\documentclass{article}
\usepackage[utf8]{inputenc}

\usepackage{graphicx}
\graphicspath{{Images/}}
\usepackage[authoryear]{natbib}
\usepackage{authblk}
\usepackage[english]{babel}
\usepackage{amsmath}
\usepackage{amssymb}
\usepackage{amsthm}
\usepackage{dsfont}
\usepackage{cases}
\usepackage{booktabs}       
\usepackage{adjustbox}
\usepackage{newtxtext,newtxmath}

\usepackage{subcaption} 
\usepackage{listings} 

\usepackage{arydshln}
\usepackage[shortlabels]{enumitem}
\usepackage{color}
\usepackage[dvipsnames]{xcolor}
\usepackage{hyperref}
\usepackage{verbatim}
\usepackage{setspace}

\usepackage{tikz}
\usetikzlibrary{arrows.meta,positioning}

\definecolor{darkgreen}{rgb}{0.1,.8,0.1}
\definecolor{darkred}{rgb}{0.8,.1,0.1}
\definecolor{darkblue}{rgb}{0.1,.1,0.8}
\definecolor{cmathieu}{rgb}{1,0,0}

\usepackage{pgfplots}
\pgfplotsset{compat=newest}

\newcommand{\nommodel}{$SIDUHR^{+/-}$~}

\def \1{\mathds{1}}

\theoremstyle{plain}

\theoremstyle{definition}

\theoremstyle{remark}

\providecommand{\keywords}[1]
{
  \small	
  \textbf{{Keywords}}: #1
}

\def \R{\mathbb{R}}

\def\drm{\mathrm{d}}

\def \red{\textcolor{red}}
\def \blue{\textcolor{blue}}
\def \fgreen{\textcolor{ForestGreen}}
\def \purple{\textcolor{Purple}}

\title{COVID-19 pandemic control:\\ balancing detection policy and \\ lockdown intervention under ICU sustainability}
\author[1]{Arthur Charpentier}
\author[2]{Romuald Elie}
\author[3]{Mathieu Lauri{\`e}re}
\author[2]{Viet Chi Tran}
\affil[1]{Universit\'e du Qu\'ebec \`a Montr\'eal (UQAM)}
\affil[2]{Universit\'e Gustave Eiffel (LAMA)} 
\affil[3]{Princeton University (ORFE)}
\date{\begin{center} This version: May 20, 2020\\[1mm] First version: April 15, 2020\end{center}}

\begin{document}

\maketitle

\begin{abstract}
We consider here an extended $SIR$ model, including several features of the recent COVID-19 outbreak: in particular the infected and recovered individuals can either be detected (+) or undetected (-) and we also integrate an intensive care unit (ICU) capacity. Our model enables a tractable quantitative analysis of the optimal policy for the control of the epidemic dynamics using both lockdown and detection intervention levers. With parametric specification based on literature on COVID-19, we investigate the sensitivities of various quantities on the optimal strategies, taking into account the subtle trade-off between the sanitary and the socio-economic cost of the pandemic, together with the limited capacity level of ICU. We identify the optimal lockdown policy as an intervention structured in 4 successive phases: First a quick and strong lockdown intervention to stop the exponential growth of the contagion; second a short transition phase to reduce the prevalence of the virus; third a long period with full ICU capacity and stable virus prevalence; finally a return to normal social interactions with disappearance of the virus. {The optimal scenario hereby avoids the second wave of infection, provided the lockdown is released sufficiently slowly. We also provide optimal intervention measures with increasing ICU capacity, as well as optimization over the effort on detection of infectious and immune individuals. Whenever massive resources are introduced to detect infected individuals, the pressure on social distancing can be released, whereas the impact of detection of immune individuals reveals to be more moderate.}  
\end{abstract}

\\

\vspace{1cm}

\thanks{The authors thank Hélène Guérin, Olivier l'Haridon, Olivier Pietquin and Gabriel Turinici  for stimulating discussions and feed-backs on early versions of that work, as well as participants of the Machine Learning Paris seminar. R.E. and V.C.T. are partly supported by the Bézout Labex, funded by ANR, reference ANR-10-LABX-58. V.C.T. is also part of ANR Cadence (ANR-16-CE32-0007), by the Chair ``Modélisation Mathématique et Biodiversité" of Veolia Environnement-Ecole Polytechnique-Museum National d'Histoire Naturelle-Fondation X. }


\section{Introduction}

Within a few weeks, the COVID-19 took most of the world by surprise, starting with Asia, then Europe and finally America (see \cite{WHO,kucharskirusselldiamondliuedmundsfunkeggo,magalwebb}). More specifically, the COVID-19 turned out to be much more contagious than expected, and official health authorities understood that letting the virus spread freely would probably saturate hospital capacities. In most countries, aggressive quarantine and lockdown measures have been taken, hopefully to stop the spread of the virus and infection, or at least to slow down its diffusion and delay the cases over time to avoid saturation of healthcare systems. The effectiveness of these non-pharmaceutical interventions have been questioned in \cite{hellewellabbottgimmabossejarvisrussellmundaykucharskiedmundsfunkeggo,kucharskigogedmunds}. It was already suggested in \cite{CDC}, but then emphasized by \cite{Imperial}, that these measures should rather aim at `{\em flattening the curve}': mitigation seems more realistic than suppression, since eliminating the threat necessarily takes time, and a massive arrival of people in need in hospitals would be dramatic.

The first control used by health authorities was based on quarantines, historically a natural technique to lower social interactions and reduce the spread of the disease, as recalled in \cite{quarantine}. Classically, infected people are isolated, to avoid further transmission, but COVID-19 has a potentially high rate of asymptomatic infected individuals, as mentioned in \cite{Al-Tawfiq,Daym1165,heetal,Nishiura} or \cite{Zhou}. This explains why most countries used indiscriminate lockdown. In order to distinguish between the susceptible, but still unaffected, individuals and the either asymptomatic infected or undetected recovered ones, testing is necessary. Piloting these interventions is a key problem, in particular for monitoring exit strategies.

In the recent literature on the COVID-19, many papers have focused on the effect of lockdown on the spread of the disease (e.g. \cite{NBERw26901,roqueskleinpapaixsarsoubeyrand}) or on the evaluation via simulations of exit-strategies (e.g. \cite{didomenicopullanosabbatiniboellecolizza,evgenioufekombovchinnikovporcherpoucholvayatis}). In this paper, we suggest a different approach for evaluating lockdown and exit scenarios based on Optimal Control theory (e.g. \cite{pontryaginboltyanskiigamkrelidzemishchenko}). Although the analysis of public health policies using controls is of clear practical value, there has not been a so dense literature on the subject (e.g. \cite{abakuks1973,Abakuks74,Behncke2000,greenhalgh,Hansen,Agusto,Sharomi}). We consider a modified $SIR$ model, incorporating all the features related to the recent COVID-19 outbreak mentioned above, and use optimization algorithms to compute the `best' level of lockdown and detection effort in testing individuals. The criterion to rank the different possible strategies takes into account: 1) the death toll of the epidemic, 2) the economic and social costs of the lockdown, 3) the required effort for detecting infected or immune individuals. Besides, we only consider strategies never exceeding the capacity level $U_{\max}$ of the ICU sanitary system, {chosen to be dedicated to COVID-19 patients}.\\
One of the originality of the model considered here is the particular focus on the outcomes of the intensive care units (ICU)  saturation, which in many countries have been put under great pressure and whose capacities are determining factors for the death toll of the disease. We assume a capacity constraints on how many individuals can be treated in ICU at a given time. Once the threshold is reached, the fatality rate for the additional patients sent to ICU rises sharply. For the COVID-19 outbreak, we are aware of some works by \cite{alvarezargentelippi,Djidjou-Demasse2020,piguillemshi}, but none of these work models the ICU capacities. In \cite{acemogluchernozhukovwerningwhinston}, no ICU capacity is explicitly introduced, but these authors choose a death rate in hospitals that depends linearly on the number of patients. The non-linearity induced by the threshold that we introduce provides very interesting phenomena and insights for the control of the disease. Furthermore, we investigate cases where the ICU capacity is increasing and evolving dynamically over time. \\
A second feature in our work is that we consider two levers for control: the quarantine ratio and {the detection effort for infectious and immune individuals}. One of our motivation is to study qualitatively the balance existing between efficient testing and strong confinement strategies. Actually, two testing procedures can be performed here. A short range one, that can detect, at time $t$ if someone is infected, or not (Polymerase chain reaction or PCR tests, that can detect whether an individual is infected and send her/him in quarantine). And a long range one, that can detect if someone is immune or not (a blood test as in \cite{Huang2020}). {For this purpose, we introduced a so-called \nommodel propagation model for the epidemic, which explicitly allows to take into account the detection ratios of immune and infected individuals.}

Our main findings consist in identifying four main phases in the optimally controlled epidemic dynamics. First, a strong and quick lockdown intervention allows to stop the exponential growth of the epidemic spreading and bring the effective reproduction number $\mathfrak{R}_t$ under level $1$. Then begins a short transition period during which $\mathfrak{R}_t$ remains below 1 as the prevalence of the virus within the population diminishes progressively until reaching a threshold $I^*$. Since the beginning of the epidemic, the number of individuals requiring ICU is rising and just reaches $U_{\max}$ when the virus prevalence attains $I^*$. Then begins a very long period during which the reproduction number $\mathfrak{R}_t$ stabilizes at level $1$ while the virus prevalence and the ICU occupation remain stable at levels $I^*$ and $U_{\max}$. Finally, when a sufficient proportion of the population has already been infected and the herd immunity level is close enough, a light final release of lockdown measure allows to retrieve a regular level of social interactions and end the epidemic phase. {Such a control strategy allows to achieve an epidemic size very close to the herd immunity level, which corresponds to the minimal size of an epidemic, in comparison to a benchmark case without intervention for example.} {Besides, given the current level of information and confidence on the COVID-19 dynamics modelling, the design of robust optimal control strategy should be seen as  a minimal requirement. For our model, the enhanced structure of the  optimally controlled epidemic dynamics reveals to be very robust to variations of the model specifications within a reasonable range.} 

{The effect of the detection effort within the population reveals to be more subtle. As explained above, two types of detections are considered. Detecting quickly infected individuals and breaking the contamination chains reveals to be key in reducing the economic and sanitary burden of the epidemic. First, it allows to slow down the exponential growth of the epidemic and decrease hereby the required epidemic size for reaching the herd immunity and the induced death toll. It also reduces the lockdown effort on the population, which helps minimize the socio-economic impact. Optimizing over the detection rate of infected individuals reveals the huge impact on both sanitary and economic burdens of early effort in detection, as well as a stable threshold level in detection. Using the detection of infectious individuals as a control lever allows to keep the reproduction number below $1$ while limiting the lockdown effort and the induced socio-economic cost. Concerning the detection of immune individuals, it allows to reduce significantly the socio-economic burden, later in time during the epidemic, whereas its impact on the sanitary outcome seems rather limited.}

The rest of the paper is organized as follows. In the `Methods' section (\ref{sec:2}), we derive our \nommodel model from a standard $SIR$ model (see Section \ref{sec:2:1}), and explain all the components of the dynamics. The infectious individuals may either: $i$) be asymptomatic or have mild symptoms that do not require hospitalization, $ii$) need hospitalization with or without intensive care. In section \ref{sec:2:2} we present the optimal control problem and objective function of public health authorities. Possible control levers are discussed in section \ref{sec:2:2:1}, with lockdown, detection efforts, and ICU capacities limitation. In section \ref{sec:2:2:2} we motivate the choice of the objective function, and discuss optimal control in sections \ref{sec:2:2:3} and \ref{sec:2:2:4}. 

In the `Discussion' section (\ref{sec:3}), we compute the optimal strategy based on a choice of parameters from the COVID-19 epidemic in France, but of course the methods and findings can be transposed to any other country or set of parameters. In section \ref{sec:3:1}, the benchmark scenario is described. Then, in section \ref{subsec:optimal:policy} we consider optimal lockdown, and discuss the 4 stage dynamics. In section 
\ref{sec:3:3}, we provide a sensitivity analysis, and discuss the {robustness of our findings in response to changes in the model specifications and parameters.} 
In sections \ref{sec_sensi_Umax} and \ref{sec:3:5}, we consider the impact of two other controls: the ICU sustainability level and detection resources.
Given the simplicity of the compartmental model considered here, our purpose is not to give detailed prediction of the epidemic but rather to highlight very interesting aspects of the interplay between different levers (lockdown, detection and ICU capacity), provide qualitative description of the outcomes in each scenarios and give rough estimates of the duration and size of the epidemics.

{Let us finally emphasize that the level on uncertainty upon the characteristics of the epidemic dynamics is still currently very high. The enhanced conclusions of this paper remain valid within the scope of the chosen model specifications chosen here, which are in line with the current understanding of the epidemic dynamics.}


\section{Methods}\label{sec:2}

\subsection{The \nommodel model and its dynamics}\label{sec:2:1}

Compartmental models, where the population is divided into classes defined by the status of the individuals with respect to the disease or how they are taken in charge by the healthcare system are very popular in mathematical epidemiology (see e.g. \cite{AndersonMay,diekmannheesterbeekbritton-book,ballbrittonlaredopardouxsirltran} for an introduction).
One of the simplest and most fundamental of all epidemiological models is the popular $SIR$ model, developed in \cite{kermack27}. The population, of fixed size, is divided into three categories: susceptible ($S$), infected ($I$) and recovered ($R$) individuals. We denote by $S_t$, $I_t$ and $R_t$ the respective proportions in the populations at time $t$ (measured in days). 
In order to adapt the $SIR$ model to the COVID-19 epidemic and {incorporate} lockdown and testing strategies with their economic and social impacts, we modify some rates and add compartments to the $SIR$ model.

We want to take into account two important features of the COVID-19 pandemic. The first one is the important proportion of asymptomatic patients, and the small use of testing in several countries. We will split the $I$ group in two: $I^-$ the non-detected, {mostly} asymptomatic individuals, but contagious, and $I^+$ the detected part of infected individuals. The second feature is the problem highlighted in the `{\em flatten the curve}' concept (discussed in \cite{RanneyNEJM}, \cite{Kissler2020} or \cite{Gudi2020}) which means that a goal in the choice of a public health intervention should be to avoid a surge of demand on the health care system and in particular in ICUs. From a modelling perspective, we introduce the proportions $H_t$ and $U_t$ of individuals using the health care system with or without the need of ICUs. There is an exogenous sustainable limit $U_{\max}$ for the ICU system, above which the death rate is significantly higher.

\paragraph{Compartments} More specifically, all possible states can be visualized on Figure \ref{fig:graphSIDHR}. The parameters for the dynamics are described in the next section.
\begin{itemize}
    \item $S$: susceptible, never tested positive, never infected and not dead (where people are by default). The can get infected by infectious and not quarantined individuals, i.e. individuals in $I^-$. When susceptible individuals get infected, they move to $I^-$.
    \item $I^-$: infected non-detected. Those individuals can be asymptomatic, not sick enough to go the the hospital, and ar{e not detected}. They can either get tested (with type-1 tests, and then move to $I^+$), get sicker and then go to the hospital ($H$) or simply recover and get immune, but still not detected ($R^-$)
    \item $I^+$: infected detected (and non-hospitalized). They can either get sicker and then go to the hospital ($H$) or recover, in which case they end up in the compartment $R^+$ since they were tested positive. Here, we assume that the individuals in $I^+$ cannot infect anyone since they are strongly isolated.
    \item $R^-$: recovered non-detected. Those can be detected using blood-type tests (also named type-2) and move to $R^+$.
    \item $R^+$: recovered detected. {We assume here that all recovered individuals (both $R^-$ and $R^+$) are immune {long enough with respect to the epidemic duration}.}
    \item $H$: hospitalized (and detected). All people entering the hospital get tested. They can either get sicker and then go to intensive care ($U$), or get immunity after recovering, and then move to $R^+$. {Note that $H$ is an intermediary state, before getting really sick, or die. Individuals are not allowed here to go from $I$ (detected or not) to $D$. Furthermore, we did not distinguish explicitly between {\em mild} and {\em severe} infected (as in many other models) {in order to emphasize the role of detection resources}, but, somehow, $H$ could correspond to {\em severe} cases.}
    \item $U$: hospitalized in ICU, for people at a more advanced stage (and detected), who need ventilators and other specialized medical care . They arrive from $H$, and they can either get immunity after recovering, and then move to $R^+$, or die ($D$).
    \item $D$: dead, from the disease.
\end{itemize}
Here, recovered are not contagious anymore and we assume that they get immunity (and therefore, can not get infected again). This is a rather strong assumption, especially on a long term horizon. \cite{choe} used a longitudinal study to show that those infected by MERS in South Korea (with a significant degree of immunity post-infection) had immunity that lasted for up to a year, as well as, more recently \cite{Aldridge}, on historical patterns of three common corona-viruses. 

\begin{figure}[h]
\centering
\begin{tikzpicture}[->,auto,node distance=2.5cm,
  thick,main node/.style={circle,draw,font=\sffamily\large\bfseries}]
  \node[main node][fill=yellow!20,rectangle,minimum size=0.8cm] (S) {$S$};
  \node[main node][fill=yellow!20,rectangle,minimum size=0.8cm] (Im) [right of=S] {$I^-$};
\node[main node][draw=none,fill=none] (O1) [below of=Im] {};
\node[main node][draw=none,fill=none] (O2) [right of=Im] {};
  \node[main node][fill=yellow!20,rectangle,minimum size=0.8cm] (Rm) [right of=O2] {$R^-$};
  \node[main node][fill=blue!10,minimum size=0.8cm] (Ip) [below of=O1] {$I^+$};
  \node[main node][draw=none,fill=none] (O3) [right of=Ip] {};
  \node[main node][fill=blue!10,minimum size=0.8cm] (H) [right of=O1] {$H$};
  \node[main node][fill=blue!10,minimum size=0.8cm] (Rp) [right of=O3] {$R^+$};
  \node[main node][fill=blue!10,minimum size=0.8cm] (U) [right of=H] {$U$};
  \node[main node][fill=blue!10,minimum size=0.8cm] (D) [right of=U] {$D$};
  \path[every node/.style={font=\sffamily\small}][color=ForestGreen]
    (S) edge node [above] {$(1-\delta)\textcolor{black}\beta$} (Im);
  \path[every node/.style={font=\sffamily\small}][color=red]
    (Im) edge node [left] {$\lambda^1$} (Ip);
      \path[every node/.style={font=\sffamily\small}][color=purple]
    (H) edge node [left] {$\gamma_{HR}$} (Rp)
    (U) edge node [left, yshift=0.1cm] {$\gamma_{UR}$} (Rp)
    (H) edge node [above] {$\gamma_{HU}$} (U)
    (U) edge node [above, xshift=0.1cm] {$\gamma_{UD}$} (D);
  \path[every node/.style={font=\sffamily\small}]
  (Im) edge node [above] {$\gamma_{IR}$} (Rm)
    (Im) edge node [above right] {$\gamma_{IH}$} (H)
      (Ip) edge node [below left] {$\gamma_{IR}$} (Rp)
    (Ip) edge node [above left] {$\gamma_{IH}$} (H);
\path[every node/.style={font=\sffamily\small}][color=blue]
    (Rm) edge[bend left] node [above left, yshift=1cm, xshift=-0.3cm] {$\lambda^2$} (Rp);
\end{tikzpicture}
\caption{The \nommodel model}\label{fig:graphSIDHR}
\end{figure}

\paragraph{Dynamics} The evolution of the sizes of each compartments is assumed to be modelled by the following system of ordinary differential equations:

\begin{equation}
\begin{cases}
    \drm S_t = - \fgreen{(1-\delta_t)}\beta I^{-} S_{t}  \drm t, &  \mbox{Susceptible} 
    \\
    \drm I^{-}_{t}  = \fgreen{(1-\delta_t)}\beta I^{-}_{t}  S_{t} \drm t - \red{\lambda^1_{t}} I^{-}_{t} \drm t - (\gamma_{IR}+\gamma_{IH}) I^{-}_{t}  \drm t,  &  \mbox{Infected undetected} \\
     \drm I^{+}_{t}  = \red{\lambda^1_{t}} I^{-}_{t} \drm t - (\gamma_{IR}+\gamma_{IH}) I^{+}_{t}  \drm t , \qquad\qquad &  \mbox{Infected detected }
    \\ 
    \drm R^{-}_{t} = \gamma_{IR} I^{-}_{t}  \drm t
                        - \blue{\lambda^2_{t}} R^{-}_{t} \drm t, & 
    \mbox{Recovered undetected}
    \\
    \drm R^{+}_{t} = \gamma_{IR} I^{+}_{t}  \drm t + \blue{\lambda^2_{t}} R^{-}_{t} \drm t + \purple{\gamma_{HR}} H_t  \drm t  +  \purple{\gamma_{UR}(U_{t})} U_t \drm t , & 
    \mbox{Recovered detected}
    \\  \drm H_{t} = \gamma_{IH} \big(I^{-}_{t}+I^{+}_{t}\big)  \drm t - \purple{(\gamma_{HR} + \gamma_{HU})} H_t \drm t,  & 
    \mbox{Hospitalized}
    \\  \drm U_{t} = \purple{\gamma_{HU}} H_t  \drm t  - (\purple{\gamma_{UR}(U_{t})} + \purple{\gamma_{UD}(U_{t})}) U_t \drm t,  & 
    \mbox{ICU}
    \\    \drm D_{t} = \purple{\gamma_{UD}(U_{t})} U_t \drm t, & 
    \mbox{Dead}    
\end{cases}
 \label{sys:SIR}
\end{equation}

Let us explain the different parameters. The transmission rate $(1-\delta_t)\beta$ is the rate at which an undetected infected individual transmits the disease to a susceptible one. Multiplying this parameter by $S_t I^{-}_t$, which is proportional to the number of pairs that we can form with an $S$ and an $I^{-}$ individual, provides the total infection rate at the level of the population and at time $t$: $(1-\delta_t)\beta S_t I^{-}_t$. Heuristically, this quantity can also be understood as the fraction of the population infected during a unit time interval at time $t$.
The infection rate $\beta$ is reduced by a time-dependent factor $(1-\delta)$ where $\delta \in [0,1]$ comes from social distancing, quarantine, isolation interventions that reduce infectious contact rates. {The strength of the lockdown intervention interprets as follows: $\delta=0$ means no intervention while $\delta=1$ means no social interaction.} We consider two ways of detecting individuals: short range tests allow to discover new infectious individuals in $I^-$ and we denote by $\lambda^1$ the rate at which each of these infectious individuals gets detected, long range test allow to find formerly infected and hereby immune individuals that are now in $R^-$ and we denote by $\lambda^2$ the corresponding rate. Here we are interested in the case where the parameters $\delta$ and $\lambda^1$ are control variables and in the sequel, these parameters are hereby time dependent.\\
The recovery rate for an infected individual is $\gamma_{IR}$. The rate at which an infected individual enters the health system is $\gamma_{IH}$. We assume here that both rates $\gamma_{IR}$ and $\gamma_{IH}$ are the same for detected or non-detected infected individuals. The individual transition rates from $H$ to $U$ or $R^+$, from $U$ to $R^+$ or to $D$ are denoted by $\gamma_{HU}$, $\gamma_{HR}$, $\gamma_{UR}(U)$ and $\gamma_{UD}(U)$ respectively. Notice that the two later rates  $\gamma_{UR}(U)$ and $\gamma_{UD}(U)$ are not constant but depend on $U_t$: when the ICU limitation is reached, the extra patients who are not taken in charge by the system can not recover and are exposed to extra death rates, as explained in the Appendix \ref{sec:parameters}.\\
These parameters are chosen following the tracks of \cite{didomenicopullanosabbatiniboellecolizza,salje-cauchemez}. In these papers, some parameters are fixed according to clinical studies and other are estimated from French data. The detail on how these parameters are computed in the present paper is given in Section \ref{sec:parameters}. However, we emphasize that although we choose French benchmarks, our input is more on methodology and can be of interest for other countries or epidemics as well.

\begin{table}[!ht]
\begin{center}
	\begin{tabular}{c|c|c}
				\hline
			Parameters & Value  & Reference\\
			\hline \hline 
			     $\mathfrak{R}_0$ & $3.3$ & based on (S)\\
				 $\beta$   & $0.436$ & based on (S)\\
				 $\gamma_{IR}$ &  $0.130$ & based on (D)\\
				 $\gamma_{IH}$ & $0.00232$ & based on (D,S)\\
				 $\gamma_{HR}$  & $0.048$  & based on (S)\\
			     $\gamma_{HU}$ & $0.091$ & based on (S)\\
			     $\gamma_{UR}(U)\times U$ & $0.078 U \wedge 1.564\ 10^{-5}$ & based on (S)\\
			     $\gamma_{UD}(U)\times U$  & $0.02\ U \wedge U_{\max} + 2(U-0.0002)^+$ & based on (S)\\
			     $U_{\max}$ & $0.0002$ & estimated\\
			     $I_0^-$ & $0.005$ & estimated\\
				 \hline 
				\end{tabular}
\caption{Parameter values for the base-scenario, referring to (D) = \cite{didomenicopullanosabbatiniboellecolizza}, (S) = \cite{salje-cauchemez} or (W) = \cite{WHO}. Individual rates are given except for the transition rates from $U$ to $R$ or $D$ where the global rates at the population level are given, to better explain the nonlinearity. Details are presented in Appendix \ref{sec:parameters}.}
\label{tab:parameters-corps}
	\end{center}
\end{table}

The system described here can be derived from individual based stochastic processes (e.g. \cite{ballbrittonlaredopardouxsirltran}) and thus, there is an underlying individual based model where all the rates have a probabilistic and statistical interpretation (see Section \ref{sec:parameters}, with connection to continuous time Markov chains. As we will discuss later on, those parameters can therefore be interpreted as inverse of transition time lengths.

{The initial time $t=0$ corresponds to the time at which policies that reduce the infection rate are put in place ($\delta>0$), or the time at which testing is started ($\lambda^1>0$). For the French COVID-19 epidemic, it would correspond to March 17th 2020, but again, we emphasize that our methodology remains of interest for other countries or diseases (see e.g. \cite{Pedersen} for country specific informations).}

\paragraph{Basic and effective reproduction numbers $\mathfrak{R}_0$ and $\mathfrak{R}_t$}

The basic reproduction number $\mathfrak{R}_0$ corresponds to the expected number of individuals directly contaminated by a typical infected individual during the early stage of the outbreak (when the proportion of susceptible individuals is close to 1), see \cite{diekmannheesterbeekmetz}. When $\mathfrak{R}_0 < 1$, the disease does not
spread quickly enough, resulting in a decay in the number of infected individuals. But when $\mathfrak{R}_0 > 1$, the infected population grows over time. 

{As recalled in \cite{trapmanballdhersintranwallingabritton}, the value of the $\mathfrak{R}_0$ is model-dependent. For the model considered here, only the $I^-$ individuals can transmit the disease to infect susceptible individuals. If we see $I^-$ as the infected class, and if we merge the classes $I^+$, $R^-$, $R^+$, $H$, $U$ and $D$ into a large `recovered' class, in this $SIR$ type of model, the dynamics of $I_t^-$ has infection rate $(1-\delta)\beta$ and removal rate $\lambda^1 + \gamma_{IR}+\gamma_{IH}$, assuming that $\delta=\delta_0$ and $\lambda^1=\lambda^1_0$  are constant over time.} In our model, the basic reproduction number is
\begin{equation}
{\mathfrak{R}_0=  \frac{(1-\delta_0)\beta}{\lambda^1_0 + \gamma_{IR}+\gamma_{IH}} }
\label{eq:R0}
\end{equation}
Without intervention (i.e. for $(\delta,\lambda^1)=(0,0)$ over time) and with our chosen parameters provided in Table \ref{tab:parameters-corps}, this corresponds to $\mathfrak{R}_0=\beta/(\gamma_{IR}+\gamma_{IH})=3.3$ which is in line with \cite{salje-cauchemez}. Sensitivity of our results with respect to variations of $\mathfrak{R}_0$ within the range $[3.3,3.6]$ is discussed in Section \ref{sec_sensi_R0}.

In the case where $\lambda$ and $\delta$ evolve with time, we end up with a dynamic reproduction number $(\mathfrak{R}_t)_t$ that is called effective reproduction number, see \cite{AndersonMay}. More specifically, {at time $t\geq 0$, let us define}
 \begin{equation}\label{eq:Rot}
{\mathfrak{R}_t=  \frac{(1-\delta_t)\beta S_t}{\lambda^1_{t} + (\gamma_{IR}+\gamma_{IH})}. }
 \end{equation}
Its interpretation is very close to the static version $\mathfrak{R}_0$. When $\mathfrak{R}_t<1$, the average number of secondary cases started from one primary case with symptom onset on day $t$, dies out quickly, resulting in a decay in the number of infected individuals in the population. But when $\mathfrak{R}_t>1$, the infected population grows over time. So we should expect to have controls on the quarantine level $\delta_t$ and on the testing rate $\lambda_t$ that will constrain $\mathfrak{R}_t$ to be reduced beyond 1, at least after some starting time.

{Another information brought by the $\mathfrak{R}_t<1$ is that, when $S$ is smaller than the so-called \emph{`herd immunity'} threshold, 
the epidemic enters into a sub-critical phase where it goes to extinction. In our case, the herd immunity threshold $S^*$ without intervention is around $30\%$ and durable lockdown and detection efforts can allow to reduce it significantly until the arrival of a vaccination solution.}


\subsection{The optimal control problem}\label{sec:2:2}

In the previous section, we described our epidemiological model in order to characterize the dynamics of the virus spreading processes. We now consider  a government acting as a global planer, who wants to control the epidemic dynamics in order to balance the induced sanitary and economic outcomes. In order to mitigate the effects of an epidemic, several parameters in the model are now interpreted as `{\em control levers}’, and we now focus on the derivation of their optimal dynamic choice.
Optimizing intervention strategies is an important policy issue for the management of infectious diseases, such as COVID-19. Optimal control in the context of pandemic models has been used since \cite{abakuks1973, Abakuks74},  \cite{BOBISUD1977165} or \cite{Sethi1978}, in the 70's.

\subsubsection{Possible control levers}\label{sec:2:2:1}

\subsubsection*{Quarantines and Lockdown}

Lockdown measures proved to be effective in controlling the COVID-19 outbreak in China, as recalled in \cite{kucharskirusselldiamondliuedmundsfunkeggo} or \cite{Lai2020}.
Quarantine is a rather old technique used to prevent the spread of diseases. It refers to the restriction of movement of anyone (not necessarily sick people), and it usually takes place at home, and may be applied at the individual level. It should be distinguished {from} isolation, which refers to the restriction of movement of {infectious individuals} who have a contagious disease. It can be done in hospitals, in dedicated facilities, or at home. Those two are seen as active controls, and are different from contact surveillance, which is passive. Modern quarantine's goal is to reduce transmission by increasing {\em social distance} between people, by reducing the number of people with whom everyone can be in contact with. This includes canceling public gatherings, closing public transportation, etc. A lockdown is intended to stop people from moving between places, and it could involve cancelling flights, closing borders, and shutting down restaurants. The idea is to reduce the flow of people to curb transmission, and we might consider here that lockdown and quarantine measures are almost equivalent.

{\cite{Sethi1978} suggested an optimal quarantine problem, in an $SIR$ model, where a proportion of infected people could be quarantined. He proved that the optimal strategy was either to quarantine all infected individuals (if so, at early stages) or none.}
In the context of the 2003 SARS pandemic in Singapore, \cite{Peng2005} recall that a {\em Home Quarantine Order} was signed, but it was more an isolation of sick people than a lockdown as the ones used to lower possible consequences of COVID-19. But options considered are different, from one country to another. \cite{NEWSC2020} mentions that several countries closesd their borders, in most European countries, schools and restaurants were closed for weeks.

\cite{Lagorio} tried to quantify the effectiveness
of a quarantine strategy, where healthy people are advised to avoid contacts with individuals that might carry the disease, using network based models. Here, the parameter has a direct interpretation, since it is related to the average rate of contact between individuals.
In \cite{Xiefei} or \cite{Ku2020}, quarantine and isolation are used as control. In our model, those controls correspond to the $\delta_t$ component. As in \cite{Chinazzieaba9757}, such control can also be related to travel bans, for a weaker form than strict quarantine. Note finally that {\cite{alvarezargentelippi} suggested that a complete lockdown was impossible to reach, and that an upper bound of $70\%$ (of the population) should be considered. In this paper, we do not incorporate any constraint (except that we cannot lockdown more than the entire population), but interestingly, in the numerical simulation, the upper bound we obtained as optimal is very close to this value.}

\subsubsection*{Testing, Tracing and Isolating}

The second most important lever that can be used is detection. As claimed by several governments (see \cite{Gostic}) in the context of the COVID-19 pandemic, testing is key to exit lockdown, and mitigate the health and economic harms of the virus. Here, we assume that health authorities can perform two kinds of {active detection actions}. 

Type-1 test, that could be an antigen test (Reverse Transcription Quantitative Polymerase Chain Reaction (RT-qPCR) - also called molecular or PCR -- Polymerase Chain Reaction -- test). This test allows to detect whether an individual is currently infected or not.  It should be used to determine who self-isolates and for contact tracing. For that test, a sample is collected – usually with a deep nasal swab (and analysed in a laboratory). This is a short-term test : when performed at time $t$ on an individual, we know whether an individual is infected or not. Together with proper detection means, the use of this test corresponds to the $\lambda^1_t$ control in our dynamics. This test allows to detect and isolate infectious individuals. In our model, they transfer from compartment $I^-$ to compartment $I^+$. It is performed as mandatory when someone arrives at the hospital. Hereby, any individual being at some point in compartment $I^+$, $H$ or $ICU$ ends up in the identified immune compartment $R^+$ if he survived the epidemic. We assume for simplicity that those tests do not have false positive or false negative and leave for further research the impact of the test sensitivity on the epidemic dynamics and optimal control. Thus, the tests are all positive on individuals in $I^-$ and negative on individuals in $S$ and $R^-$. Other individuals will not be tested.

{Type-2 test is an antibody test (using serological immunoassays that detects viral-specific antibodies -- Immunoglobin M (IgM) and G (IgG) -- also called serology or immunity test).} It allows to test whether an individual can now be considered immune to the virus. It could potentially be used to issue immunity certificates in order to help restarting the economy quicker. {It also reduces the overall uncertainty and potential fear of individuals to be infected, and has a positive impact on the economic consumption.} It is also useful for contact tracing purpose as it helps identifying individuals that can not become infectious again.  For such a test, a blood or saliva sample is applied to a strip that identifies presence of antibodies. We assume here that immunity lasts long enough with respect to the epidemic duration, and $R^+$ regroups the collection of immune detected individuals. It is a long-term one : when performed at time $t$ on an individual, we know if that person {\em had} the disease before. The detection using those tests identifies in our model to the control variable  $\lambda^2$. It allows to transfer individuals from compartment  $R^-$ to compartment $R^+$, but again, people in $S$ will be tested also, as well as people in $I^-$. As for Type-1 tests, we will suppose here that these tests do not make any wrong identification. 

For type-2 tests, a small sample of a patient’s blood -- for instance via a pin prick -- and the test looks for two specific types of antibody: IgM and IgG. IgM are the first antibodies to be produced by the immune system. They have a half-life of around five days, and they usually appear within five to seven days of infection and peak at around 21 days. Detection of these antibodies suggests the person has existent or a recent infection. IgG  antibodies are more numerous and can be detected around 10 to 14 days after infection. The presence of these antibodies indicates a person has recovered from the virus and is now immune.

\subsubsection*{`Raising the line' and ICU Capacities}

So far, we have assumed that the sustainable capacity of ICUs was not a control, and should be considered as an exogeneous fixed boundary (denoted $U_{\max}$).
As discussed in \cite{NATION}, {\em preventing a health care system from being overwhelmed requires a society to do two things: 'flatten the curve' -- that is, slow the rate of infection so there aren't too many cases that need hospitalization at one time -- and 'raise the line' -- that is, boost the hospital system's capacity to treat large numbers of patients}" (see also \cite{VOX}). Therefore, a natural control variable is the level of {\em the line}, or  $U_{\max}$ as we named it.

Nevertheless, we will not consider $U_{\max}$ as a control variable that can be optimized (such as the strength of the lockdown, or the effort in detection), since it is difficult to assess the cost of raising it. Nevertheless, in Section \ref{sec_sensi_Umax}, we consider the case where health authorities choose to increase that capacity, by making ventilators and other medical material necessary in the context of the pandemic. But it cannot be increased indefinitely since ICU also require trained personal and even if adjustments can be considered, we assume that it is not realistic to assume that $U_{\max}$ can be increased by more than $50\%$.

\subsubsection{Objective function}\label{sec:2:2:2}

We now turn to the design of the objective function, trying to take into consideration the sanitary and socio-economic outcomes of the lockdown and detection policy. In the context of planning vaccination campaigns, \cite{daCruz2011} and \cite{1470088} suggest a convex quadratic cost function by minimizing both the number of infected individuals in a time horizon and the cost to implement the control policy. In \cite{KIM201774}, a model for 2009 A/H1N1 influenza in Korea is considered: the goal is to minimize the number of infected individuals and the cost of implementing the control measures, and the cost is taken to be a nonlinear quadratic function. Quarantine and vaccination are considered as control variables in \cite{Iacoviello2008}, and the optimal control is obtained by minimizing a quadratic cost function. Inspired by those approaches, we will also consider quadratic cost functions. Before going further, let first introduce some relevant quantities characterizing the epidemic phase that will reveal useful in the upcoming analysis.

\paragraph{Quarantined individuals}
 Let denote by $Q$ the quantity of people concerned by the lockdown policy and defined by:
 \begin{equation}\label{eq_def_Q}
 Q_t := R^-_t + I^-_t + S_t \;, \quad t\ge 0.
 \end{equation}
 All the individuals in $Q_t$ are having the level of contact rate $\delta_t$ with the population at time $t$. This proportion of individuals identifies to the people for which we can not say if they already contracted the virus or not. Therefore, they are all concerned with non targeted lockdown strategies as well as possibly randomized testing trials. {Recall that, on the opposite, we} suppose that infected detected individuals in $I^+$ are isolated and have no social interactions with the population, whereas detected immune individuals suffer no mobility constraints and can reach a level of local interactions similar to the one they had before the beginning of the pandemic. 
 
 \paragraph{Global level of social interactions} On the other hand, the global level of social interactions among the population is denoted $W_t$ and given by: 
 \begin{equation}\label{eq_def_W}
 W_t := (1- \delta_t) Q_t + R^+_t \;, \quad t\ge 0.
 \end{equation}
 The quantity $W_t$ represents the proportion of social interactions in the population and can be interpreted as a macroscopic labour force level for the economy at time $t$. 
 
 \paragraph{Testing resources.} Finally, in order to measure the detection efforts of infected and immune individuals, we introduce the following metrics; which identifies to the number of virologic and anti-body tests done in a (somehow quite unrealistic) fully randomized detection trial:  
 \begin{equation}\label{eq_def_N1_N2}
 N^1_t  
 := \lambda^1_t Q_t + \gamma_{IH} I^-_t  \qquad \mbox{ and }\qquad N^2_t  
 := \lambda^2_t Q_t   \;, \quad t\ge 0.
 \end{equation}
Observe that the different formulation between both metrics is due to the fact that individual are automatically tested using virologic detection methods, when admitted at the hospital. 

\paragraph{Vaccination arrival date}
All health authorities hope for the development of a vaccine for Covid19 in a close future. In order to encompass uncertainty concerns upon the arrival date of the vaccine, we consider the arrival date of the vaccination solution to be a random time with exponential distribution of parameter $\alpha$, {denoted $\tau$}. According to recent studies, \cite{JiangVaccine,Cohen14}, the creation of such vaccine for a sufficient quantity of individuals can presumably be assumed to require around $500$ days. We shall consider a parameter $\alpha$ equal either to $0$, $1/250$ or $1/500$ for our numerical experiments. {For convenience, we assume here that both a vaccine and a cure simultaneously appear at time $\tau$}. 
{\cite{alvarezargentelippi} also use such a discount approach, on top of some economic discount rate (assuming that the vaccine will arrive at an expected date of 18 months).}

\paragraph{Sanitary cost} 

 For the sanitary cost, it seems rather natural to simply consider a death count metric given by $D_T$. Nevertheless, \cite{Olivier2013} legitimates the use of an appropriate discounting of health outcomes. About the kind of discount we should use, \cite{vanderPol2002} discusses the use of hyperbolic discount (opposed to classical exponential discount), in the case of social choice regarding health outcome. The hyperbolic discount allows for asymmetry in discounting, and non-stationarity, in the sense that postponing in two years now can be substantially different from postponing in three years (from now) next year. On the other hand, \cite{vanderPol2002} claims that is not necessarily a realistic assumption, so that using a standard time-consistant exponential discount makes sense in terms of health outcomes. Besides, the longer we can postpone deaths, the more likely a vaccine will be found before. Adding up a exponential discounting factor reveals to be also perfectly consistent with the consideration of random development time of a vaccine solution at large scale, with exponential distribution. Hence, we will consider a sanitary cost of the form:
\begin{equation}
    C_{\text{sanitary}} := E[D_{\tau}] = \int_0^\infty  e^{-\alpha t}  \drm D_t \;,
\end{equation}
recalling that the random arrival time of the vaccine follows an exponential distribution with parameter $\alpha$.

\paragraph{Economic and social cost}

{ The quantification of the socio-economic induced by lockdown restrictions is not easy to quantify properly. More involved modelling may wish to take into account a feedback effect between labor and consumption levels together with the epidemic propagation characterized by the global rate of social interactions within the population. Such approach goes beyond the scope of this paper and we solely intend to quantify the impact of the epidemic prevalence in terms of reduction of social interactions.  Here, we would like to capture the loss of productivity and well-being among the population due to lockdown situation. In a very general setting, the production function is a function of capital input ($K$, that usually corresponds to machinery, equipment, buildings, etc.) and labor ($W$, that represents the total number of people working): the output produced in a given period of time is classically given by a Cobb-Douglas production function function (as introduced in \cite{CobbDouglas}) $K^{1-a}W^{a}$, with $a\in(0,1)$. Power coefficients identify to the output elasticities. Here, the function display constant returns to scale, in the sense that doubling the use of capital $K$ and labor $W$ will also double produced output. Assuming that, on the short term (say less than a year) capital remains constant, it means that if at time $t$, only a proportion $W_t$ of individuals can be economically active, the productivity loss is proportional to $1-W_t^{a}$. Thus, the inter-temporal economic cost at time $t$ is proportional to $\displaystyle{(1-W_t^{a})}$, or (up to an additive constant) $\displaystyle{-W_t^{a}}$. Nevertheless, several other metrics have been used on the literature in order to account for the economic cost induced by the lockdown situation: \cite{Guerrieri} or \cite{Eichenbaum} considered a linear function, while \cite{piguillemshi}, \cite{Bernstein} or \cite{Djidjou-Demasse2020} considered a convex function. As for the other cost structures, we choose for simplicity to keep a quadratic structure quantifying the distance between the current global rate of social interactions $W$ given in \eqref{eq_def_W} (and identified as a metric for welfare, well-being or working-force) and its optimal value $1$.}
Hereby, the social and economic cost will be given by: 
\begin{equation}\label{eq_cost_eco}
    C_{\text{econ}} :=  E\left[ \int_0^\tau (1-W_t)^2 \drm t \right]= \int_0^\infty e^{-\alpha t}  (1-W_t)^2 \drm t \;,
\end{equation}
 {where we recall that the discounting rate $\alpha$ encompass the distribution of arrival time of a vaccination solution solution.}

\paragraph{Detection cost}

The quantification of efforts put into the detection of infectious individuals is a difficult task, as outlined in \cite{Weiyong} in the case of children, but several detection means are possible. As mentioned in \cite{lshtm4656563} and \cite{Hellewell}, testing is only part of a strategy. The World Health Organization recommends a combination of measures: rapid diagnosis and immediate isolation of cases, rigorous tracking and precautionary self-isolation of close contacts. \cite{Cohen1287} compares various techniques used to detect infected people. Recall that we assume that a virologic testing is performed for all individuals entering into the hospital. Besides, {proper tracing of contamination chains together with potential}  random screening of the population should be conducted on the non detected individuals, i.e. on the sub-population $Q_t$. This group contains the susceptible $S_t$, the undetected infected $I_t^-$ as well as the undetected immune $R_t^-$. As antibodies detection should also be performed on the same group $Q_t$, it seems important to take into account the impact of the size of this group within the population. Besides, as testing resources are sparse, the metric quantifying the effort of detection must take into account the increasing difficulty for scaling massively testing resources. For all these reasons, we pick a cost for detecting infected or immune individuals of quadratic form as follows: 
\begin{eqnarray}\label{eq_cost_prevalence}
    C_{\text{prevalence}} &:=& E\left[\int_0^\tau |N^1_t|^2 \drm t \right] = \int_0^\infty e^{-\alpha t} |N^1_t|^2 \drm t   \quad\\ \mbox{ and } \quad    C_{\text{immunity}} &:=& E\left[\int_0^\tau |N^2_t|^2 \drm t \right] = \int_0^\infty e^{-\alpha t} |N^2_t|^2 \drm t\;,\label{eq_cost_immunity}
\end{eqnarray}
 where we recall that $N^1_t$ and $N^2_t$ are given in \eqref{eq_def_N1_N2}. Again, we consider here a detection cost, not a testing cost.

\subsubsection{Optimal control under ICU capacity constraint}\label{sec:2:2:3}

We are now in position to turn to the design of the global objective function. {We are looking towards the societal optimum chosen  by an omniscient government acting as a global planer and is facing a multi-objective  control problem.} As in \cite{KUMAR2017334}, we shall consider a weighted sum of different costs:
\begin{itemize}
      \item the sanitary cost described in \eqref{eq_cost_prevalence} directly related to the mortality rate of the virus;
      \item An economic and social cost given in \eqref{eq_cost_eco} due to the reduction of social interactions within the population;
      \item A detection cost for identifying infectious and immune individuals; which will be mainly discussed in Section \ref{sec:3:5} below. 
\end{itemize}
 Combining these effects, the global objective function is given by: 
 \begin{eqnarray}\label{eq_Global_cost}
     J(\delta, \lambda^1,\lambda^2) &:=& w_{\text{sanitary}} C_{\text{sanitary}} + w_{\text{econ}} C_{\text{econ}} + w_{\text{prevalence}} C_{\text{prevalence}} + w_{\text{immunity}} C_{\text{immunity}}\nonumber \\
     &=& w_{\text{sanitary}} \int_0^\infty e^{-\alpha t}  \drm D_t + w_{\text{econ}} \int_0^\infty e^{-\alpha t} (1-W_t)^2  \drm t\\ &+& w_{\text{prevalence}} \int_0^\infty e^{-\alpha t} |N^1_t|^2 \drm t + w_{\text{immunity}} \int_0^\infty e^{-\alpha t} |N^2_t|^2 \drm t\,,\nonumber
 \end{eqnarray} 
 where the $w_{i}$ terms are positive weights associated to each marginal cost. {The choice of proper weights $w_{i}$ is a hard ethical task, that we choose to avoid as out main purpose is to highlight the main patterns of the optimally controlled epidemic dynamics. We pick some weight levels for our numerical experiments and study in Section \ref{sec_sensi_weight} the sensitivity of our findings to strong variations of these weights.}
 
{In our approach, the global planer tries to minimize the  global cost function given in \eqref{eq_Global_cost}} using the lockdown control lever $\delta$ together with the detection lever $\lambda=(\lambda^1;\lambda^2)$. Nevertheless, we did not yet take into account the risks and outcomes induced by the saturation of ICU hospital care facilities. The dynamics of the \nommodel model already takes into account the strong impact of ICU saturation on the mortality rate on individuals with severe symptoms. Whenever the proportion of individual in $U$ exceeds the upper bound level $U_{\max}$, patients can not be treated correctly so that patients are more likely to transfer to $D$ instead of $R^+$, in comparison to calmer times where $U<U_{\max}$. With such feature, the optimal lockdown and detection strategies should undoubtedly try to limit the duration of periods where the maximal ICU capacity is exceeded. Nevertheless, as we believe that exceeding capacities of ICUs would have a strong negative impact on the hospital staff and even the entire population, we rather choose to impose the non-saturation of ICU as a required state constraint on the system. Hence, we try are looking for a robust solution to  the following control problem: 
\begin{eqnarray}\label{Control_Problem}
 \inf_{(\delta,\lambda^1,\lambda^2)\in\mathcal{A} ~ \mbox{s.t.}~  U \le U_{\max}} \big\lbrace J(\delta,\lambda^1,\lambda^2)\big\rbrace\;,
\end{eqnarray}
where $\mathcal{A}$ denotes the set of admissible strategies driving the dynamics of the \nommodel model and is given by 
\begin{eqnarray}
  \mathcal{A} &:=& \left\{ (\delta,\lambda^1,\lambda^2) : [0,T] \rightarrow [0,1]^3\,,\quad  (\delta,\lambda^1,\lambda^2) \;\;\mbox{measurable }  \right\}\;.
\end{eqnarray}

\subsubsection{Numerical approximation of the solution}\label{sec:2:2:4}

From a mathematical point of view, the optimal control problem  of interest falls into the class of control for deterministic dynamical system over infinite horizon  in the presence of additional state constraint. The detailed presentation of the numerical resolution of this problem is presented in Section \ref{Sec_algo} below. We briefly recall here the main underlying approximations used in the numerical approach. 

In order to optimize the impact on the dynamical structure of the underlying \nommodel of the chosen control, we make use of Pontryagin maximum principle. For numerical purposes, we restrict our analysis to the consideration of a control problem with finite horizon. We thus pick a maturity $T$ large enough in order to have a very small remaining level of infected individuals at time $T$. {Besides, whenever the level of infected is small enough, $SIR$ type dynamics are not fully reliable anymore, limiting the impact of such approximation on the derived optimal strategy}. In practice, the numerical experiments are indeed not sensitive to the choice of maturity $T$, as soon as it is picked large enough. 
For the graphs shown here, we used $T=700$ days (almost two years){, corresponding to a scenario where a vaccination solution is available within 2 years}. We did run computations up to $T=900$ looking for a potential impact on the interpretation of the optimal strategy. But since we did not observe relevant ones, we restricted our attention to $T=700$ in order to simplify the overall presentation of the results. Besides, in order to encompass the state constraint into the maximum principle based algorithm, we simply represent it under the form of a penalization cost, whenever the ICU constraint is not satisfied. The penalization is strong enough to discourage crossing this threshold for most scenarios. See Appendix \ref{sec_appendix_math} for more details on the implementation. 

The algorithm then relies on an iterative procedure. Time is discretized so that the search for an optimal control reduces to the search for one vector per control variable ($\delta, \lambda^1,\lambda^2$), with one value per time step. Starting from an initial guess, at each iteration the approximation for the optimal control is updated based on the expression of the gradient in terms of the solution to a forward-backward system of ordinary differential equations. In order to ensure that the controls remain between $0$ and $1$, a projection of the new controls on the interval $[0,1]$ is performed at each iteration.


\section{Discussion}\label{sec:3}

\subsection{Benchmark scenario without  intervention}\label{sec:3:1}

\begin{figure}[h]
	\begin{subfigure}{.5\columnwidth}
		\centering
		\includegraphics[width=\columnwidth]{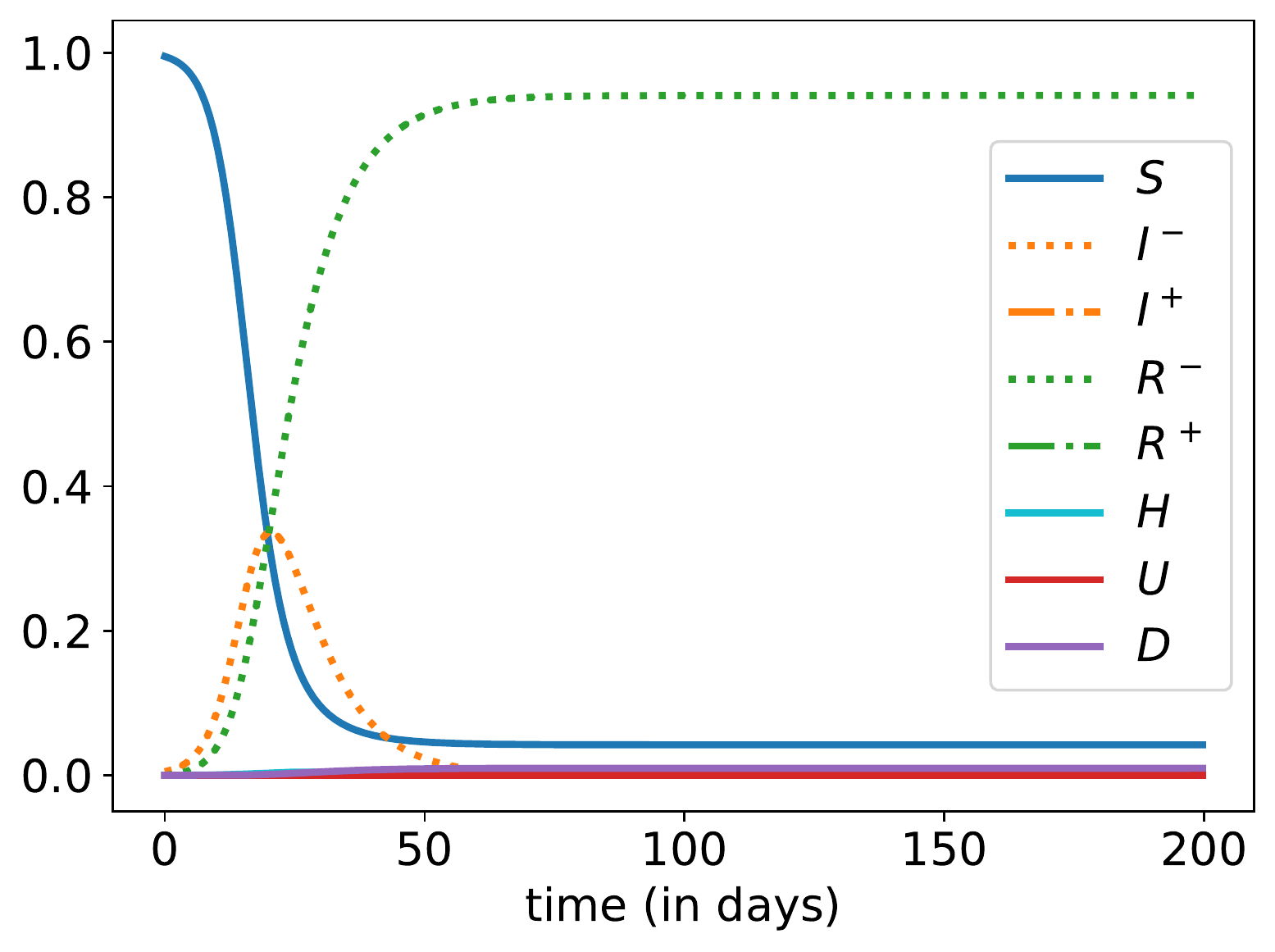}
		\caption{\, All states}
	\end{subfigure}%
	\begin{subfigure}{.5\columnwidth}
		\centering 
		\includegraphics[width=\columnwidth]{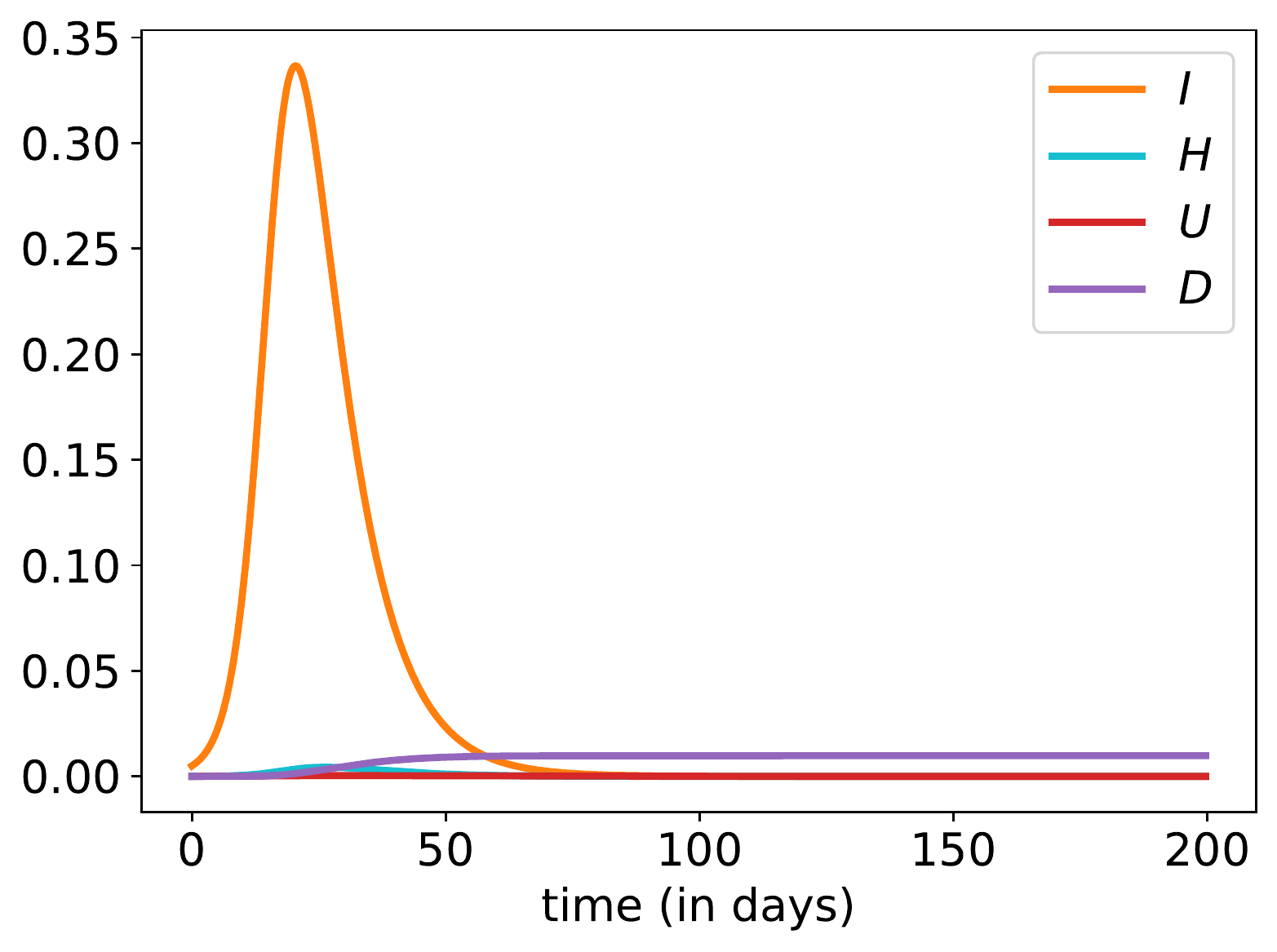}
		\caption{\, States $I=I^-+I^+, H, U, D$ only}
	\end{subfigure}
	\caption{Evolution of states without control, over 200 days.}
 	\label{fig:evol-nocontrol}
\end{figure}

Let us first investigate the outcomes of the epidemic dynamics when no lockdown, detection nor quarantine policies are available. {As commonly done in the current literature, we do not take into account the reaction of individuals in response to the virus prevalence dynamics, as discussed e.g. in \cite{elie2020contact}. Anticipating realistic  individuals behaviors in such context is still difficult, as no available data allows to tackle this phenomenon properly.} 
We suppose that the virus dynamics has no impact on the level of social interactions of individuals,{which remains constant equal to $\beta_0$}.
Such analysis provides a very useful no-intervention benchmark scenario for the rest of our study. The dynamics of each compartment is provided in Figure \ref{fig:evol-nocontrol} while Table \ref{tab:XT-nocontrol} provides important epidemic characteristics computed at terminal date $T=700$ days\footnote{In all other figures of this article, but that one, we consider a 700 day horizon. Here, the epidemic phase is over within two months so we shorten the time frame on Figure \ref{fig:evol-nocontrol} for ease of readability.}. 

As no intervention interferes with the epidemic dynamics, the overall duration of the epidemic is very short as the number of susceptible vanishes exponentially fast. The initial reproduction number $\mathfrak{R}_0=3.3$ induces an epidemic lasting around 80 days. The virus prevalence is quite important over that period, leading to the contamination of almost $96\%$ of the population, way above the herd immunity threshold, $S^*\sim 71\%$ in our example. The epidemic peak happens after a one month period, and as much as one third of the population is infected at that date. Around $7\%$ of the population is hospitalized due to the virus, leading to a highly severe overwhelming of the healthcare capacity  system for around 60 days. The epidemic induces in total a mortality of almost $1\%$ of the population, {while half of deaths occur while the ICU capacity is saturated}. 

It is worth noticing that our model, with the parameter values of Table \ref{tab:parameters-corps} that were calibrated on the French epidemic of COVID-19 (\cite{didomenicopullanosabbatiniboellecolizza,salje-cauchemez}), gives predictions that are in accordance with the ones made by \cite{rouxmassonnaudcrepey} in absence of interventions and neglecting the ICU capacity  constraint (see Appendix \ref{sec:parameters} for details).

\subsection{Optimal lockdown policy without detection effort}\label{subsec:optimal:policy}

\begin{figure}[h]
	\begin{subfigure}{.33\columnwidth}
		\centering
		\includegraphics[width=\columnwidth]{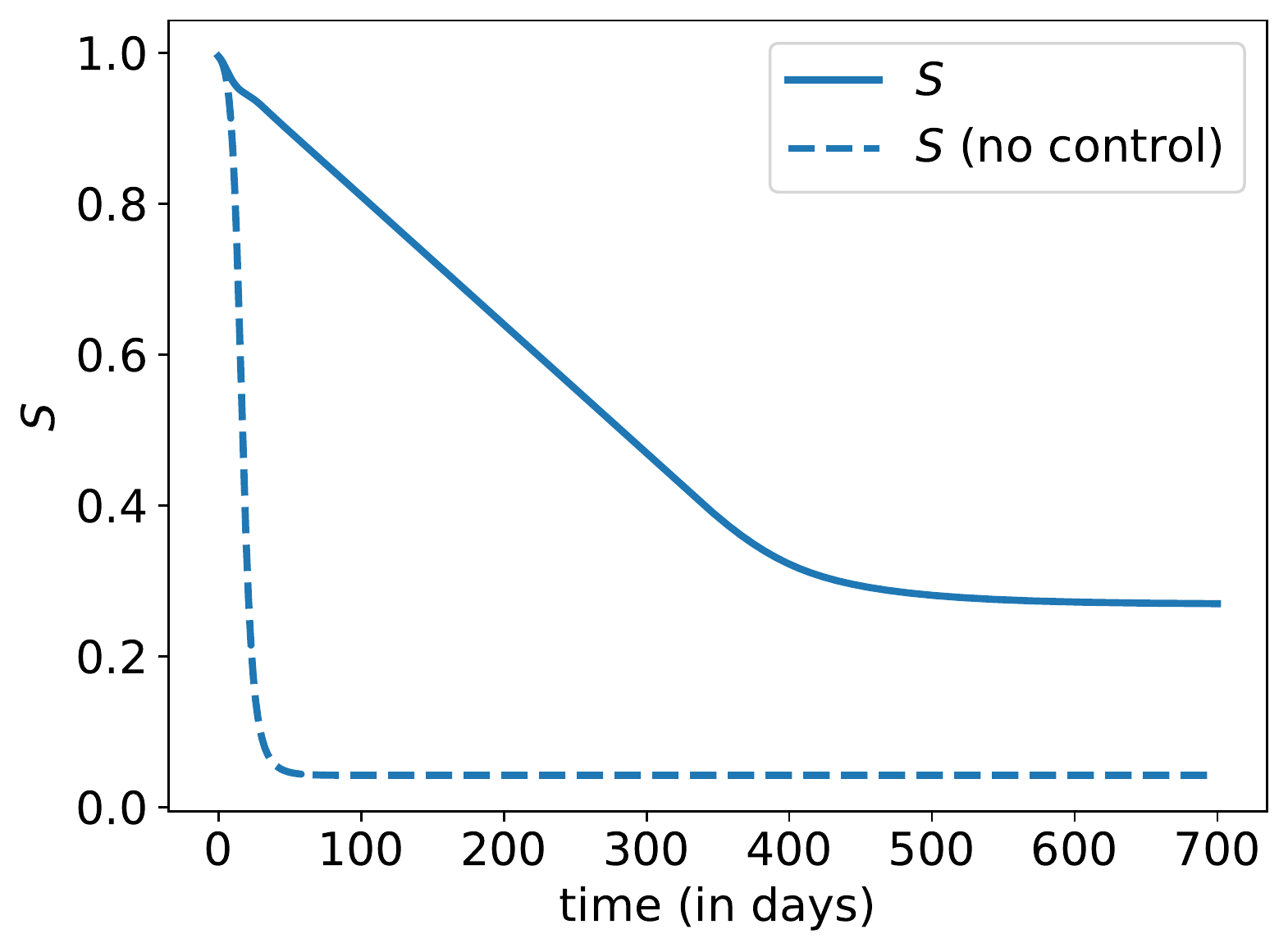}
	\end{subfigure}%
	\begin{subfigure}{.33\columnwidth}
		\centering 
		\includegraphics[width=\columnwidth]{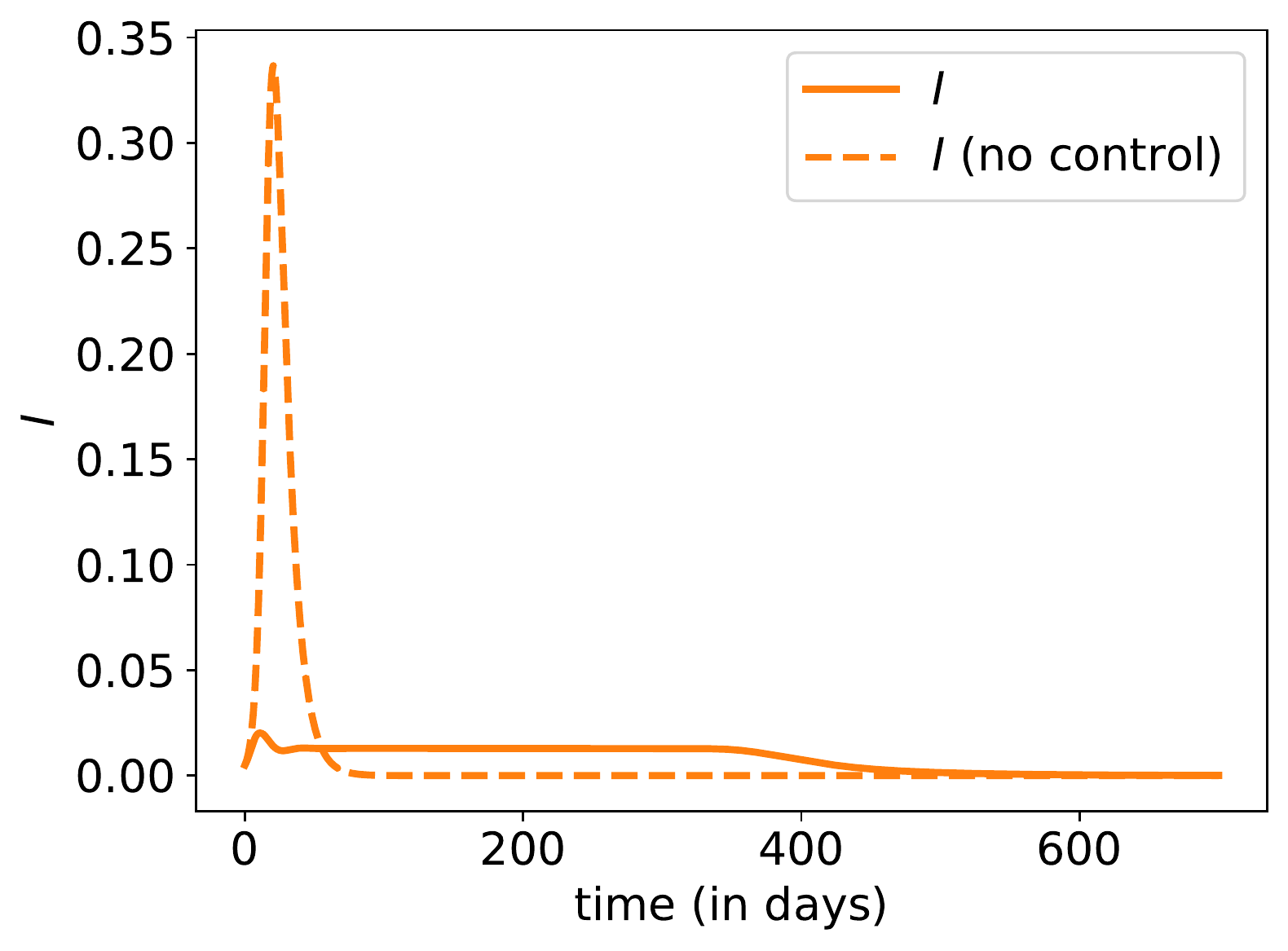}
	\end{subfigure}
	\begin{subfigure}{.33\columnwidth}
		\centering 
		\includegraphics[width=\columnwidth]{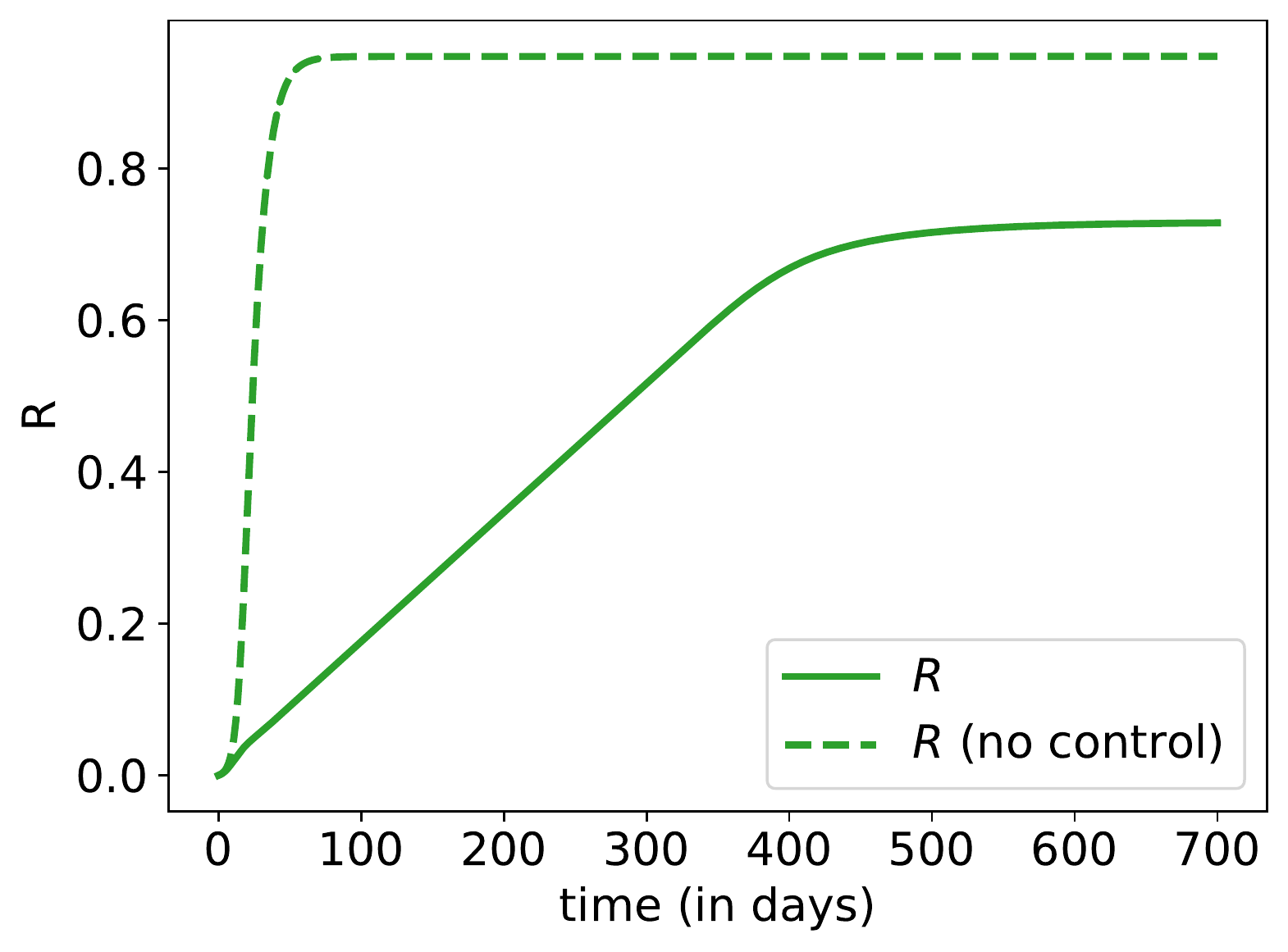}
	\end{subfigure}
	
	\begin{subfigure}{.33\columnwidth}
		\centering
		\includegraphics[width=\columnwidth]{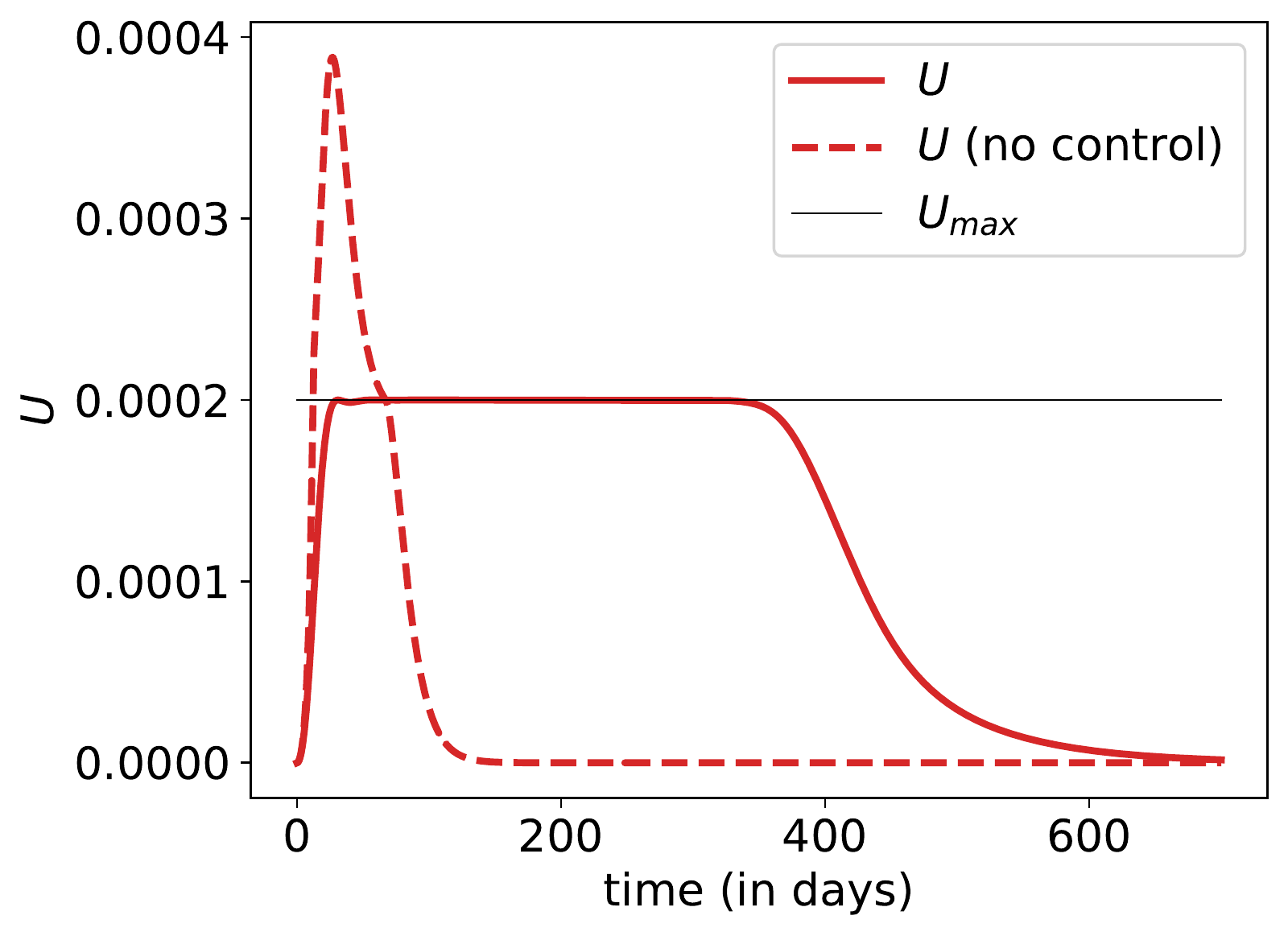}
	\end{subfigure}%
	\begin{subfigure}{.33\columnwidth}
		\centering 
		\includegraphics[width=\columnwidth]{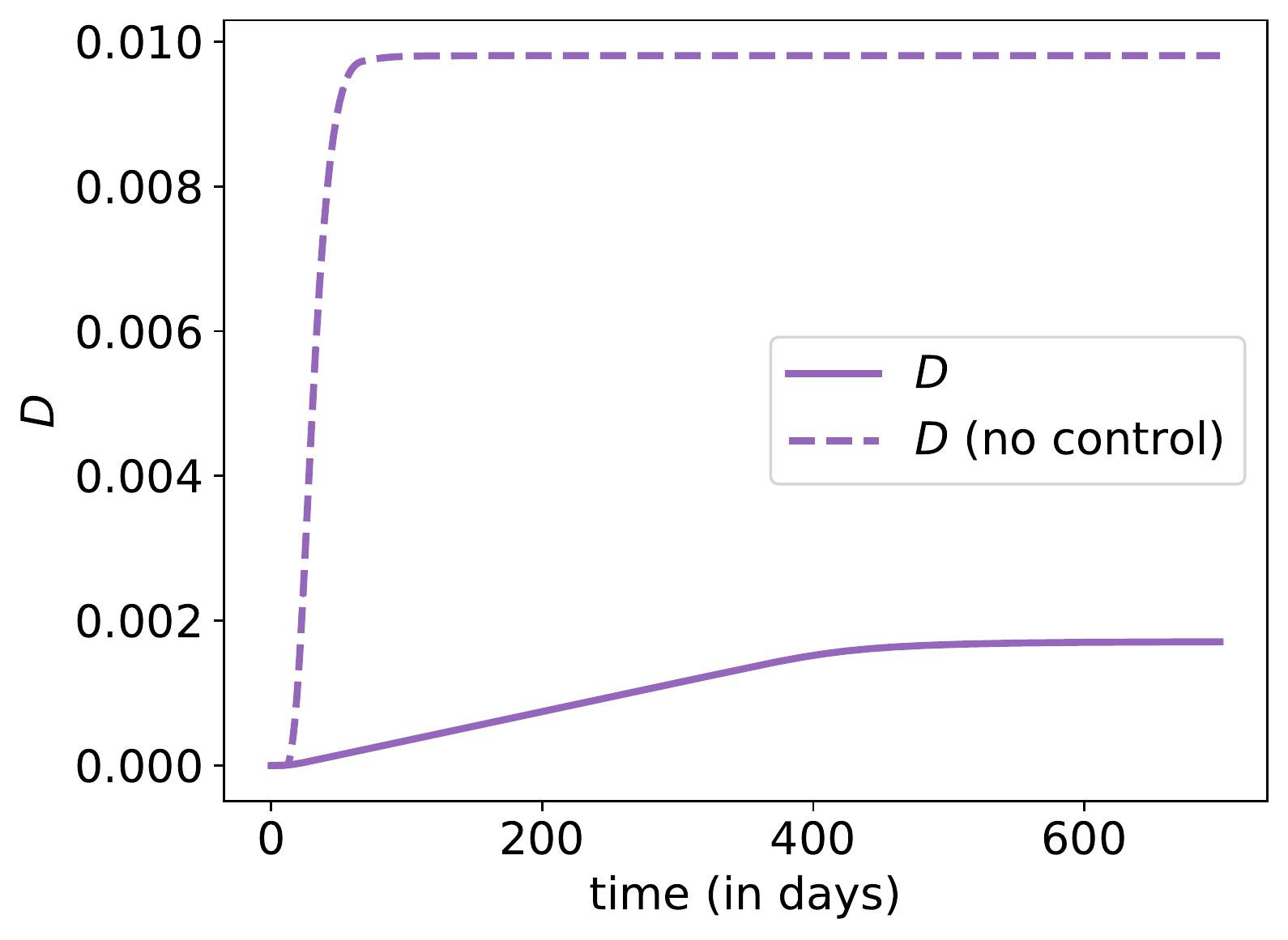}
	\end{subfigure}
	\begin{subfigure}{.33\columnwidth}
		\centering 
		\includegraphics[width=\columnwidth]{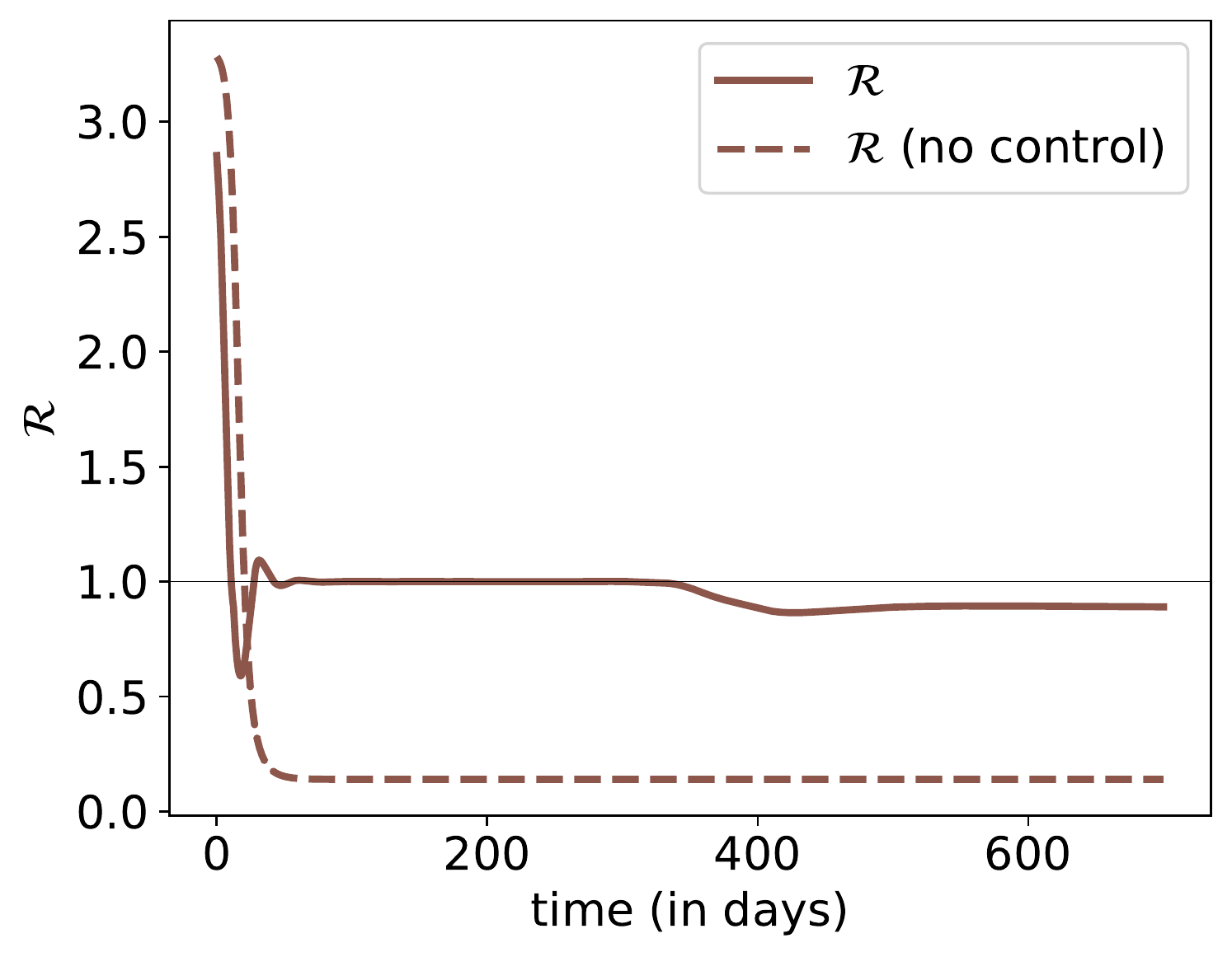}
	\end{subfigure}
	
	\begin{subfigure}{.33\columnwidth}
		\centering
		\includegraphics[width=\columnwidth]{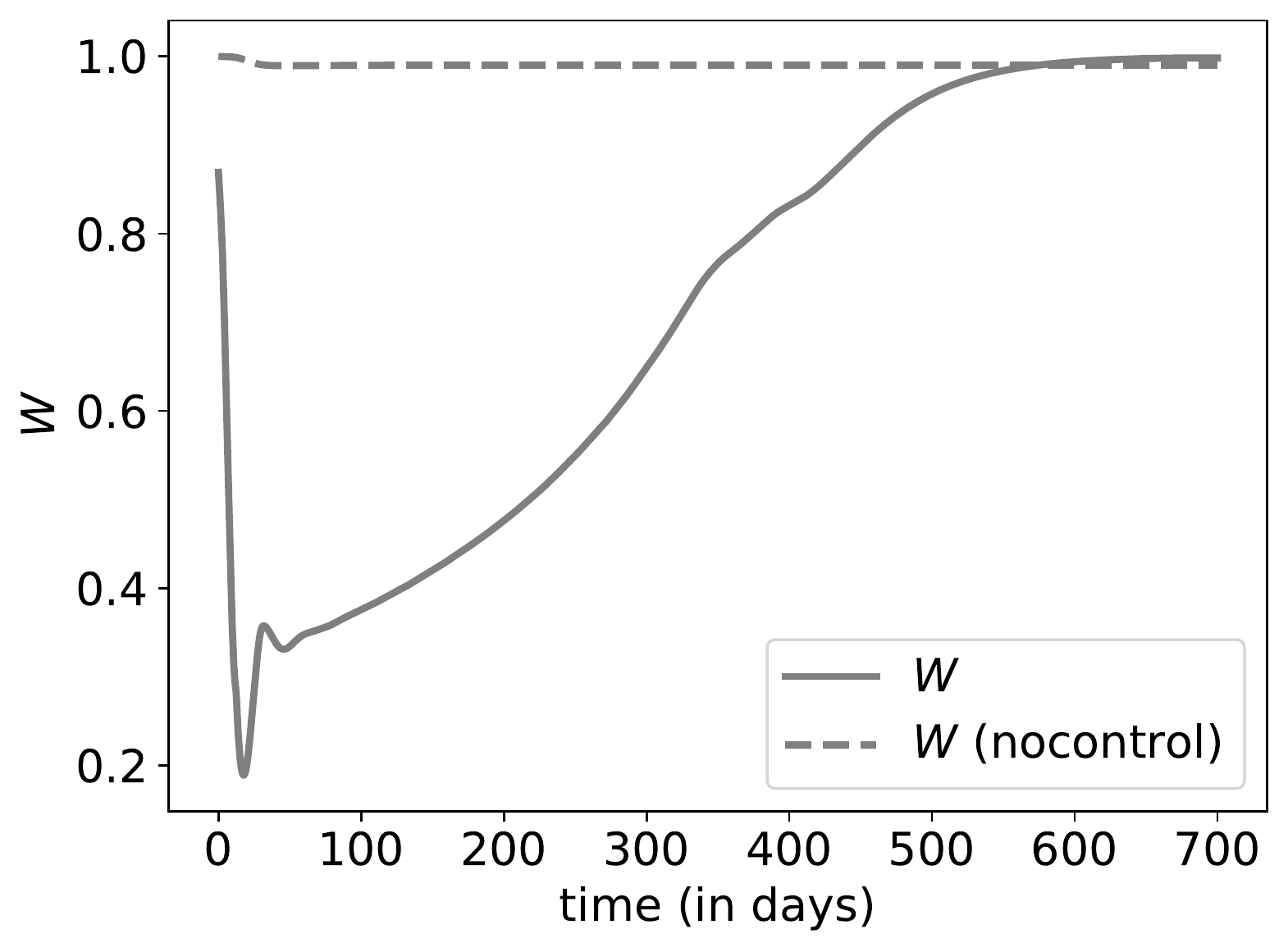}
	\end{subfigure}%
	\begin{subfigure}{.33\columnwidth}
		\centering 
		\includegraphics[width=\columnwidth]{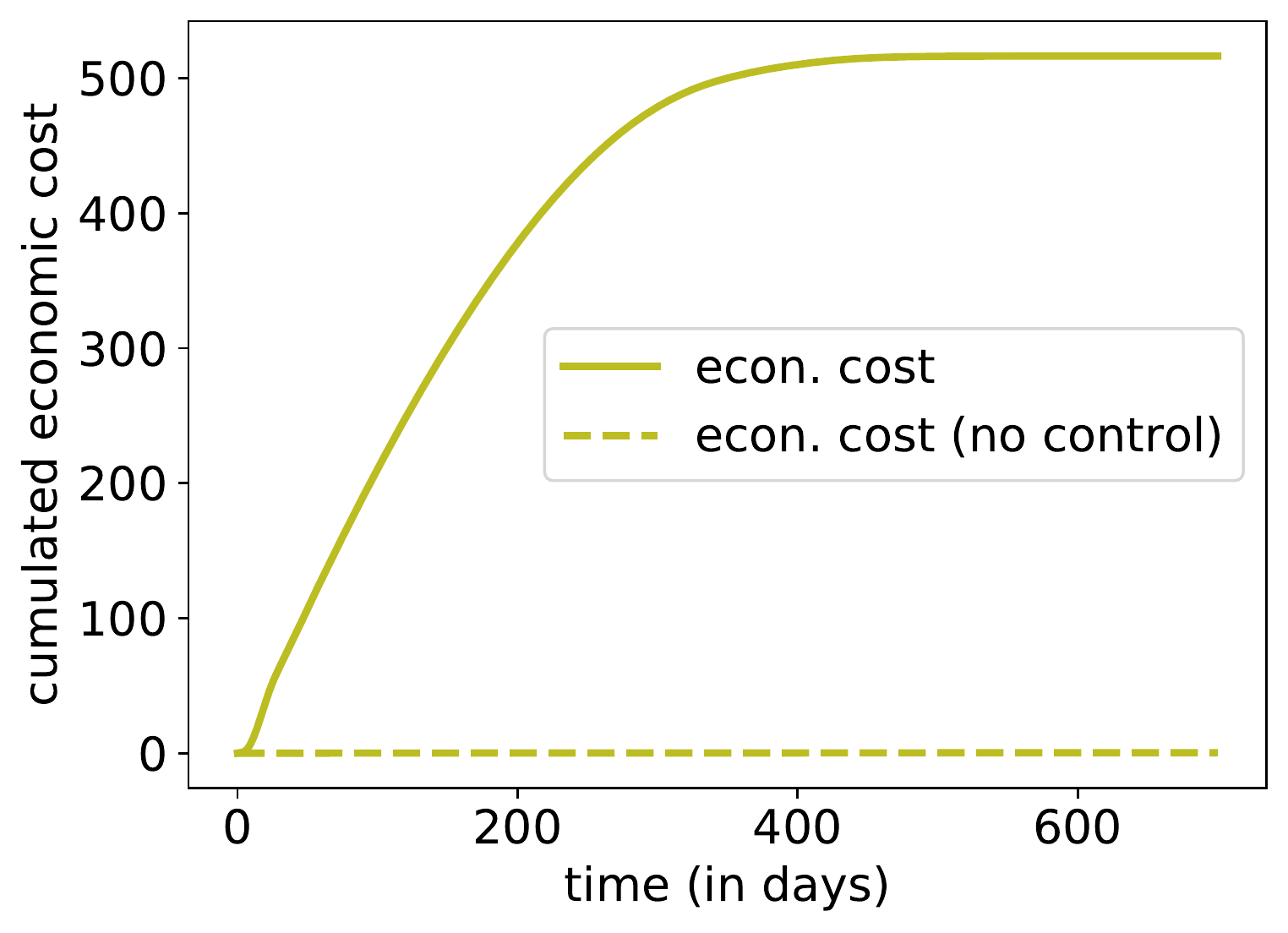}
	\end{subfigure}
	\begin{subfigure}{.33\columnwidth}
		\centering 
		\includegraphics[width=\columnwidth]{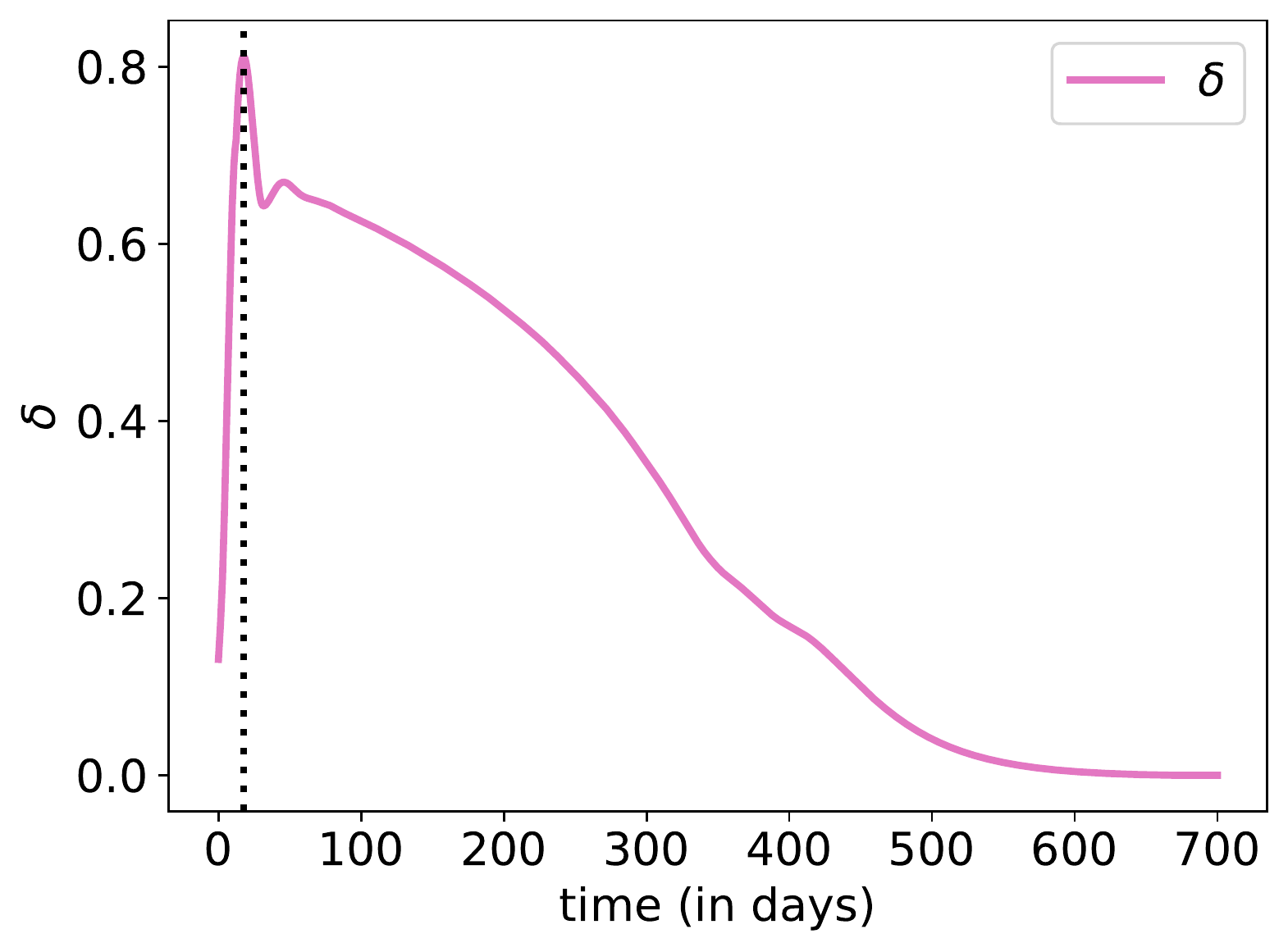}
	\end{subfigure}
	\caption{Evolution of state proportions -- $S_t$ (susceptible), $I_t$ (infected, detected or not, $I_t^-+I_t^+$), $R_t$ (recovered and immune, detected or not, $R_t^-+R_t^+$), $U_t$ (hospitalized in ICU), $D_t$ (dead), $\mathfrak{R}_t$ (dynamic version of the reproduction number), $W_t$ (labor force available), $Q_t$ (susceptible to be quanrantines, $S_t+I_t^-+R_t^-$) and $\delta_t$ (lockdown intensity) -- with optimal control $\delta$ (plain line) and without control (dashed line). States vary here from $t=0$ (beginning of the pandemic, 
	or at least of possible measures, with $\delta\geq 0$)
	to $t=700$, and computations are based on a $\drm t=1/5$ days.}
 	\label{fig:Scenario_control_benchmark}
\end{figure}

{In order to emphasize the marginal role of each component in the optimal control problem of interest, we first focus on the optimal lockdown strategy and corresponding epidemic trajectory induced without any detection effort. This does not mean that we do not test anyone, we simply assume that only the people entering in the hospital are tested with type-1 tests. No testing is done on infected individuals without severe symptoms ($\lambda^1=0$) and  there no testing is done  to detect recovered people after being asymptomatically sick (type-2 test, $\lambda^2=0$). 

Hereby, we are starting with a weaker form of the optimal control problem, where we control only the lockdown intervention policy:
\begin{eqnarray}
 \inf_{\delta\in\mathcal{A}_{\delta} \quad \mbox{s.t.}~  U \le U_{\max}} \big\lbrace J(\delta,0,0)\big\rbrace\;,
\end{eqnarray}
with the following choice of parameters and $w_{\text{sanitary}}=100000$, $w_{\text{eco}}=1$ and $\alpha=0$. The sensitivity of the enhanced results to this ad-hoc parametric choice is discussed in Sections \ref{sec_sensi_weight} and  \ref{sec_sensi_alpha}. Besides,
once we will have a better understanding about the derived  lockdown strategy and its robustness, we will focus on the impact of detection strategies in Section \ref{sec:3:5}. The optimal lockdown strategy together with the main characteristics of the pandemic dynamics are presented in Figure \ref{fig:Scenario_control_benchmark} and Table \ref{tab:XT-nocontrol}. The optimal control problem has been solved using the numerical algorithm described in Appendix \ref{Sec_algo} and numerical approximation errors undoubtedly remain in the results presented here.}  \\

\begin{table}[h!]
	\begin{center}
		\caption{Epidemic characteristics}
			\begin{tabular}{r|c|c|c|c|c}
				\hline 
				\hline
				  & $S_T$    & $I_T$ & $R_T$ & $D_T$ & $\max_t(I_t)$ 
				 \\
				\hline
				Without control & $4.2\%$    & $0\%$ & $94.8\%$ & $9.8$\textperthousand & $33.7\%$ 
				\\
				\hline
				With optimal $\delta$ & $27\%$    & $\sim 0\%$ & $72.9\%$ & $1.7$\textperthousand & $2\%$ 
				 \\
				\hline 
				\hline
			\end{tabular}
		\label{tab:XT-nocontrol}
	\end{center}
\end{table}

When we follow the lockdown intervention ($\delta_t$) proposed by the optimal response policy, four distinct consecutive phases can be observed:
 \begin{enumerate}
 \item \textbf{Quick activation of a strong lockdown}: The optimal strategy consists in activating as soon as possible a strong lockdown measure on the entire population as no targeting or detection is considered or possible at that stage. This strategy has been widely used in March and April 2020 as half of the world population was under lockdown. With our parameters, the contact rate over the population is reduced by $80\%$ over a couple of days, in order to stop immediately the exponential growth of infected and infectious individuals. Hereby, the size of the epidemic peak is highly reduced and, after less than one month, only $2\%$ of the population is contaminated at the peak of the epidemic (instead of $34\%$ in the no-intervention benchmark case). Of course, at the exact same date, the effective reproduction number $\mathfrak{R}_t$ goes below $1$. During this initial short period, the effort in economic activity reduction has been very important: {the welfare level $W$} went down by $80\%$, while at the same time the occupation rate in ICU hospital care system has been constantly rising.
 \item \textbf{Light lockdown release and   prevalence decrease:} During the short second phase, the intensity of the lockdown remains intense (above $60\%$) in order to keep $\mathfrak{R}_t$ below $1$. The prevalence of the virus decreases slowly, while the level of ICU admissions is still rising. After the previous phase where urgent decisions had to be taken by the global planner, a short transition period of a couple of days now starts. During this period, the proportion of infectious individuals decreases strongly in order reach an optimal prevalence level $I^*$  in the population, required for the very long next phase.
 \item \textbf{Long period with stable  prevalence and ICU sustainable capacity}: The third phase lasts around one year during which the effective reproduction number $\mathfrak{R}_t$ remains stable at level $1$. Hereby, the prevalence of the virus remains stable while the ICU capacity remains at the sustainable level $U_{\max}$. During this one year period, the level of contact rate within the population is slowly growing at a regular constant rate, moving gradually from $40\%$ to $80\%$. The working force is rising back to more normal levels and the economy is slowly restarting: one year after the beginning of the pandemic, around $70\%$ of regular social interactions level is already attained  within the population. The mortality rate over this period remains stable, and the death toll grows linearly, almost reaching its terminal value at the end of this period. 
 \item \textbf{Terminal slow progressive release of the lockdown:} The last phase is a couple of months long and consists in bringing progressively back to normal the level of social interactions, while the occupation level in ICU is decreasing and the prevalence of the virus is heading  towards $0$. The effective reproduction number $\mathfrak{R}_t$ is smaller than $1$. At some point, the deterministic equations \eqref{sys:SIR} are no more valid and the system becomes stochastic, but the extinction of the disease occurs with probability 1.
 \end{enumerate}
 
 \begin{figure}[h]
		\centering
		\includegraphics[width=8cm]{{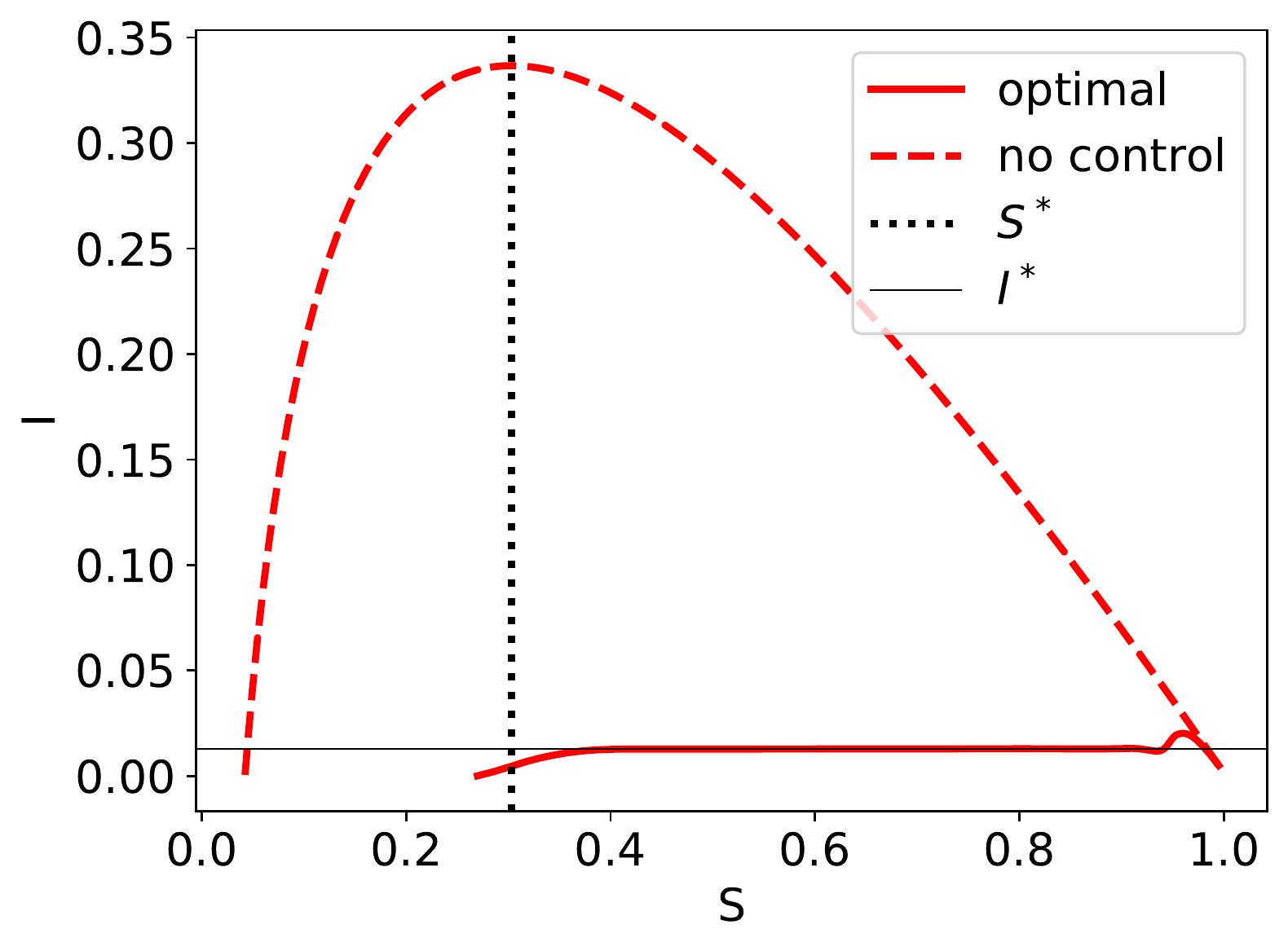}}
		\caption{\, Phase diagram $S_t$ (susceptible) versus $I_t$ (infected, detected or not, $I^-_t+I^+_t$), together with the herd immunity threshold $S^*$ and the level $I^*$ of infected corresponding to $U=U_{\max}$.  }
 	\label{fig:phase-diagram}
\end{figure}

 The optimal structure in four steps for controlling the pandemic dynamics is quite robust and very natural.  It is worth observing that although the dynamic reproduction number is not part of the criterion of interest, reaching its optimal trajectory with four successive patterns is key for controlling the dynamic level of admissions in ICU. Using the optimal control policy, the infections by the virus in the population have been spread out over 500 days, providing a sustainable level of admissions in ICU for the sanitary system. In comparison to the no-intervention benchmark case, the death toll has been divided by almost 100, thanks to a strong reduction in the level of social interactions. The size of the pandemic only represents $70\%$ of the population in comparison to $96\%$ in the no-intervention scenario. 

A nice visualisation of the epidemic dynamics is provided by looking at the phase diagram presented in Figure \ref{fig:phase-diagram}. In the no-intervention scenario, the epidemic dynamics starts from the point $(S,I)=(1,0)$ and moves up and left until reaching the herd immunity threshold $S^*$ corresponding to the time of epidemic peak. Then, $I$ starts decreasing until the end of the epidemic, point where $S$ is valued around $4\%$ in our example: hence, the final epidemic size is almost $96\%$ of the population. In the optimally controlled scenario, we can clearly identify the 4 phases described above: $I$ is still rising during the first phase until reaching a small epidemic peak level, it starts decreasing during the second transition phase until reaching the optimal level $I^*$ corresponding to the level $U_{\max}$ of ICU capacity. It remains stable there during the long third phase and finally decreases in the last phase until reaching a susceptible proportion close to the herd immunity level $S^*$.

Of course, both dynamics depend on the value of the parameters, such as the initial virus prevalence $I_0^-$, the starting basic reproductive number $\mathfrak{R}_0$, expectations about possible arrival of vaccine and cure (through parameter $\alpha$ or weights considered in the objective function). But as we can see in Appendix \ref{appendix:b}, and as we will discuss more now, the enhanced results are robust to the parameter specifications.


\subsection{Sensitivity of the optimally controlled epidemic dynamics to the model specifications}\label{sec:3:3}

 The calibration of $SIR$ type dynamics to current available data is currently a challenging task, as the quality of available mortality and hospitalization figures is still questionable and incomplete. It is worth noting that one of the main advantages of the \nommodel dynamics is the possibility of calibrating its dynamics to mortality, infected detected, hospitalized and ICU data points. Still, such inverse calibration problem remains  ill-posed and the precision range over  the calibrated parameters is not satisfying. 
 
 In order to demonstrate the robustness of our findings in such context, we provide a sensitivity analysis of our results to modifications in the model specifications. More specifically, we analyse the impact of reasonable  variations with respect to the initial prevalence $I_0^-$ in Appendix \ref{sec_sensi_I0}, the basic reproduction number $\mathfrak{R}_0$ in Appendix \ref{sec_sensi_R0}, the balance between sanitary and socio-economic cost in  Appendix \ref{sec_sensi_weight}, and the arrival date of a vaccination solution in Appendix \ref{sec_sensi_alpha}. Let now present the main outputs of those numerical experiments{, while keeping in mind that numerical errors due to the optimization approximation scheme can still alter some  numerical results presented in this study.}

{On Figure \ref{fig:Sensi_I0}, we can visualize the impact of a change in $I_0^-$, with a smaller (five times) and larger (also five times) value than the one used in benchmark computations (we used $I_0^-=5$\textperthousand). The lower value was suggested from \cite{Sachdeva}, with $1$\textperthousand, five times lower than our estimate, and an upper value five times upper, of $2.5\%$, close to the situation two weeks after according to \cite{Sachdeva}. A smaller value means that measures were taken earlier, but overall, it looks like the overall impact would have been rather small. Note that we observe here that a smaller $I_0^-$ might cause more deaths ($D_T$), because of some balance with economic gains (economics losses are smaller with a lower $I_0^-$). Observe also that variations in $I_0^-$ have almost no impact on the ICU saturation dynamics: it might start earlier with a large $I_0^-$, but the duration remains more or less the same.}

{On Figure \ref{fig:Sensi_Rzero}, we can visualize the impact of a change in $\mathfrak{R}_0$, with a smaller ($-0.3$ i.e. $-10\%$) and larger ($+0.3$ i.e. $+10\%$) value than the one used in benchmark computations ($\mathfrak{R}_0=3.3$). Such variation in $\mathfrak{R}_0$ is compatible with the literature, see \cite{salje-cauchemez,didomenicopullanosabbatiniboellecolizza}. The overall 4-phases global patterns of the optimally controlled epidemics remain identical. The lockdown effort increases slightly but significantly  with $\mathfrak{R}_0$, together with the ICU saturation phase duration, the global death toll and global economic cost. Note that a decrease of $\mathfrak{R}_0$ from 3.6 to 3.0 {(attained for example through extra hygiene habits)}  would decrease ICU saturation time length by almost 25\%.   }

{On Figure \ref{fig:Sensi_wsanitary}, we can visualize the impact of a change of the ratio between the sanitary weight  $w_{\text{sanitary}}$ and the socio-economic one $w_{\text{eco}}$, with a smaller (half) and larger (twice) value than the one used in benchmark computations. The global patterns in the epidemic dynamics remain globally similar. A larger sanitary weight induces a smaller death toll, a smaller epidemic peak together with a shorter ICU saturation period. The economic gain or loss seems to be mostly sensible during the second half of the epidemic duration.}

Figure \ref{fig:Sensi_alpha} provides the shapes of numerically approximated  optimal epidemic curves for different beliefs about the expected arrival time of a vaccine (and a cure). {In comparison to the benchmark case where no vaccine potential creation is taken into account, the arrival date $\tau$ of the vaccine is supposed to follow an exponential distribution with expectation $100$, $250$ or $500$ days, while the optimization is occurring on the time interval $[0,T\wedge\tau]$} . The epidemic curves are very similar {when the anticipated time for vaccine availability is long enough}, confirming the robustness of the approach and main findings. {Whenever a vaccination procedure is supposed to arise rather early, the optimal strategy consists in postponing the mortality while keeping a stronger and longer lockdown effort, hoping for a quick vaccination solution. The sanitary burden is hereby smaller during the early stage of the epidemics but rises more strongly at the terminal phase.} \\

 Our main overall observation is that the dynamics of the optimally controlled epidemic dynamics  always present the same patterns and enlights the robustness of our findings. It divides into four successive phases in a similar way to the one described in the previous main scenario analysis. Besides, the effective reproduction number $\mathfrak{R}_t$  together with the optimal lockdown intervention policy also have the exact similar patterns. Of course small variations in the specifications still have an influence on the exact figures characterizing the balance between sanitary and economic outcomes, the total duration of the epidemic or the pressure on the ICU hospital system, but the overall characteristics remain identical. These observations confirm the robustness of our findings when dealing the the optimal control problem at hand, under specification uncertainty.  The more specific dependence to the ICU care system capacity is studied in more details and discussed in the next section.

\subsection{On the impact of additional ICU capacities}\label{sec_sensi_Umax}

As discussed earlier, most countries have been urged to take public health measures to `{\em flatten the curve}'. An important motivation was to avoid that hospital systems get overwhelmed, and more specifically, social distancing was introduced to ensure that the occupation of ICUs with ventilators remain within the capacities of hospitals, dedicated to COVID-19 patients. Previously, we integrated an upper limit to ICU -- denoted by $U_{\max}$ -- and in this section, we discuss the impact of a potential increase of that capacity. Of course, increasing capacity does not simply mean getting more masks, beds, ventilators, or medicines for the serious cases. It also means diverting resources,  training more staff to work in the ICU, and probably introduce telemedicine solutions for non-COVID-19 patients (who do not require hospitalization) to free up beds.

\begin{figure}[h]
	\begin{subfigure}{.33\columnwidth}
		\centering
		\includegraphics[width=\columnwidth]{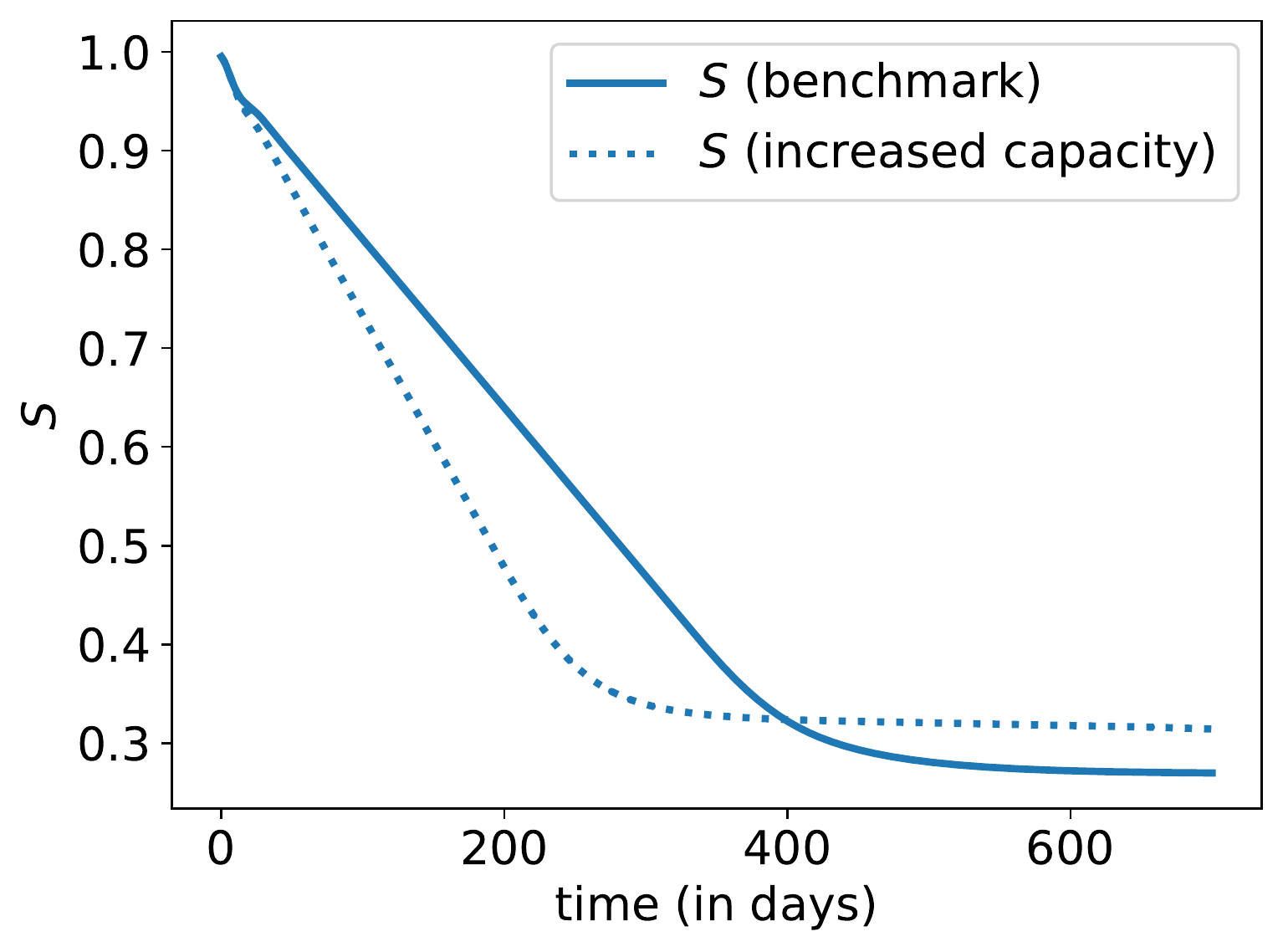}
	\end{subfigure}%
	\begin{subfigure}{.33\columnwidth}
		\centering 
		\includegraphics[width=\columnwidth]{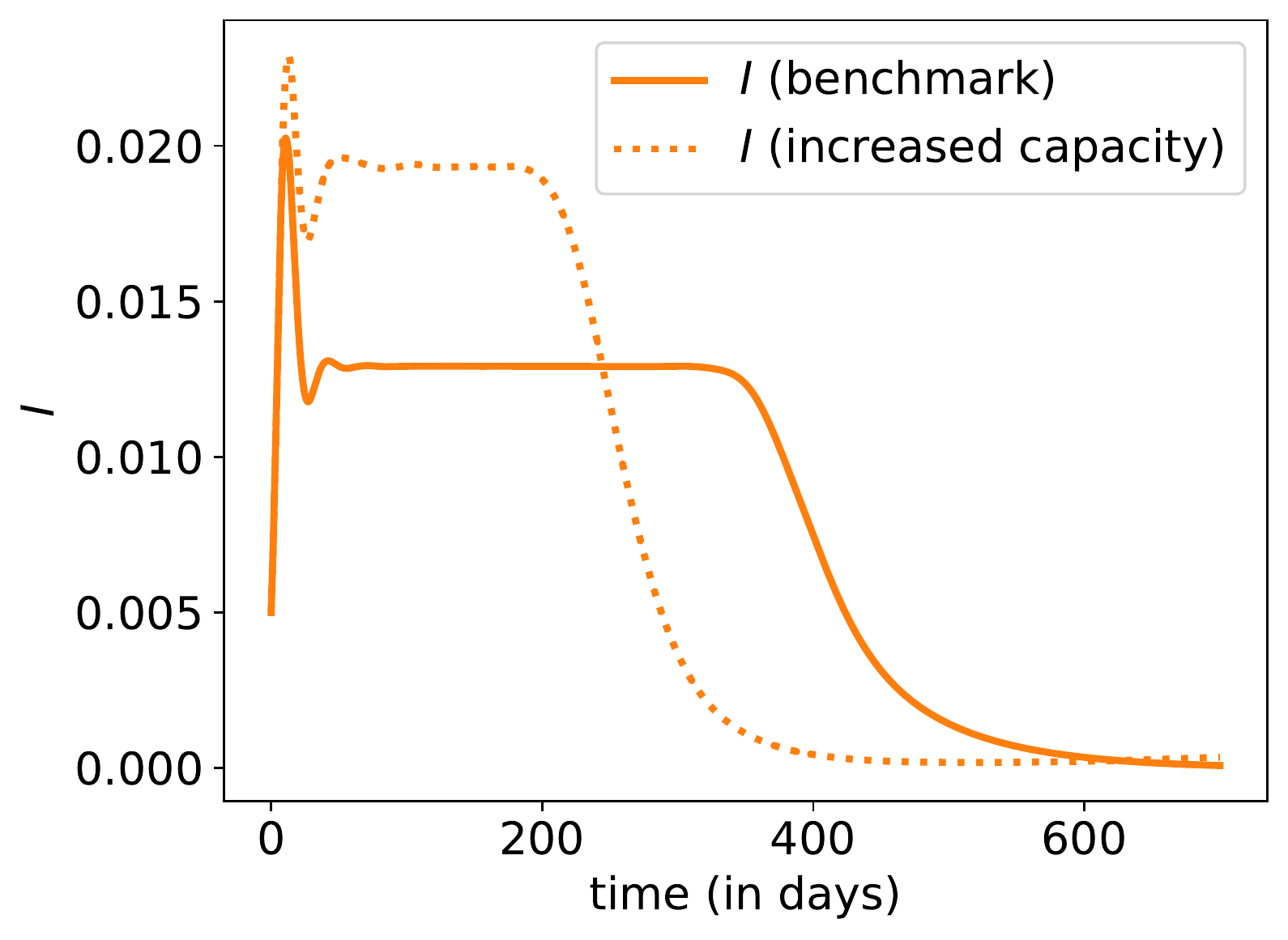}
	\end{subfigure}
	\begin{subfigure}{.33\columnwidth}
		\centering 
		\includegraphics[width=\columnwidth]{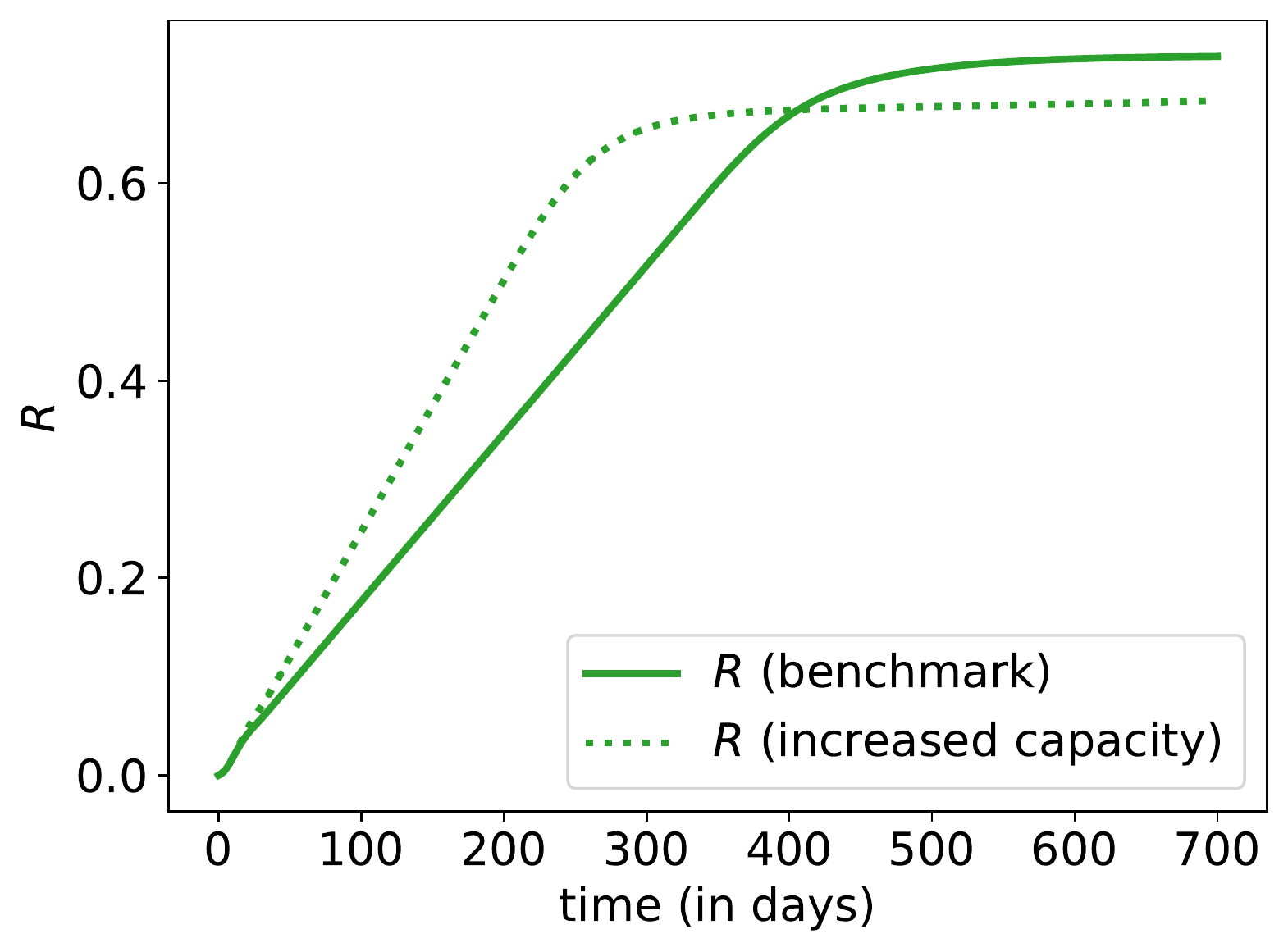}
	\end{subfigure}
	
	\begin{subfigure}{.33\columnwidth}
		\centering
		\includegraphics[width=\columnwidth]{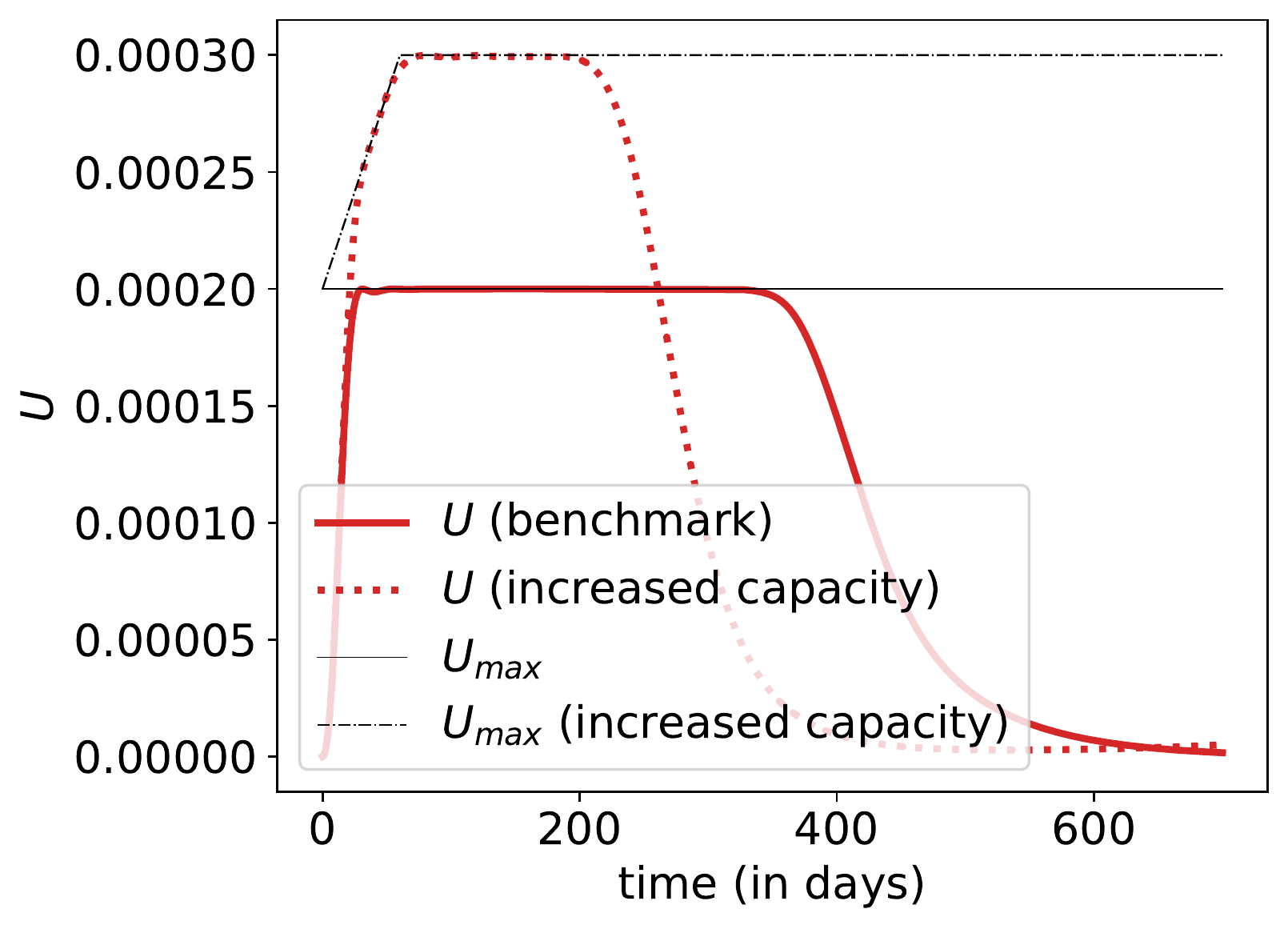}
	\end{subfigure}%
	\begin{subfigure}{.33\columnwidth}
		\centering 
		\includegraphics[width=\columnwidth]{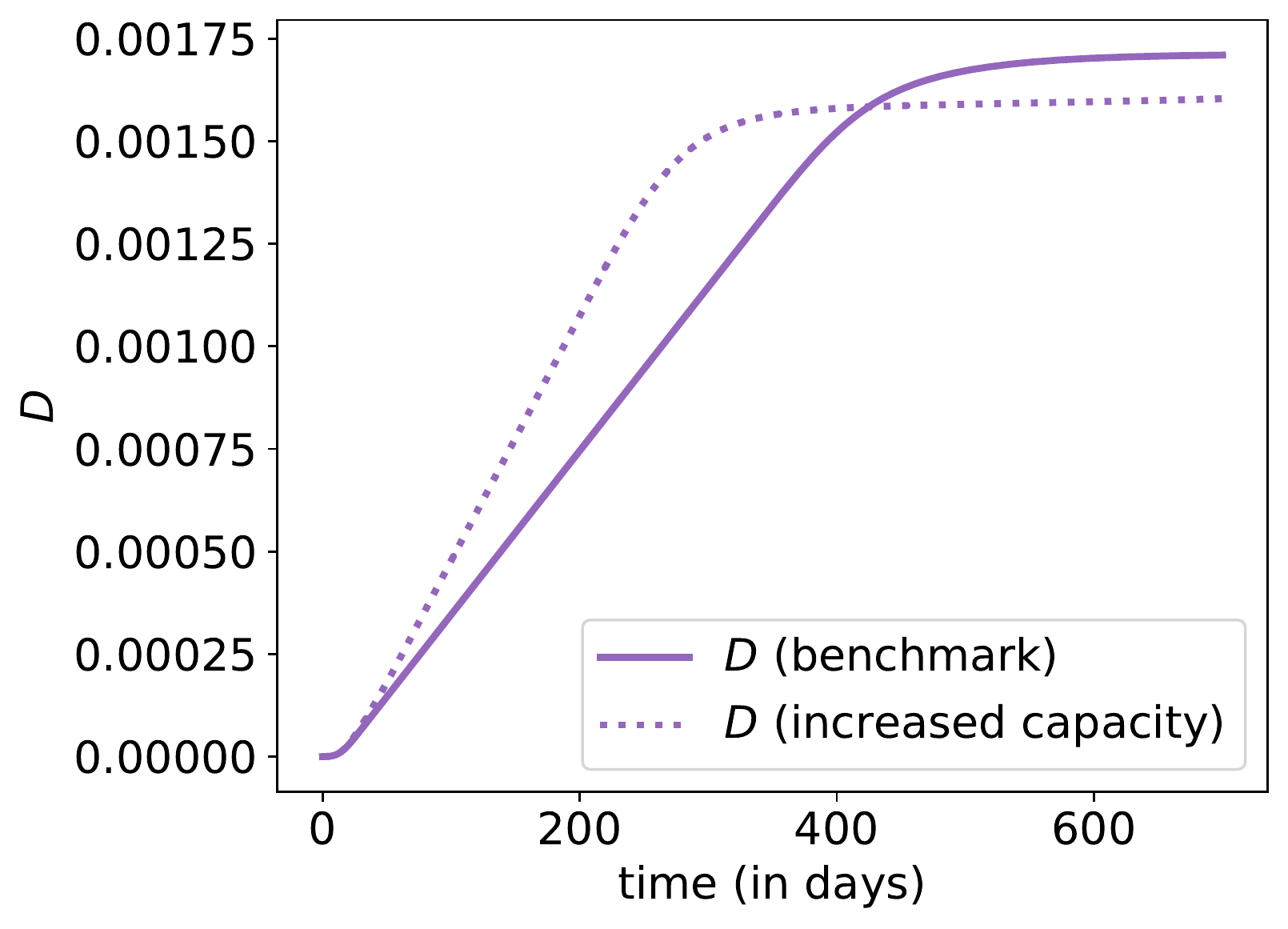}
	\end{subfigure}
	\begin{subfigure}{.33\columnwidth}
		\centering 
		\includegraphics[width=\columnwidth]{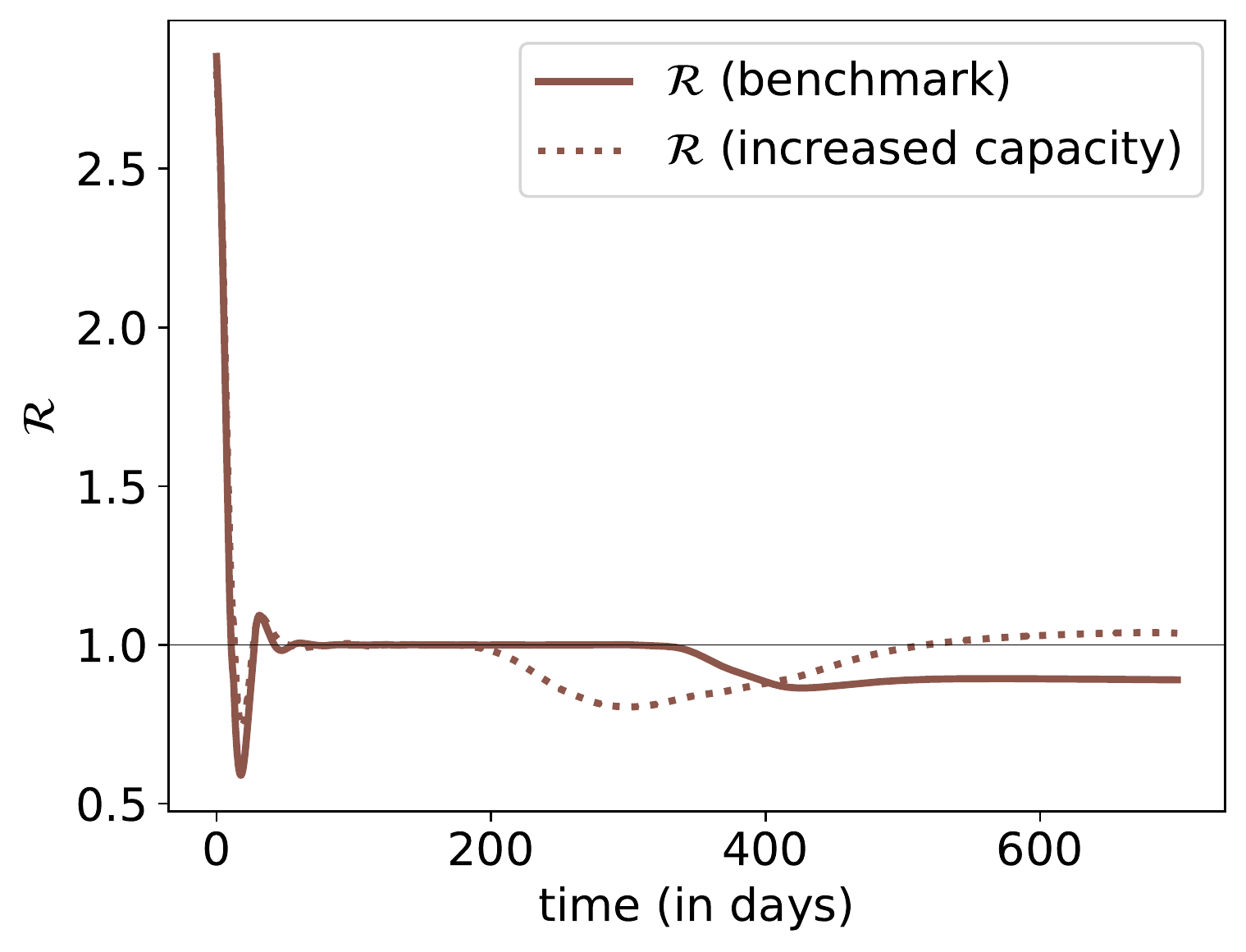}
	\end{subfigure}
	
	\begin{subfigure}{.33\columnwidth}
		\centering
		\includegraphics[width=\columnwidth]{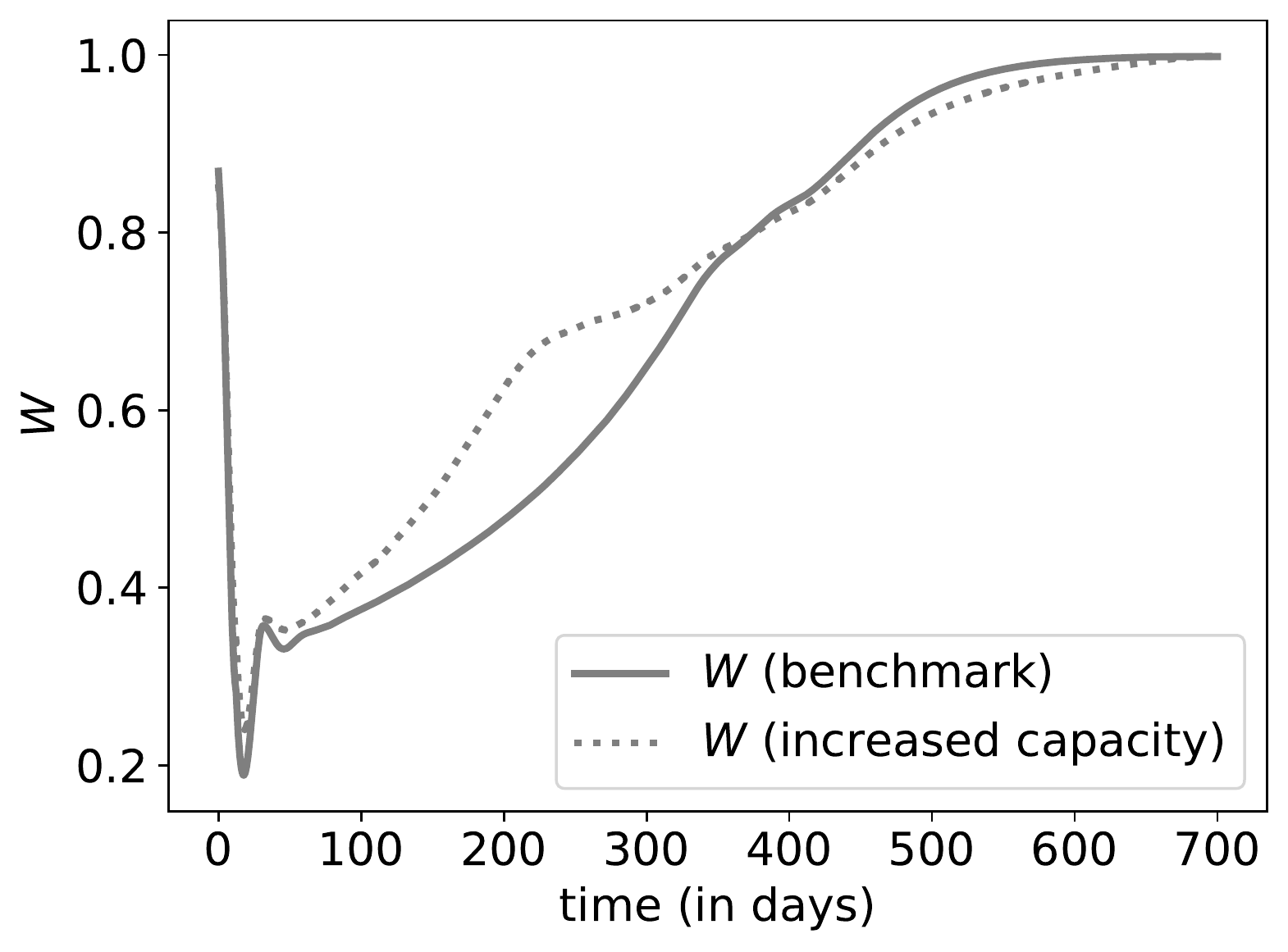}
	\end{subfigure}%
	\begin{subfigure}{.33\columnwidth}
		\centering 
		\includegraphics[width=\columnwidth]{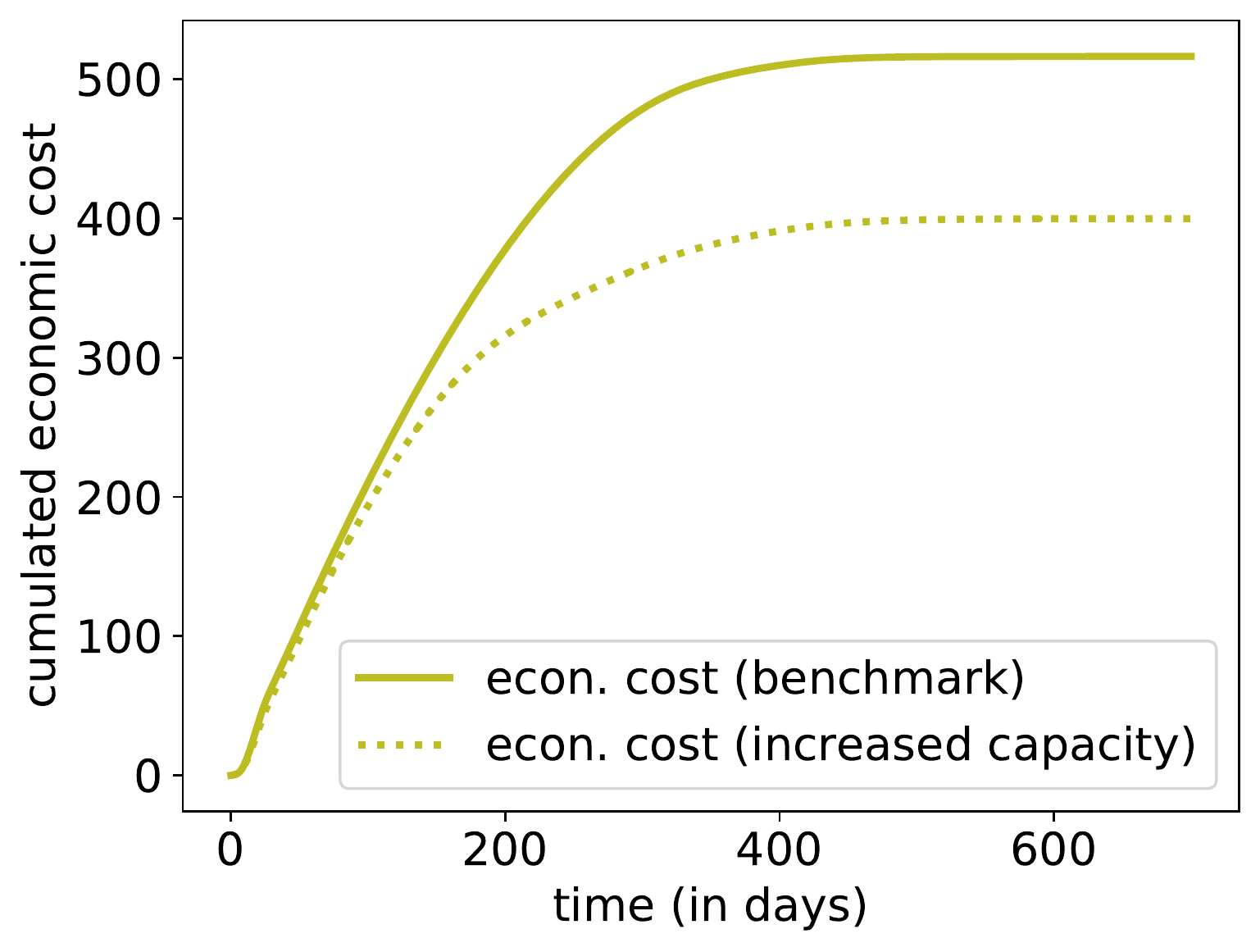}
	\end{subfigure}
	\begin{subfigure}{.33\columnwidth}
		\centering 
		\includegraphics[width=\columnwidth]{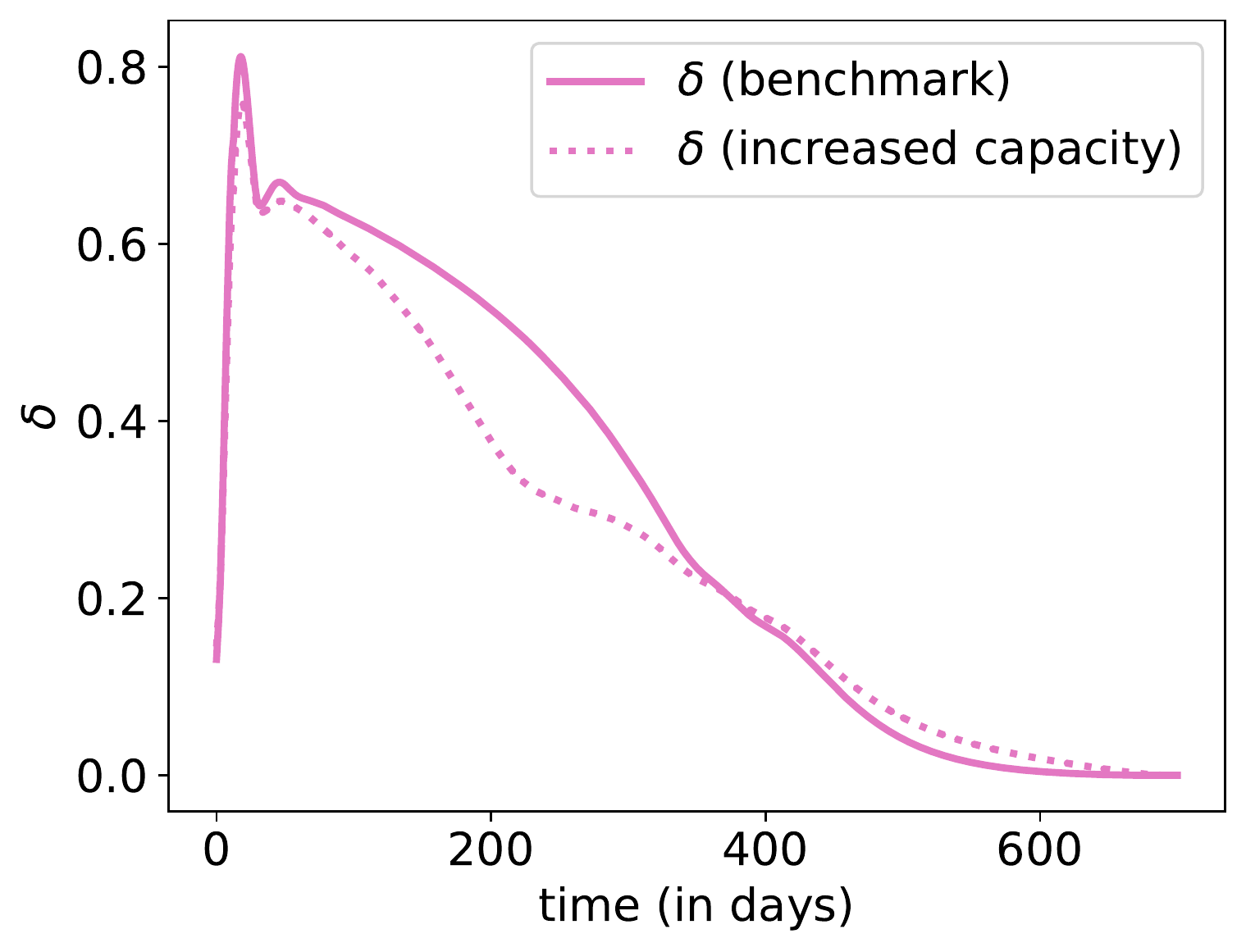}
	\end{subfigure}
	\caption{Evolution of states with optimal control $\delta$ when `raising the line' (increase of $U_{\max}$ by $50\%$). The plain line is the benchmark scenario discussed in Section \ref{subsec:optimal:policy}; the dashed line corresponds to the scenario with a capacity of $U_{\max}$ which increases linearly from time $t=0$ until $t=60$ days up to $1.5$ times the initial capacity and then stays at this level. }
 	\label{fig:Sensi_UMAX}
\end{figure}

On Figure \ref{fig:Sensi_UMAX}, we consider a change in $U_{\max}$, where we assume that health authorities are able to increase ICU capacities by $50\%$ over a 2 months period and remain at that level. We can imagine an increase in the number of ventilators or back-ups with staff coming from other regions that are spared by the epidemics, as was done in China or France for instance. Another possible interpretation is that the medical staff on the front line might be infected in early periods leading to an initial shortage of personal at the beginning followed by a return to {\em normal} within a few weeks. Here, we consider a regular (linear) increase of ICU capacities over two months, until reaching a new limit, 50\% higher than the original one. 

This scenario has a major impact on all quantities. First, we observe that the derived optimal lockdown strategy is able to let the ICU admission level bind the sustainable constraint over time, although it now grows dynamically with time. {This corresponds to a longer second phase duration until the stable in prevalence and ICU capacity starts. } The final number of dead people is overall  smaller than in the original scenario, as the expansion of ICU capacity allows to treat more patients simultaneously. Therefore, the social distancing can be much weaker during the long ICU saturation phase, which also happens to be way shorter. Furthermore, ICU saturation duration is much shorter with a major positive impact on the economy from day 100 until day 400, as the lockdown intervention remains much weaker (after 200 days, the global level of social interactions is around 65\% in comparison to less than 50\% in our benchmark scenario). This reduces the socio-economic impact of the epidemic and hereby allows to reach a smaller mortality count at time $T$. The overall gain at terminal date is massive  (even if here, we do not include a cost for increasing capacities). Nevertheless, the augmentation of ICU capacity implies also a higher daily dynamic death toll, so that an early non-anticipated vaccination solution at an intermediate date would induce a higher sanitary cost in comparison to the scenario without ICU capacity rising.

    
\subsection{On the impact of detection resources }\label{sec:3:5}

We now focus our analysis on the impact and optimization over the detection efforts {$\lambda^1$ and $\lambda^2$}  of infectious or immune individuals. {As stated in \cite{Petom1163}, {\em mass testing could end the epidemic rapidly}}, and we now try to emphasize the impact of light or mass testing upon the lockdown optimal intervention policy. The numerical results are presented in Appendix B for ease of readability and we present here the main outputs of our experiments.

{\subsubsection*{Effort in virologic detection}}
{We first quantify on Figure \ref{fig:Sensi_lambda1B5} the impact of introducing  additional virologic detection resources with fixed success rate $\lambda^1$. Namely, we consider a constant daily detection level $\lambda^1$ of $1$\textperthousand, $1\%$, $5\%$ or $10\%$.} 
{Of course, the more effort we put into detecting infected people, the better the output. For instance we observe a smaller death toll, a weaker economic impact as well as less ICU saturation. Up to remaining numerical errors, moving the detection capacity $\lambda^1$ from $1\%$ to $10\%$ has a strong impact on the overall outcomes of the epidemic: The epidemic size drops from $70\%$ to $30\%$, the remaining number of susceptible individuals $S_T\simeq 70\%$ corresponds to the herd immunity obtained with $\lambda^1=0.1$ and $\delta\simeq 0.25$, the duration of ICU capacity saturation is divided by a factor larger than $2$, and the overall sanitary and economic costs are divided by $2$. The huge detection of infected individuals has a strong impact on the socio-economic cost for two reasons: First, detection of infected individuals allows to slow down the epidemic propagation. Second, detected infected individuals are later on identified as immune people, i.e. are in compartment $R^+$, so that they are not impacted by social relations limitation, which reduces the overall economic cost. Finally, observe that massive testing has a major impact on most output (and must be {\em massive} to really impact, as suggested in \cite{Petom1163}). Observe that $S_T$ remains rather large, but hopefully any  epidemic restart is avoided by the detection effort $\lambda^1$ which must not be released (the patterns of some figures close to terminal date $T$ are due to the numerical approximation of the infinite horizon problem by a finite one). }

 {Another experiment presented in Figure \ref{fig:Sensi_lambda1B5_linear}  considers the case where the detection resources are rising linearly from $0\%$ to $20\%$ aver $700$ days. In comparison to the corresponding case where detection resources of level $\lambda^1=10\%$ are constantly available, the sanitary death toll induced by a rising $\lambda^1$ is way higher. This confirms the intuition that intense detection resources are required from the beginning of the epidemic. More specifically, we now focus  on the optimal dispatch over time of allocation resources: Figure \ref{fig:Sensi_lambda1B6} provides the optimal policy for the detection of infected  whenever no lockdown measures are in place, while Figure \ref{fig:Sensi_lambda1B7} provides the optimal combination of detection policy and lockdown intervention.}
For both cases, the weight $w_{\text{prevalence}}$ has been set up equal to $1$. 
 {As we can observe on both Figure \ref{fig:Sensi_lambda1B6} and Figure \ref{fig:Sensi_lambda1B7}, and as stated clearly in \cite{Petom1163}, massive testing and detection strategies could have been used to end the epidemic rapidly. The optimal strategy consists in having extremely massive and early type-1 detection resources, in order to isolate infected individuals and break the contamination chains. 
 {Herd immunity is ensured while we keep the detection active so that $\mathfrak{R}_t$ remains below 1, allowing to epidemic size to remain very small, while everyone simply waits for for a vaccination cure. }

 With that strategy, the peak on ICU is 25\% of the upper limit $U_{\max}$, $D_T$ is almost null, with no real impact on the economy (there is still an impact since, in order to be safe, we still quarantine a significant proportion of people, above 10\%). Note here that the optimal $\lambda_1$, once we passed the peak, exceeds 25\%, which is higher than the scenarios considered earlier. Observe also that  massive detection tools induce a constant  optimal lockdown level.}\\

{\subsubsection*{Effort in immunity  detection}}

{We now turn to the impact of detecting immune individuals, as  found by the optimally controlled epidemic dynamics. First, we consider on Figure \ref{fig:Sensi_lambda2} the introduction of fixed type-2 detection strategy (or testing) and measure the impact on the optimal lockdown policy. Such detection resources allow to transfer individuals from compartment $R^-$ to $R^+$. These additional resources have no significant impact on the optimal lockdown strategy and death toll dynamics of the epidemic. The only impact concerns the socio-economic cost of the pandemic, since identified recovered (and immune) individuals will be allowed to have a regular level of social interactions and won't be quarantined.}

{On Figure \ref{fig:Optim_delta_lambda2}, we look towards the optimal detection strategy ( with $ w_{\text{immunity}}=1$) whenever no lockdown policy is in place, while Figure \ref{fig:Optim_delta_lambda2B10} focuses on the optimization on both lockdown intervention $\delta$ and immunity detection effort $\lambda^2$. As expected, additional optimization over $\lambda^2$ has few impact on the sanitary burden, but can be used to lower the economic cost. The optimal detection strategy consists in identifying recovered individuals as soon as possible, and requires a large amount of effort in the first part of the epidemic phase. As a consequence, in the presence of immunity detection policy, the level of social interactions during the last phase of the epidemic is significantly higher, thus reducing the induced economic cost.}

{Finally, on Figure \ref{fig:Optim_delta_lambda2B11}, we  optimize over all possible control levers: lockdown measure, virologic detection effort $\lambda^1$ as well as the immunity detection one $\lambda^2$. We observe an optimal policy and induced epidemic dynamics rather similar to the one obtained in Figure \ref{fig:Sensi_lambda1B7}, whenever no immunity detection mean was available. The optimal $\lambda^2$ helps in reducing the overall economic costs by detecting immune people.}

\section{Conclusion}

The present work investigates, using an approach from Optimal Control theory, what could be optimal interventions for limiting the spread of a disease like the COVID-19 epidemic. We consider first the case where only the lockdown lever is operated, as was done in France when massive testing was still not an option. Then, in the other scenarios, optimization is done on both lockdown and {detection capacity of infected and immune individuals}.\\
The notion of `best scenario' is of course very much impacted by the choice of objective function that is made. The optimization takes into account the trade-off between lowering the number of deaths and minimizing the economic and social costs. To our knowledge, taking into consideration these diverse aspects into the objective function has not been studied much in the recent literature (see e.g. \cite{acemogluchernozhukovwerningwhinston}){, although the literature on the topic is growing rapidly.}\\

Another originality of our approach is the consideration of a strict constraint {on the ICU capacity level dedicated to COVID-19 patients.} We observe that in the solutions, the lockdown and detection controls are set so that the occupation of ICU remains a high as possible under $U_{\max}$ (due to the economic and social costs). It is really this constraint that shapes the decisions leading to a curve that is `\emph{flattened}' sufficiently to ensure the sustainability of the health system. This is particularly visible in the scenario where $U_{\max}$ is increased of 50\% within 2 months at the beginning of the epidemic: we see a relaxation of the lockdown that is permitted by the increased capacity of ICUs. \\

The best scenarios that we obtain are structured into 4 phases: 1) quick and strong measures to recover the control of the epidemic, 2) relaxation of the epidemic once under control to reach the fluxes imposed by the sustainability of the health care system, 3) the fluxes obtained after the period 2 are kept as long as possible to flatten the curve in order to avoid overwhelming the ICUs, 4) once a herd immunity is reached, the lockdown and testing controls can be lowered.\\
Without being an explicit target in the control problem, the evolution of the effective reproduction number $\mathfrak{R}_t$ is extremely interesting, and consistent with existing epidemiological literature. Again, without monitoring that indicator, we obtain in all the optimal strategies, $\mathfrak{R}_t$ becomes controlled below 1 after a few days (after the phase 1, and at the price of an important initial lockdown), and remains just below that critical value\footnote{On some graphs, we have a long term value of $\mathfrak{R}_t$ above 1, but it is obtained whenever close to the terminal date $T$ and due to the numerical approximation of the infinite horizon control problem by a finite horizon one.}. It is allowed to reach and remain stable at level $1$ in order to stabilize the ICU occupancy at their sustainable level $U_{\max}$. \\

Of course, the model considered here is schematic because we wanted : 1) to put the methodology in evidence, 2) to explore easily the different scenarios. Generalizations can be carried in a quite straightforward manner.
First, the major drawback of our approach is the necessity to fully observe some underlying latent variables and parameters of the epidemic dynamics, such as the prevalence of the virus. But we observed that the shape of the optimal control is quite robust to a variation of the initial prevalence $I_0^-$, and more generally to variations upon lots of parameters in the model. 
An important issue about that COVID-19 pandemic was the poor quality of data in the beginning, especially when trying to compare among countries to get robust estimators of various quantities regarding the disease. In our model, we tuned parameters using values from existing literature, and most of the results we obtained were actually extremely robust to moderate changes of those values.\\

Second, most individuals will not experience severe symptoms whenever contaminated by the virus. Such analysis calls for the use of differential quarantine strategies based on an estimation of the own risk of each individual. In our approach we use immunity detection  procedure for a similar purpose, but less accurate estimates of the risk of each individuals are also available depending on their current and past health characteristics. As discussed in \cite{evgenioufekombovchinnikovporcherpoucholvayatis}, the key ingredient for considering and using a differential lockdown strategy is a proper handling of the risk of bad specification of the  group associated to some individuals. A dynamic optimisation over both the dynamic lockdown intervention together with the risk level associated to each misspecification is left for further research. Similarly the risk of wrong identification of infected or immune  individuals using parameters $\lambda^1$ and $\lambda^2$ 
should be also taken into account in the design of a proper and efficient detection strategy. Also, in view of more practical applications for health policy guidance, a vector based version of \nommodel should probably be derived, as in \cite{acemogluchernozhukovwerningwhinston} for instance. Categories can be splited depending on the age of individuals. Hence, $S_t$ will then be $(S^{\text{c}}_t,S^{\text{a}}_t,S^{\text{s}}_t)$, with children, adults and seniors, for instance.
Most of the equations of the dynamics become multivariate, and parameters will be vectors, except $\beta$ that becomes a WAIFW ({\em or Who
Acquires Infection From Whom}) matrix, as in \cite{Wallinga2006} or \cite{Mossong2008}. The main interest would be to have target and age-specific controls, with testing and lockdown that can be per age group. The difficult component in this multivariate extension is on the objective function, where the economic component should probably be more related to well-being cost of being quarantined, especially for people who do not belong to labor groups. \\

{Finally let us notice that in this paper, as always in epidemic modelling, an important feature is the $-/+$ problem, with undetected individuals. The huge impact of  infected detection resources $\lambda^1$ in our numerical experiments emphasize the critical role of tracing and testing resources upon the dynamic of the epidemic. 
As suggested in \cite{eameskeeling-CT,arazozaclemencontran,housekeeling,kissgreenkao} (among many others), tracing is clearly a natural tool that should be optimized, to lower the socio-economic cost of the disease without endangering the entire population (and increase the sanitary cost of the pandemic). This paves the way for further research topics.}

\bibliographystyle{chicago}
\bibliography{bibliography}

\begin{thebibliography}{}

\bibitem[\protect\citeauthoryear{Abakuks}{Abakuks}{1973}]{abakuks1973}
Abakuks, A. (1973).
\newblock An optimal isolation policy for an epidemic.
\newblock {\em Journal of Applied Probability\/}~{\em 10\/}(2), 247–262.

\bibitem[\protect\citeauthoryear{Abakuks}{Abakuks}{1974}]{Abakuks74}
Abakuks, A. (1974).
\newblock Optimal immunisation policies for epidemics.
\newblock {\em Advances in Applied Probability\/}~{\em 6\/}(3), 494--511.

\bibitem[\protect\citeauthoryear{Acemoglu, Chernozhukov, Werning, and
  Whinston}{Acemoglu et~al.}{2020}]{acemogluchernozhukovwerningwhinston}
Acemoglu, D., V.~Chernozhukov, I.~Werning, and M.~Whinston (2020).
\newblock A multi-risk sir model with optimally targeted lockdown.
\newblock {\em NBER Working Paper\/}~{\em 27102}, 1--38.
\newblock JEL No. I18.

\bibitem[\protect\citeauthoryear{Agusto and Adekunle}{Agusto and
  Adekunle}{2014}]{Agusto}
Agusto, F. and A.~Adekunle (2014).
\newblock Optimal control of a two-strain tuberculosis-hiv/aids co-infection
  model.
\newblock {\em Biosystems\/}~{\em 119}, 20 -- 44.

\bibitem[\protect\citeauthoryear{Al-Tawfiq}{Al-Tawfiq}{2020}]{Al-Tawfiq}
Al-Tawfiq, J.~A. (2020).
\newblock Asymptomatic coronavirus infection: Mers-cov and sars-cov-2
  (covid-19).
\newblock {\em Travel Medicine and Infectious Disease\/}, 101608.

\bibitem[\protect\citeauthoryear{Aldridge, Lewer, Beale, Johnson, Zambon,
  Hayward, Fragaszy, and null}{Aldridge et~al.}{2020}]{Aldridge}
Aldridge, R., D.~Lewer, S.~Beale, A.~Johnson, M.~Zambon, A.~Hayward,
  E.~Fragaszy, and n.~null (2020).
\newblock Seasonality and immunity to laboratory-confirmed seasonal
  coronaviruses (hcov-nl63, hcov-oc43, and hcov-229e): results from the flu
  watch cohort study.
\newblock {\em Wellcome Open Research\/}~{\em 5\/}(52).

\bibitem[\protect\citeauthoryear{Alvarez, Argente, and Lippe}{Alvarez
  et~al.}{2020}]{alvarezargentelippi}
Alvarez, F., D.~Argente, and F.~Lippe (2020).
\newblock A simple planning problem for {COVID-19} lockdown.
\newblock {\em the National Bureau of Economic Research\/}~(26981), 1--35.

\bibitem[\protect\citeauthoryear{Anderson and May}{Anderson and
  May}{1991}]{AndersonMay}
Anderson, R. and R.~May (1991).
\newblock {\em Infectious Diseases of Humans: dynamics and Control}.
\newblock Oxford: Oxford University Press.

\bibitem[\protect\citeauthoryear{Ball, Britton, Lar\'edo, Pardoux, Sirl, and
  Tran}{Ball et~al.}{2019}]{ballbrittonlaredopardouxsirltran}
Ball, F., T.~Britton, C.~Lar\'edo, E.~Pardoux, D.~Sirl, and V.~Tran (2019).
\newblock {\em Stochastic Epidemic Models with Inference\/} (T. Britton and E.
  Pardoux ed.).
\newblock Lecture Notes in Mathematics. Springer.

\bibitem[\protect\citeauthoryear{Barclay}{Barclay}{2020}]{VOX}
Barclay, E. (2020).
\newblock The us doesn't just need to flatten the curve. it needs to "raise the
  line".
\newblock {\em
  https://www.vox.com/2020/4/7/21201260/coronavirus-usa-chart-mask-shortage-ventilators-flatten-the-curve\/}.
\newblock 2020-04-07.

\bibitem[\protect\citeauthoryear{Behncke}{Behncke}{2000}]{Behncke2000}
Behncke, H. (2000).
\newblock Optimal control of deterministic epidemics.
\newblock {\em Optimal Control Applications and Methods\/}~{\em 21}, 269 --
  285.

\bibitem[\protect\citeauthoryear{Berger, Herkenhoff, and Mongey}{Berger
  et~al.}{2020}]{NBERw26901}
Berger, D.~W., K.~F. Herkenhoff, and S.~Mongey (2020).
\newblock An seir infectious disease model with testing and conditional
  quarantine.
\newblock {\em National Bureau of Economic Research\/}~{\em 26901}.

\bibitem[\protect\citeauthoryear{Bernstein, Richter, and
  Throckmorton}{Bernstein et~al.}{2020}]{Bernstein}
Bernstein, J., A.~W. Richter, and N.~Throckmorton (2020).
\newblock Covid-19: A view from the labor market.
\newblock {\em Federal Reserve Bank of Dallas Working Paper\/}~{\em 2010}.

\bibitem[\protect\citeauthoryear{Bobisud}{Bobisud}{1977}]{BOBISUD1977165}
Bobisud, L. (1977).
\newblock Optimal control of a deterministic epidemic.
\newblock {\em Mathematical Biosciences\/}~{\em 35\/}(1), 165 -- 174.

\bibitem[\protect\citeauthoryear{Chinazzi, Davis, Ajelli, Gioannini, Litvinova,
  Merler, Pastore~y Piontti, Mu, Rossi, Sun, Viboud, Xiong, Yu, Halloran,
  Longini, and Vespignani}{Chinazzi et~al.}{2020}]{Chinazzieaba9757}
Chinazzi, M., J.~T. Davis, M.~Ajelli, C.~Gioannini, M.~Litvinova, S.~Merler,
  A.~Pastore~y Piontti, K.~Mu, L.~Rossi, K.~Sun, C.~Viboud, X.~Xiong, H.~Yu,
  M.~E. Halloran, I.~M. Longini, and A.~Vespignani (2020).
\newblock The effect of travel restrictions on the spread of the 2019 novel
  coronavirus (covid-19) outbreak.
\newblock {\em Science\/}.

\bibitem[\protect\citeauthoryear{Choe, Perera, Park, Song, Bang, Kim, and
  Oh}{Choe et~al.}{2017}]{choe}
Choe, P., R.~Perera, W.~Park, K.~Song, J.~Bang, E.~Kim, and M.~Oh (2017).
\newblock Mers-cov antibody responses 1 year after symptom onset, south korea,
  2015.
\newblock {\em Emerging Infectious Diseases\/}~{\em 23\/}(7), 1079--1084.

\bibitem[\protect\citeauthoryear{Cl\'{e}men\c{c}on, Tran, and
  Arazoza}{Cl\'{e}men\c{c}on et~al.}{2008}]{arazozaclemencontran}
Cl\'{e}men\c{c}on, S., V.~Tran, and H.~D. Arazoza (2008).
\newblock A stochastic {S}{I}{R} model with contact-tracing: large population
  limits and statistical inference.
\newblock {\em Journal of Biological Dynamics\/}~{\em 2\/}(4), 391--414.

\bibitem[\protect\citeauthoryear{Cobb and Douglas}{Cobb and
  Douglas}{1928}]{CobbDouglas}
Cobb, C.~W. and P.~H. Douglas (1928).
\newblock A theory of production.
\newblock {\em The American Economic Review\/}~{\em 18\/}(1), 139--165.

\bibitem[\protect\citeauthoryear{Cohen}{Cohen}{2020}]{Cohen14}
Cohen, J. (2020).
\newblock Vaccine designers take first shots at covid-19.
\newblock {\em Science\/}~{\em 368\/}(6486), 14--16.

\bibitem[\protect\citeauthoryear{Cohen and Kupferschmidt}{Cohen and
  Kupferschmidt}{2020}]{Cohen1287}
Cohen, J. and K.~Kupferschmidt (2020).
\newblock Countries test tactics in {\textquoteleft}war{\textquoteright}
  against covid-19.
\newblock {\em Science\/}~{\em 367\/}(6484), 1287--1288.

\bibitem[\protect\citeauthoryear{da~Cruz, Cardoso, and Takahashi}{da~Cruz
  et~al.}{2011}]{daCruz2011}
da~Cruz, A.~R., R.~T.~N. Cardoso, and R.~H.~C. Takahashi (2011).
\newblock Multiobjective dynamic optimization of vaccination campaigns using
  convex quadratic approximation local search.
\newblock In R.~H.~C. Takahashi, K.~Deb, E.~F. Wanner, and S.~Greco (Eds.),
  {\em Evolutionary Multi-Criterion Optimization}, Berlin, Heidelberg, pp.\
  404--417. Springer Berlin Heidelberg.

\bibitem[\protect\citeauthoryear{Day}{Day}{2020}]{Daym1165}
Day, M. (2020).
\newblock Covid-19: identifying and isolating asymptomatic people helped
  eliminate virus in italian village.
\newblock {\em British Medical Journal\/}~{\em 368}.

\bibitem[\protect\citeauthoryear{Diekmann, Heesterbeek, and Britton}{Diekmann
  et~al.}{2012}]{diekmannheesterbeekbritton-book}
Diekmann, O., H.~Heesterbeek, and T.~Britton (2012).
\newblock {\em Mathematical Tools for Understanding Infectious Disease
  Dynamics}.
\newblock Princeton Series in Theoretical and Computational Biology. New
  Jersey: Princeton University Press.

\bibitem[\protect\citeauthoryear{Diekmann, Heesterbeek, and Metz}{Diekmann
  et~al.}{1990}]{diekmannheesterbeekmetz}
Diekmann, O., J.~Heesterbeek, and J.~Metz (1990).
\newblock On the definition and the computation of the basic reproduction ratio
  $r_0$ in models for infectious diseases in heterogeneous populations.
\newblock {\em Journal of Mathematical Biology\/}~{\em 28}, 365--382.

\bibitem[\protect\citeauthoryear{Djidjou-Demasse, Michalakis, Choisy, Sofonea,
  and Alizon}{Djidjou-Demasse et~al.}{2020}]{Djidjou-Demasse2020}
Djidjou-Demasse, R., Y.~Michalakis, M.~Choisy, M.~T. Sofonea, and S.~Alizon
  (2020).
\newblock Optimal covid-19 epidemic control until vaccine deployment.
\newblock {\em medRxiv\/}~{\em 2020.04.02.20049189}.

\bibitem[\protect\citeauthoryear{Domenico, Pullano, Sabbatini, Bo\"elle, and
  Colizza}{Domenico et~al.}{2020}]{didomenicopullanosabbatiniboellecolizza}
Domenico, L.~D., G.~Pullano, C.~Sabbatini, P.-Y. Bo\"elle, and V.~Colizza
  (2020).
\newblock Expected impact of lockdown in {I}le-de-france and possible exit
  strategies.
\newblock {\em medrxiv\/}~{\em 2020.04.13.20063933}.

\bibitem[\protect\citeauthoryear{Eames and Keeling}{Eames and
  Keeling}{2003}]{eameskeeling-CT}
Eames, K. and M.~Keeling (2003).
\newblock Contact tracing and disease control.
\newblock {\em Proceedings of the Royal Society London B\/}~{\em 270},
  2565--2571.

\bibitem[\protect\citeauthoryear{Eichenbaum, Rebelo, and Trabandt}{Eichenbaum
  et~al.}{2020}]{Eichenbaum}
Eichenbaum, M.~S., S.~Rebelo, and M.~Trabandt (2020).
\newblock The macroeconomics of epidemics.
\newblock Working Paper 26882, National Bureau of Economic Research.

\bibitem[\protect\citeauthoryear{Elie}{Elie}{2008}]{elie2008finite}
Elie, R. (2008).
\newblock Finite time merton strategy under drawdown constraint: a viscosity
  solution approach.
\newblock {\em Applied Mathematics and Optimization\/}~{\em 58\/}(3), 411--431.

\bibitem[\protect\citeauthoryear{Elie, Hubert, and Turinici}{Elie
  et~al.}{2020}]{elie2020contact}
Elie, R., E.~Hubert, and G.~Turinici (2020).
\newblock Contact rate epidemic control of covid-19: an equilibrium view.
\newblock {\em arXiv preprint arXiv:2004.08221\/}.

\bibitem[\protect\citeauthoryear{Evgeniou, Fekom, Ovchinnikov, Porcher,
  Pouchol, and Vayatis}{Evgeniou
  et~al.}{2020}]{evgenioufekombovchinnikovporcherpoucholvayatis}
Evgeniou, T., M.~Fekom, A.~Ovchinnikov, R.~Porcher, C.~Pouchol, and N.~Vayatis
  (2020).
\newblock Epidemic models for personalised {COVID-19} isolation and exit
  policies using clinical risk predictions.
\newblock {\em medRxiv\/}~{\em 2020.04.21.20074054}.

\bibitem[\protect\citeauthoryear{Ferguson, Laydon, Nedjati-Gilani, Imai,
  Ainslie, Baguelin, Bhatia, Boonyasiri, Cucunubá, Cuomo-Dannenburg, Dighe,
  Dorigatti, Fu, Gaythorpe, Green, Hamlet, Hinsley, Okell, van Elsland, and
  Ghani}{Ferguson et~al.}{2020}]{Imperial}
Ferguson, N., D.~Laydon, G.~Nedjati-Gilani, N.~Imai, K.~Ainslie, M.~Baguelin,
  S.~Bhatia, A.~Boonyasiri, Z.~Cucunubá, G.~Cuomo-Dannenburg, A.~Dighe,
  I.~Dorigatti, H.~Fu, K.~Gaythorpe, W.~Green, A.~Hamlet, W.~Hinsley, L.~C.
  Okell, S.~van Elsland, and A.~C. Ghani (2020).
\newblock Impact of non-pharmaceutical interventions (npis) to reduce covid-19
  mortality and healthcare demand.
\newblock {\em Imperial College COVID-19 Response Team\/}~{\em 9}.

\bibitem[\protect\citeauthoryear{Flaxman, Mishra, Gandy, Unwin, Coupland,
  Thomas A~Mellan, Berah, Ghani, Donnelly, Riley, Okell, Vollmer, Ferguson, and
  Bhatt}{Flaxman et~al.}{2020}]{report13}
Flaxman, S., S.~Mishra, A.~Gandy, J.~Unwin, H.~Coupland, a.~Z. Thomas A~Mellan,
  T.~Berah, A.~Ghani, C.~A. Donnelly, S.~Riley, L.~C. Okell, M.~A.~C. Vollmer,
  N.~M. Ferguson, and S.~Bhatt (2020).
\newblock Estimating the number of infections and the impact of
  non-pharmaceutical interventions on covid-19 in 11 european countries.
\newblock {\em Imperial College COVID-19 Response Team\/}~{\em 13}.

\bibitem[\protect\citeauthoryear{Gelardi}{Gelardi}{2020}]{NATION}
Gelardi, C. (2020).
\newblock Colonialism made puerto rico vulnerable to coronavirus catastrophe.
\newblock {\em
  https://www.thenation.com/article/politics/puerto-rico-coronavirus/\/}.
\newblock 2020-04-09.

\bibitem[\protect\citeauthoryear{Gostic, Gomez, Mummah, Kucharski, and
  Lloyd-Smith}{Gostic et~al.}{2020}]{Gostic}
Gostic, K., A.~C. Gomez, R.~O. Mummah, A.~J. Kucharski, and J.~O. Lloyd-Smith
  (2020).
\newblock Estimated effectiveness of symptom and risk screening to prevent the
  spread of covid-19.
\newblock {\em eLife\/}~{\em 9}, e55570.

\bibitem[\protect\citeauthoryear{Greenhalg}{Greenhalg}{1988}]{greenhalgh}
Greenhalg, D. (1988).
\newblock Some results on optimal control applied to epidemics.
\newblock {\em Mathematical Biosciences\/}~{\em 88}, 125--158.

\bibitem[\protect\citeauthoryear{Gudi, Undela, Venkataraman, Mateti, Chhabra,
  Nyamagoud, and Tiwari}{Gudi et~al.}{2020}]{Gudi2020}
Gudi, S.~K., K.~Undela, R.~Venkataraman, U.~V. Mateti, M.~Chhabra,
  S.~Nyamagoud, and K.~K. Tiwari (2020).
\newblock Knowledge and beliefs towards universal safety precautions to flatten
  the curve during novel coronavirus disease (ncovid-19) pandemic among general
  public in india: Explorations from a national perspective.
\newblock {\em medRxiv\/}~{\em 2020.03.31.20047126}.

\bibitem[\protect\citeauthoryear{Guerrieri, Lorenzoni, Straub, and
  Werning}{Guerrieri et~al.}{2020}]{Guerrieri}
Guerrieri, V., G.~Lorenzoni, L.~Straub, and I.~Werning (2020).
\newblock Macroeconomic implications of covid-19: Can negative supply shocks
  cause demand shortages?
\newblock Working Paper 26918, National Bureau of Economic Research.

\bibitem[\protect\citeauthoryear{Hansen and Day}{Hansen and Day}{2011}]{Hansen}
Hansen, E. and T.~Day (2011).
\newblock Optimal control of epidemics with limited resources.
\newblock {\em Journal of Mathematical Biology\/}~{\em 62\/}(3), 423--451.

\bibitem[\protect\citeauthoryear{He, Lau, Wu, Deng, Wang, and Hao}{He
  et~al.}{2020}]{heetal}
He, X., E.~Lau, P.~Wu, X.~Deng, J.~Wang, and X.~Hao (2020).
\newblock Temporal dynamics in viral shedding and transmissibility of covid-19.
\newblock {\em Nature Medicine\/}.

\bibitem[\protect\citeauthoryear{Hellewell, Abbott, Gimma, Bosse, Jarvis,
  Russell, Munday, Kucharski, Edmunds, Group, Funk, and Eggo}{Hellewell
  et~al.}{2020}]{hellewellabbottgimmabossejarvisrussellmundaykucharskiedmundsfunkeggo}
Hellewell, J., S.~Abbott, A.~Gimma, N.~Bosse, C.~Jarvis, T.~Russell, J.~Munday,
  A.~Kucharski, J.~Edmunds, C.~C.-.~W. Group, S.~Funk, and R.~Eggo (2020).
\newblock Feasibility of controlling covid-19 outbreaks by isolation of cases
  and contacts.
\newblock {\em Lancet Global Health\/}~{\em 8}, 488--496.

\bibitem[\protect\citeauthoryear{Hellewell, Abbott, Gimma, Bosse, Jarvis,
  Russell, Munday, Kucharski, Edmunds, Sun, Flasche, Quilty, Davies, Liu,
  Clifford, Klepac, Jit, Diamond, Gibbs, [van Zandvoort], Funk, and
  Eggo}{Hellewell et~al.}{2020}]{Hellewell}
Hellewell, J., S.~Abbott, A.~Gimma, N.~I. Bosse, C.~I. Jarvis, T.~W. Russell,
  J.~D. Munday, A.~J. Kucharski, W.~J. Edmunds, F.~Sun, S.~Flasche, B.~J.
  Quilty, N.~Davies, Y.~Liu, S.~Clifford, P.~Klepac, M.~Jit, C.~Diamond,
  H.~Gibbs, K.~[van Zandvoort], S.~Funk, and R.~M. Eggo (2020).
\newblock Feasibility of controlling covid-19 outbreaks by isolation of cases
  and contacts.
\newblock {\em The Lancet Global Health\/}~{\em 8\/}(4), e488 -- e496.

\bibitem[\protect\citeauthoryear{House and Keeling}{House and
  Keeling}{2010}]{housekeeling}
House, T. and M.~Keeling (2010).
\newblock The impact of contact tracing in clustered populations.
\newblock {\em PLoS Computational Biology\/}~{\em 6\/}(3), e1000721.

\bibitem[\protect\citeauthoryear{Huang, Shi, Gong, Jiang, Liu, Yang, Tang, You,
  Jiang, Long, Zeng, Luo, Zeng, Zeng, Wang, Yang, and Yang}{Huang
  et~al.}{2020}]{Huang2020}
Huang, L., Y.~Shi, B.~Gong, L.~Jiang, X.~Liu, J.~Yang, J.~Tang, C.~You,
  Q.~Jiang, B.~Long, T.~Zeng, M.~Luo, F.~Zeng, F.~Zeng, S.~Wang, X.~Yang, and
  Z.~Yang (2020).
\newblock Blood single cell immune profiling reveals the interferon-mapk
  pathway mediated adaptive immune response for covid-19.
\newblock {\em medRxiv\/}~{\em 2020.03.15.20033472}.

\bibitem[\protect\citeauthoryear{Iacoviello and Liuzzi}{Iacoviello and
  Liuzzi}{2008}]{Iacoviello2008}
Iacoviello, D. and G.~Liuzzi (2008).
\newblock Optimal control for sir epidemic model: A two treatments strategy.
\newblock {\em 2008 Mediterranean Conference on Control and Automation -
  Conference Proceedings, MED'08\/}, 842--847.

\bibitem[\protect\citeauthoryear{Jiang, Deng, and Zhang}{Jiang
  et~al.}{2020}]{JiangVaccine}
Jiang, F., L.~Deng, and L.~Zhang (2020).
\newblock Review of the clinical characteristics of coronavirus disease 2019
  (covid-19).
\newblock {\em Journal of General Internal Medicine\/}~{\em 35}, 1545–1549.

\bibitem[\protect\citeauthoryear{Kermack and McKendrick}{Kermack and
  McKendrick}{1927}]{kermack27}
Kermack, W. and A.~McKendrick (1927).
\newblock A contribution to the mathematical theory of epidemics.
\newblock {\em Proceedings of the Royal Society of London\/}~{\em A 115},
  700--721.

\bibitem[\protect\citeauthoryear{Kim, Lee, and Jung}{Kim
  et~al.}{2017}]{KIM201774}
Kim, S., J.~Lee, and E.~Jung (2017).
\newblock Mathematical model of transmission dynamics and optimal control
  strategies for 2009 a/h1n1 influenza in the republic of korea.
\newblock {\em Journal of Theoretical Biology\/}~{\em 412}, 74 -- 85.

\bibitem[\protect\citeauthoryear{Kiss, Green, and Kao}{Kiss
  et~al.}{2013}]{kissgreenkao}
Kiss, I., D.~Green, and R.~Kao (2013).
\newblock Infectious disease control using contact tracing in random and
  scale-free networks.
\newblock {\em Journal of the Royal Society Interface\/}~{\em 3\/}(6), 55--62.

\bibitem[\protect\citeauthoryear{Kissler, Tedijanto, Lipsitch, and
  Grad}{Kissler et~al.}{2020}]{Kissler2020}
Kissler, S.~M., C.~Tedijanto, M.~Lipsitch, and Y.~Grad (2020).
\newblock Social distancing strategies for curbing the covid-19 epidemic.
\newblock {\em medRxiv\/}~{\em 2020.03.22.20041079}.

\bibitem[\protect\citeauthoryear{Ku, Ng, and Lin}{Ku et~al.}{2020}]{Ku2020}
Ku, C.~C., T.-C. Ng, and H.-H. Lin (2020).
\newblock Epidemiological benchmarks of the covid-19 outbreak control in china
  after wuhan’s lockdown: A modelling study with an empirical approach.
\newblock {\em SSRN Electronic Journal\/}~{\em 3543589}.

\bibitem[\protect\citeauthoryear{Kucharski, Klepac, Conlan, Kissler, Tang, Fry,
  Gog, Edmunds, and working group}{Kucharski
  et~al.}{2020}]{kucharskigogedmunds}
Kucharski, A., P.~Klepac, A.~Conlan, S.~Kissler, M.~Tang, H.~Fry, J.~Gog,
  J.~Edmunds, and C.~C.-. working group (2020).
\newblock Effectiveness of isolation, testing, contact tracing and physical
  distancing on reducing transmission of sars-cov-2 in different settings.
\newblock {\em medRxiv\/}~{\em 2020.04.23.20077024}.

\bibitem[\protect\citeauthoryear{Kucharski, Russell, Diamond, Liu, Edmunds,
  Funk, and Eggo}{Kucharski
  et~al.}{2020}]{kucharskirusselldiamondliuedmundsfunkeggo}
Kucharski, A., T.~Russell, C.~Diamond, Y.~Liu, J.~Edmunds, S.~Funk, and R.~Eggo
  (2020).
\newblock Early dynamics of transmission and control of covid-19: a
  mathematical modelling study.
\newblock {\em Lancet Infectious Disease\/}~{\em 20}, 553--58.

\bibitem[\protect\citeauthoryear{Kumar and Srivastava}{Kumar and
  Srivastava}{2017}]{KUMAR2017334}
Kumar, A. and P.~K. Srivastava (2017).
\newblock Vaccination and treatment as control interventions in an infectious
  disease model with their cost optimization.
\newblock {\em Communications in Nonlinear Science and Numerical
  Simulation\/}~{\em 44}, 334 -- 343.

\bibitem[\protect\citeauthoryear{Lagorio, Dickison, Vazquez, Braunstein, Macri,
  Migueles, Havlin, and Stanley}{Lagorio et~al.}{2011}]{Lagorio}
Lagorio, C., M.~Dickison, F.~Vazquez, L.~A. Braunstein, P.~A. Macri, M.~V.
  Migueles, S.~Havlin, and H.~E. Stanley (2011).
\newblock Quarantine-generated phase transition in epidemic spreading.
\newblock {\em Physical Review E\/}~{\em 83\/}(2), 026102.

\bibitem[\protect\citeauthoryear{Lai, Ruktanonchai, Zhou, Prosper, Luo, Floyd,
  Wesolowski, Santillana, Zhang, Du, Yu, and Tatem}{Lai et~al.}{2020}]{Lai2020}
Lai, S., N.~W. Ruktanonchai, L.~Zhou, O.~Prosper, W.~Luo, J.~R. Floyd,
  A.~Wesolowski, M.~Santillana, C.~Zhang, X.~Du, H.~Yu, and A.~J. Tatem (2020).
\newblock Effect of non-pharmaceutical interventions for containing the
  covid-19 outbreak in china.
\newblock {\em medRxiv\/}~{\em 2020.03.03.20029843}.

\bibitem[\protect\citeauthoryear{Liu, Zhang, Chen, Xiang, Song, Shu, Chen,
  Liang, Zhou, You, Wu, Zhang, Lu, Xia, Huang, Yang, Liu, Semple, Cowling, Lan,
  Sun, Yu, and Liu}{Liu et~al.}{2020}]{Weiyong}
Liu, W., Q.~Zhang, J.~Chen, R.~Xiang, H.~Song, S.~Shu, L.~Chen, L.~Liang,
  J.~Zhou, L.~You, P.~Wu, B.~Zhang, Y.~Lu, L.~Xia, L.~Huang, Y.~Yang, F.~Liu,
  M.~G. Semple, B.~J. Cowling, K.~Lan, Z.~Sun, H.~Yu, and Y.~Liu (2020).
\newblock Detection of covid-19 in children in early january 2020 in wuhan,
  china.
\newblock {\em New England Journal of Medicine\/}~{\em 382\/}(14), 1370--1371.

\bibitem[\protect\citeauthoryear{Magal and Webb}{Magal and
  Webb}{2020}]{magalwebb}
Magal, P. and G.~Webb (2020).
\newblock Predicting the number of reported and unreported cases for the
  {C}{O}{V}{I}{D}-{1}{9} epidemic in {S}outh {K}orea, {I}taly, {F}rance and
  {G}ermany.
\newblock medRxiv, doi:10.1101/2020.03.21.20040154.

\bibitem[\protect\citeauthoryear{Mossong, Hens, Jit, Beutels, Auranen, and
  Mikolajczyk}{Mossong et~al.}{2008}]{Mossong2008}
Mossong, J., N.~Hens, M.~Jit, P.~Beutels, K.~Auranen, and R.~Mikolajczyk
  (2008).
\newblock Social contacts and mixing patterns relevant to the spread of
  infectious diseases.
\newblock {\em PLoS Med\/}~{\em 5\/}(3), e74.

\bibitem[\protect\citeauthoryear{Nishiura, Kobayashi, Suzuki, S-Mok, Hayashi,
  Kinoshita, Yang, Yuan, Akhmetzhanov, Linton, and Miyama}{Nishiura
  et~al.}{2020}]{Nishiura}
Nishiura, H., T.~Kobayashi, A.~Suzuki, J.~S-Mok, K.~Hayashi, R.~Kinoshita,
  Y.~Yang, B.~Yuan, A.~Akhmetzhanov, N.~Linton, and T.~Miyama (2020).
\newblock Estimation of the asymptomatic ratio of novel coronavirus infections
  (covid-19).
\newblock {\em International Journal of Infectious Diseases\/}.

\bibitem[\protect\citeauthoryear{Oliver}{Oliver}{2013}]{Olivier2013}
Oliver, A. (2013).
\newblock A normative perspective on discounting health outcomes.
\newblock {\em Journal of Health Services Research \& Policy\/}~{\em 18\/}(3),
  186--189.

\bibitem[\protect\citeauthoryear{Ooi, Lim, and Chew}{Ooi
  et~al.}{2005}]{Peng2005}
Ooi, P.~L., S.~Lim, and S.~K. Chew (2005).
\newblock Use of quarantine in the control of sars in singapore.
\newblock {\em American Journal of Infection Control\/}~{\em 33\/}(5), 252 --
  257.

\bibitem[\protect\citeauthoryear{Pedersen and Meneghini}{Pedersen and
  Meneghini}{2020}]{Pedersen}
Pedersen, M.~G. and M.~Meneghini (2020).
\newblock A simple method to quantify country-specific effects of covid-19
  containment measures.
\newblock {\em medRxiv\/}~{\em 2020.04.07.20057075}.

\bibitem[\protect\citeauthoryear{Peto}{Peto}{2020}]{Petom1163}
Peto, J. (2020).
\newblock Covid-19 mass testing facilities could end the epidemic rapidly.
\newblock {\em BMJ\/}~{\em 368}.

\bibitem[\protect\citeauthoryear{Piguillem and Shi}{Piguillem and
  Shi}{2020}]{piguillemshi}
Piguillem, F. and L.~Shi (2020).
\newblock Optimal {COVID-19} quarantine and testing policies.
\newblock {\em EIEF Working Papers Series\/}~(2004).

\bibitem[\protect\citeauthoryear{Pontryagin, Boltyanskii, Gamkrelidze, and
  Mishchenko}{Pontryagin
  et~al.}{1964}]{pontryaginboltyanskiigamkrelidzemishchenko}
Pontryagin, L., G.~Boltyanskii, R.~Gamkrelidze, and E.~Mishchenko (1964).
\newblock {\em Mathematical Theory of Optimal Processes}.
\newblock New York.

\bibitem[\protect\citeauthoryear{Qualls, Levitt, and Kanade}{Qualls
  et~al.}{2017}]{CDC}
Qualls, N., A.~Levitt, and N.~Kanade (2017).
\newblock Community mitigation guidelines to prevent pandemic influenza —
  united states.
\newblock {\em Morbidity and Mortality Weekly Report\/}~{\em 66}, 1--34.

\bibitem[\protect\citeauthoryear{Ranney, Griffeth, and Jha}{Ranney
  et~al.}{2020}]{RanneyNEJM}
Ranney, M.~L., V.~Griffeth, and A.~K. Jha (2020).
\newblock Critical supply shortages — the need for ventilators and personal
  protective equipment during the covid-19 pandemic.
\newblock {\em New England Journal of Medicine\/}.

\bibitem[\protect\citeauthoryear{Roques, Klein, Papaix, Sar, and
  Soubeyrand}{Roques et~al.}{2020}]{roqueskleinpapaixsarsoubeyrand}
Roques, L., E.~Klein, J.~Papaix, A.~Sar, and S.~Soubeyrand (2020).
\newblock Effect of a one-month lockdown on the epidemic dynamics of {COVID-19}
  in {F}rance.
\newblock {\em medRxiv\/}~{\em 2020.04.21.20074054}.

\bibitem[\protect\citeauthoryear{Roux, Massonnaud, and Cr\'epey}{Roux
  et~al.}{2020}]{rouxmassonnaudcrepey}
Roux, J., C.~Massonnaud, and P.~Cr\'epey (2020).
\newblock C{O}{V}{I}{D}-19: One-month impact of the french lockdown on the
  epidemic burden.
\newblock
  https://www.ehesp.fr/wp-content/uploads/2020/04/Impact-Confinement-EHESP-20200322v1-1.pdf.

\bibitem[\protect\citeauthoryear{Sachdeva and Sheth}{Sachdeva and
  Sheth}{2020}]{Sachdeva}
Sachdeva, A. and A.~Sheth (2020).
\newblock Covid-19, panic now!! a call to action because the numbers are
  deceptive.
\newblock {\em SSRN\/}~{\em 3563419}.

\bibitem[\protect\citeauthoryear{Salath{\'e}, Althaus, Neher, Stringhini,
  Hodcroft, Fellay, Zwahlen, Senti, Battegay, Wilder-Smith, Eckerle, Egger, and
  Low}{Salath{\'e} et~al.}{2020}]{lshtm4656563}
Salath{\'e}, M., C.~L. Althaus, R.~Neher, S.~Stringhini, E.~Hodcroft,
  J.~Fellay, M.~Zwahlen, G.~Senti, M.~Battegay, A.~Wilder-Smith, I.~Eckerle,
  M.~Egger, and N.~Low (2020, March).
\newblock Covid-19 epidemic in switzerland: on the importance of testing,
  contact tracing and isolation.
\newblock {\em Swiss medical weekly\/}~{\em 150\/}(11-12), w20225--.
\newblock add\_research\_centre.php.

\bibitem[\protect\citeauthoryear{Salje, Kiem, Lefrancq, Courtejoie, Bosetti,
  Paireau, Andronico, Hoze, Richet, Dubost, Strat, Lessler, Bruhl, Fontanet,
  Opatowski, Boelle, and Cauchemez}{Salje et~al.}{2020}]{salje-cauchemez}
Salje, H., C.~T. Kiem, N.~Lefrancq, N.~Courtejoie, P.~Bosetti, J.~Paireau,
  A.~Andronico, N.~Hoze, J.~Richet, C.-L. Dubost, Y.~L. Strat, J.~Lessler,
  D.~L. Bruhl, A.~Fontanet, L.~Opatowski, P.-Y. Boelle, and S.~Cauchemez
  (2020).
\newblock Estimating the burden of sars-cov-2 in france.
\newblock https://hal-pasteur.archives-ouvertes.fr/pasteur-02548181.

\bibitem[\protect\citeauthoryear{Sethi}{Sethi}{1978}]{Sethi1978}
Sethi, S.~P. (1978).
\newblock Optimal quarantine programmes for controlling an epidemic spread.
\newblock {\em Journal of the Operational Research Society\/}~{\em 29\/}(3),
  265--268.

\bibitem[\protect\citeauthoryear{Sharomi and Malik}{Sharomi and
  Malik}{2017}]{Sharomi}
Sharomi, O. and T.~Malik (2017).
\newblock Optimal control in epidemiology.
\newblock {\em Annals of Operation Research\/}~{\em 251}, 55--71.

\bibitem[\protect\citeauthoryear{Tognotti}{Tognotti}{2013}]{quarantine}
Tognotti, E. (2013).
\newblock Lessons from the history of quarantine, from plague to influenza a.
\newblock {\em Emerging Infectious Diseases\/}~{\em 19\/}(2), 254--259.

\bibitem[\protect\citeauthoryear{Trapman, Ball, Dhersin, Tran, Wallinga, and
  Britton}{Trapman et~al.}{2016}]{trapmanballdhersintranwallingabritton}
Trapman, P., F.~Ball, J.-S. Dhersin, V.~Tran, J.~Wallinga, and T.~Britton
  (2016).
\newblock Inferring r0 in emerging epidemics-the effect of common population
  structure is small.
\newblock {\em Journal of the Royal Society Interface\/}~{\em 13}, 20160288.

\bibitem[\protect\citeauthoryear{van~der Pol and Cairns}{van~der Pol and
  Cairns}{2002}]{vanderPol2002}
van~der Pol, M. and J.~Cairns (2002).
\newblock A comparison of the discounted utility model and hyperbolic
  discounting models in the case of social and private intertemporal
  preferences for health.
\newblock {\em Journal of Economic Behavior \& Organization\/}~{\em 49\/}(1),
  79 -- 96.

\bibitem[\protect\citeauthoryear{{Verriest}, {Delmotte}, and
  {Egerstedt}}{{Verriest} et~al.}{2005}]{1470088}
{Verriest}, E., F.~{Delmotte}, and M.~{Egerstedt} (2005).
\newblock Control of epidemics by vaccination.
\newblock In {\em Proceedings of the 2005, American Control Conference, 2005.},
  pp.\  985--990 vol. 2.

\bibitem[\protect\citeauthoryear{Wallinga, Teunis, and Kretzschmar}{Wallinga
  et~al.}{2006}]{Wallinga2006}
Wallinga, J., P.~Teunis, and M.~Kretzschmar (2006).
\newblock {Using Data on Social Contacts to Estimate Age-specific Transmission
  Parameters for Respiratory-spread Infectious Agents}.
\newblock {\em American Journal of Epidemiology\/}~{\em 164\/}(10), 936--944.

\bibitem[\protect\citeauthoryear{Wong, Vaughan, Quilty-Harper, and
  Liverpool}{Wong et~al.}{2020}]{NEWSC2020}
Wong, S., A.~Vaughan, C.~Quilty-Harper, and L.~Liverpool (2020).
\newblock Covid-19 news: Us not involved in global who plan to tackle pandemic.
\newblock {\em New Scientist\/}~{\em 24 April}.

\bibitem[\protect\citeauthoryear{World Health~Organization}{World
  Health~Organization}{2020}]{WHO}
World Health~Organization, W. (2020).
\newblock Coronavirus disease 2019 (covid-19).
\newblock https://www.who.int/emergencies/diseases/novel-coronavirus-2019.

\bibitem[\protect\citeauthoryear{Yan and Zou}{Yan and Zou}{2008}]{Xiefei}
Yan, X. and Y.~Zou (2008).
\newblock Optimal and sub-optimal quarantine and isolation control in sars
  epidemics.
\newblock {\em Mathematical and Computer Modelling\/}~{\em 47\/}(1–2),
  235–245.

\bibitem[\protect\citeauthoryear{Zhou, Li, Li, and Zhang}{Zhou
  et~al.}{2020}]{Zhou}
Zhou, X., Y.~Li, T.~Li, and W.~Zhang (2020).
\newblock Follow-up of the asymptomatic patients with sars-cov-2 infection.
\newblock {\em Clinical Microbiology and Infection\/}.

\end{thebibliography}

\clearpage

\appendix


\section{Parameters of the epidemic dynamics}\label{sec:parameters}

First, let us notice, as mentioned in the main text, that there is an underlying individual based model for the system \eqref{sys:SIR}, in which the rates have a probabilistic interpretation, and can be interpreted as inverse of mean times (e.g. \cite{ballbrittonlaredopardouxsirltran}), possibly with thinning when the corresponding events only occur for a fraction of the population.  
Based on these remarks, we can compute the rates in our model  with the parameters provided in the French COVID-19 literature (see  \cite{didomenicopullanosabbatiniboellecolizza,salje-cauchemez}). 

\paragraph{Obtaining the transition rates from the sojourn times and transition probabilities}

First, we introduce some variables on which the mechanistic construction of our rates will depend:
\begin{itemize}
    \item $p_a$: conditioned on being infected, the probability of having light symptoms or no symptoms;
    \item $p_{H}$: conditioned on being mild/severely ill, the probability of needing hospitalization;
    \item $p_{U}$: conditioned on needing hospitalization, the probability of needing ICU;
    \item $p_{d}$: conditioned on \emph{being admitted to ICU}, the probability of dying,
\end{itemize}
and for the durations:
\begin{itemize}
    \item $N^{(a)}_R$: if asymptomatic, number of days until recovery; 
    \item $N^{(s)}_R$: if symptomatic, number of days until recovery without hospital; 
    \item $N_H$: (if severe symptomatic) number of days until hospitalization;
    \item $N_{HU}$: if in H, number of days until ICU ;
    \item $N_{HR}$: if hospitalized but not in ICU, the number of days until recovery; 
    \item $N_{UD}=10$: if in ICU, number of days until death;
    \item $N_{UR}=20$: if in ICU, number of days until recovery.
\end{itemize}
These variables can be measured quite easily on the data. Then, once these parameters are chosen, we can propose the following transition rates for our model:

\begin{itemize}
    \item $\gamma_{IR}$, the daily individual transition rate from $I$ to $R$, is assumed to be of the form:
    $$\gamma_{IR} = (1-p_a) \cdot (1-p_{H}) \cdot \displaystyle\frac{1}{N^{(s)}_{R}} + p_a \cdot \displaystyle\frac{1}{N^{(a)}_{R}}, 
$$
where on the right hand side, the first term is associated to mild symptomatic while the second term is associated to severe ones. 
\item $\gamma_{IH}$, the daily rate from $I$ to $H$, is assumed to be
$$
    \gamma_{IH} = (1-p_a) \cdot p_{H} \cdot \displaystyle\frac{1}{N_{IH}}.
$$ 
\item $\gamma_{HU}$, the daily rate from $H$ to $U$, is given by
$$
\gamma_{HU} = p_{U} \cdot \displaystyle\frac{1}{N_{HU}} .
$$
\item $\gamma_{HR}$, the daily rate from $H$ to $R$, is given by
$$
        \gamma_{HR} = (1-p_{U}) \cdot \displaystyle\frac{1}{N_{HR}} .
$$
\item For the transition rates from $U$ to $R$ or $D$, the nonlinearity induced by the constraint on the ICU capacity $U_{\max}$ implies that it is easier to give these rates at the population level rather than at the individual level. For the transition from $U$ to $R$, the individual rate is denoted by $\gamma_{UR}(U) $ and the population rate is then $\gamma_{UR}(U) \cdot U$. We assume here that:

\begin{align*}
    \gamma_{UR}(U) \times U
    &=
    \begin{cases}
        (1-p_{d})\cdot\displaystyle\frac{1}{N_{UR}}U , & \hbox{ if } U < U_{\max},
        \\
        (1-p_{d})\cdot\displaystyle\frac{1}{N_{UR}}U_{\max}, & \hbox{ if } U \ge U_{\max}.
    \end{cases}
\end{align*}
This means that when the ICU capacity $U_{\max}$ is reached, new patients can not be taken in charge any more and since they correspond to individuals with severe forms of the disease, they can not recover until the ICU occupation is lowered. 
\item As for the transition rate from $U$ to $R$, the transition rate from $U$ to $D$ is nonlinear. We denote by $\gamma_{UD}(U)$ the daily death rate of an individual needing ICU, and by $\gamma_{UD}(U)\times U$ the transition at the population level, that we assume is of the form:
\begin{align}
    \gamma_{UD}(U)\times U
    &=
    \begin{cases}
        p_{d}\cdot\displaystyle\frac{1}{N_{UD}}U , & \hbox{ if } U < U_{\max}, 
        \\
        p_{d}\cdot\displaystyle\frac{1}{N_{UD}}U_{\max} +  \frac{20}{N_{UD}}(U-U_{\max}), & \hbox{ if } U \ge U_{\max}.
    \end{cases}\label{eq:gammaUR-annexe}
\end{align}
The hypothesis is that when ICU gets saturated, the patients who are not taken in charge experience an extra death rate. 
\end{itemize}

\paragraph{Parameter values}

For the infection parameter $\beta$, we fix the value of $\mathcal{R}_0$ and then compute the corresponding value of $\beta$ using \eqref{eq:R0}. \cite{salje-cauchemez} use $\mathcal{R}_0=3.3$ (in absence of lockdown intervention and of testing policy), which we also recover from the early French dynamics (\cite{WHO}) using the methods described in \cite{trapmanballdhersintranwallingabritton} (not shown). 

The other transition rates appearing on Fig. \ref{fig:graphSIDHR} can be obtained once the probabilities and durations detailed above are fixed. For this, we choose our parameters on the basis of the works by \cite{didomenicopullanosabbatiniboellecolizza,salje-cauchemez} for the French COVID-19 epidemics. We emphasize that despite of this choice, the methodology developed here is general and can apply to any country of disease.

\begin{table}[!ht]
\centering
{
\footnotesize
		\caption{Parameters. (a): parameters obtained from Di Domenico et al. \cite{didomenicopullanosabbatiniboellecolizza}. (b): parameters obtained from Salje et al. \cite{salje-cauchemez}. (c): parameters used in the benchmark scenario in this paper}  \label{table:parameters}
			\begin{tabular}{c|c|c|c|c|c|c|c|c|c|c|c}
				
				\hline
				
				& \multicolumn{4}{|c|}{Probabilities}
				&\multicolumn{7}{c}{Durations (days)} 
				\\
				
				\hline \hline
				
				 & $p_a$    & $p_{H}$ & $p_{U}$ & $p_{d}$ 
				     & $N^{(a)}_R$ & $N^{(s)}_R$ & $N_{IH}$ & $N_{HU}$ &  $N_{HR}$ &$N_{UD}$ & $N_{UR}$ 
				 \\
				
				\hline
				 
				 \hline
				 
				 (a) & 0.52 & 0.17 & 0.25 & 0.20 & 7.5 & 7.5 &  &  &  &  &  \\
				
				\hline
				(b) &  &  & 0.182 & 0.20 &  &  & 11 & 2 & 17.15 & 10 & 10.23 \\

				\hline
				(c) & 0.85 & 0.17   & 0.182 & 0.20 & 7.5 & 7.5  & 11 & 2 & 17.15 & 10 & 10.23 
				
				 \\
				
				\hline \hline
			\end{tabular}
		\label{tab:parameters}
}
\end{table}

The line (a) in the Table \ref{tab:parameters} can be obtained as follows: the model of \cite{didomenicopullanosabbatiniboellecolizza} for Ile-de-France is structured into 3 age-classes: children (25\%), adults (60\%) and seniors (15\%). Using the age-specific transition probabilities given in this reference and averaging over the age classes, we obtain the values announced:
\begin{align*}
    p_a= &  0.2 + 0.8 \big(0.25  \times 1 + 0.75 \times 0.2\big)=0.52\\
    p_H = & \frac{0.60 \times 0.10 + 0.15\times 0.20}{0.60 \times (0.70+0.10) + 0.15\times (0.6+ 0.20)}=0.17\\
    p_U= & 0.60 \times 0.36+ 0.15\times 0.2=0.246\\
    p_d= & \frac{0.60 \times 0.0074+ 0.15 \times 0.029}{060 \times (0.0074+0.05)+ 0.15 \times (0.029+0.036)}=0.20\\
\end{align*}

However, it appears in our simulations that the probability $p_a=0.52$ to be asymptomatic or only with light symptoms does not provide realistic orders for the proportion of dead individuals. In \cite{salje-cauchemez}, they estimate the proportion of being hospitalized when infected at 2.6\%. This proportion corresponds to $(1-p_a)\, p_H$ in our model, yielding a probability $p_a=0.85$ when we use the estimate $p_H=0.17$ of \cite{didomenicopullanosabbatiniboellecolizza}.\\

In France, at the beginning of the lockdown on March 15th 2020, there was roughly 12,000 beds in ICUs, which provides an estimate for the value of $U_{\max}$. 
The additional mortality rate for individuals in ICUs when $U_t>U_{\max}$ is chosen 20 times more than the usual death rate in \eqref{eq:gammaUR-annexe}. This factor is calibrated from the observation by \cite{salje-cauchemez} that the mean time to death in hospitals is bi-modal, with a group dying quickly after 0.67 days on average and a second group who die after longer time periods of mean 13.2 days. Extrapolating we assumed that `urgent' cases die 20 times faster than `usual cases'. \\

For the value of $I_0^-$, \cite[Table 1]{report13} report a total number of infected of 3\%, with a 95\% credible interval [1.1\%;7.8\%], as of 28th March 2020 (which is two week after our $t=0$). Because of the exponential increase, a crude estimate would be below $1\%$ two weeks before. The number of reported cases in France on March 15 was around 5,000. According to \cite{Sachdeva}, when Italy had 5,000 reported cases, an estimation of 60,000 cases was suggested, which would imply a $I_0^-$ close to $1$\textperthousand. 
In our benchmark scenario, we kept that $5$\textperthousand~ value for $I_0^-$.\\

To check the validity of our parameters, we simulated our model in absence of intervention ($\delta=0$, $\lambda^1\equiv 0$ and $\lambda^2 \equiv 0$) and compare the results with the prediction of \cite{rouxmassonnaudcrepey}. Because in the latter reference they do not model the ICU saturation, we also removed the threshold $U_{\max}$ and the associated additional mortality (which amounts to considering 
$U_{\max}=+\infty$). Doing this we obtain similar results as those of \cite{rouxmassonnaudcrepey}.

\newpage 


\section{Sensitivities to the model specifications}\label{appendix:b}

In this appendix, we provide sensitivity graphs, where parameters are changed, to assess the robustness of our conclusions. Do not forget that numerical approximations errors can affect a too precise interpretation of the following numerical illustrations. 

{We consider the following benchmark scenario. The parameters for the dynamics are specified in Table~\ref{tab:parameters-corps}. In the control problem, the benchmark situation is the one described in Section \ref{subsec:optimal:policy}. To be precise, for the numerical implementation, we use the cost function is the one defined in~\eqref{eq_Global_cost_penalizationICU} with the following choice of parameters: $\alpha = 0$, $w_{\text{sanitary}} = 10^5$,  
$w_{\text{econ}} = 1$,
$w_{\text{prevalence}} = 1$,
$w_{\text{immunity}} = 1$, 
$w_{\text{ICU}} = 5.10^4$ and $T=700$. Hence the optimal control problem for which we compute an approximate solution numerically is:
\begin{eqnarray}
 \inf_{\delta\in\tilde{\mathcal{A}}} \big\lbrace \tilde J_T(\delta,0,0)\big\rbrace\;,
\end{eqnarray}
with $\tilde{\mathcal{A}}$ the set of measurable functions from $[0,T]$ to $[0,1]$.

}

\subsection{Impact of the initial virus prevalence $I_0^-$}\label{sec_sensi_I0}

\begin{figure}[h]
	\begin{subfigure}{.33\columnwidth}
		\centering
		\includegraphics[width=\columnwidth]{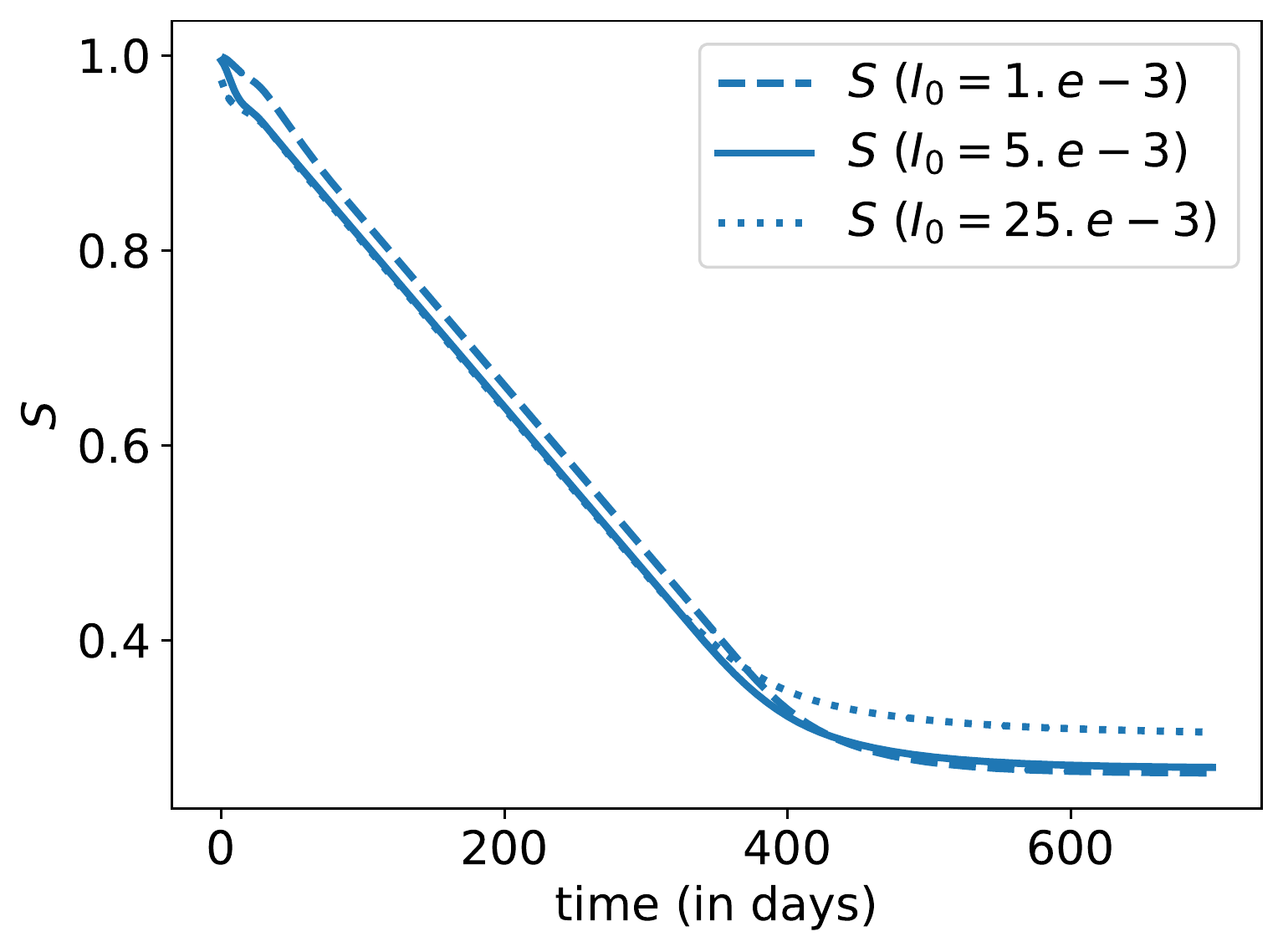}
	\end{subfigure}%
	\begin{subfigure}{.33\columnwidth}
		\centering 
		\includegraphics[width=\columnwidth]{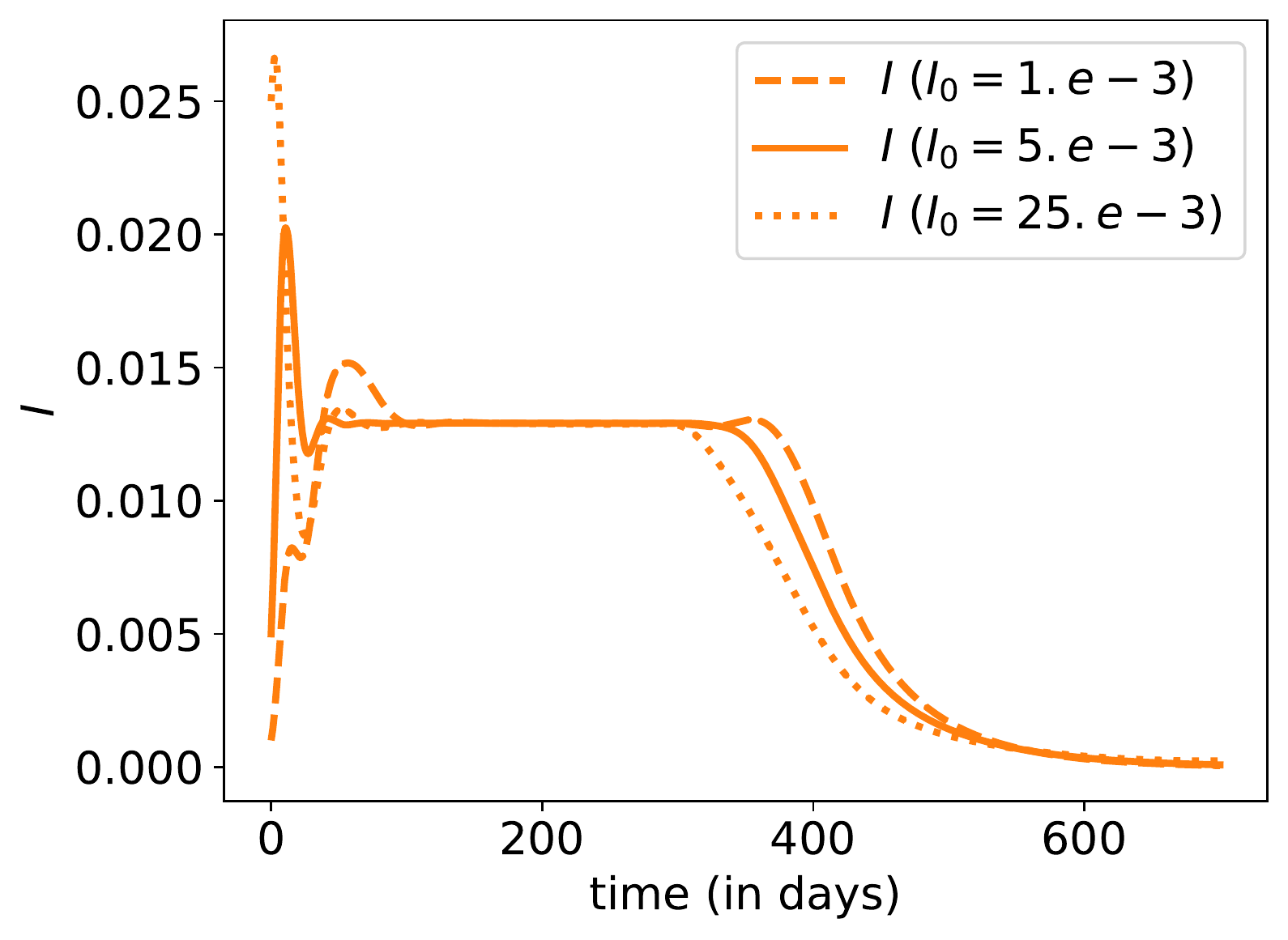}
	\end{subfigure}
	\begin{subfigure}{.33\columnwidth}
		\centering 
		\includegraphics[width=\columnwidth]{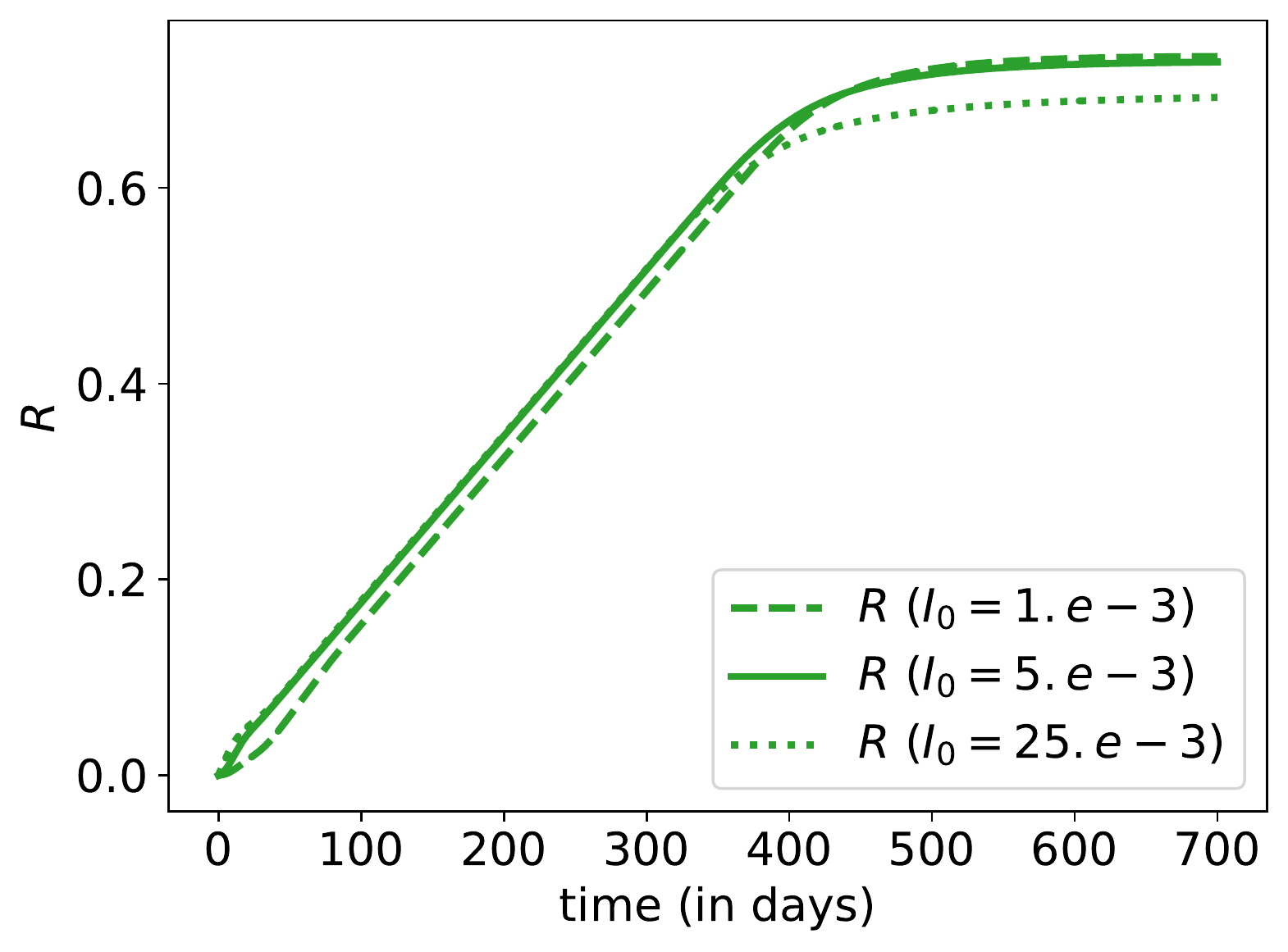}
	\end{subfigure}
	
	\begin{subfigure}{.33\columnwidth}
		\centering
		\includegraphics[width=\columnwidth]{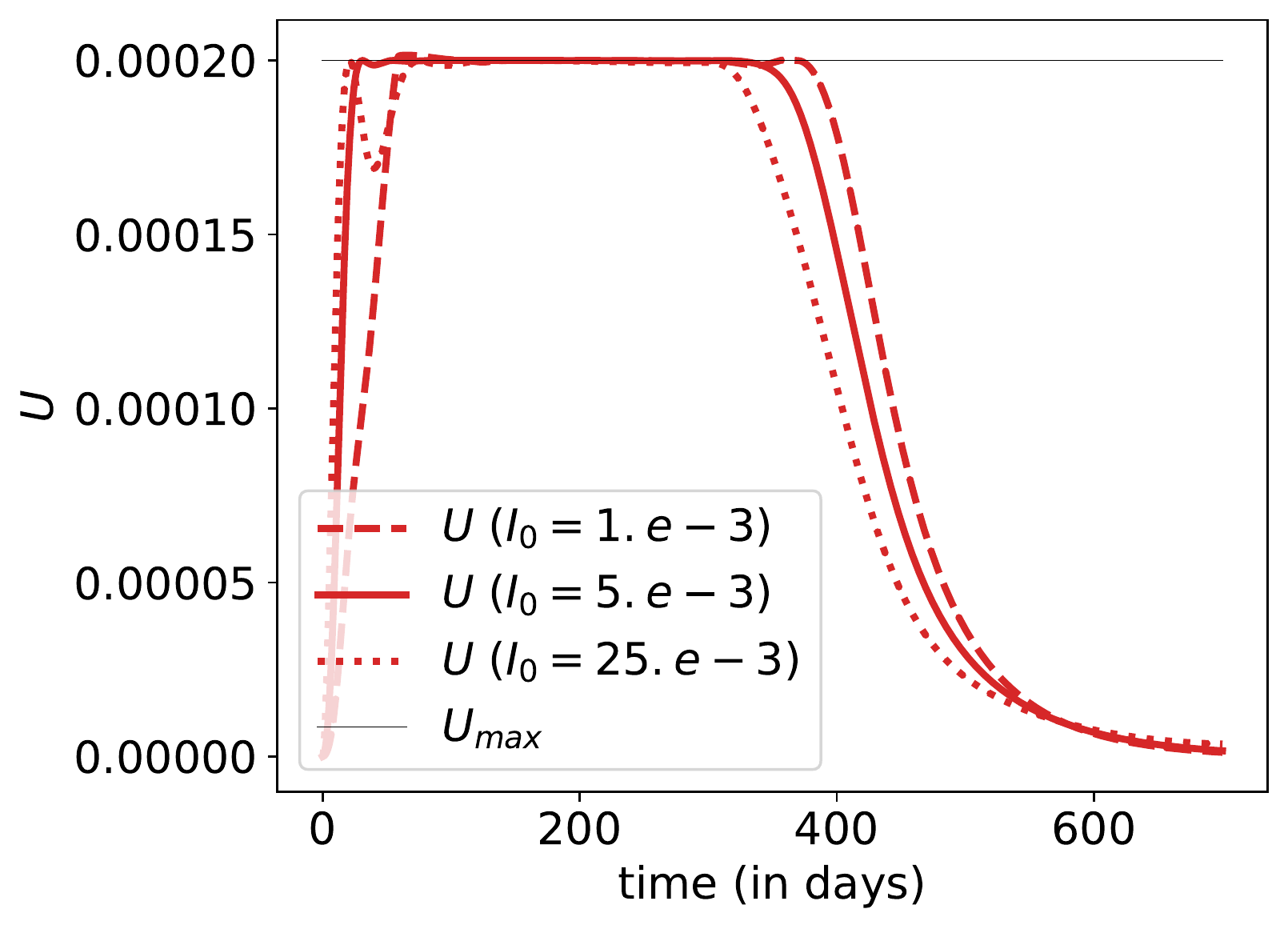}
	\end{subfigure}%
	\begin{subfigure}{.33\columnwidth}
		\centering 
		\includegraphics[width=\columnwidth]{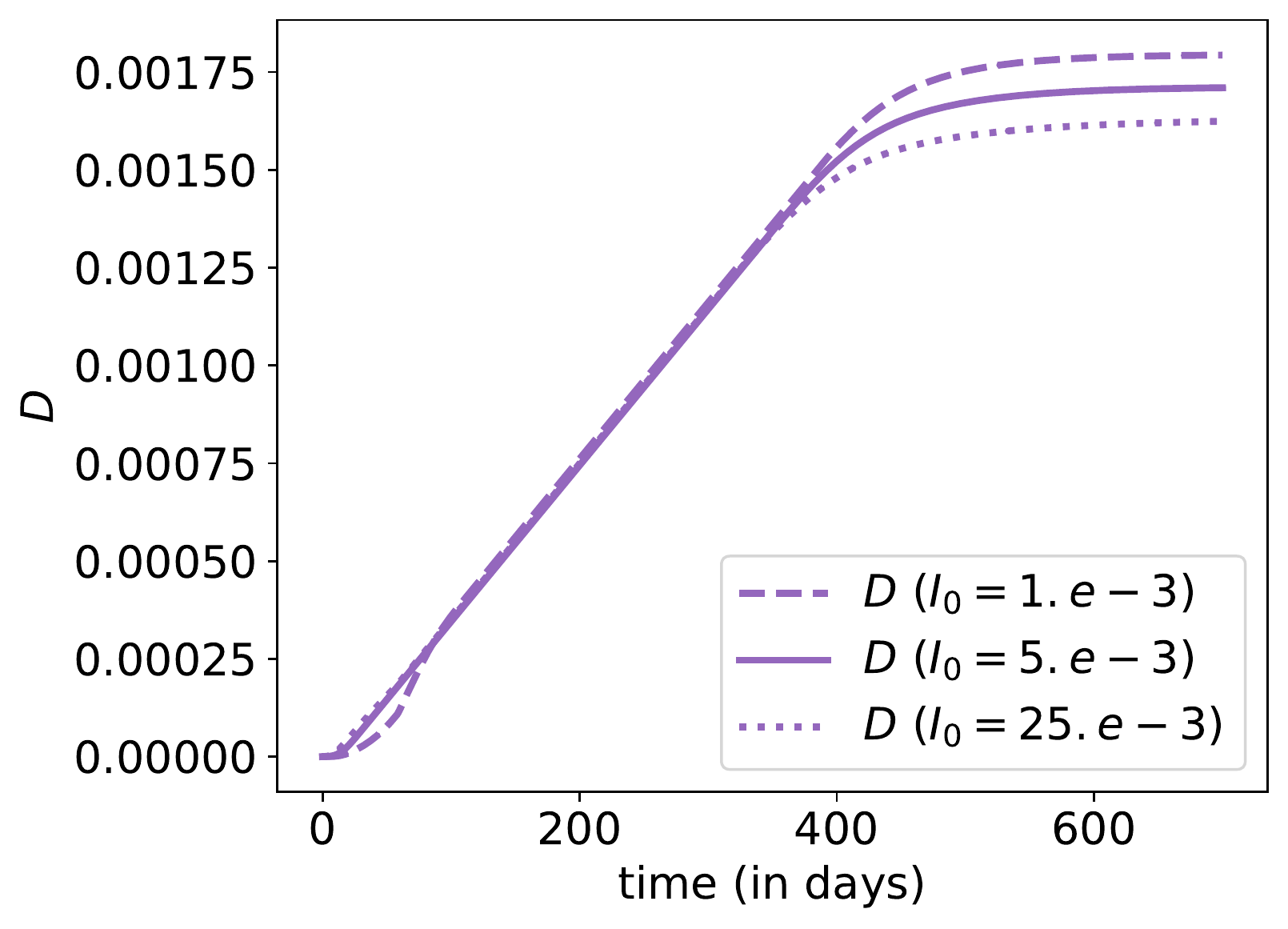}
	\end{subfigure}
	\begin{subfigure}{.33\columnwidth}
		\centering 
		\includegraphics[width=\columnwidth]{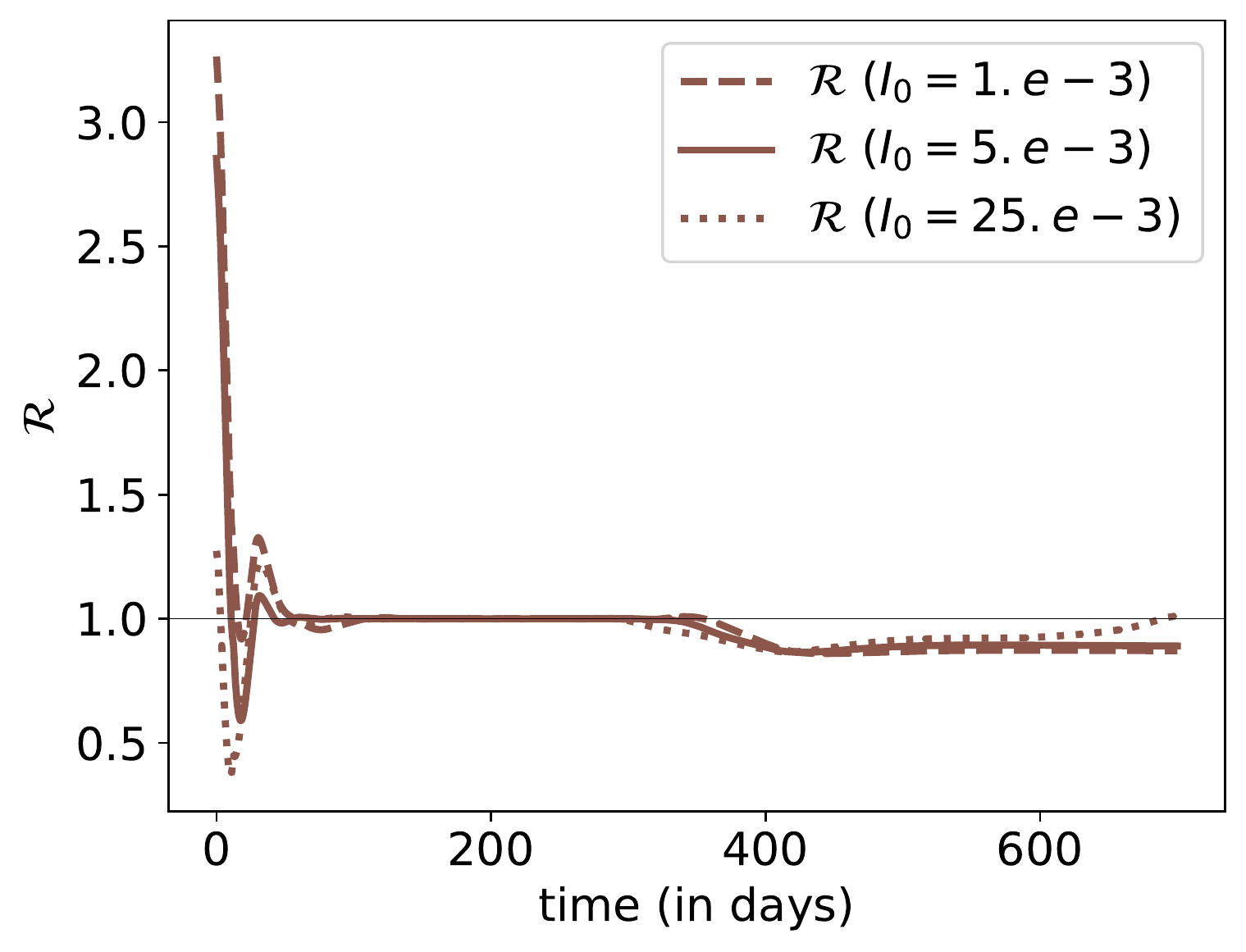}
	\end{subfigure}
	
	\begin{subfigure}{.33\columnwidth}
		\centering
		\includegraphics[width=\columnwidth]{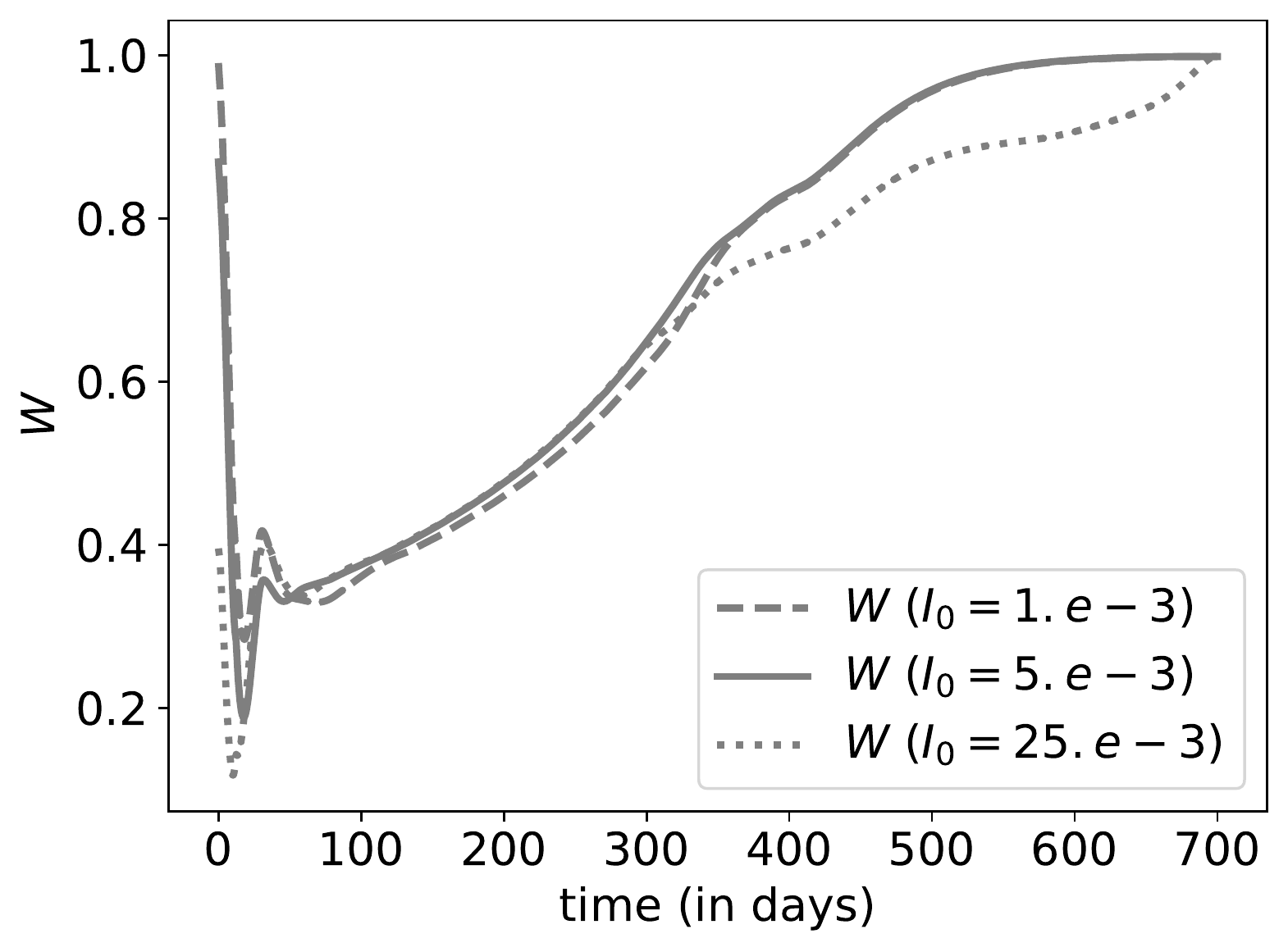}
	\end{subfigure}%
	\begin{subfigure}{.33\columnwidth}
		\centering 
		\includegraphics[width=\columnwidth]{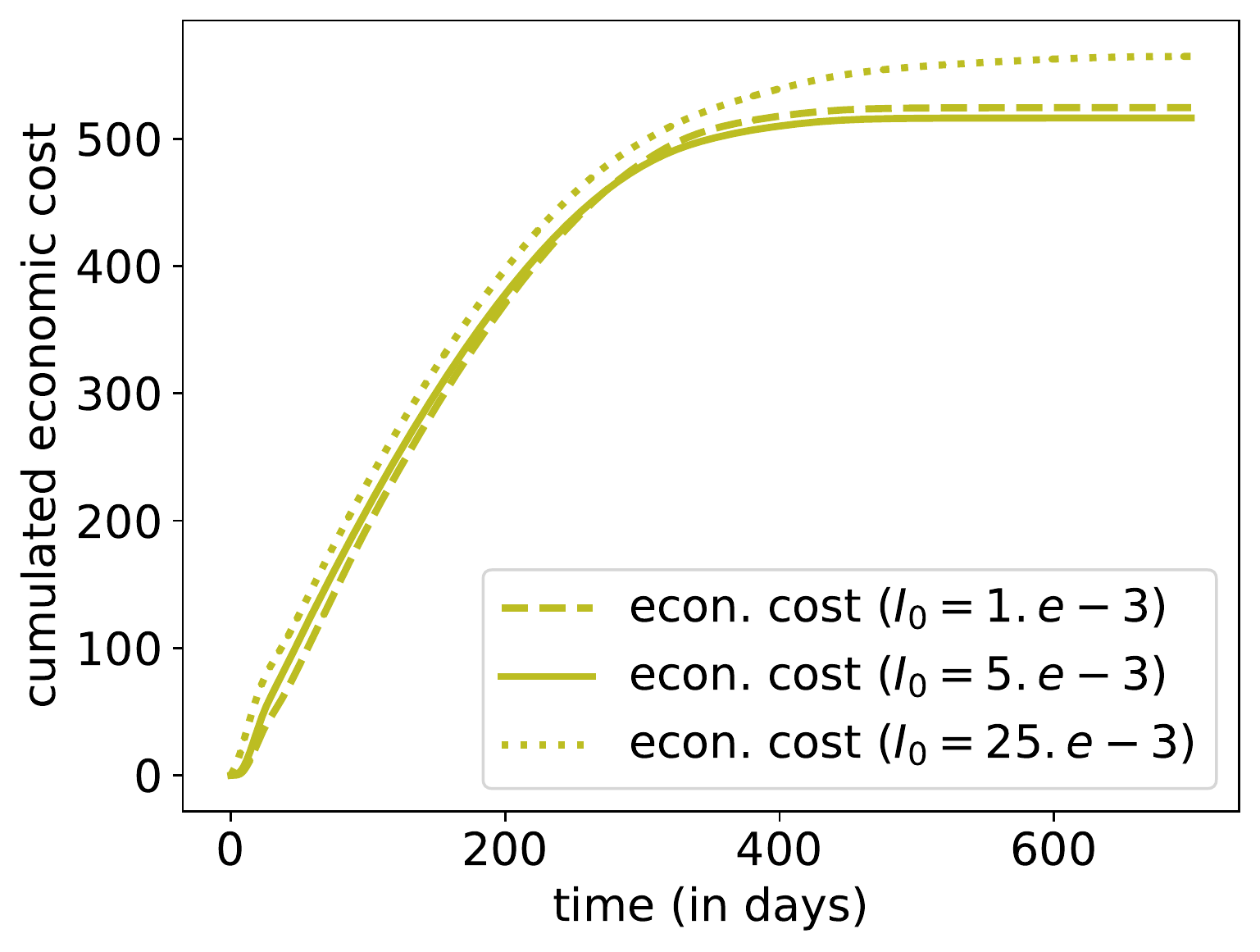}
	\end{subfigure}
	\begin{subfigure}{.33\columnwidth}
		\centering 
		\includegraphics[width=\columnwidth]{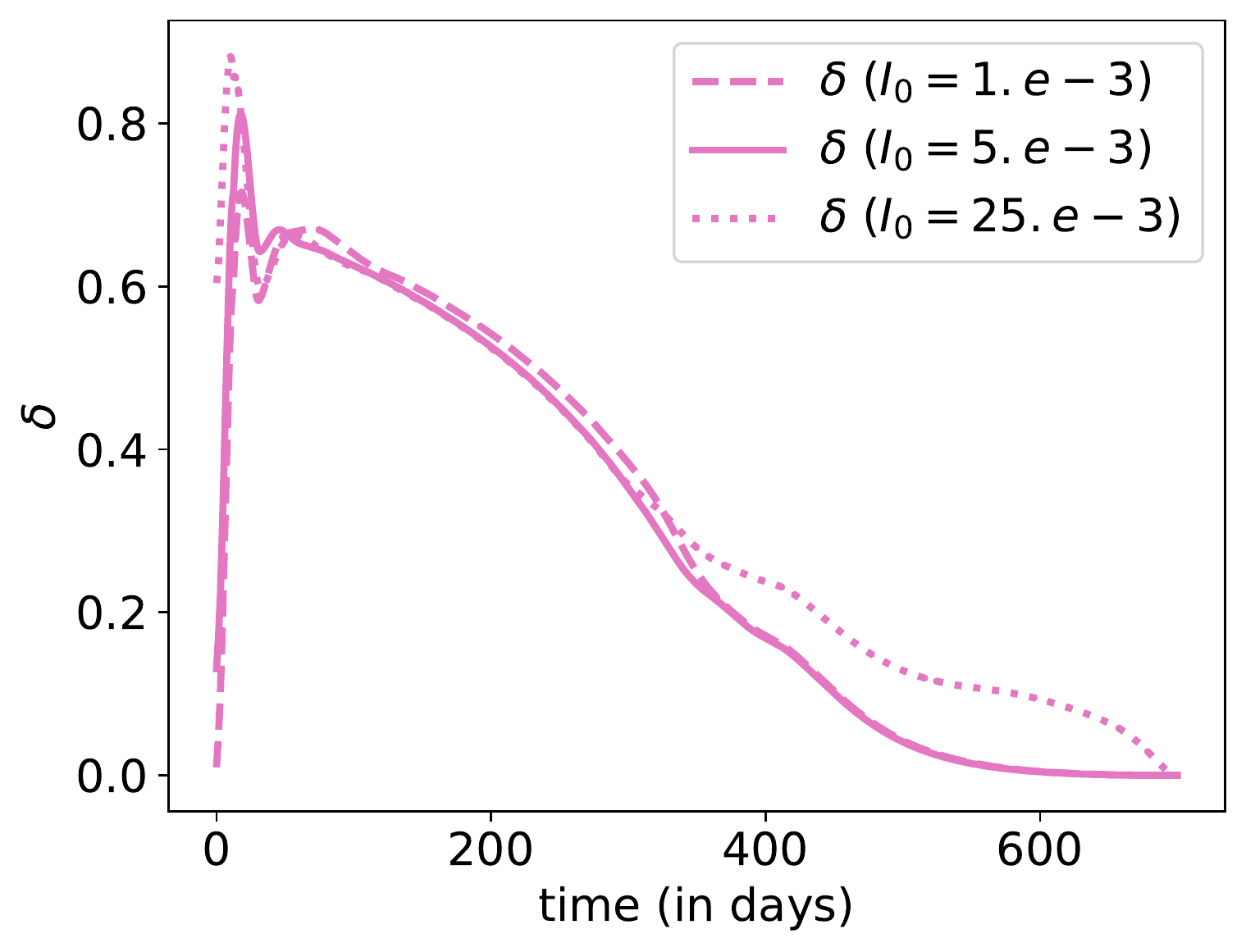}
	\end{subfigure}
	\caption{Evolution of states with optimal controls $(\delta,0,0)$ -- without any testing -- for three different values of $I_0^-$. The plain line is the benchmark scenario discussed in Section \ref{subsec:optimal:policy}; the dashed line corresponds to a smaller weight for $I_0^-$ while the dotted line corresponds to a larger $I_0^-$. }
 	\label{fig:Sensi_I0}
\end{figure}

\newpage

\subsection{Impact of the basic reproductive number $\mathfrak{R}_0$}\label{sec_sensi_R0}

\begin{figure}[h]
	\begin{subfigure}{.33\columnwidth}
		\centering
		\includegraphics[width=\columnwidth]{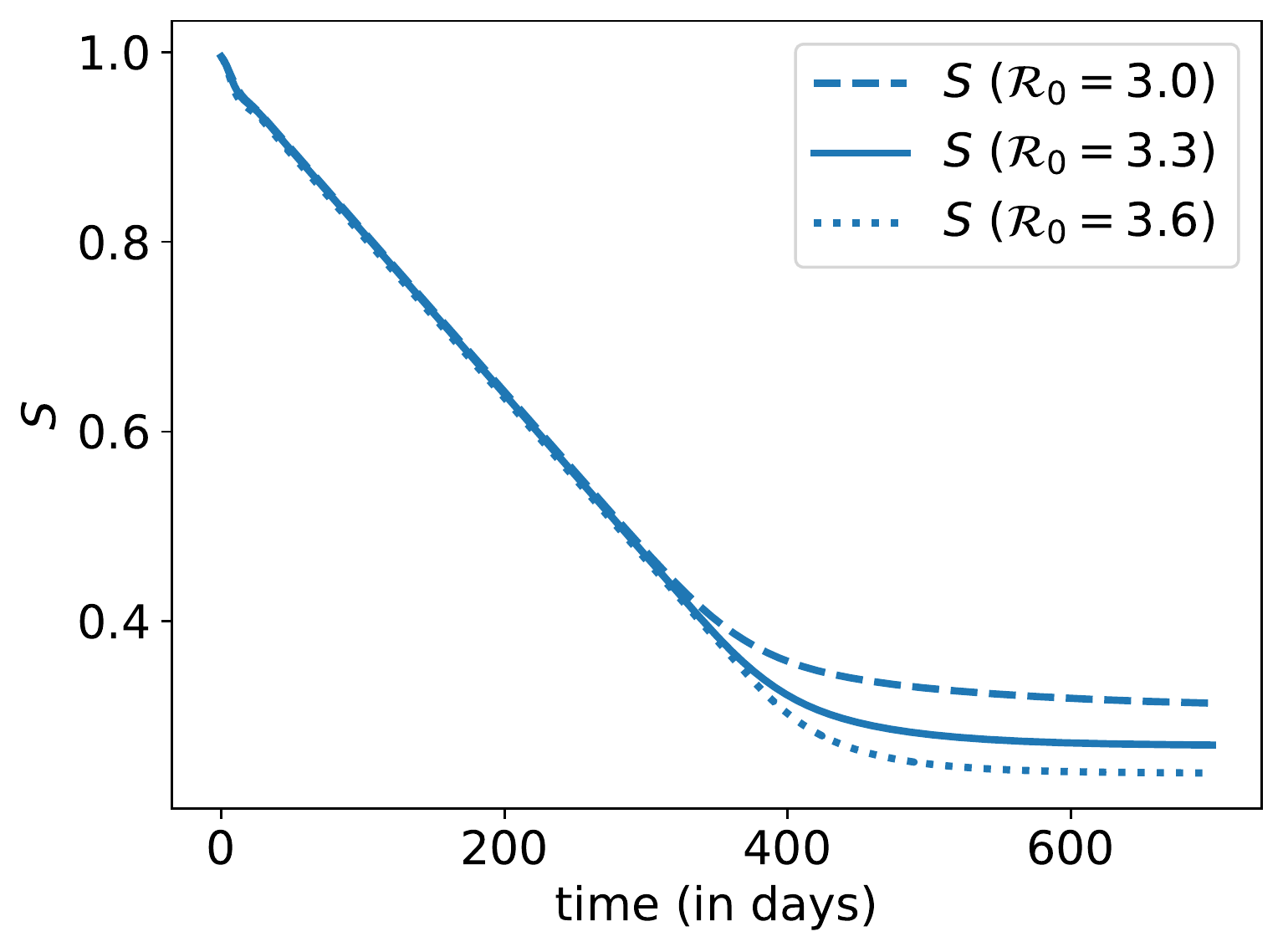}
	\end{subfigure}%
	\begin{subfigure}{.33\columnwidth}
		\centering 
		\includegraphics[width=\columnwidth]{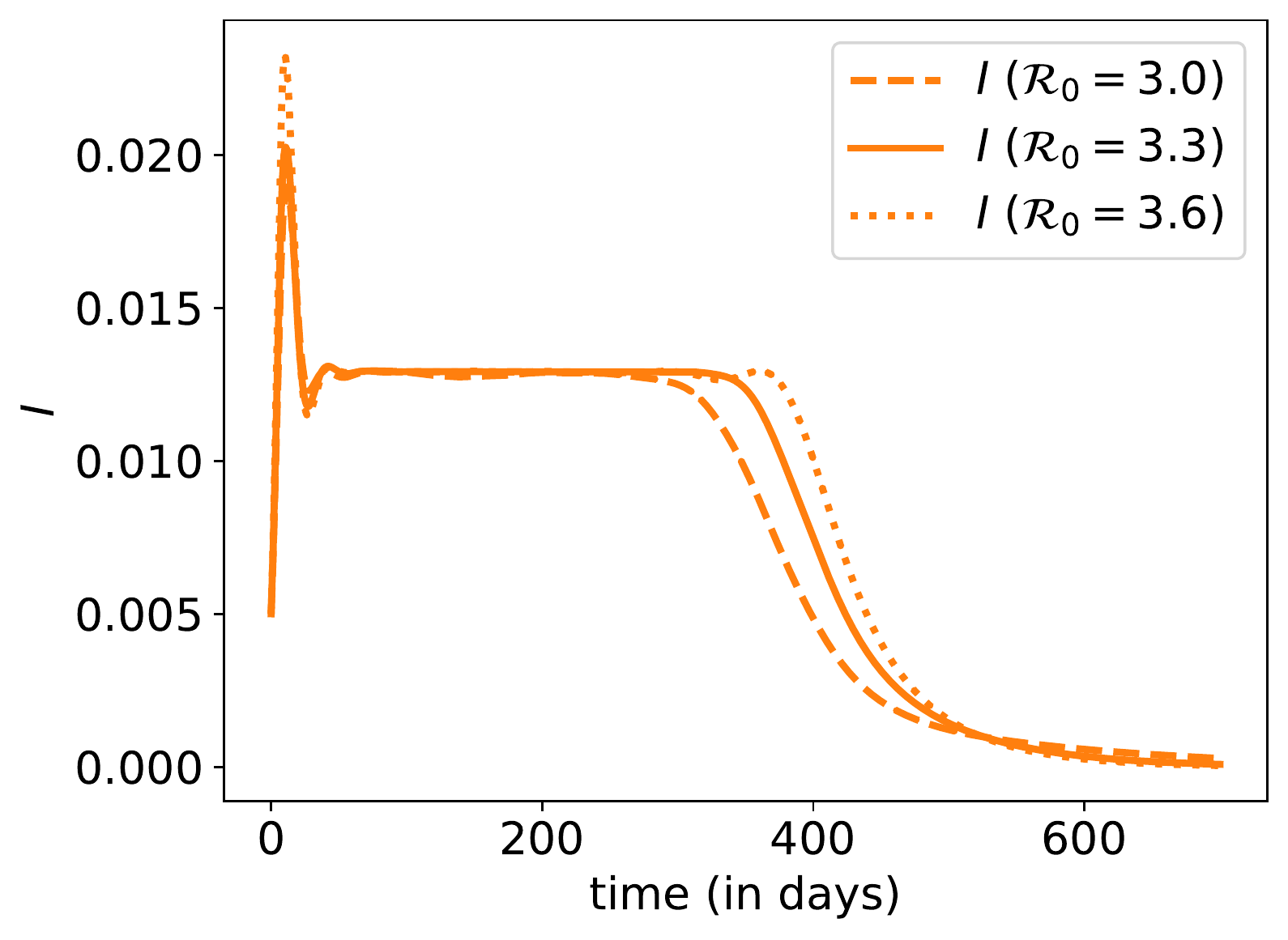}
	\end{subfigure}
	\begin{subfigure}{.33\columnwidth}
		\centering 
		\includegraphics[width=\columnwidth]{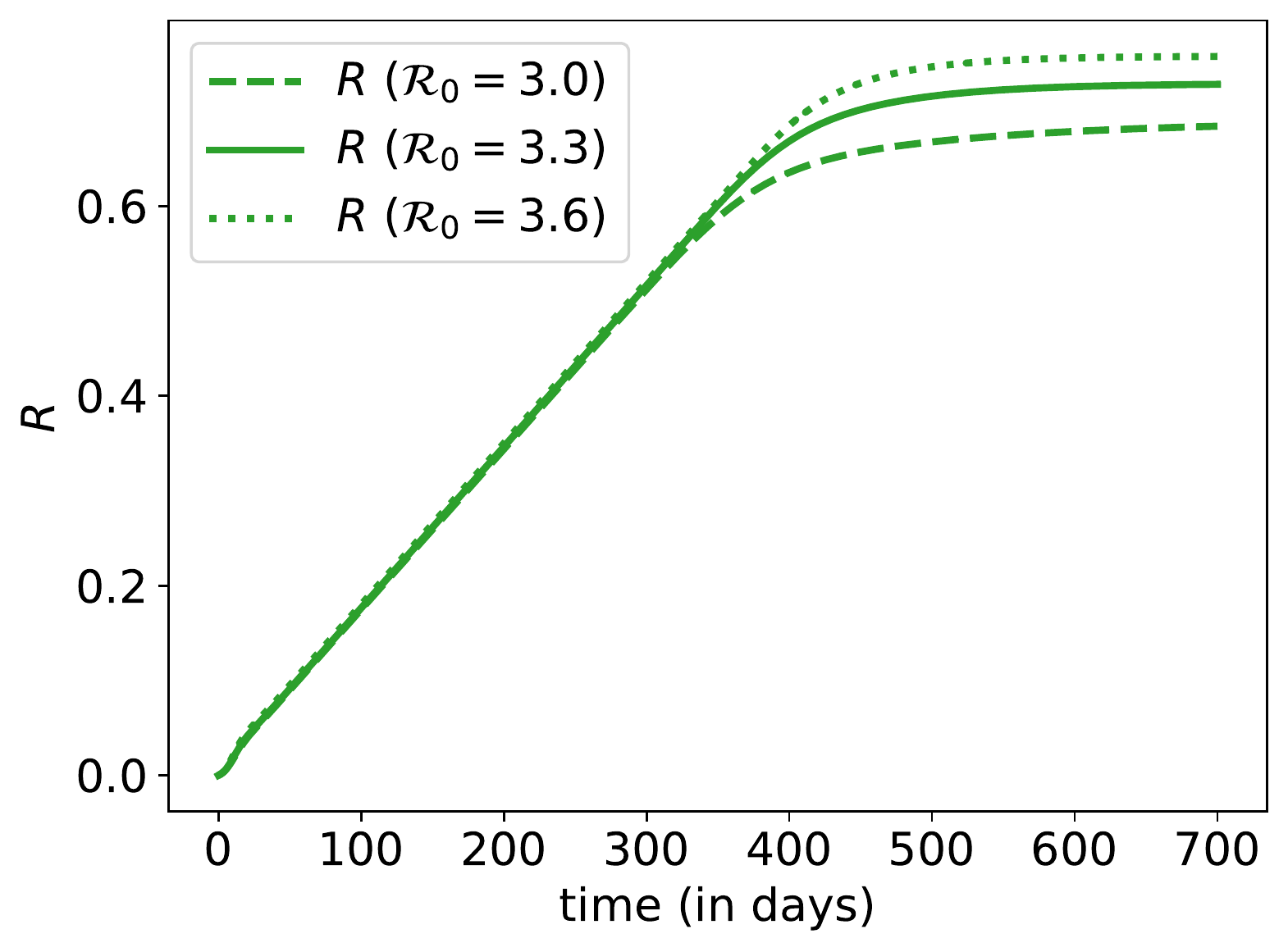}
	\end{subfigure}
	
	\begin{subfigure}{.33\columnwidth}
		\centering
		\includegraphics[width=\columnwidth]{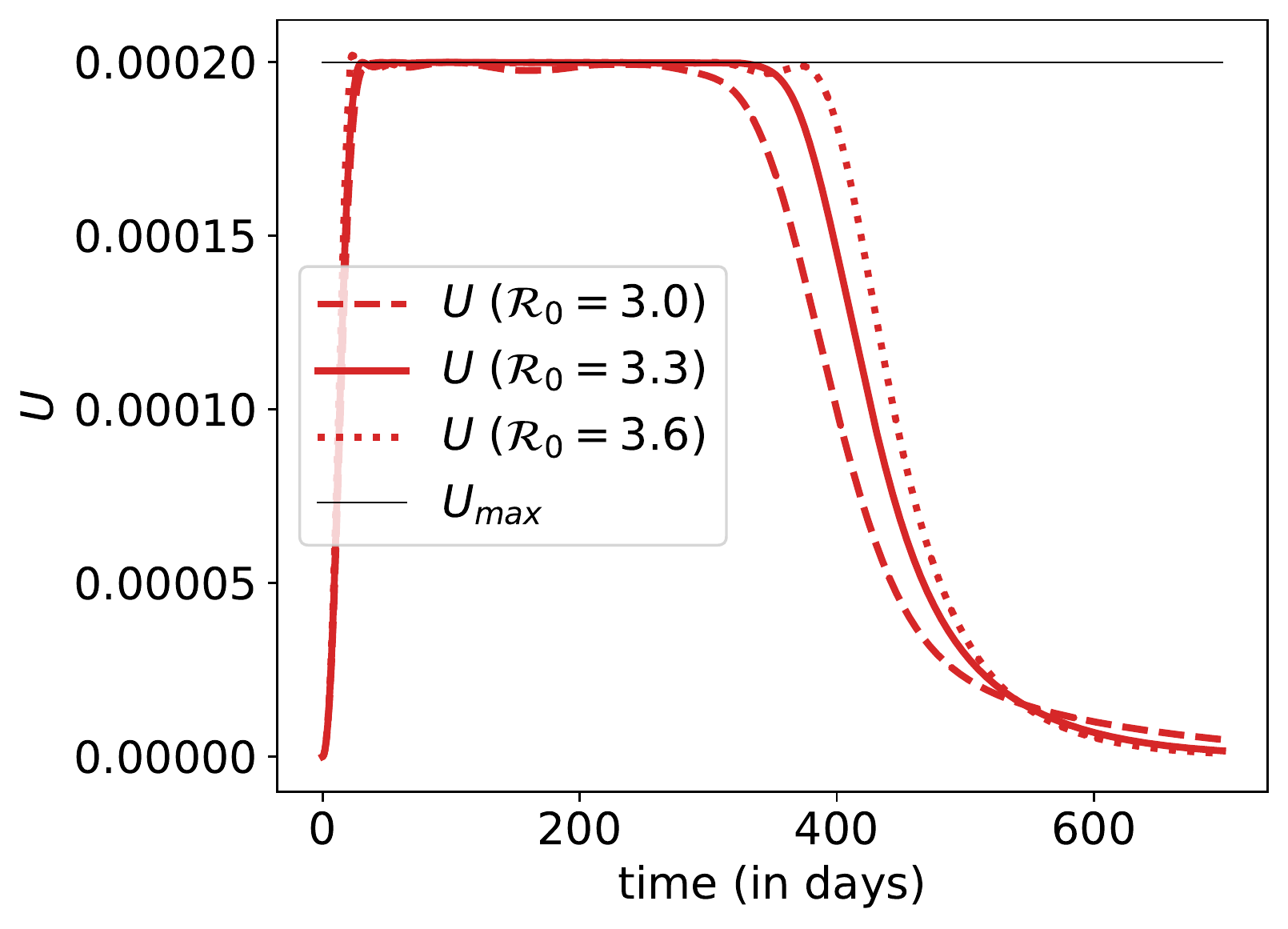}
	\end{subfigure}%
	\begin{subfigure}{.33\columnwidth}
		\centering 
		\includegraphics[width=\columnwidth]{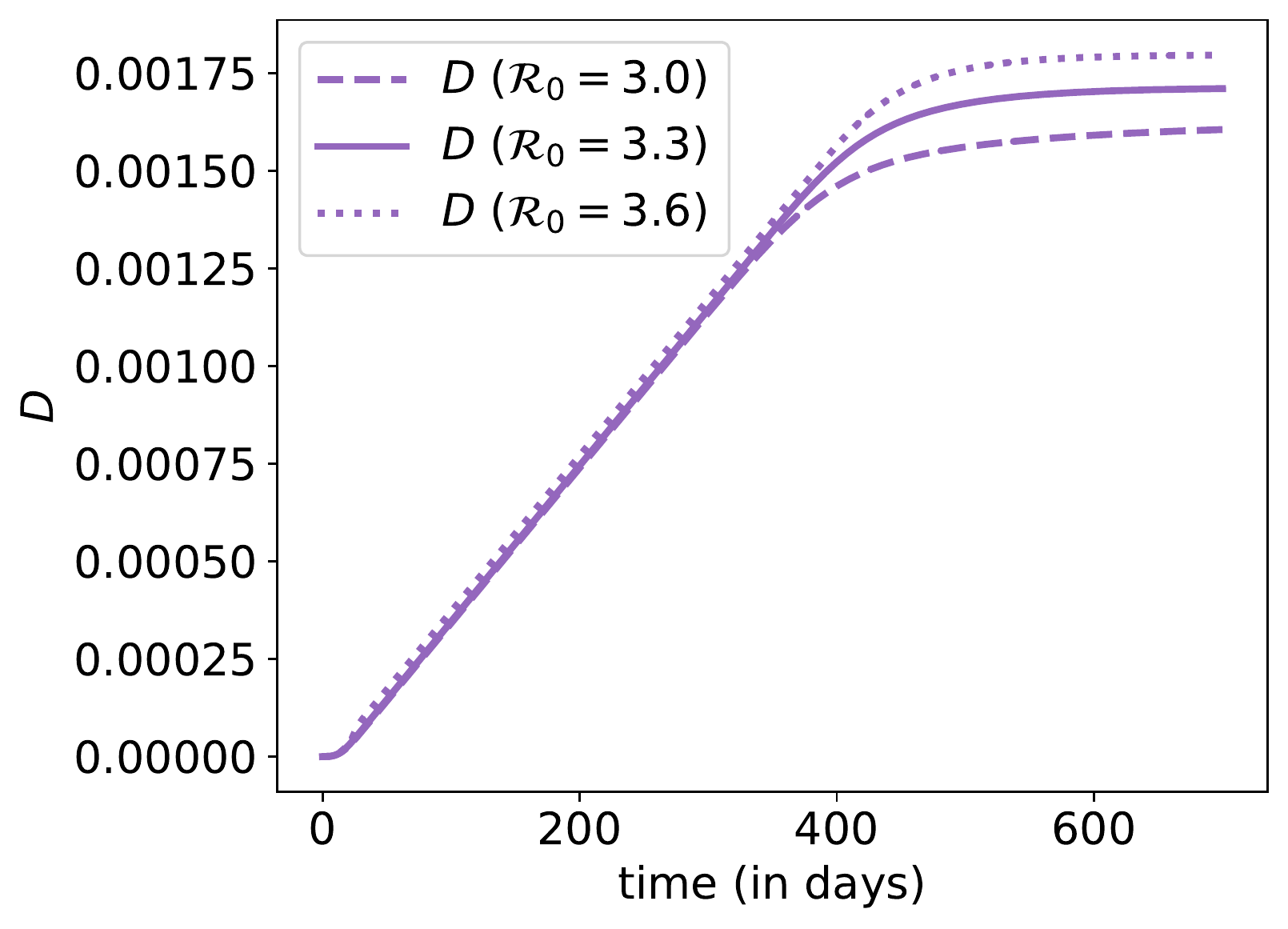}
	\end{subfigure}
	\begin{subfigure}{.33\columnwidth}
		\centering 
		\includegraphics[width=\columnwidth]{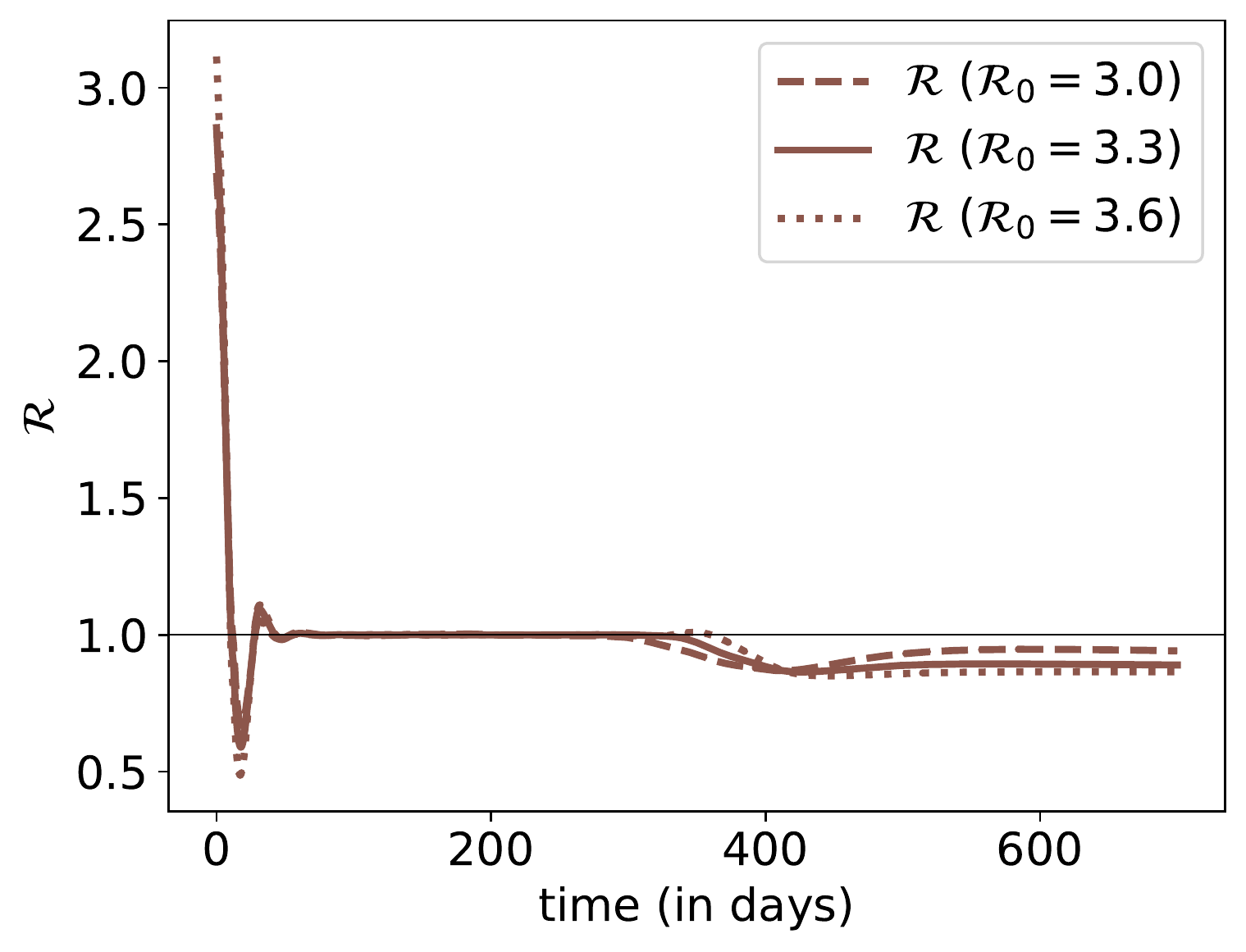}
	\end{subfigure}
	
	\begin{subfigure}{.33\columnwidth}
		\centering
		\includegraphics[width=\columnwidth]{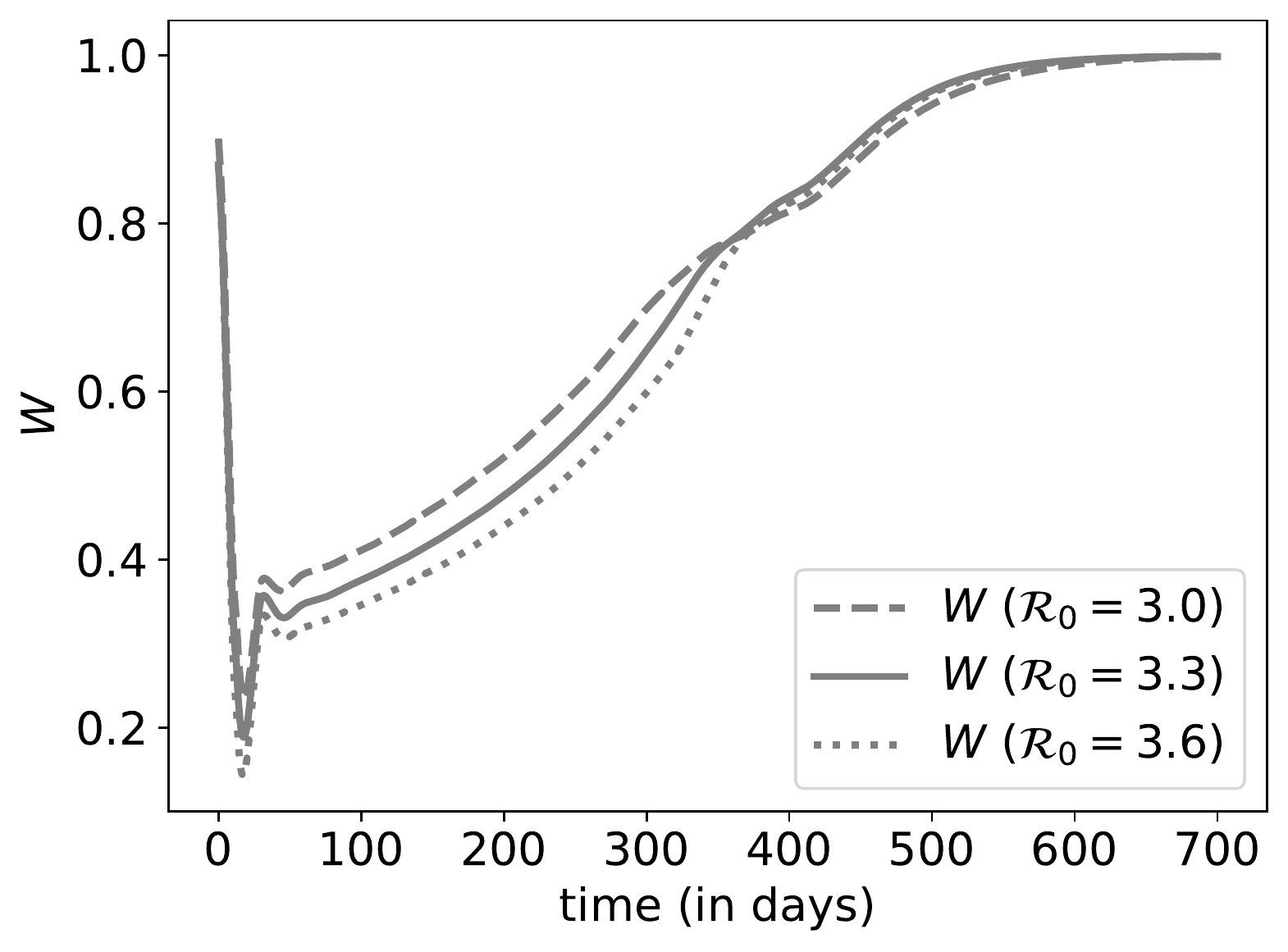}
	\end{subfigure}%
	\begin{subfigure}{.33\columnwidth}
		\centering 
		\includegraphics[width=\columnwidth]{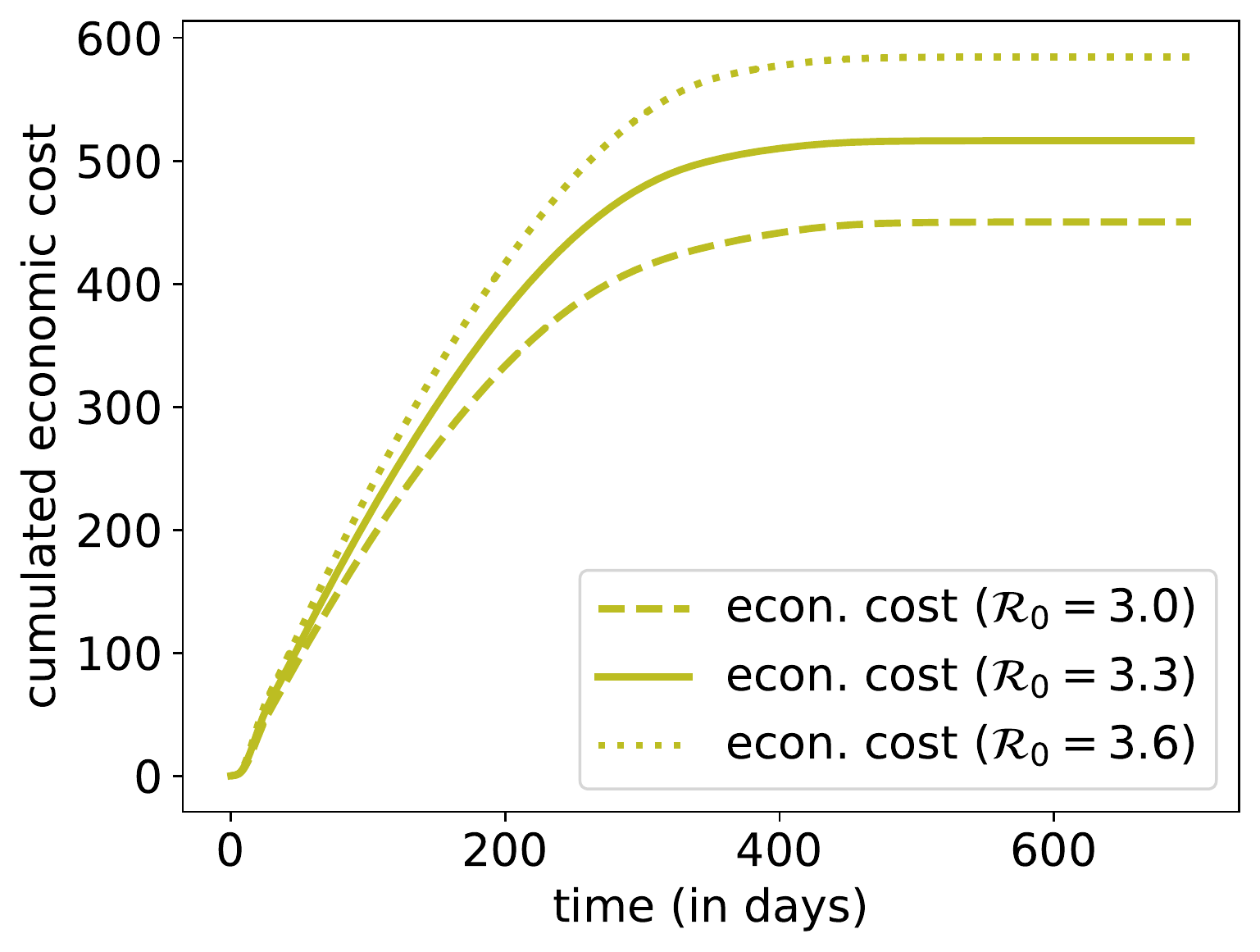}
	\end{subfigure}
	\begin{subfigure}{.33\columnwidth}
		\centering 
		\includegraphics[width=\columnwidth]{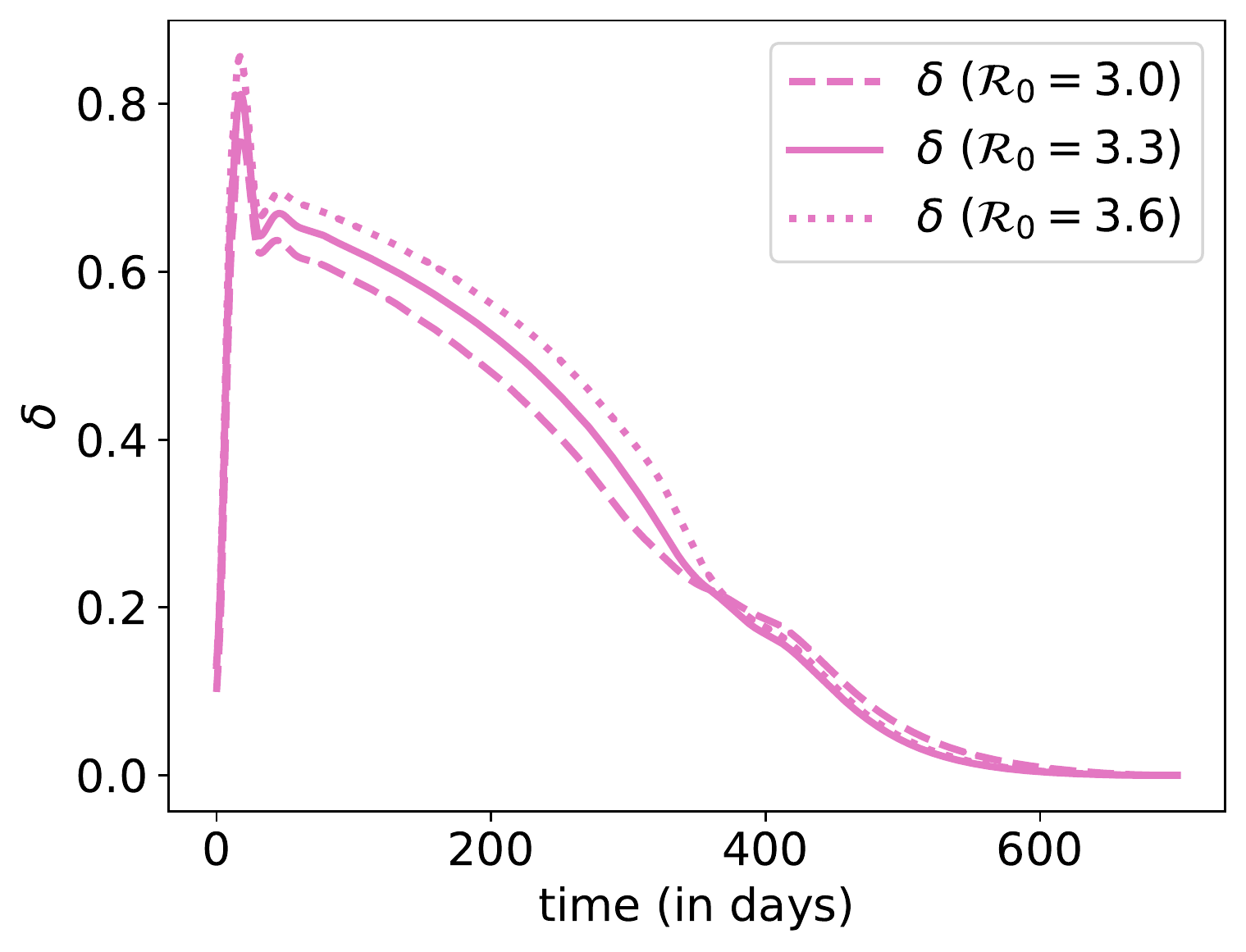}
	\end{subfigure}
	\caption{Evolution of states with optimal controls $(\delta,0,0)$ -- without any testing -- for three different values of $\mathfrak{R}_0$. The plain line is the benchmark scenario discussed in Section \ref{subsec:optimal:policy}; the dashed line corresponds to a smaller value of $\mathfrak{R}_0$ ($-0.3$) while the dotted line corresponds to a larger value of $\mathfrak{R}_0$ ($+0.3$). 
	}
 	\label{fig:Sensi_Rzero}
\end{figure}

\newpage

\subsection{Impact of balancing sanitary and socio-economic costs}\label{sec_sensi_weight}

\begin{figure}[h]
	\begin{subfigure}{.33\columnwidth}
		\centering
		\includegraphics[width=\columnwidth]{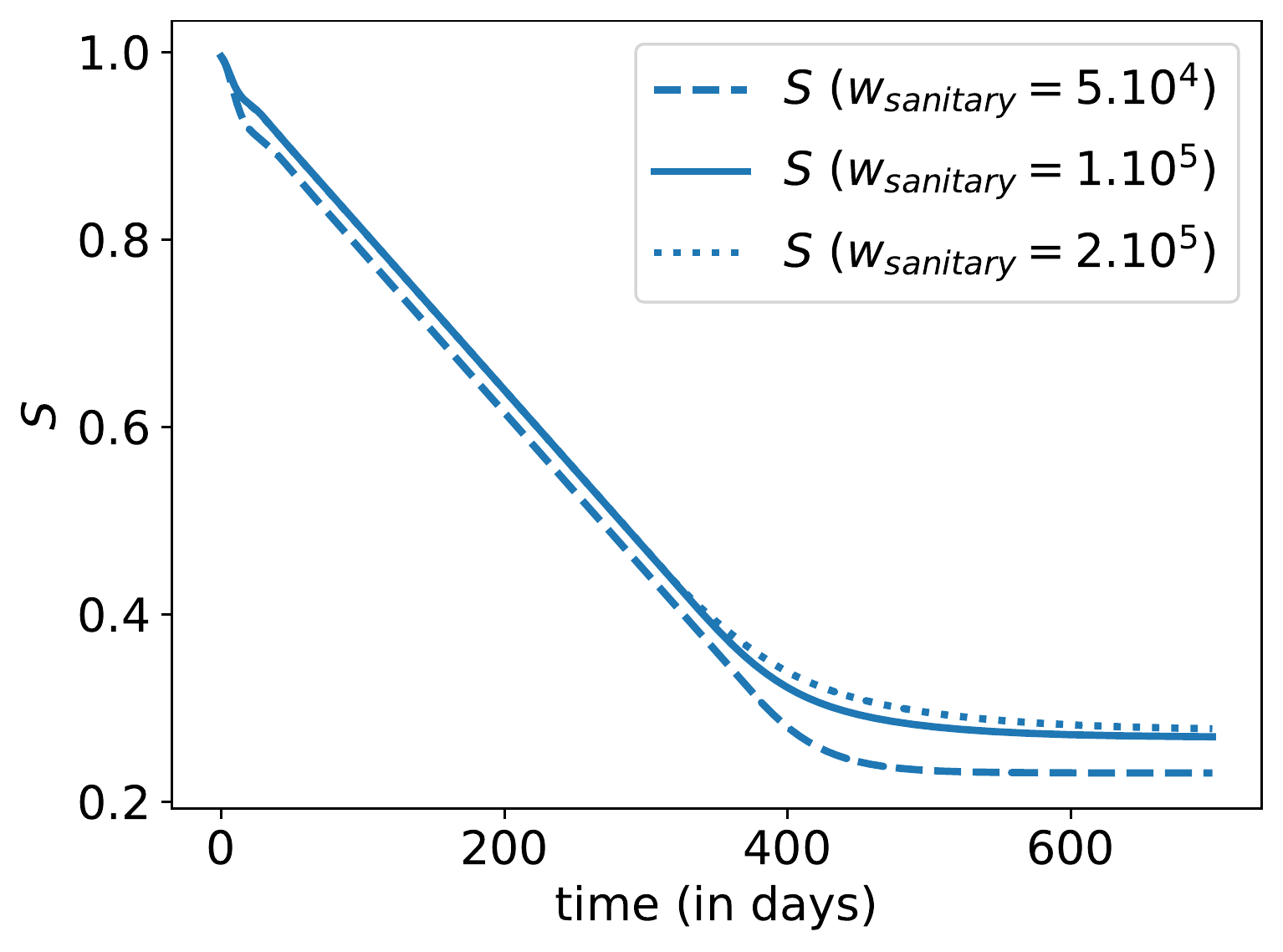}
	\end{subfigure}%
	\begin{subfigure}{.33\columnwidth}
		\centering 
		\includegraphics[width=\columnwidth]{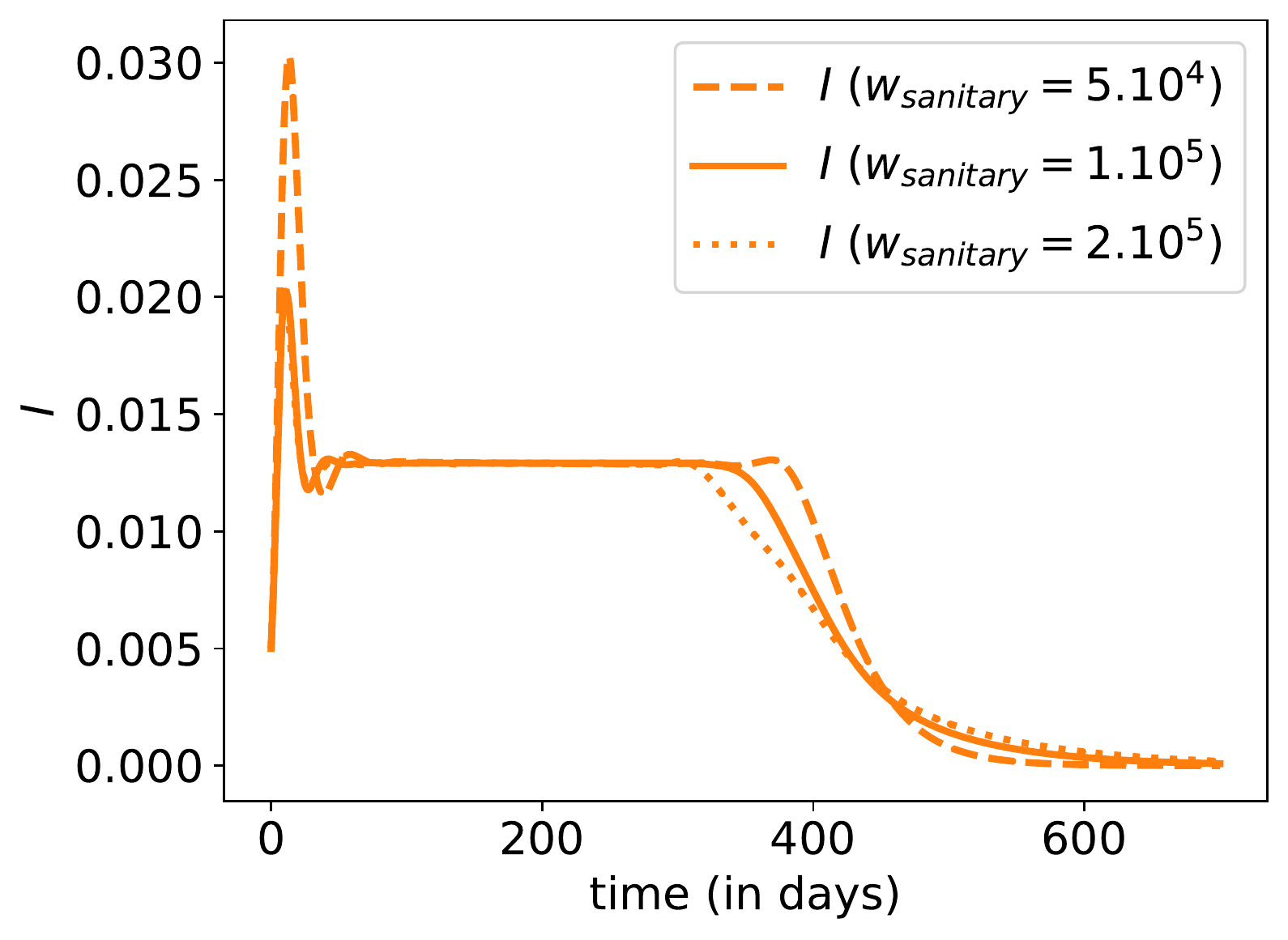}
	\end{subfigure}
	\begin{subfigure}{.33\columnwidth}
		\centering 
		\includegraphics[width=\columnwidth]{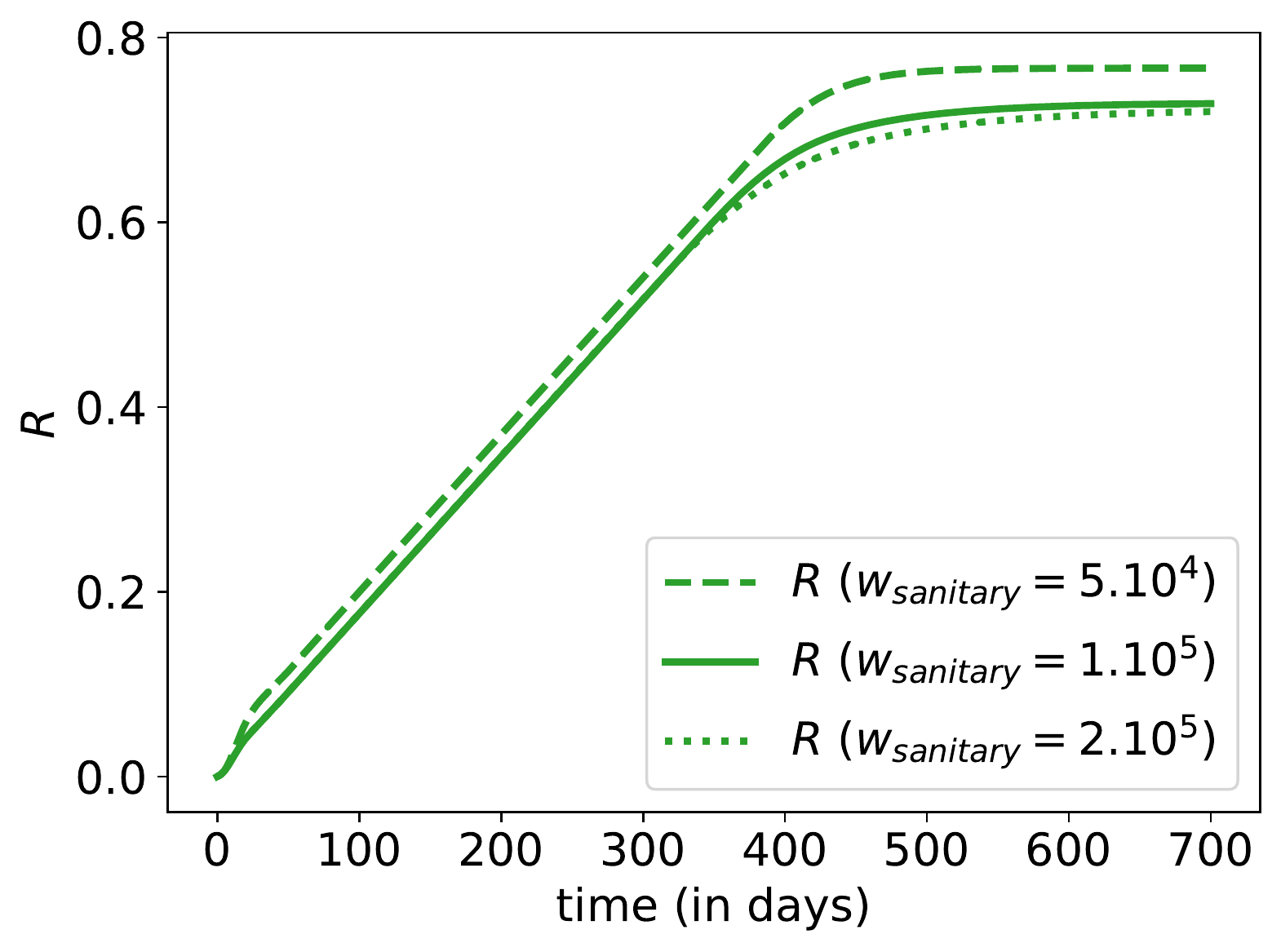}
	\end{subfigure}
	
	\begin{subfigure}{.33\columnwidth}
		\centering
		\includegraphics[width=\columnwidth]{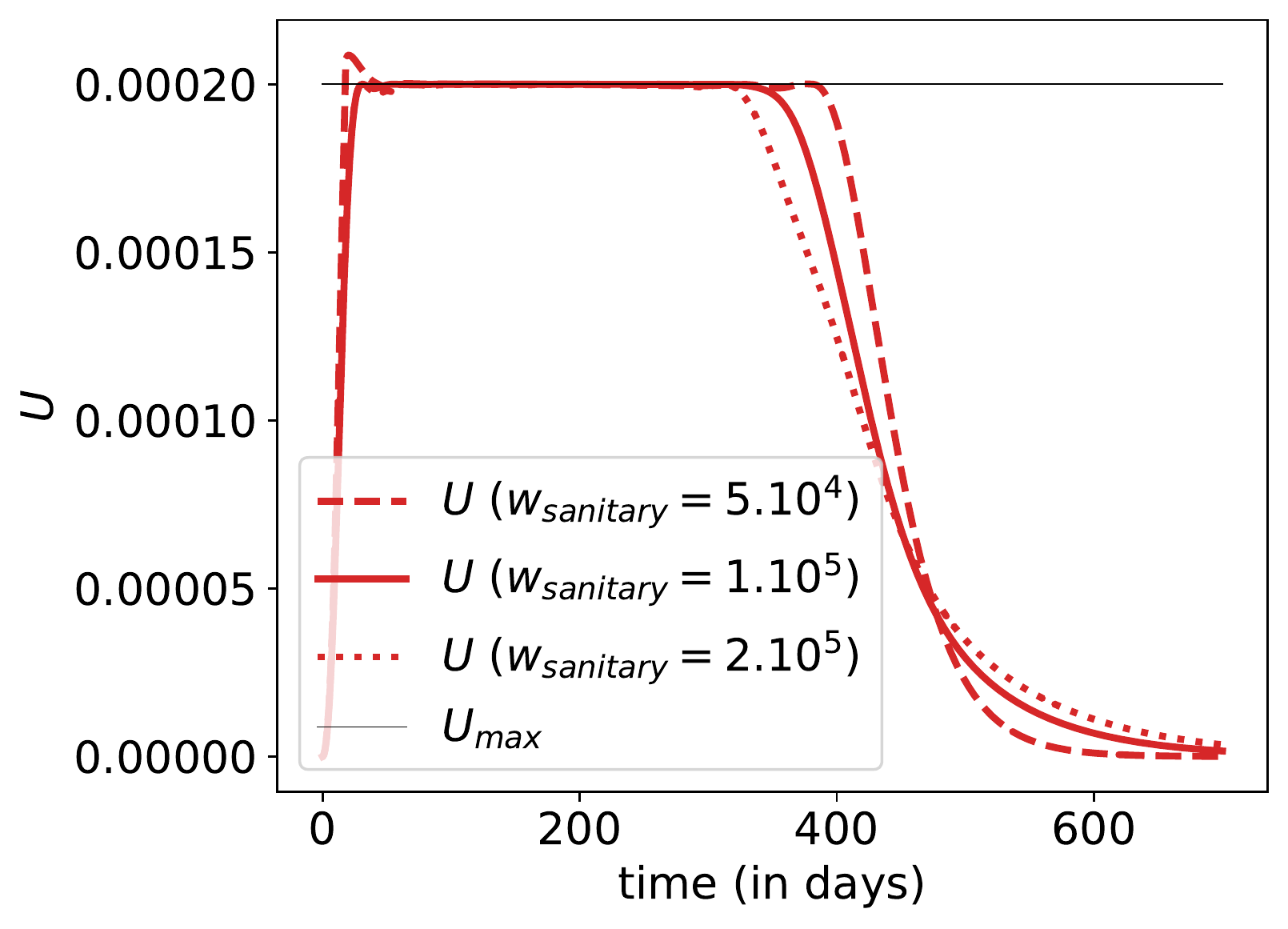}
	\end{subfigure}%
	\begin{subfigure}{.33\columnwidth}
		\centering 
		\includegraphics[width=\columnwidth]{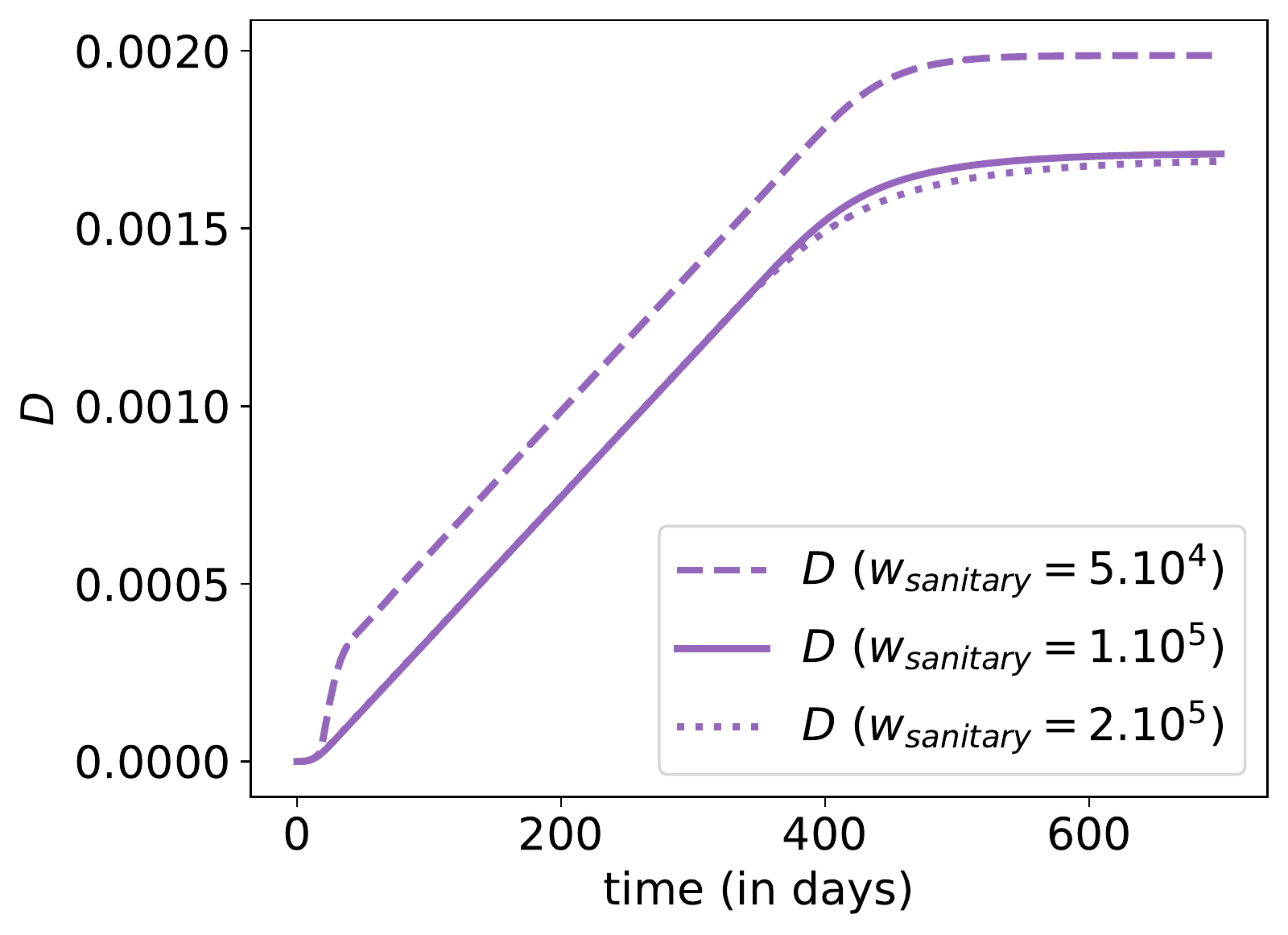}
	\end{subfigure}
	\begin{subfigure}{.33\columnwidth}
		\centering 
		\includegraphics[width=\columnwidth]{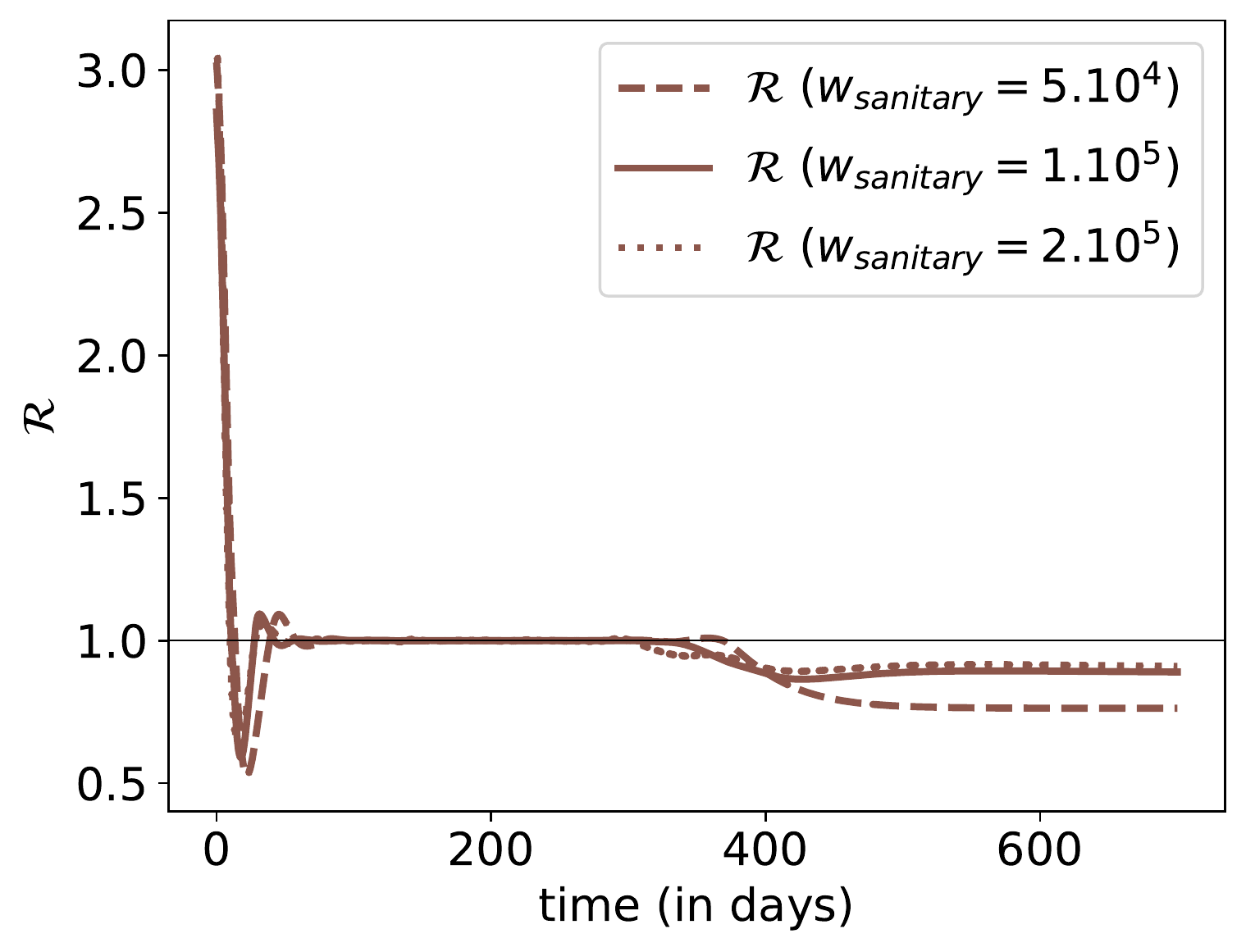}
	\end{subfigure}
	
	\begin{subfigure}{.33\columnwidth}
		\centering
		\includegraphics[width=\columnwidth]{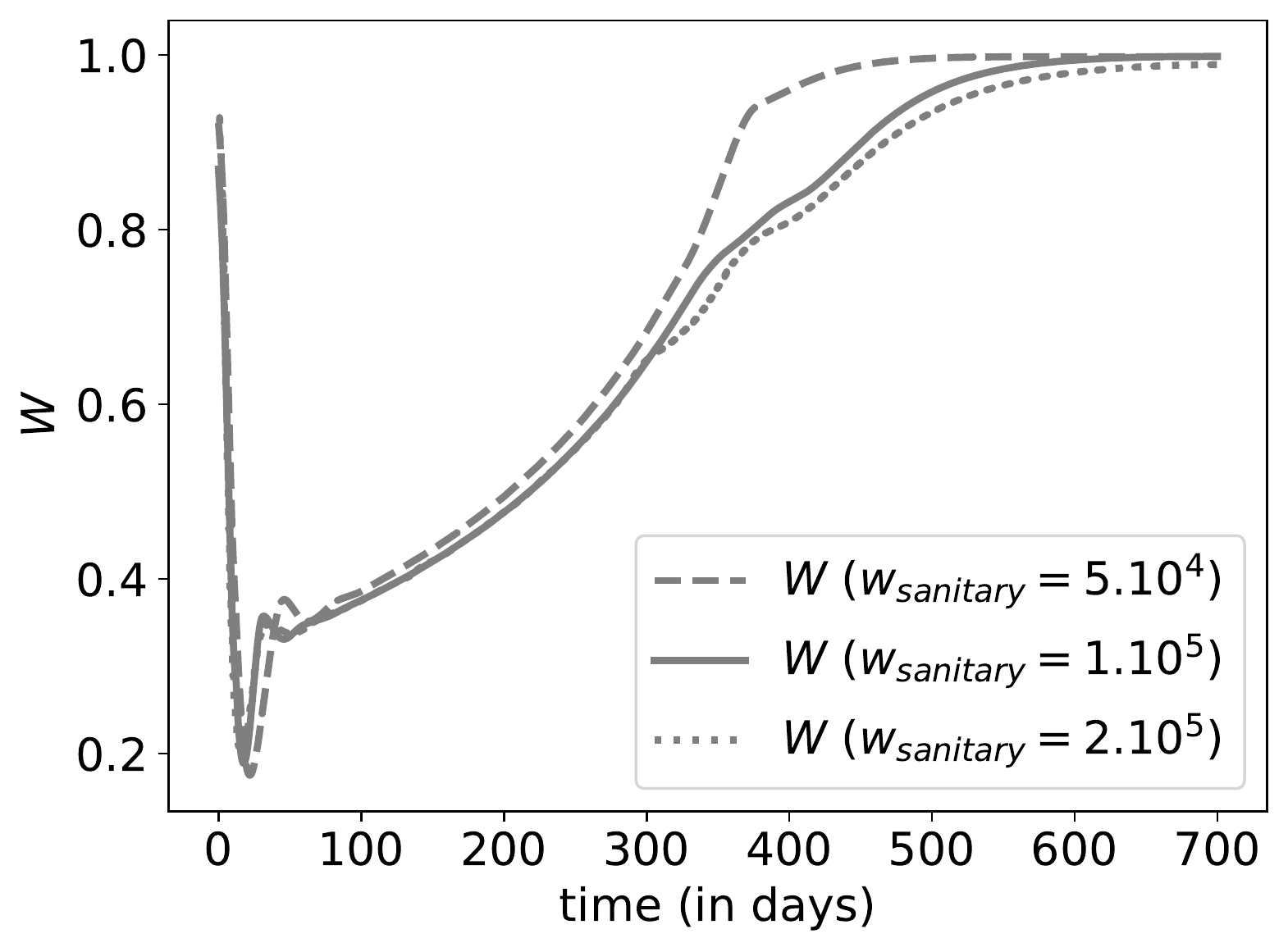}
	\end{subfigure}%
	\begin{subfigure}{.33\columnwidth}
		\centering 
		\includegraphics[width=\columnwidth]{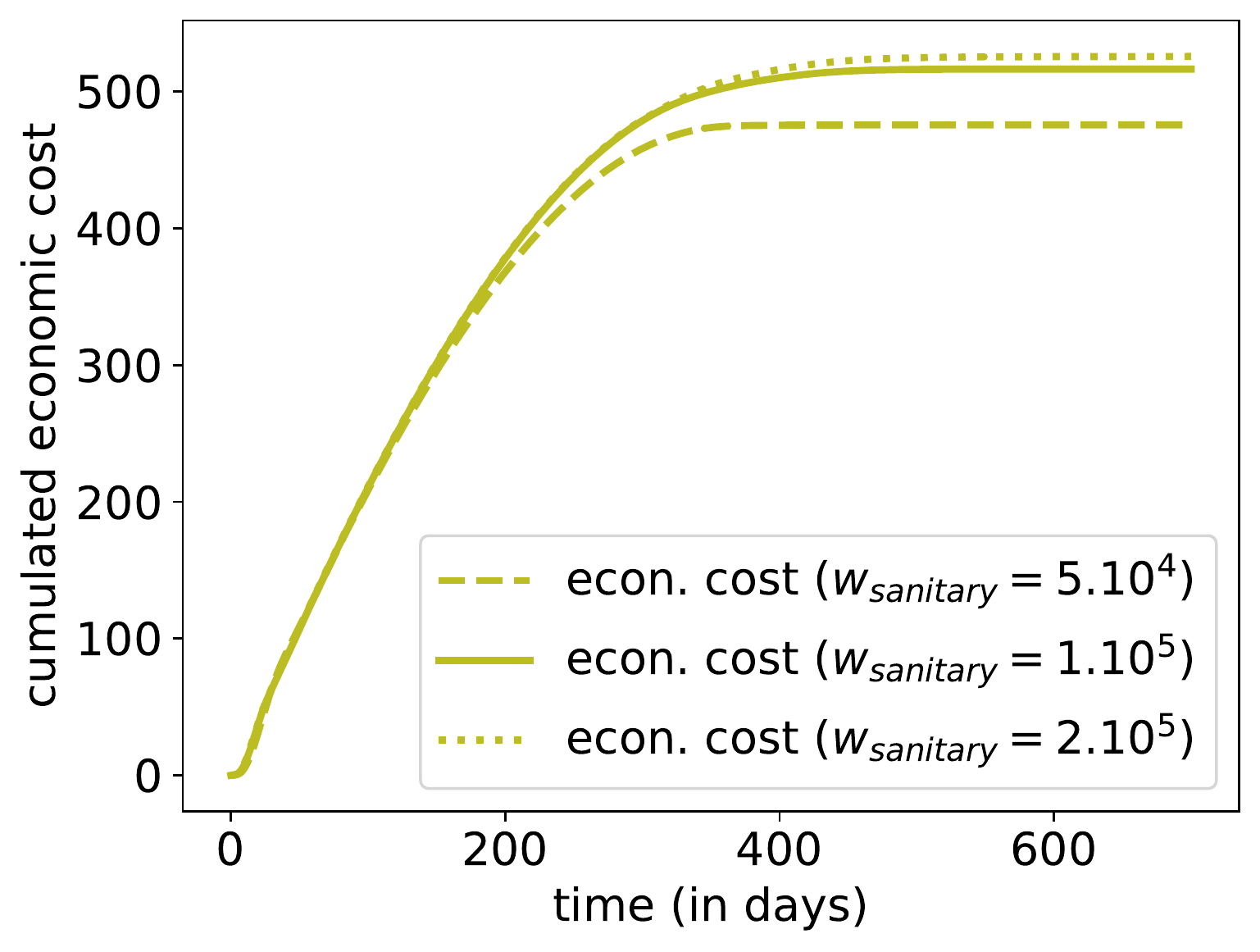}
	\end{subfigure}
	\begin{subfigure}{.33\columnwidth}
		\centering 
		\includegraphics[width=\columnwidth]{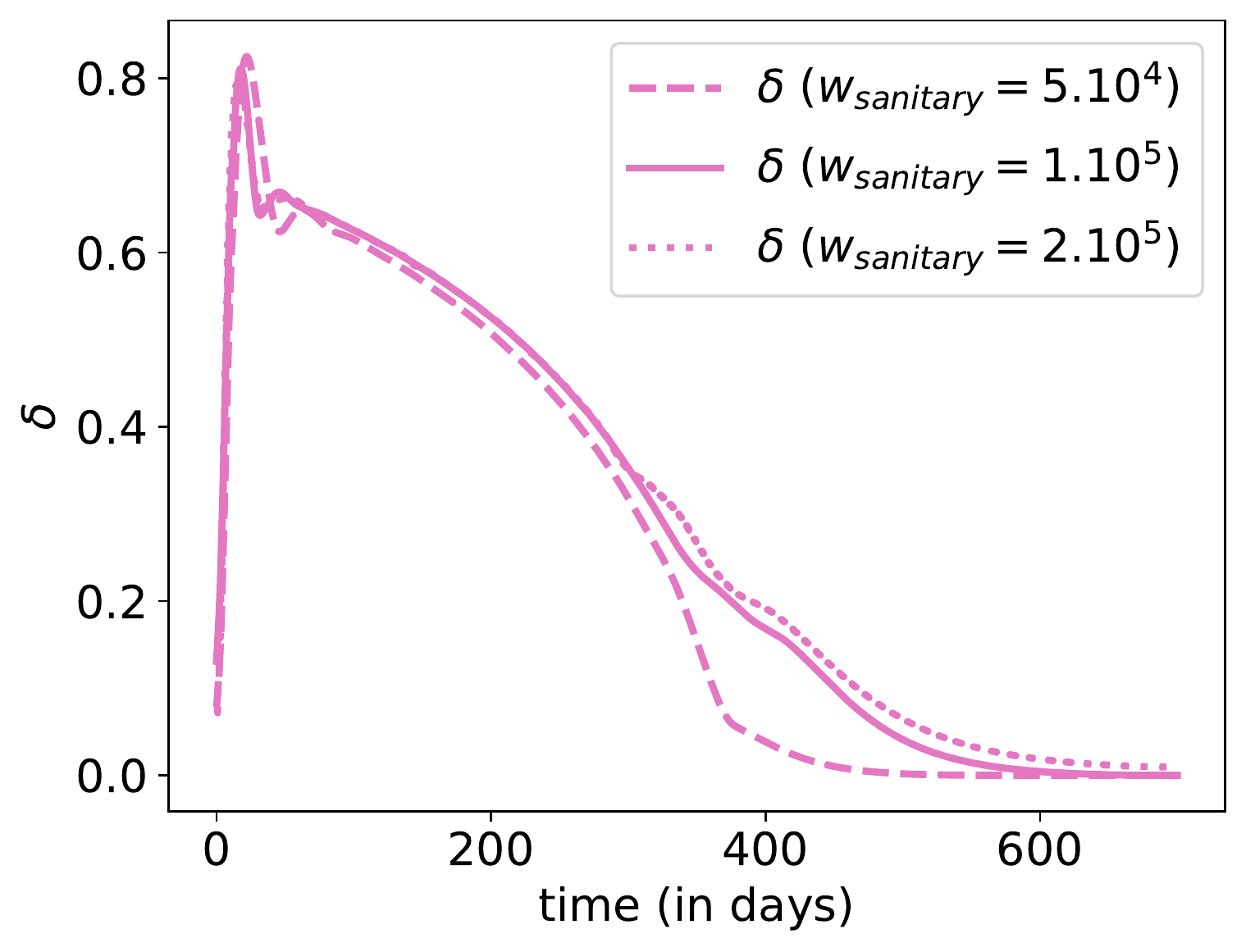}
	\end{subfigure}
	\caption{Evolution of states with optimal controls $(\delta,0,0)$ -- without any testing --  for three different values of $w_{\text{sanitary}}$. The plain line is the benchmark scenario discussed in Section \ref{subsec:optimal:policy}; the dashed line corresponds to a smaller weight for $w_{\text{sanitary}}$ (half) while the dotted line corresponds to a larger weight (twice). }
 	\label{fig:Sensi_wsanitary}
\end{figure}

\newpage

\subsection{Impact of the vaccination arrival date anticipation $\alpha$}\label{sec_sensi_alpha}

\begin{figure}[h]
	\begin{subfigure}{.33\columnwidth}
		\centering
		\includegraphics[width=\columnwidth]{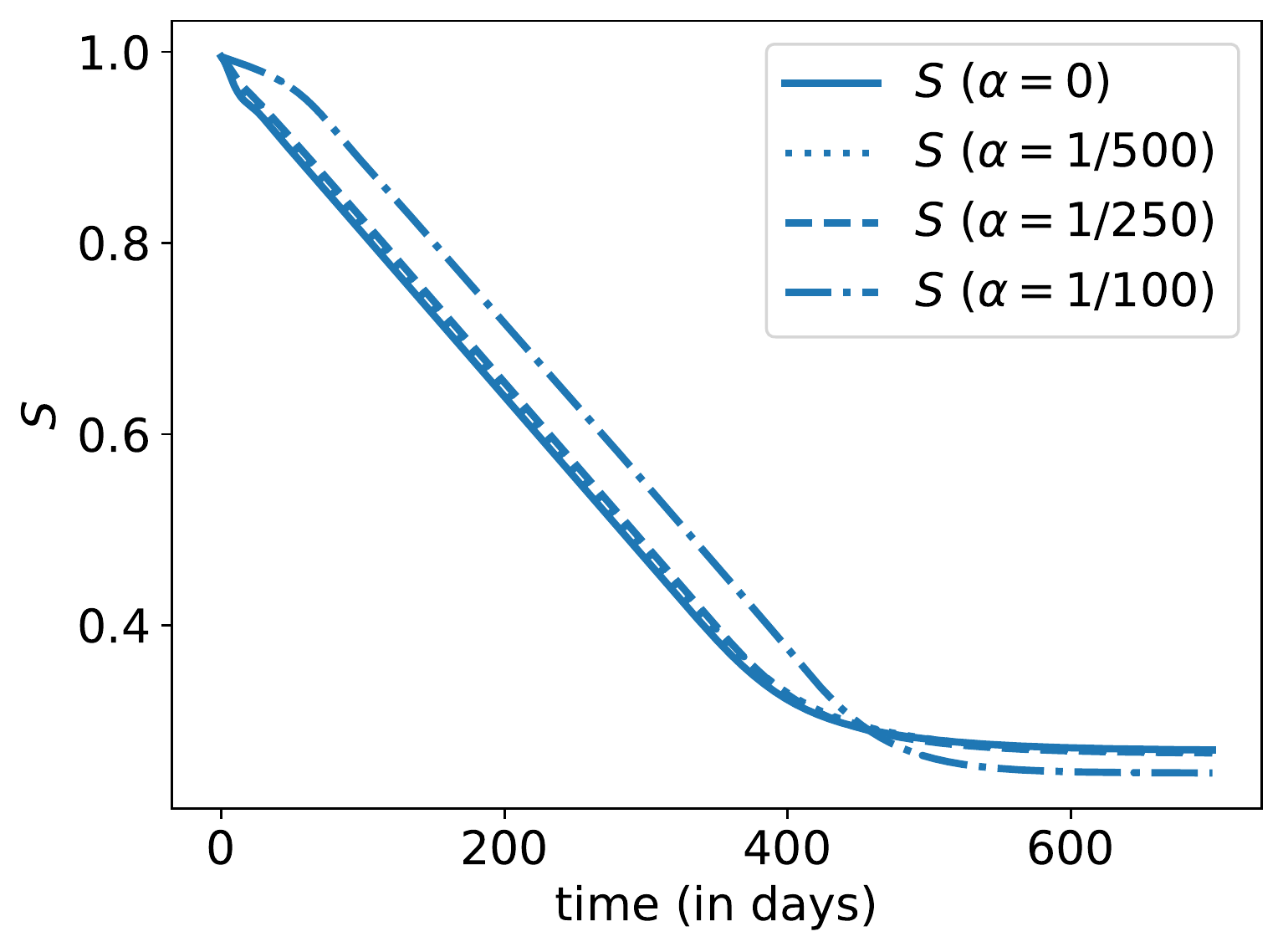}
	\end{subfigure}%
	\begin{subfigure}{.33\columnwidth}
		\centering 
		\includegraphics[width=\columnwidth]{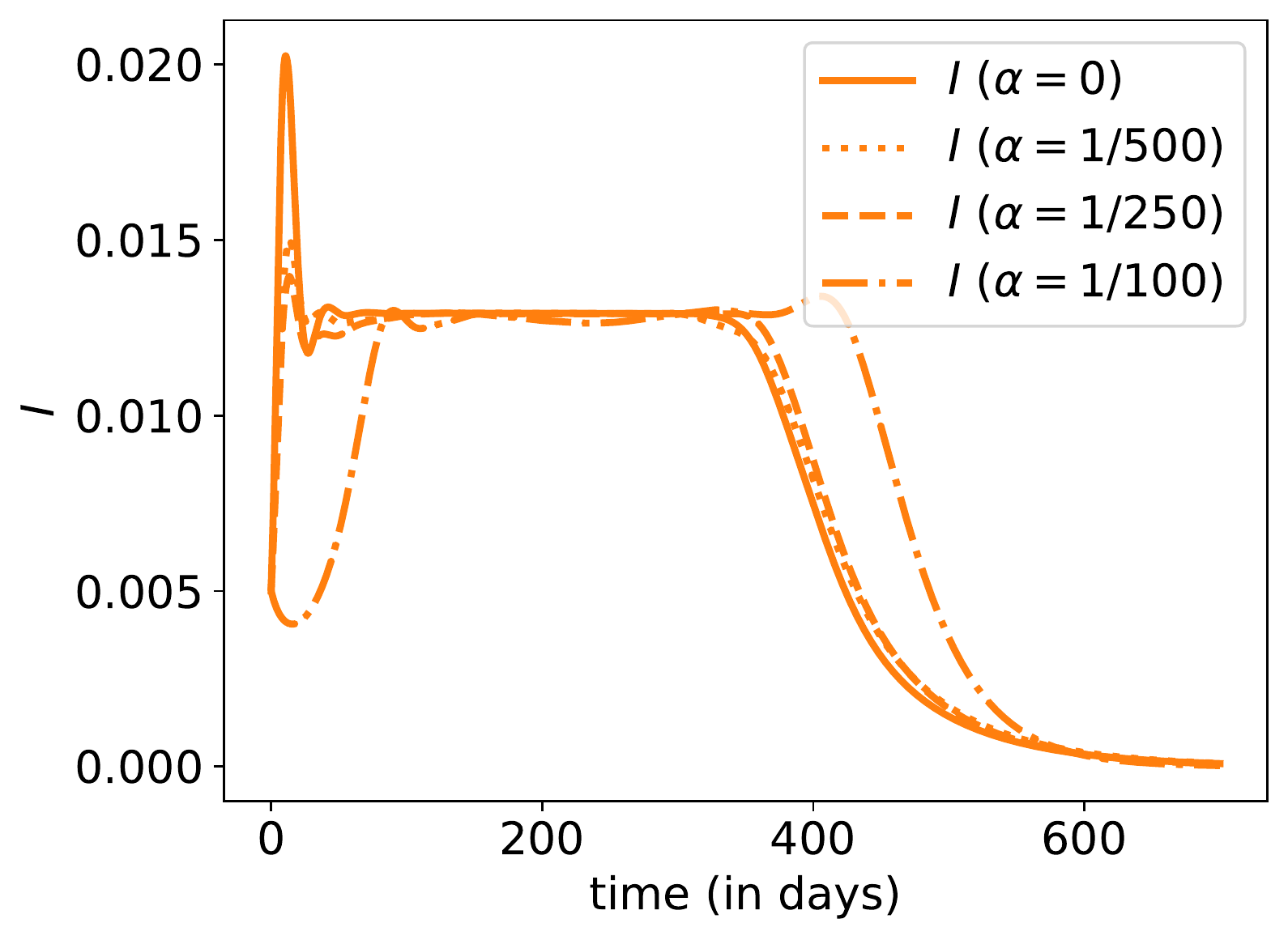}
	\end{subfigure}
	\begin{subfigure}{.33\columnwidth}
		\centering 
		\includegraphics[width=\columnwidth]{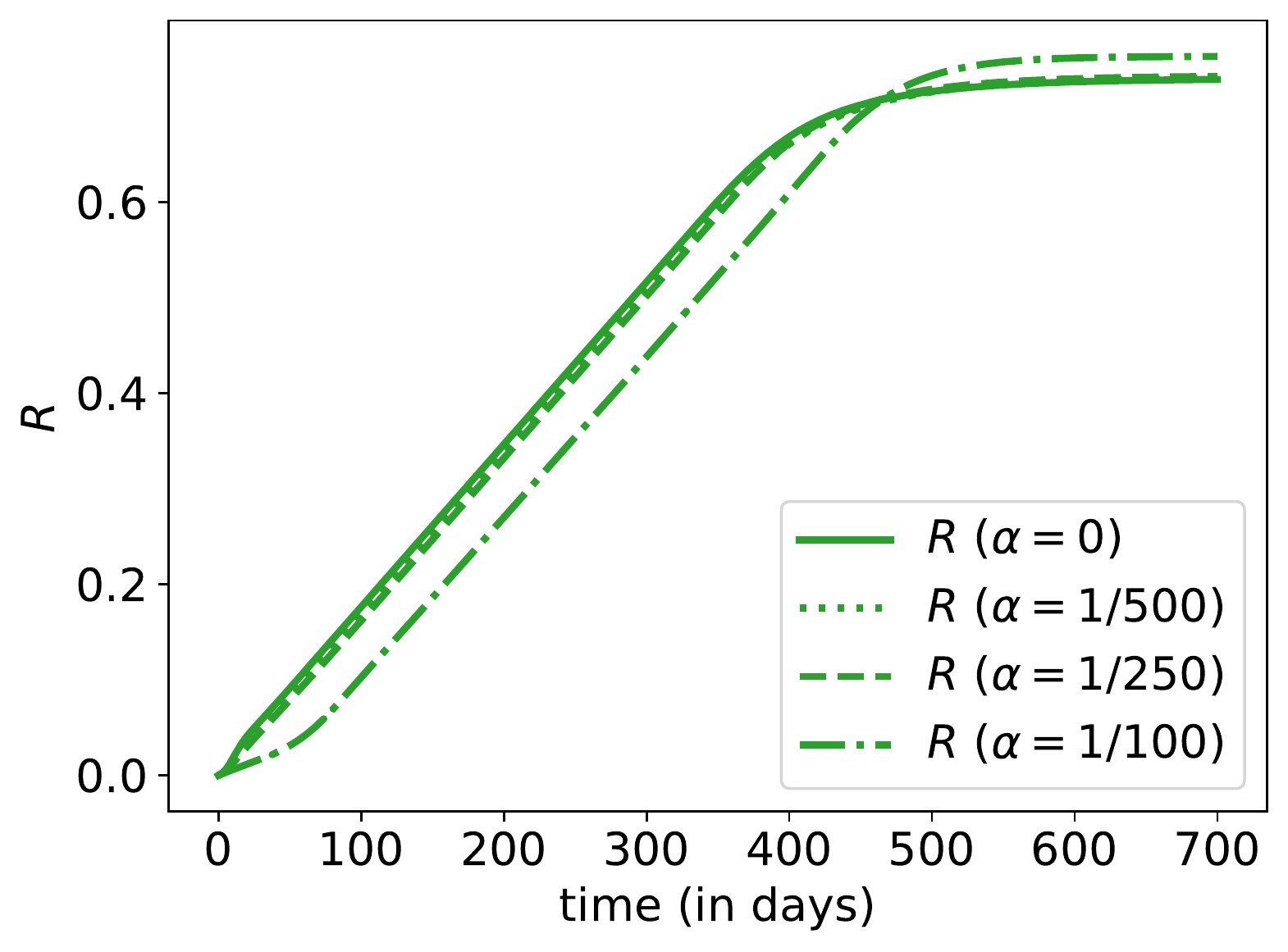}
	\end{subfigure}
	
	\begin{subfigure}{.33\columnwidth}
		\centering
		\includegraphics[width=\columnwidth]{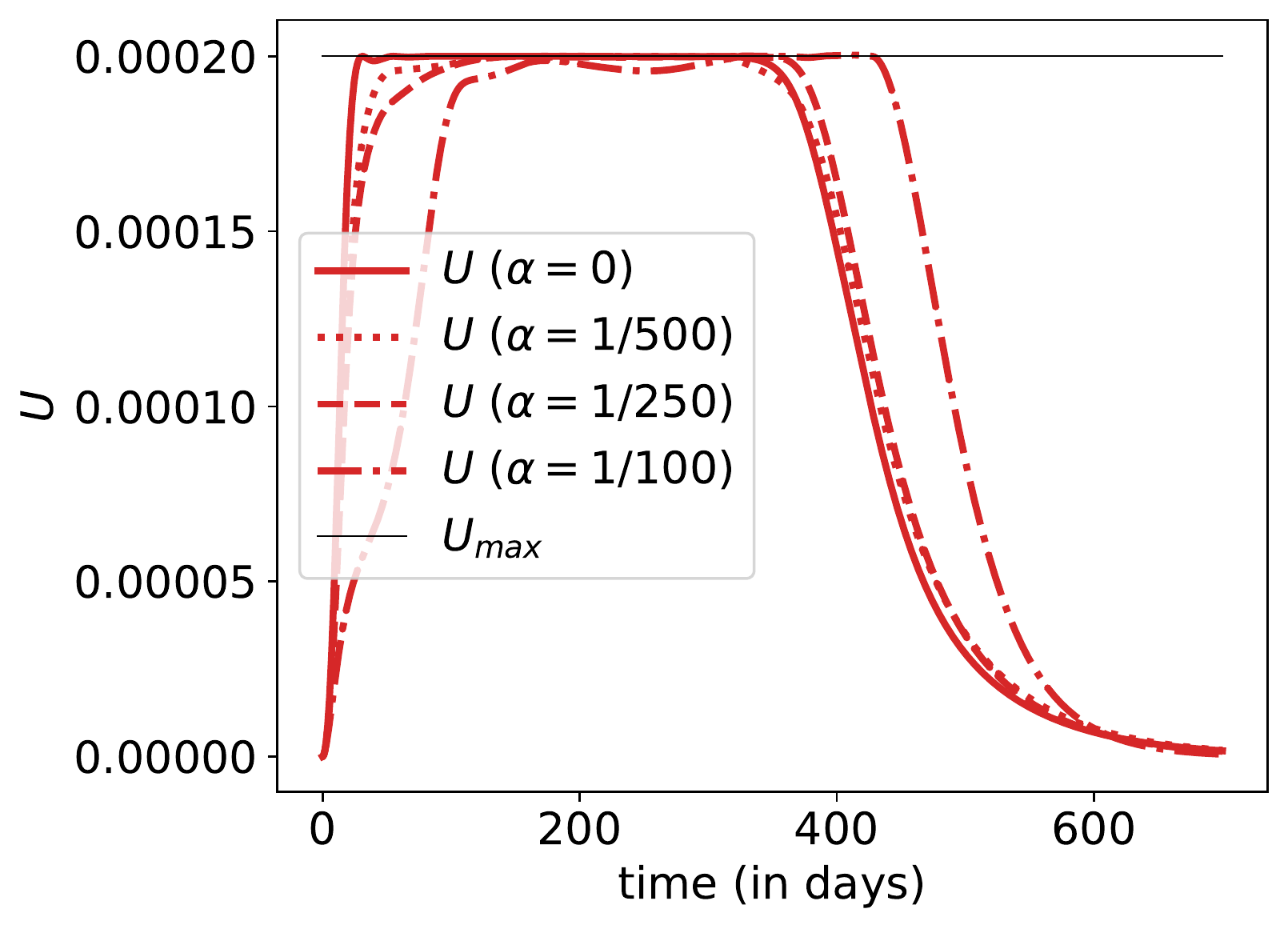}
	\end{subfigure}%
	\begin{subfigure}{.33\columnwidth}
		\centering 
		\includegraphics[width=\columnwidth]{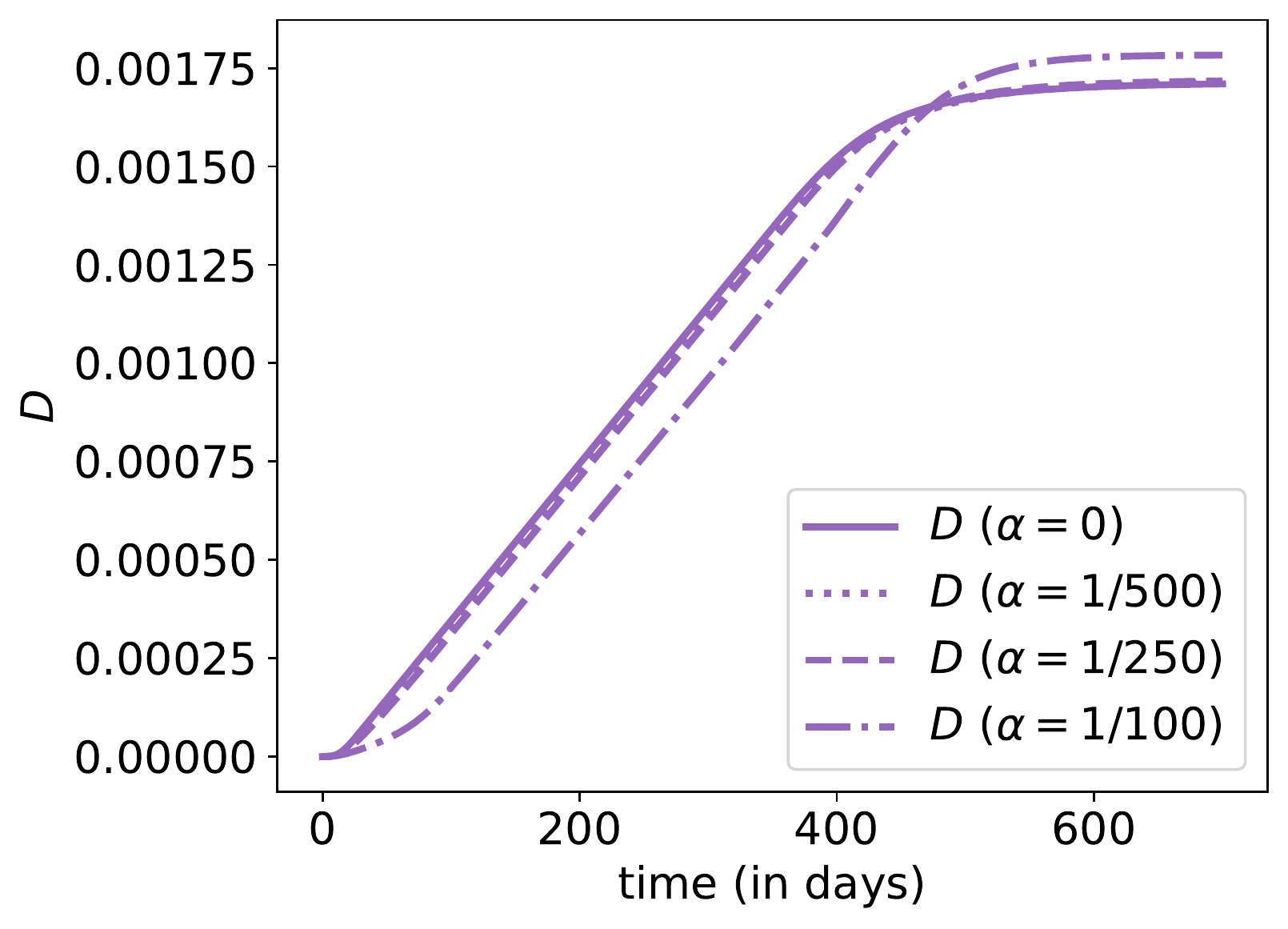}
	\end{subfigure}
	\begin{subfigure}{.33\columnwidth}
		\centering 
		\includegraphics[width=\columnwidth]{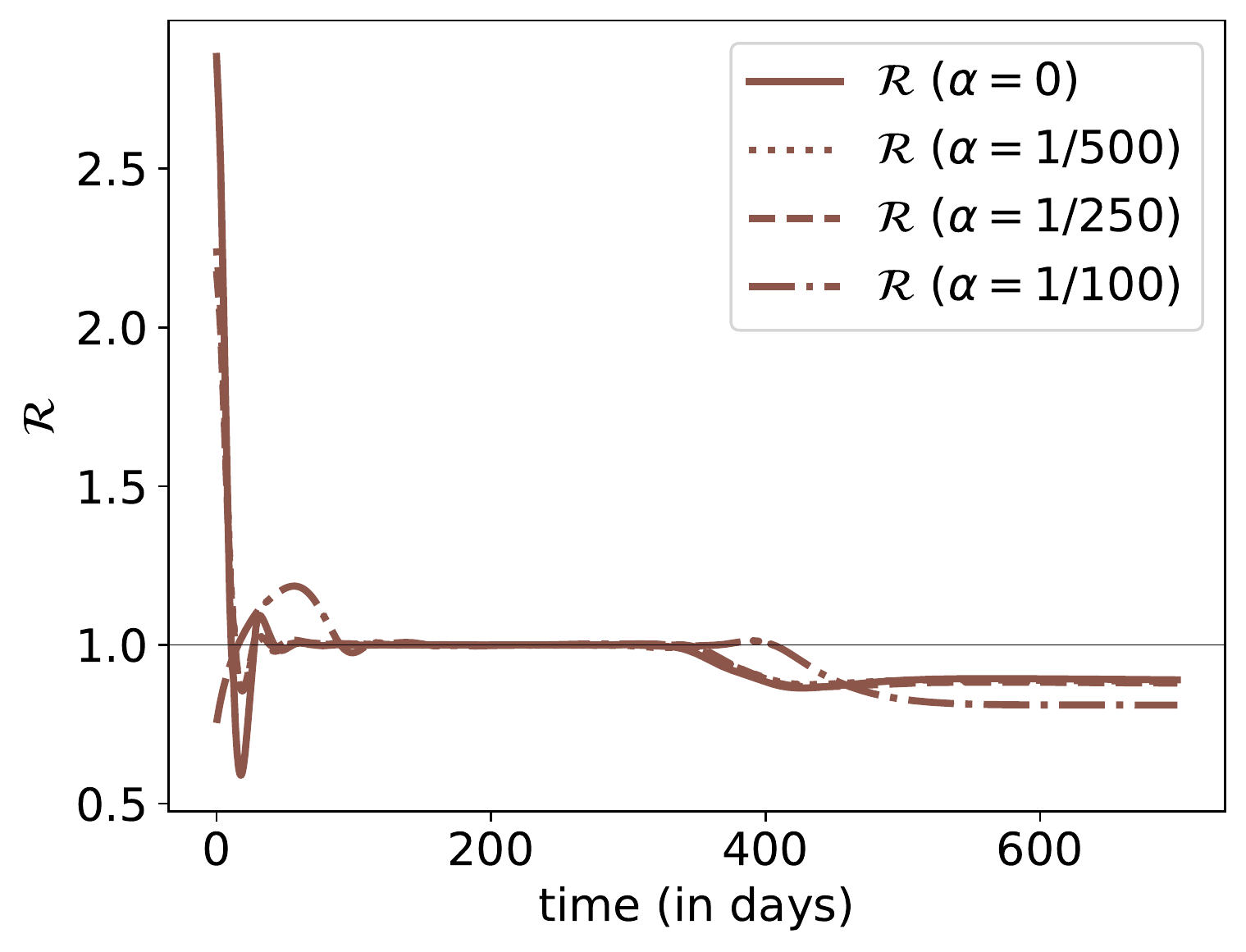}
	\end{subfigure}
	
	\begin{subfigure}{.33\columnwidth}
		\centering
		\includegraphics[width=\columnwidth]{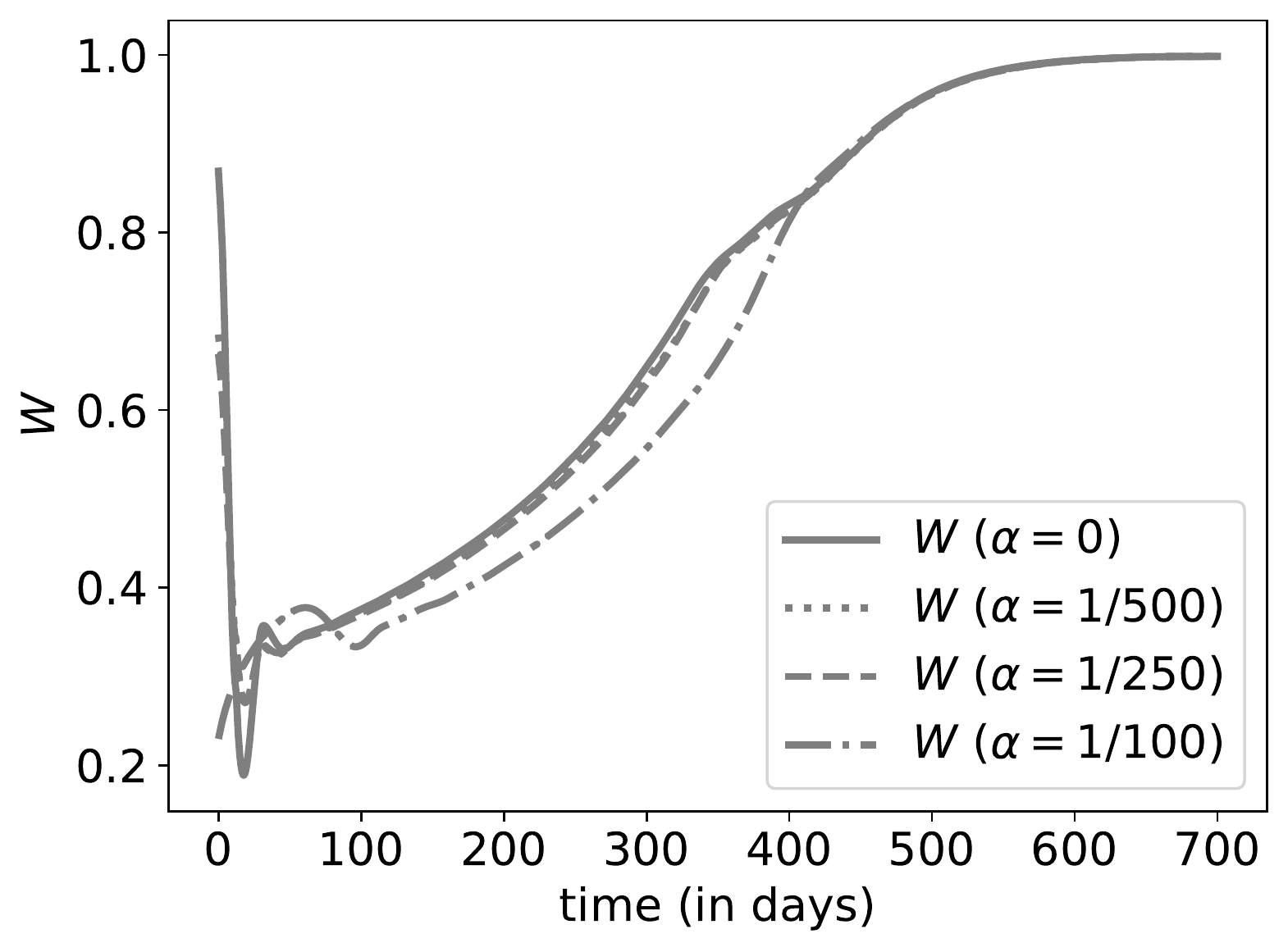}
	\end{subfigure}%
	\begin{subfigure}{.33\columnwidth}
		\centering 
		\includegraphics[width=\columnwidth]{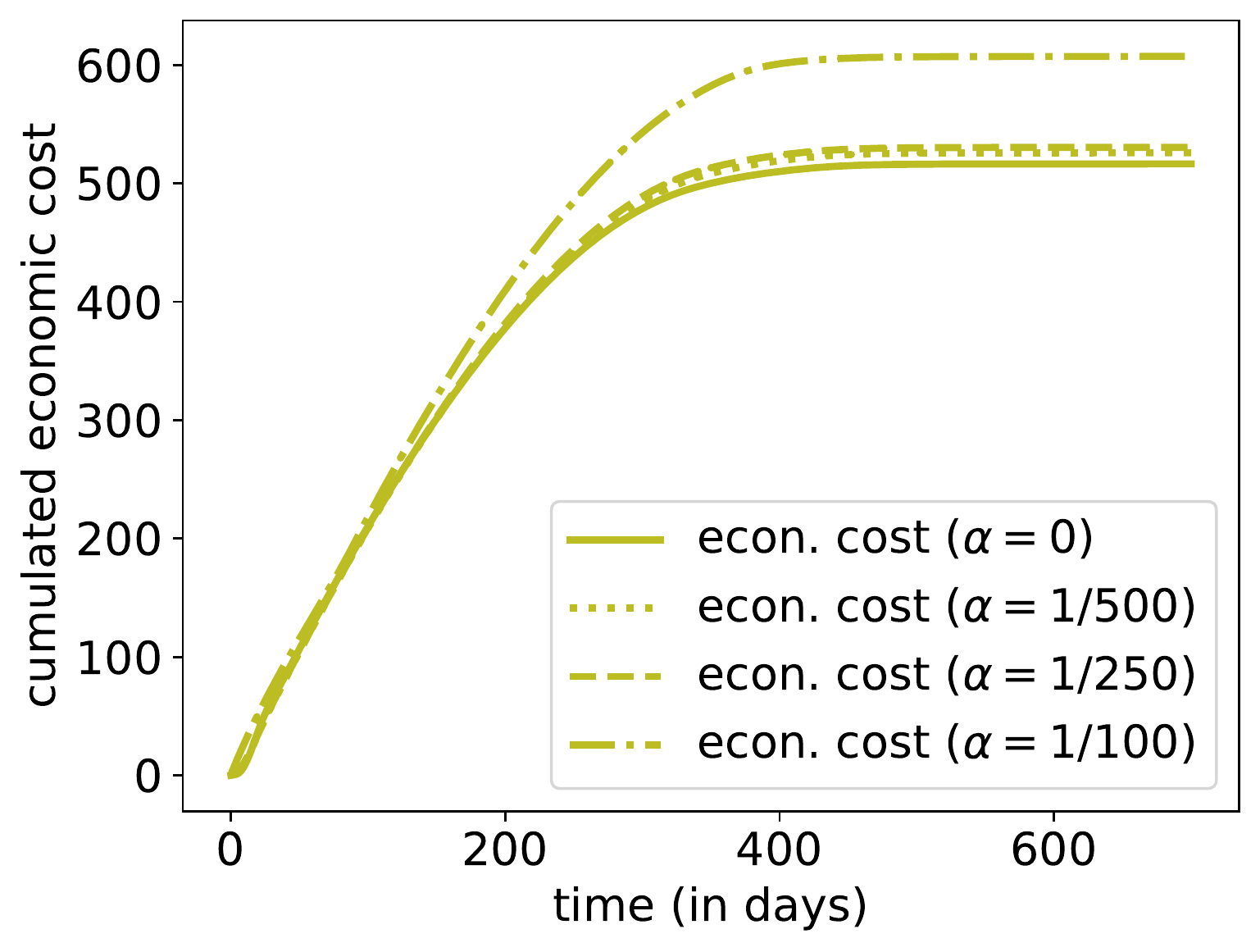}
	\end{subfigure}
	\begin{subfigure}{.33\columnwidth}
		\centering 
		\includegraphics[width=\columnwidth]{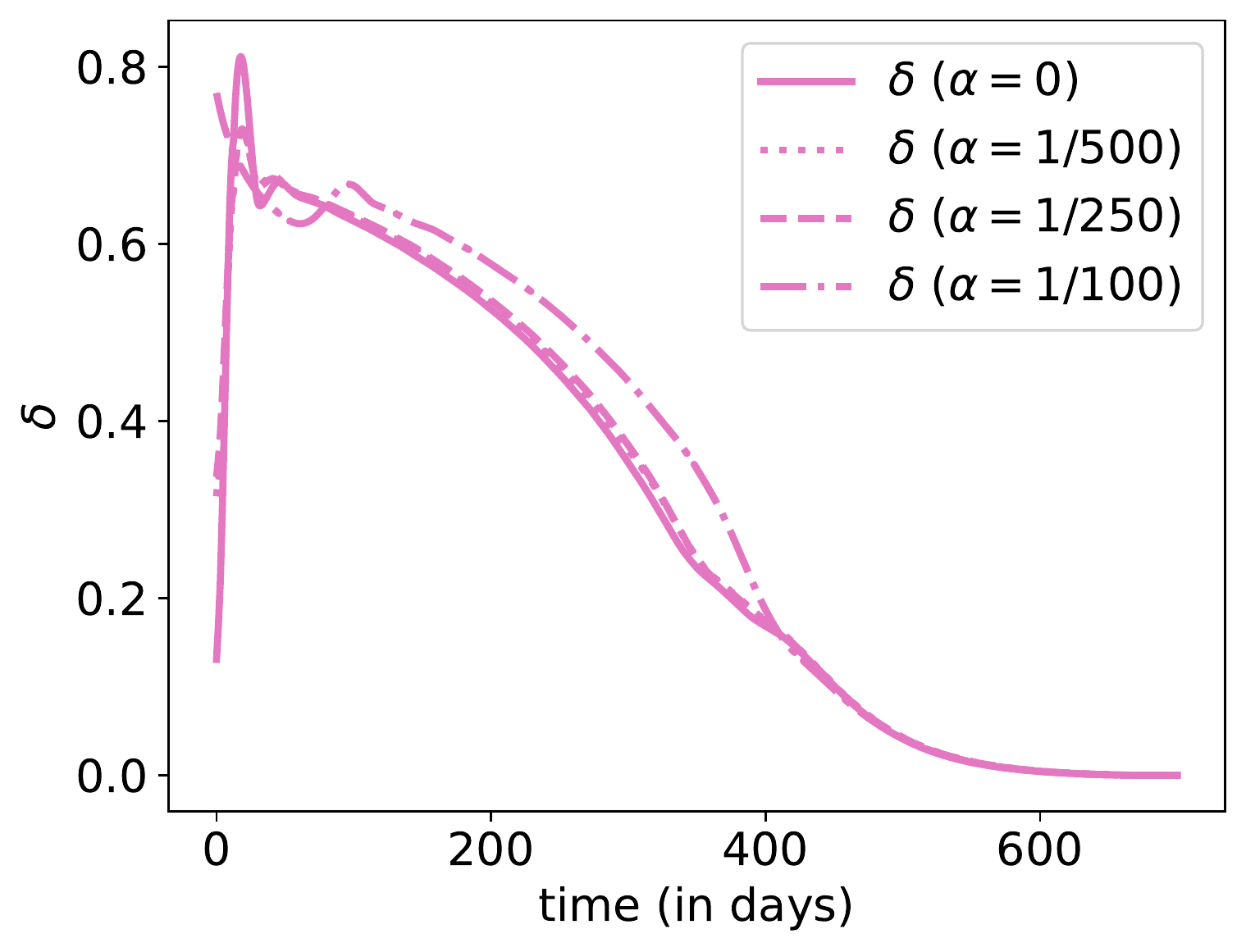}
	\end{subfigure}
	\caption{Evolution of states with optimal controls $(\delta,0,0)$ -- without any testing --  for three different values of $\alpha$, the discount rate that reflects the expected arrival time of a vaccine (and a cure). The plain line is the benchmark scenario discussed in Section \ref{subsec:optimal:policy}; the dotted line corresponds to $\alpha=1/500$, dashed line corresponds to $\alpha=1/250$ and the mixed line corresponds to $\alpha=1/100$.}
 	\label{fig:Sensi_alpha}
\end{figure}

\newpage

\subsection{Impact of additional constant effort in virologic detection $\lambda^1$}\label{sec_sensi_lambda1}

{
Here, we add virologic effort for a constant value of $\lambda^1$ and we compare the result of the optimization over $\delta$ for several values of $\lambda^1$. The optimal control problem for which we compute an approximate solution numerically is:
\begin{eqnarray}
 \inf_{\delta\in\tilde{\mathcal{A}}} \big\lbrace \tilde J_T(\delta,\lambda^1,0)\big\rbrace\;,
\end{eqnarray}
with $\tilde{\mathcal{A}}$ the set of measurable functions from $[0,T]$ to $[0,1]$.
}

\begin{figure}[h]
	\begin{subfigure}{.33\columnwidth}
		\centering
		\includegraphics[width=\columnwidth]{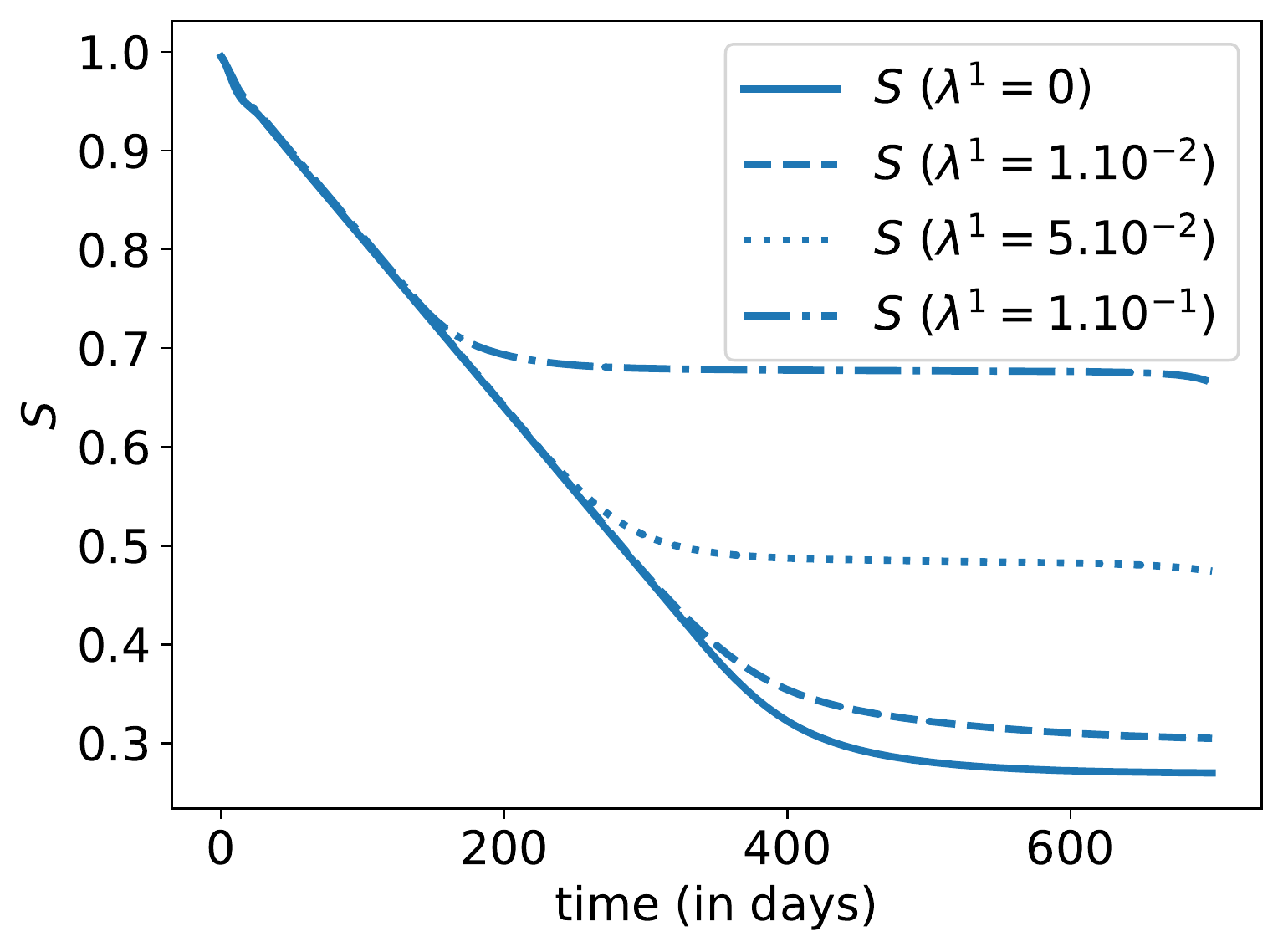}
	\end{subfigure}%
	\begin{subfigure}{.33\columnwidth}
		\centering 
		\includegraphics[width=\columnwidth]{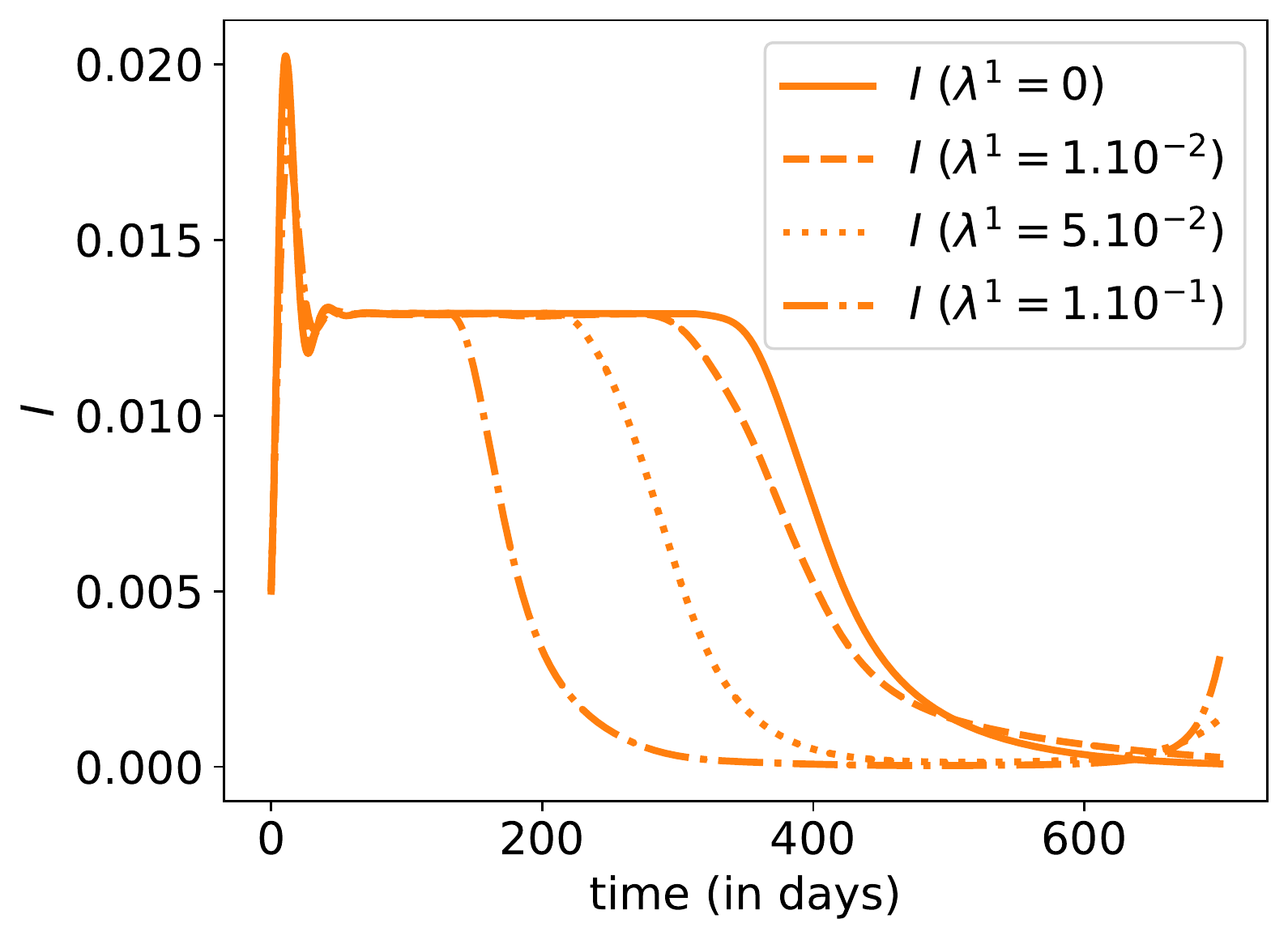}
	\end{subfigure}
	\begin{subfigure}{.33\columnwidth}
		\centering 
		\includegraphics[width=\columnwidth]{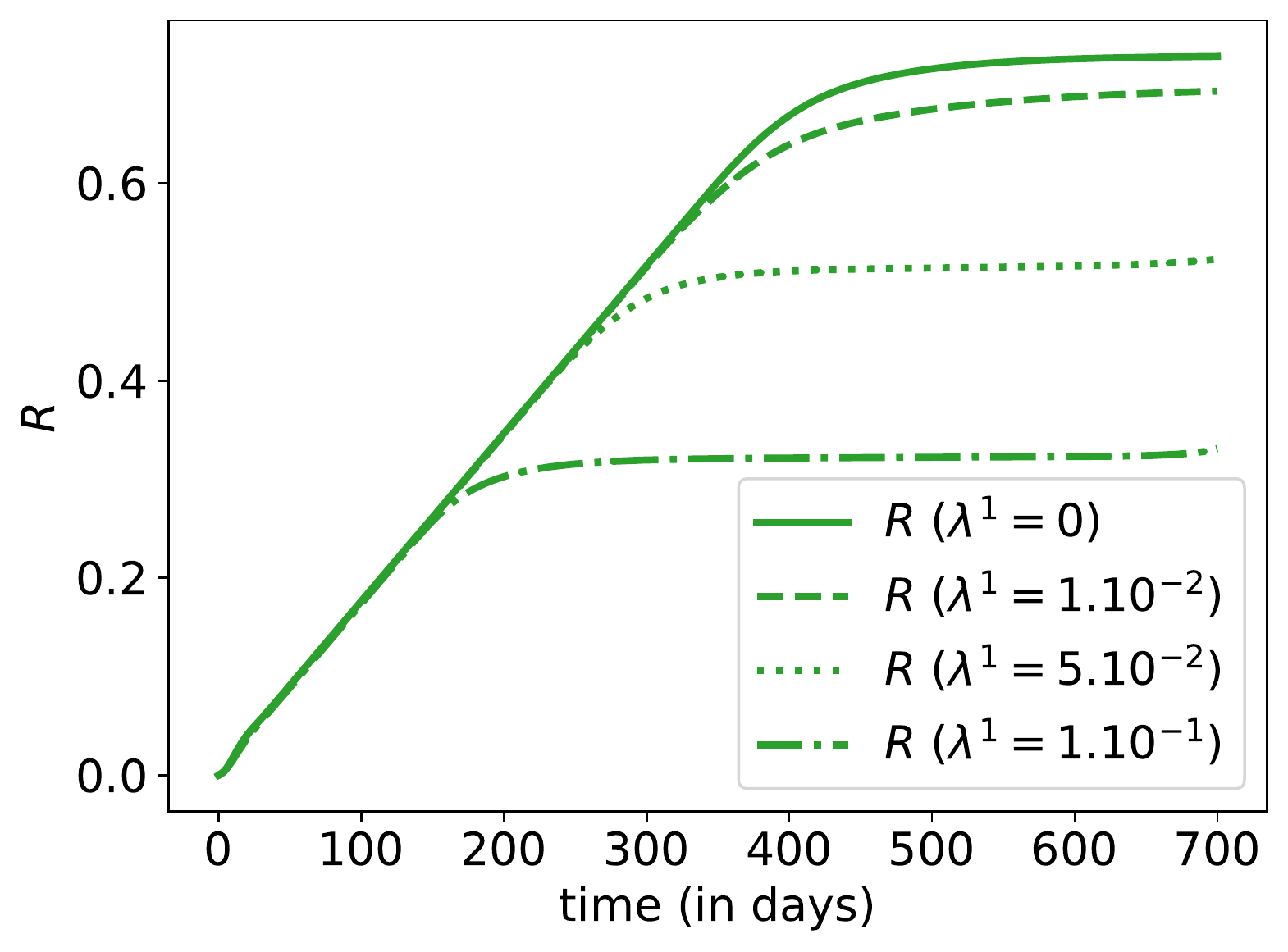}
	\end{subfigure}
	
	\begin{subfigure}{.33\columnwidth}
		\centering
		\includegraphics[width=\columnwidth]{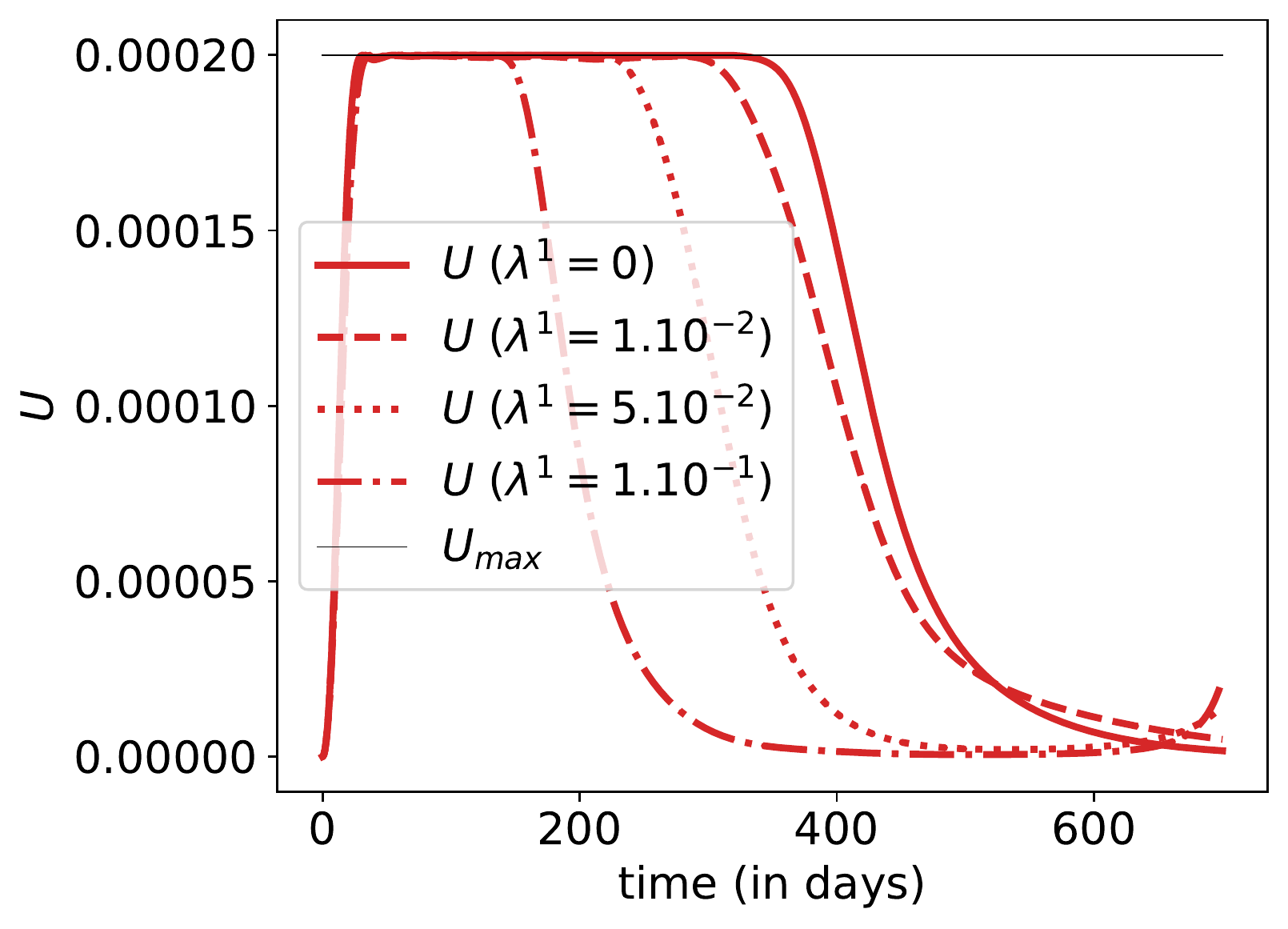}
	\end{subfigure}%
	\begin{subfigure}{.33\columnwidth}
		\centering 
		\includegraphics[width=\columnwidth]{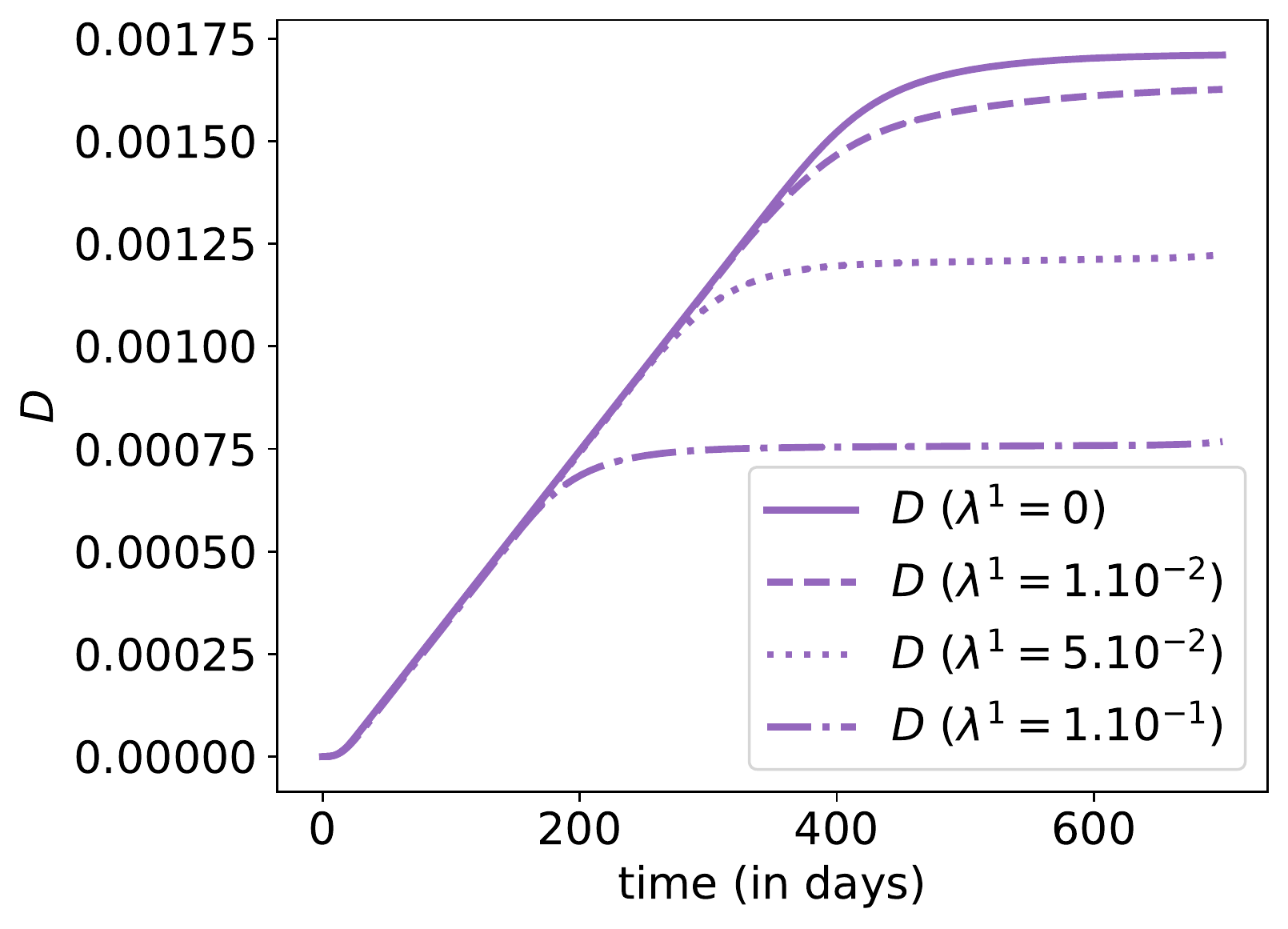}
	\end{subfigure}
	\begin{subfigure}{.33\columnwidth}
		\centering 
		\includegraphics[width=\columnwidth]{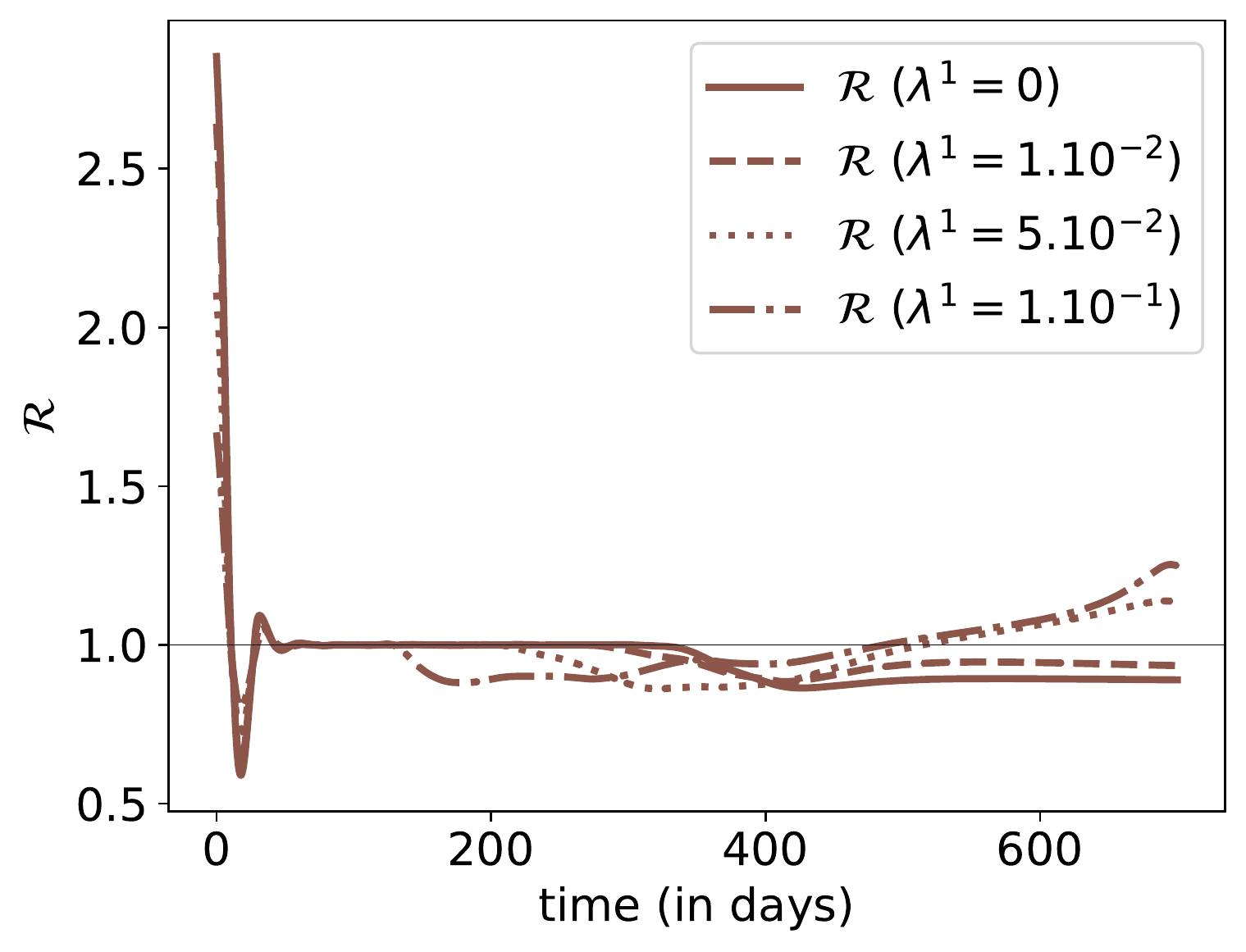}
	\end{subfigure}
	
	\begin{subfigure}{.33\columnwidth}
		\centering
		\includegraphics[width=\columnwidth]{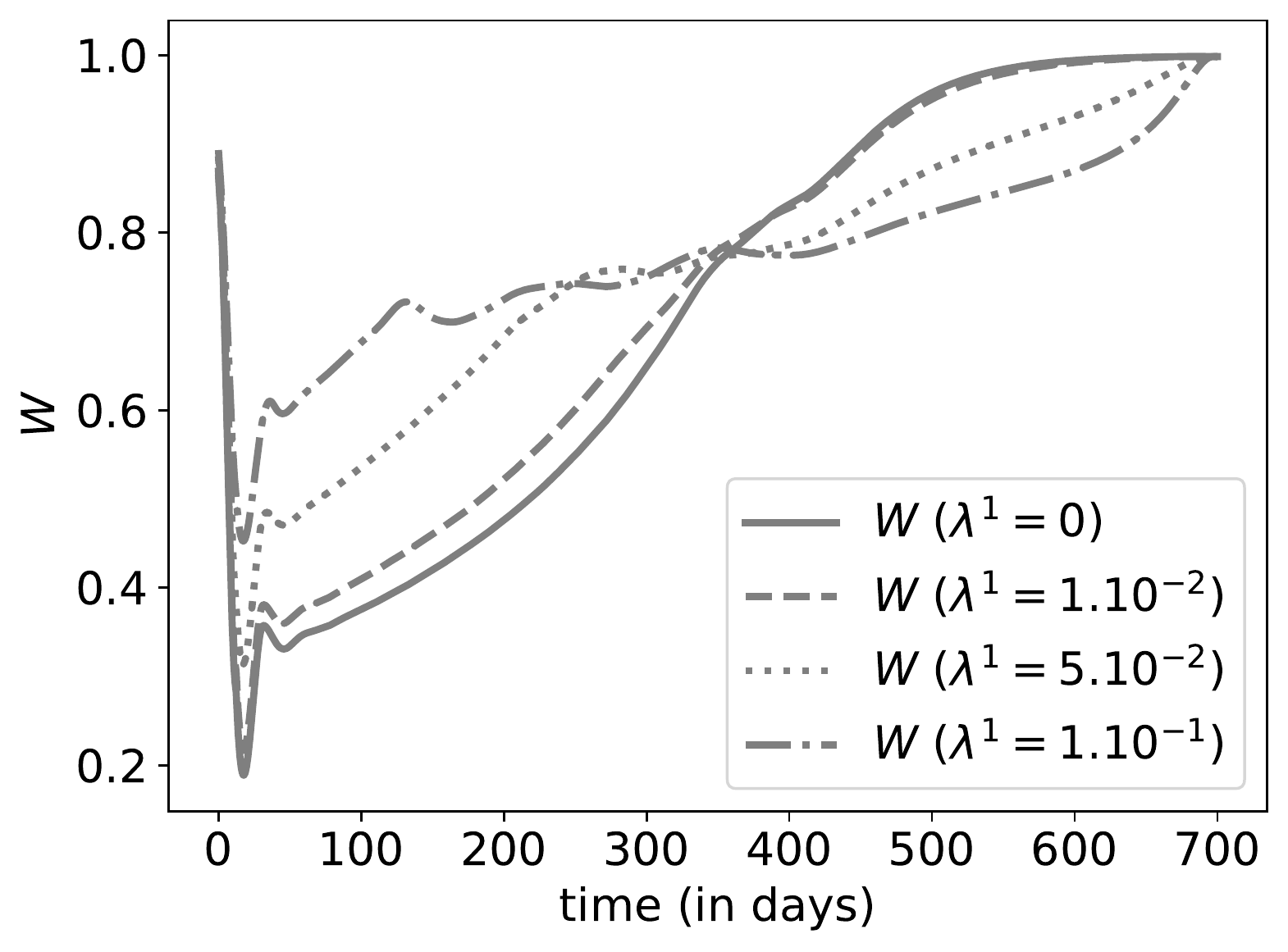}
	\end{subfigure}%
	\begin{subfigure}{.33\columnwidth}
		\centering 
		\includegraphics[width=\columnwidth]{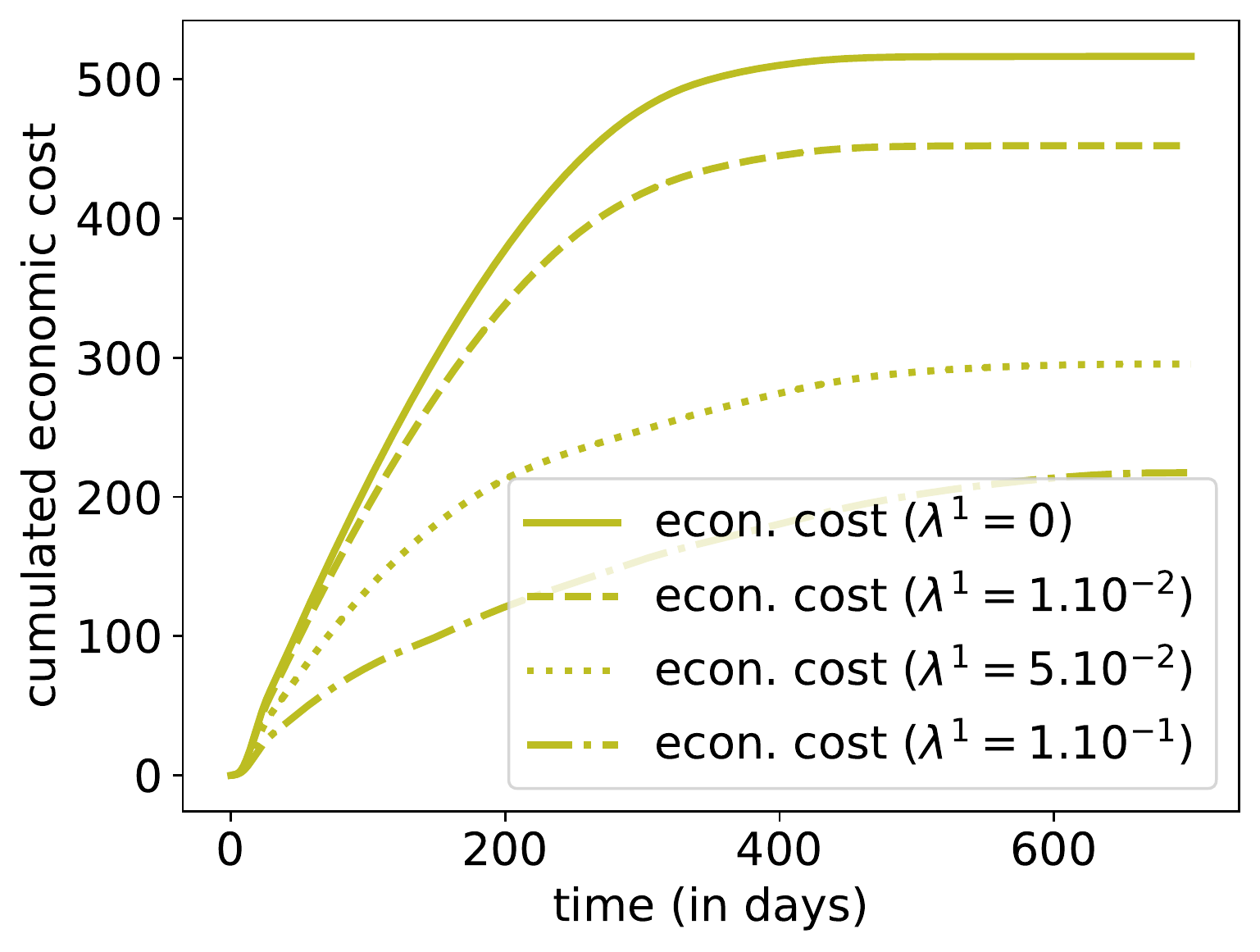}
	\end{subfigure}
	\begin{subfigure}{.33\columnwidth}
		\centering 
		\includegraphics[width=\columnwidth]{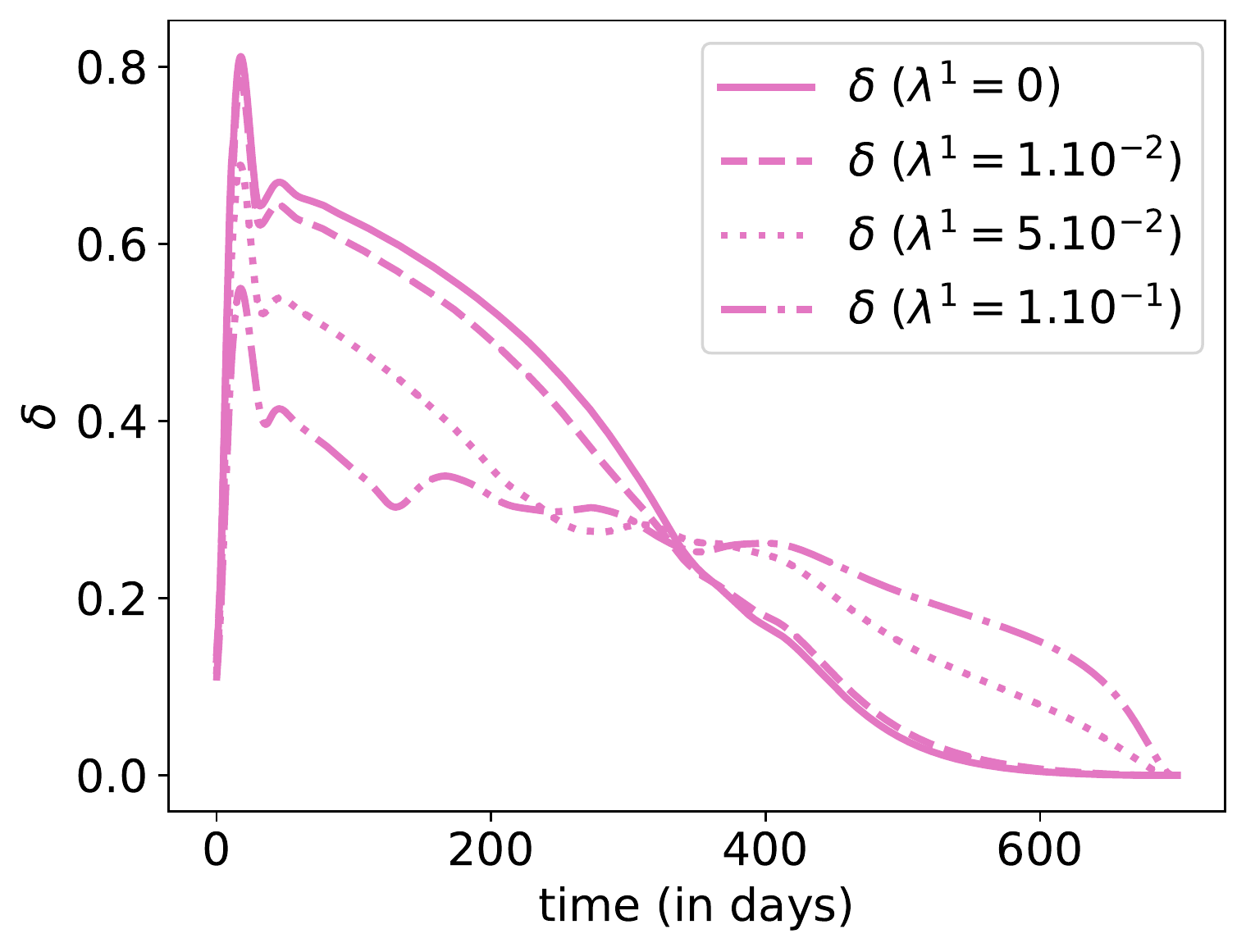}
	\end{subfigure}
	
	\begin{subfigure}{.33\columnwidth}
		\centering
		\includegraphics[width=\columnwidth]{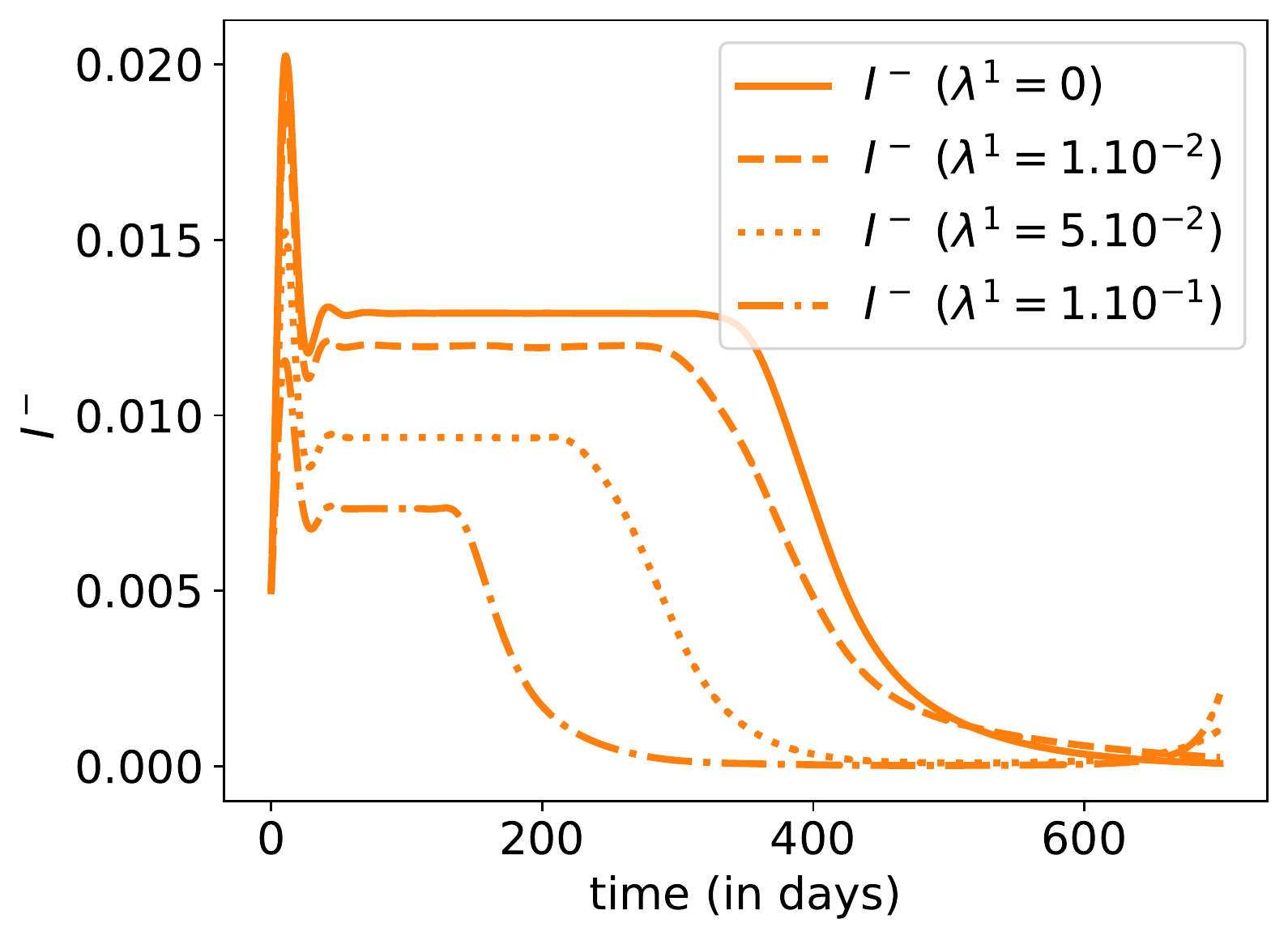}
	\end{subfigure}%
	\begin{subfigure}{.33\columnwidth}
		\centering 
		\includegraphics[width=\columnwidth]{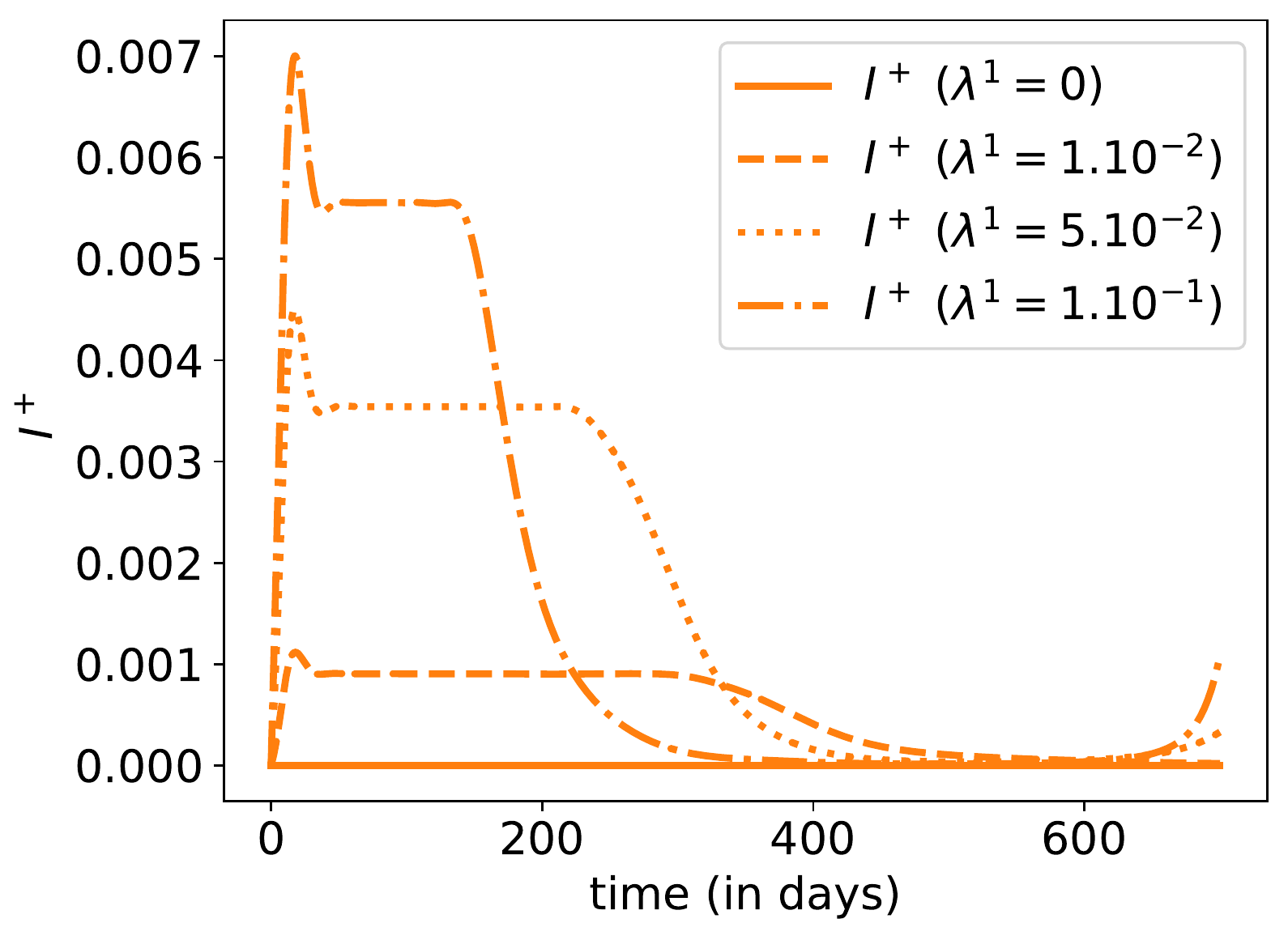}
	\end{subfigure}
	\begin{subfigure}{.33\columnwidth}
		\centering 
		\includegraphics[width=\columnwidth]{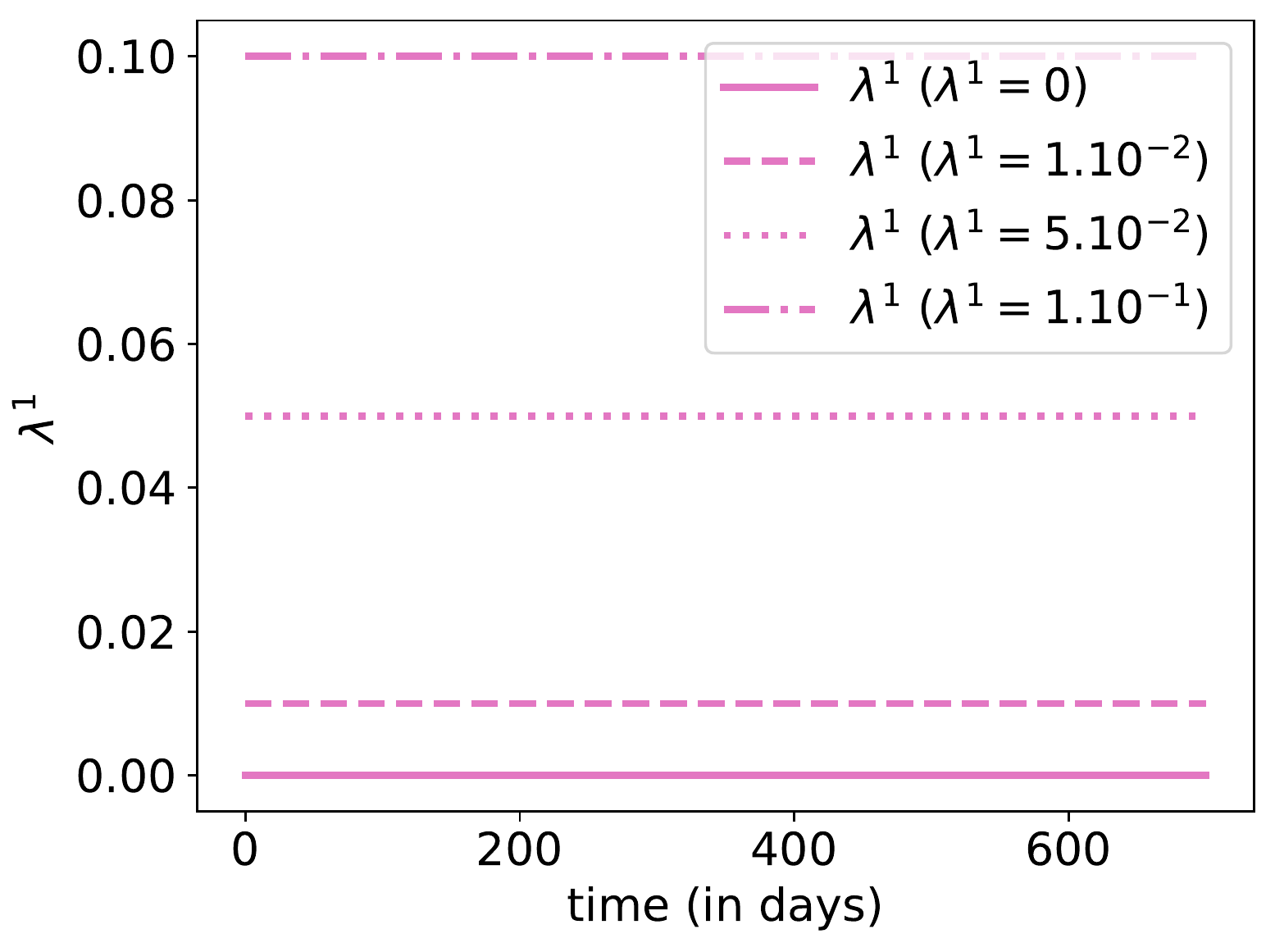}
	\end{subfigure}
	\caption{Evolution of states with optimal control $\delta$ for three different values of $\lambda^1$, or $(\delta^*,\lambda^1,0)$. The plain line is the benchmark scenario discussed in Section \ref{subsec:optimal:policy}, with $\lambda^1=0$; and then three scenarios, with $\lambda^1=1\%$~(dashed line),  $\lambda^1=5\%$ (dotted line) and $\lambda^1=10\%$ (mixed line).}
 	\label{fig:Sensi_lambda1B5}
\end{figure}

\newpage

\subsection{Impact of additional rising effort in virologic detection $\lambda^1$}\label{sec_sensi_lambda1_2}

{
Here, we add virologic effort for a value of $\lambda^1$ which increases linearly in time, $\tilde \lambda^1: t \mapsto 0.2 t/T$. The optimal control problem for which we compute an approximate solution numerically is:
\begin{eqnarray}
 \inf_{\delta\in\tilde{\mathcal{A}}} \big\lbrace \tilde J_T(\delta,\tilde \lambda^1,0)\big\rbrace\;,
\end{eqnarray}
with $\tilde{\mathcal{A}}$ the set of measurable functions from $[0,T]$ to $[0,1]$.
}

\begin{figure}[h]
	\begin{subfigure}{.33\columnwidth}
		\centering
		\includegraphics[width=\columnwidth]{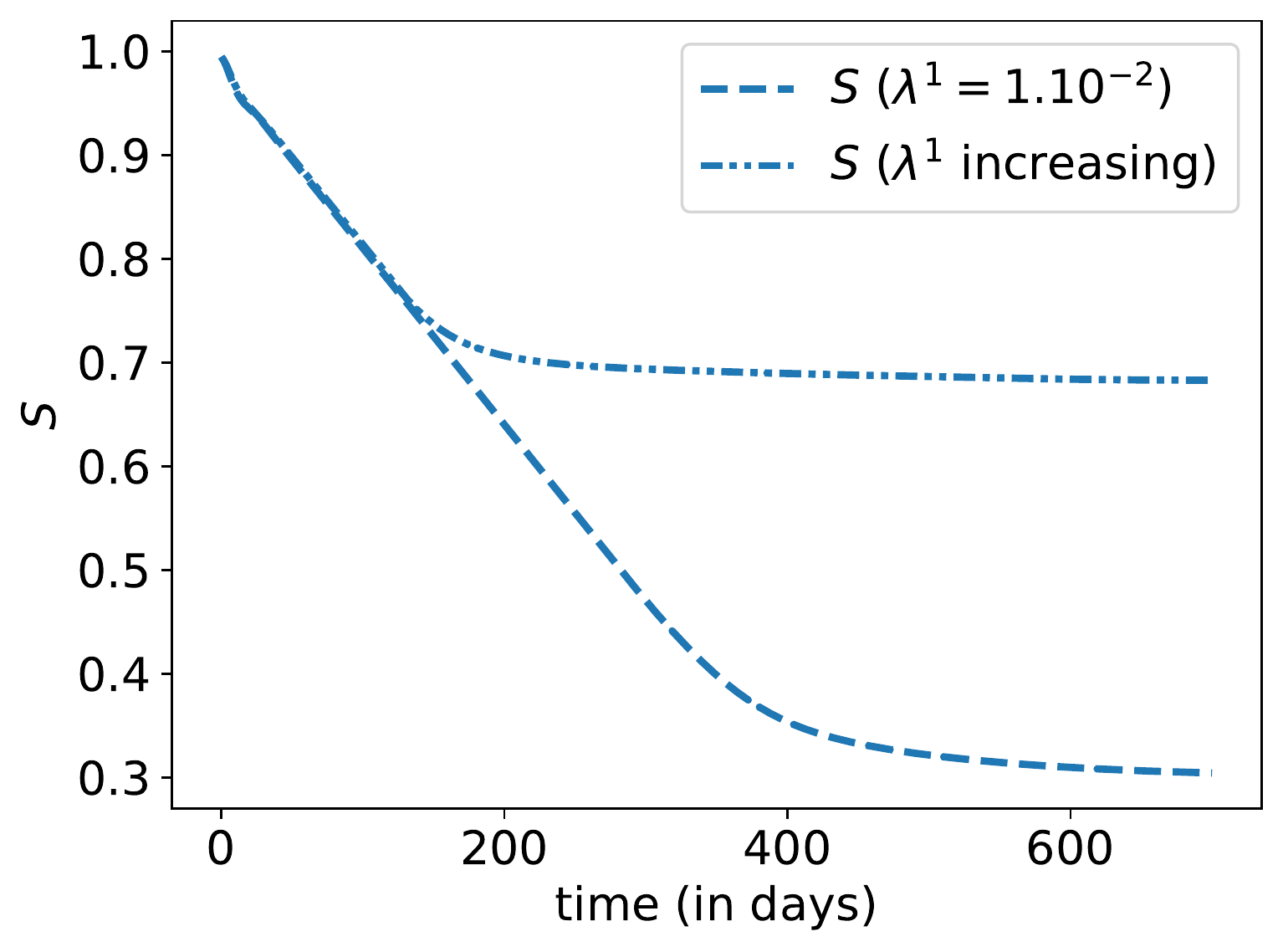}
	\end{subfigure}%
	\begin{subfigure}{.33\columnwidth}
		\centering 
		\includegraphics[width=\columnwidth]{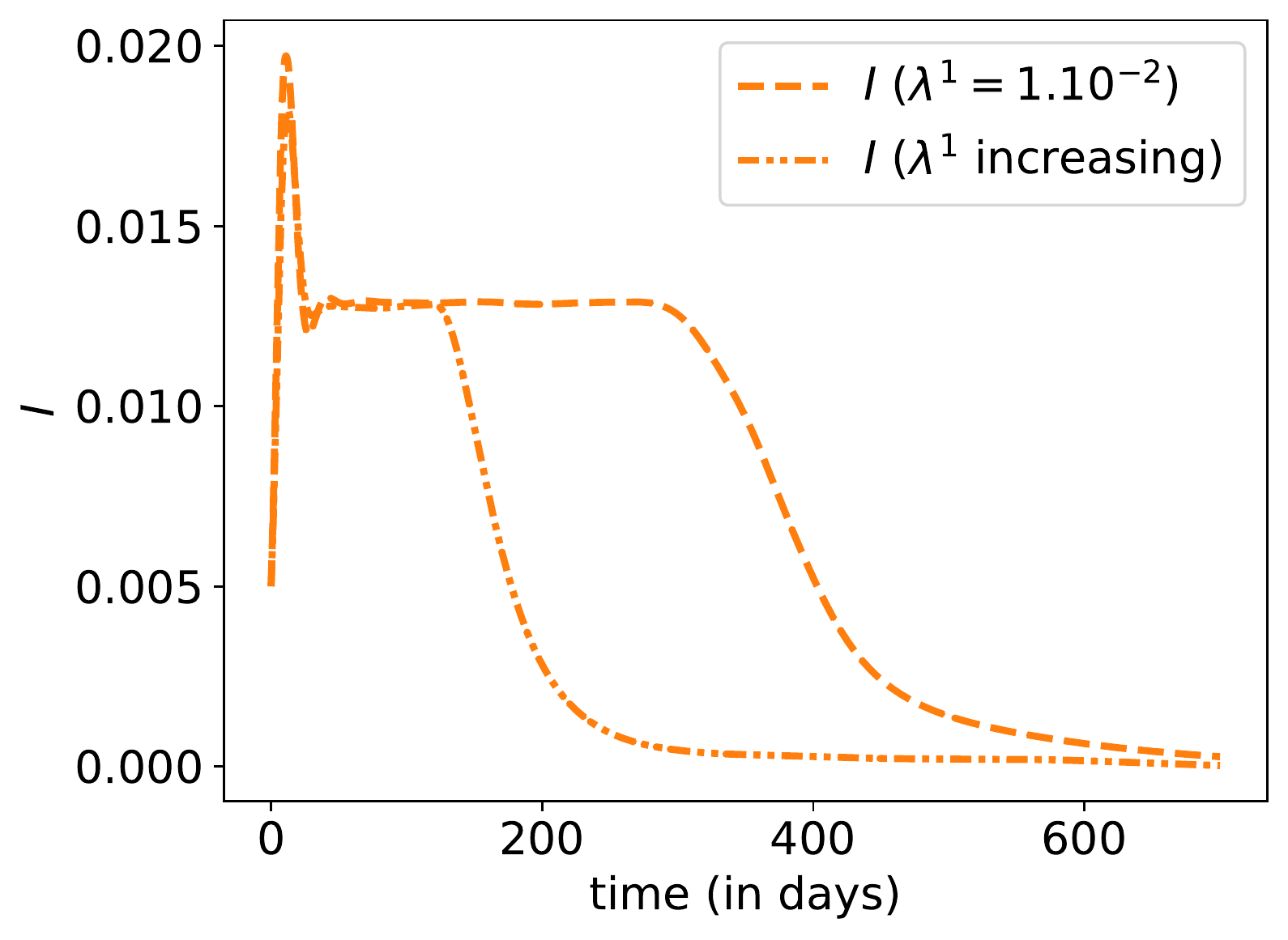}
	\end{subfigure}
	\begin{subfigure}{.33\columnwidth}
		\centering 
		\includegraphics[width=\columnwidth]{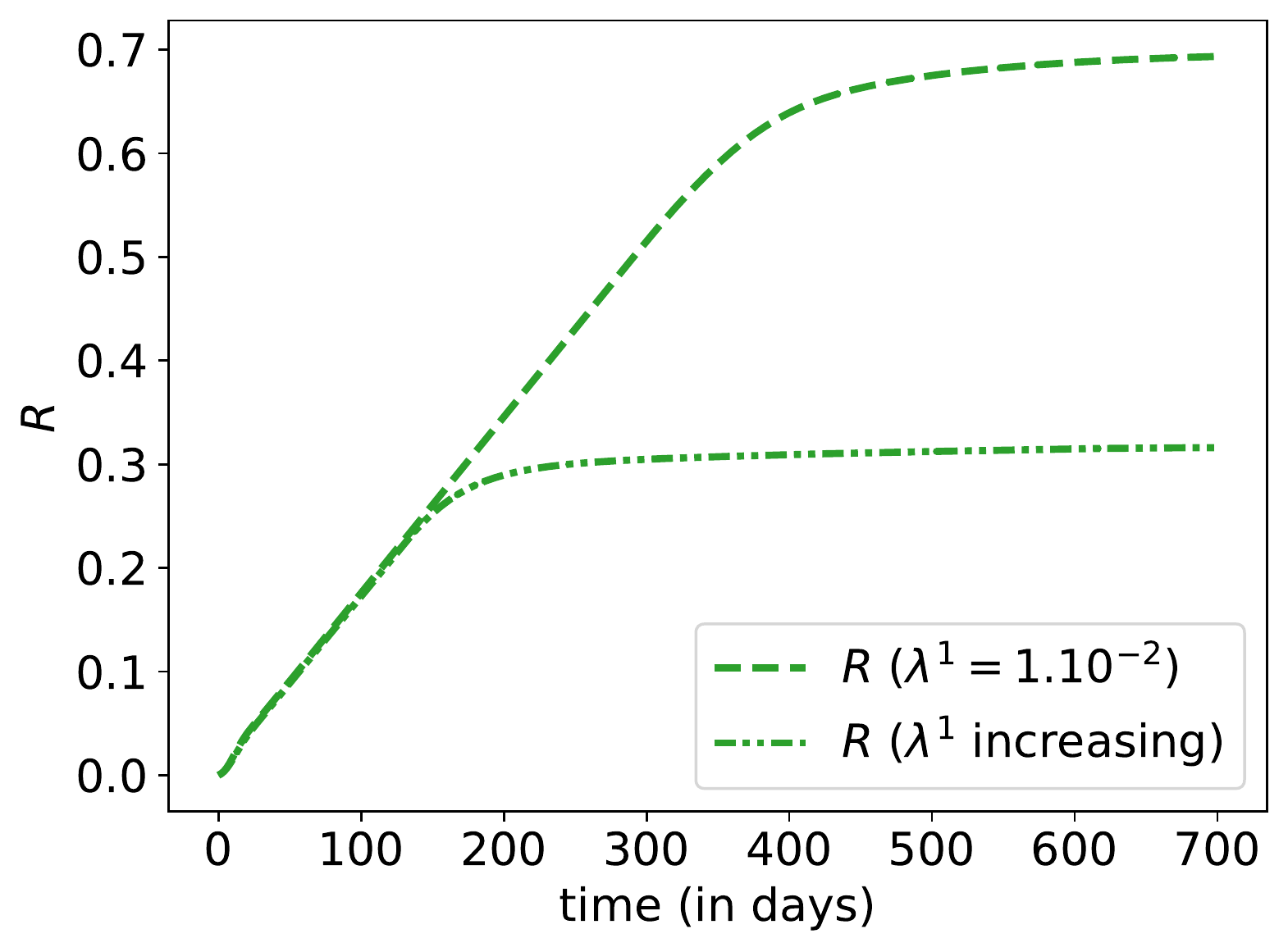}
	\end{subfigure}
	
	\begin{subfigure}{.33\columnwidth}
		\centering
		\includegraphics[width=\columnwidth]{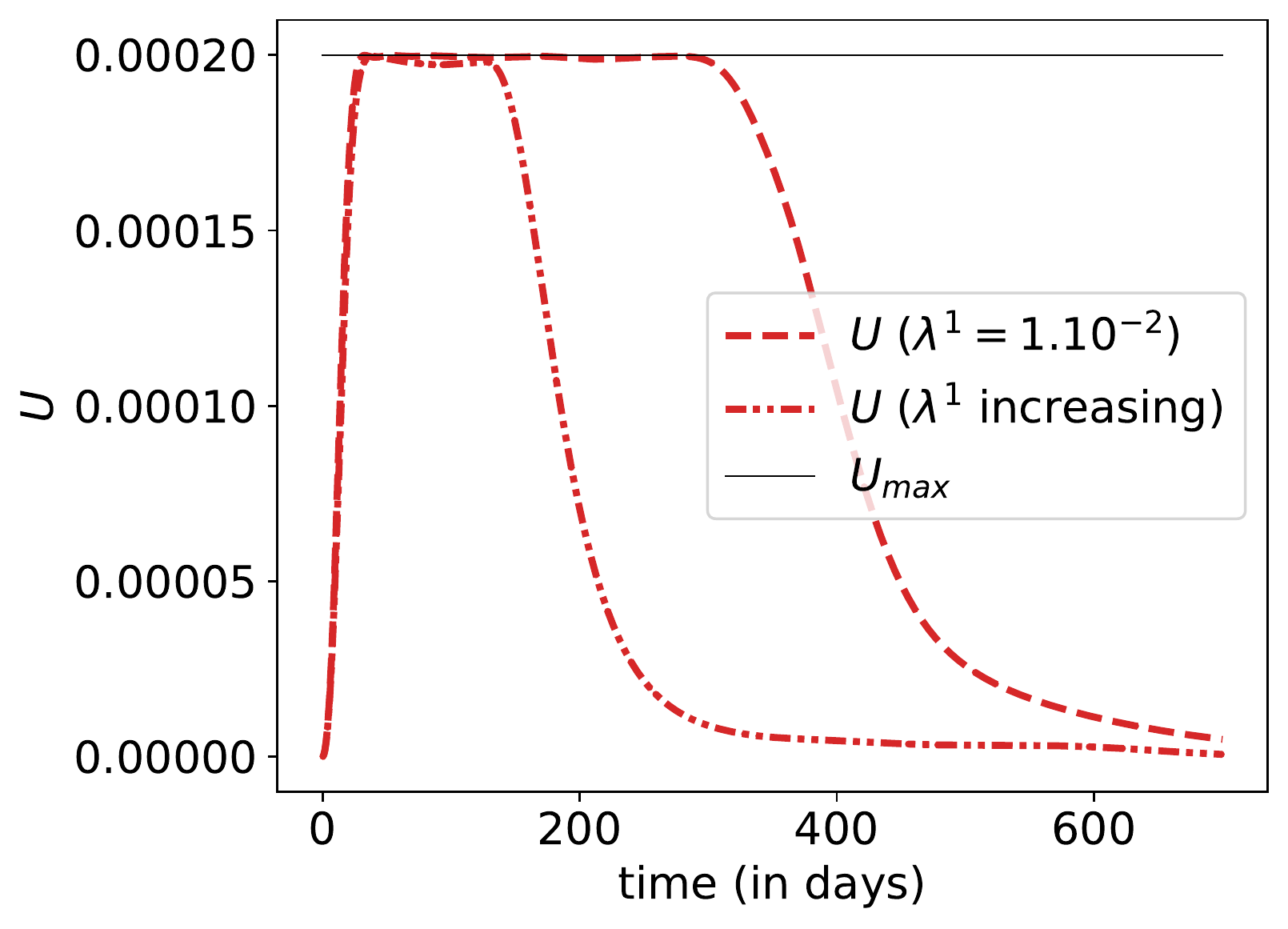}
	\end{subfigure}%
	\begin{subfigure}{.33\columnwidth}
		\centering 
		\includegraphics[width=\columnwidth]{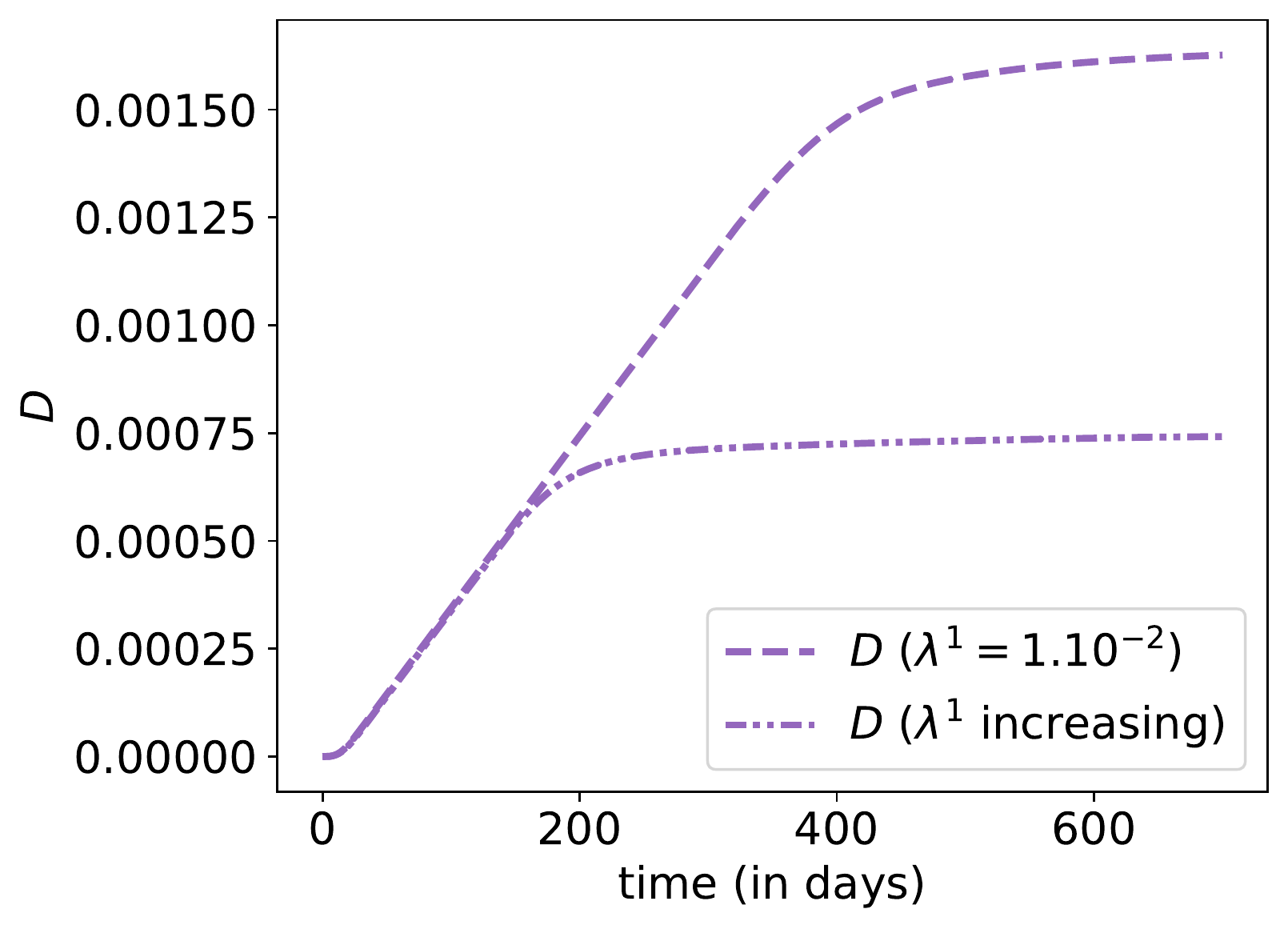}
	\end{subfigure}
	\begin{subfigure}{.33\columnwidth}
		\centering 
		\includegraphics[width=\columnwidth]{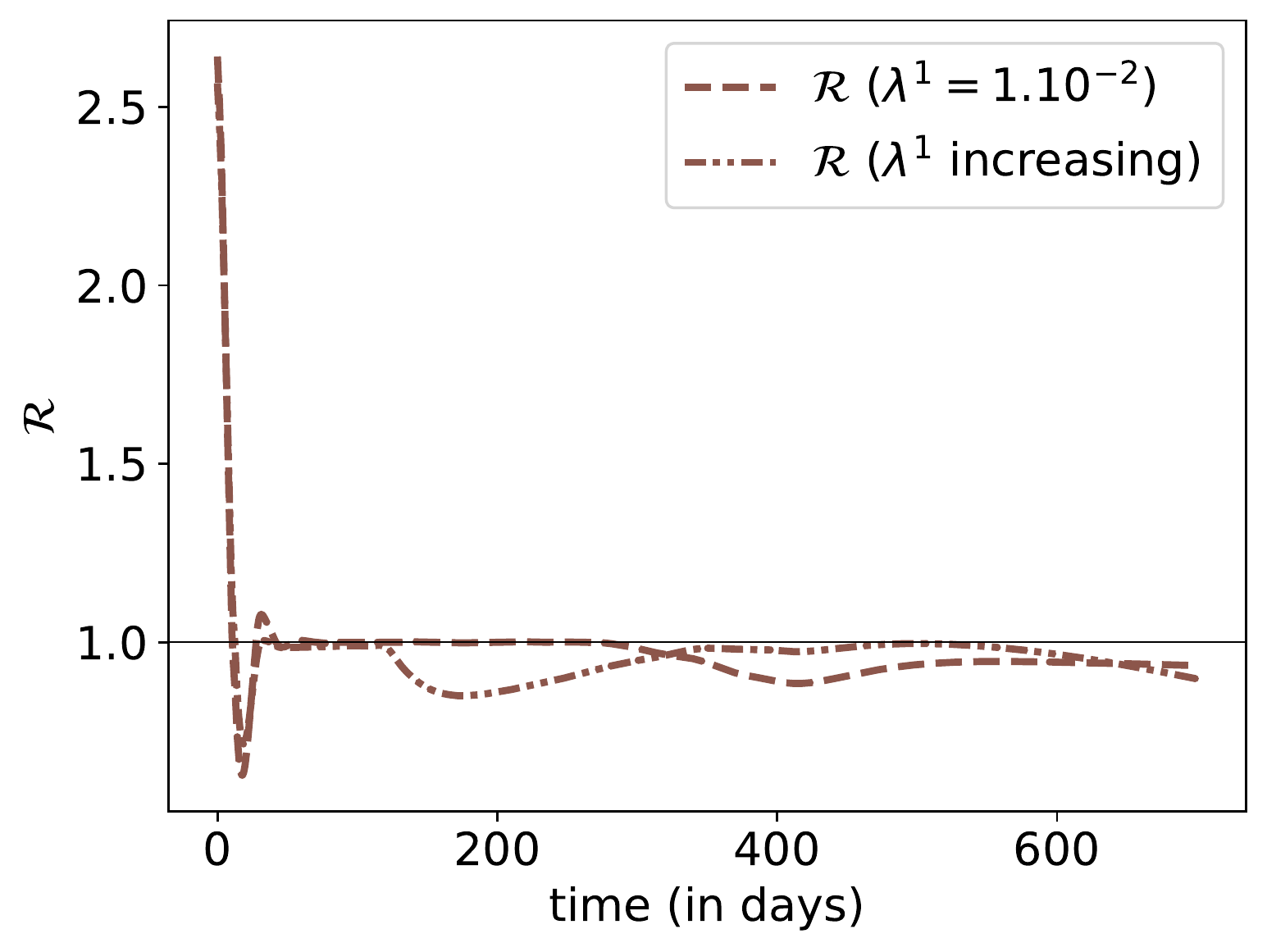}
	\end{subfigure}
	
	\begin{subfigure}{.33\columnwidth}
		\centering
		\includegraphics[width=\columnwidth]{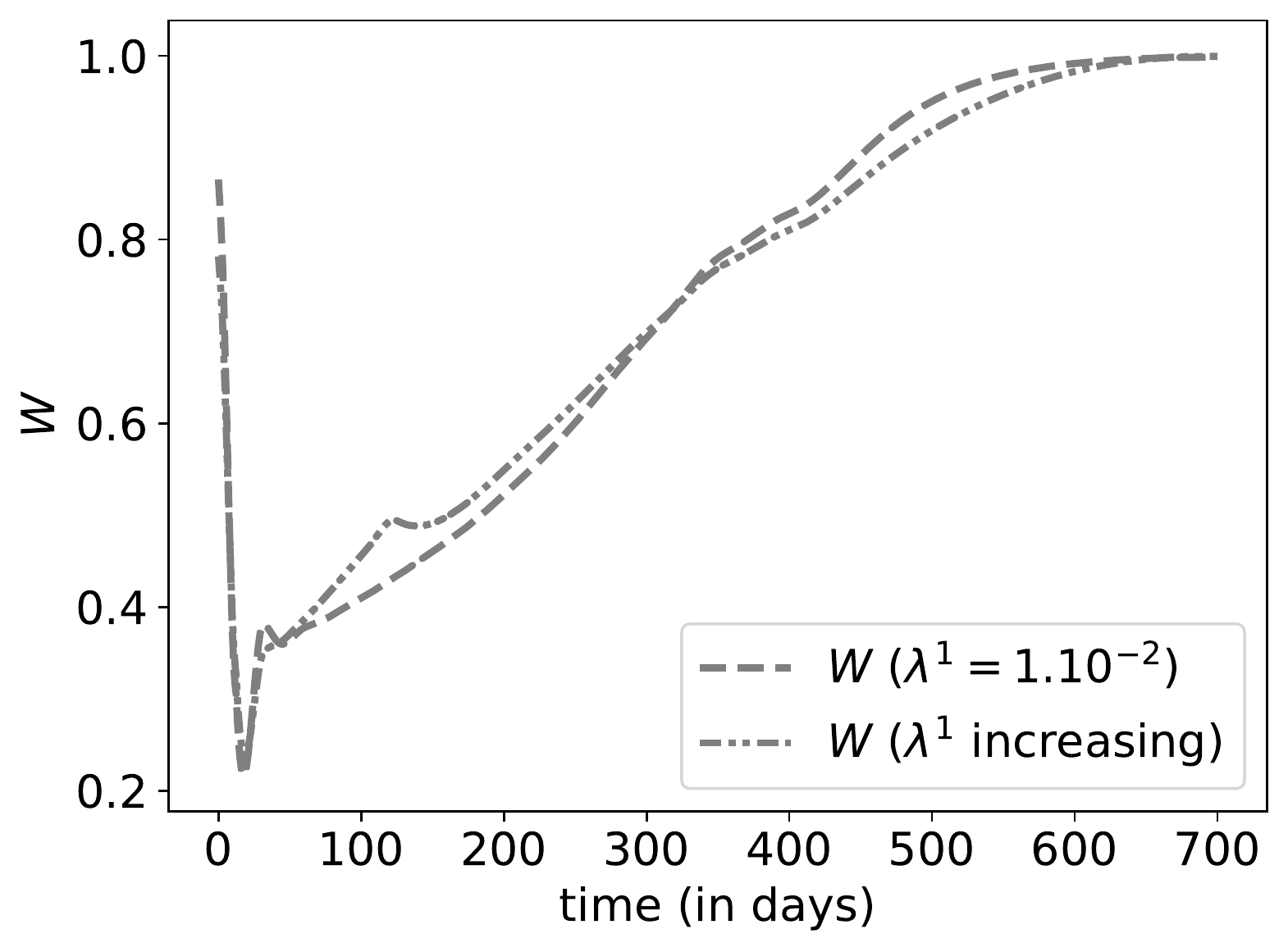}
	\end{subfigure}%
	\begin{subfigure}{.33\columnwidth}
		\centering 
		\includegraphics[width=\columnwidth]{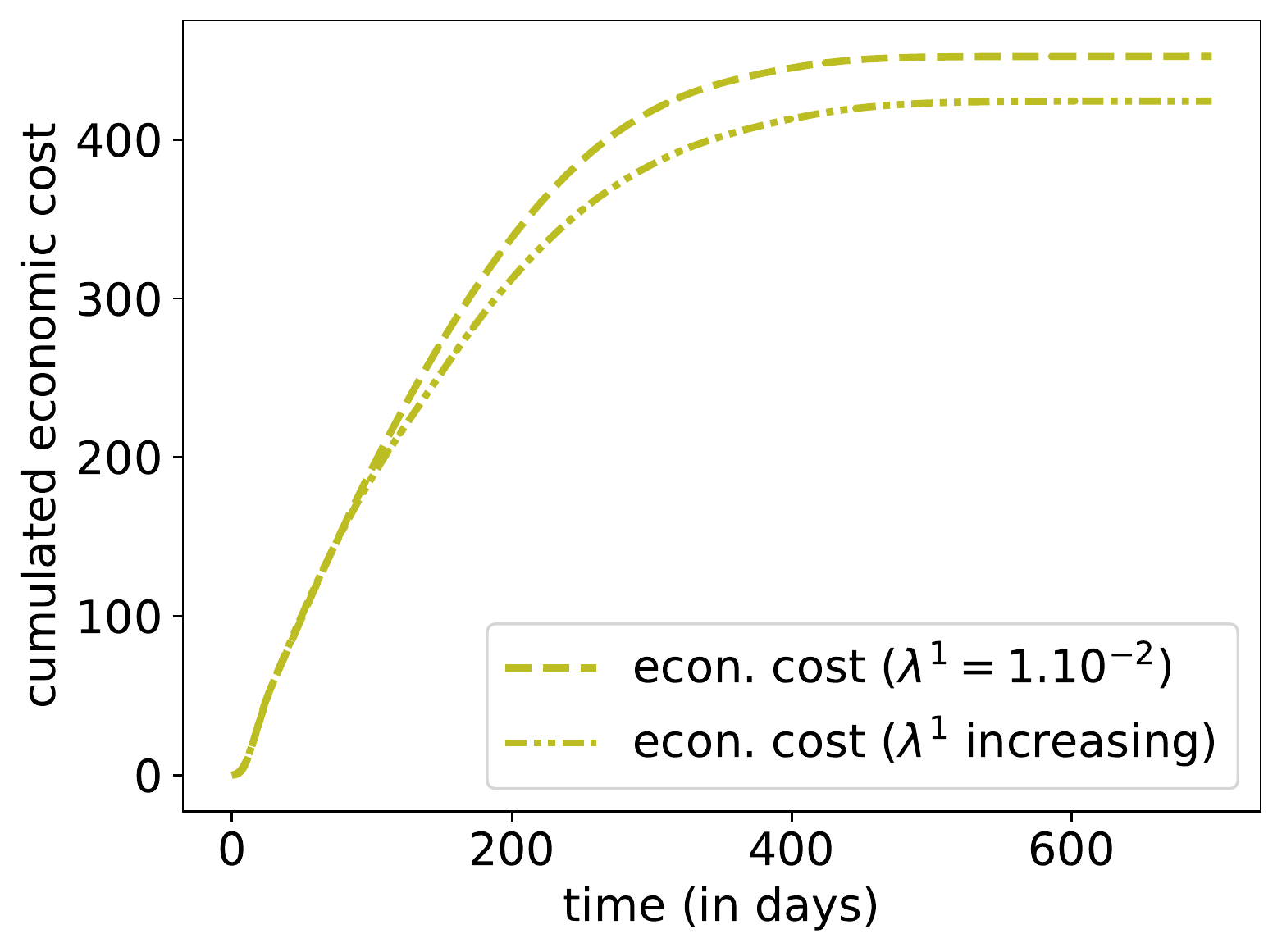}
	\end{subfigure}
	\begin{subfigure}{.33\columnwidth}
		\centering 
		\includegraphics[width=\columnwidth]{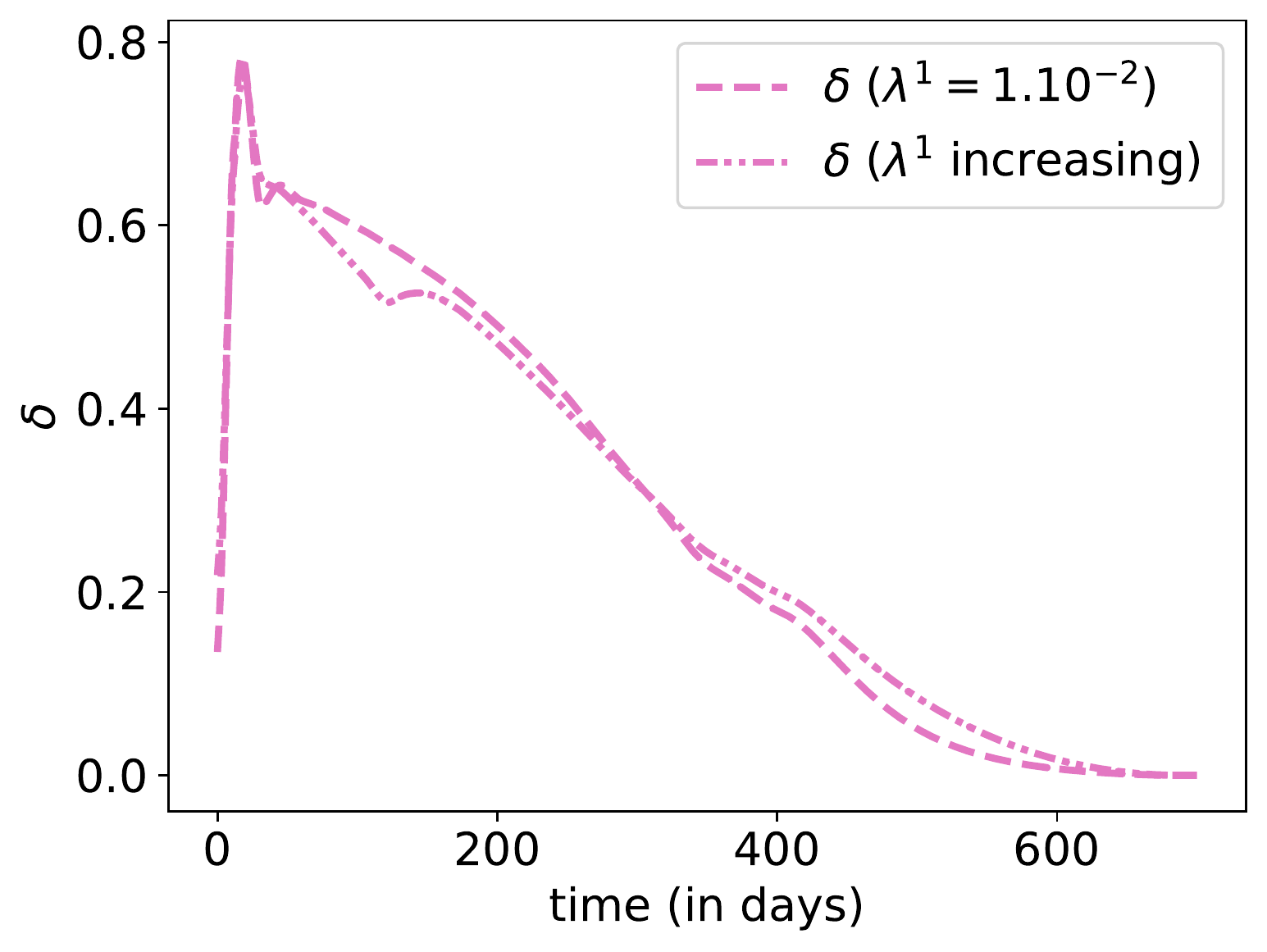}
	\end{subfigure}
	
	\begin{subfigure}{.33\columnwidth}
		\centering
		\includegraphics[width=\columnwidth]{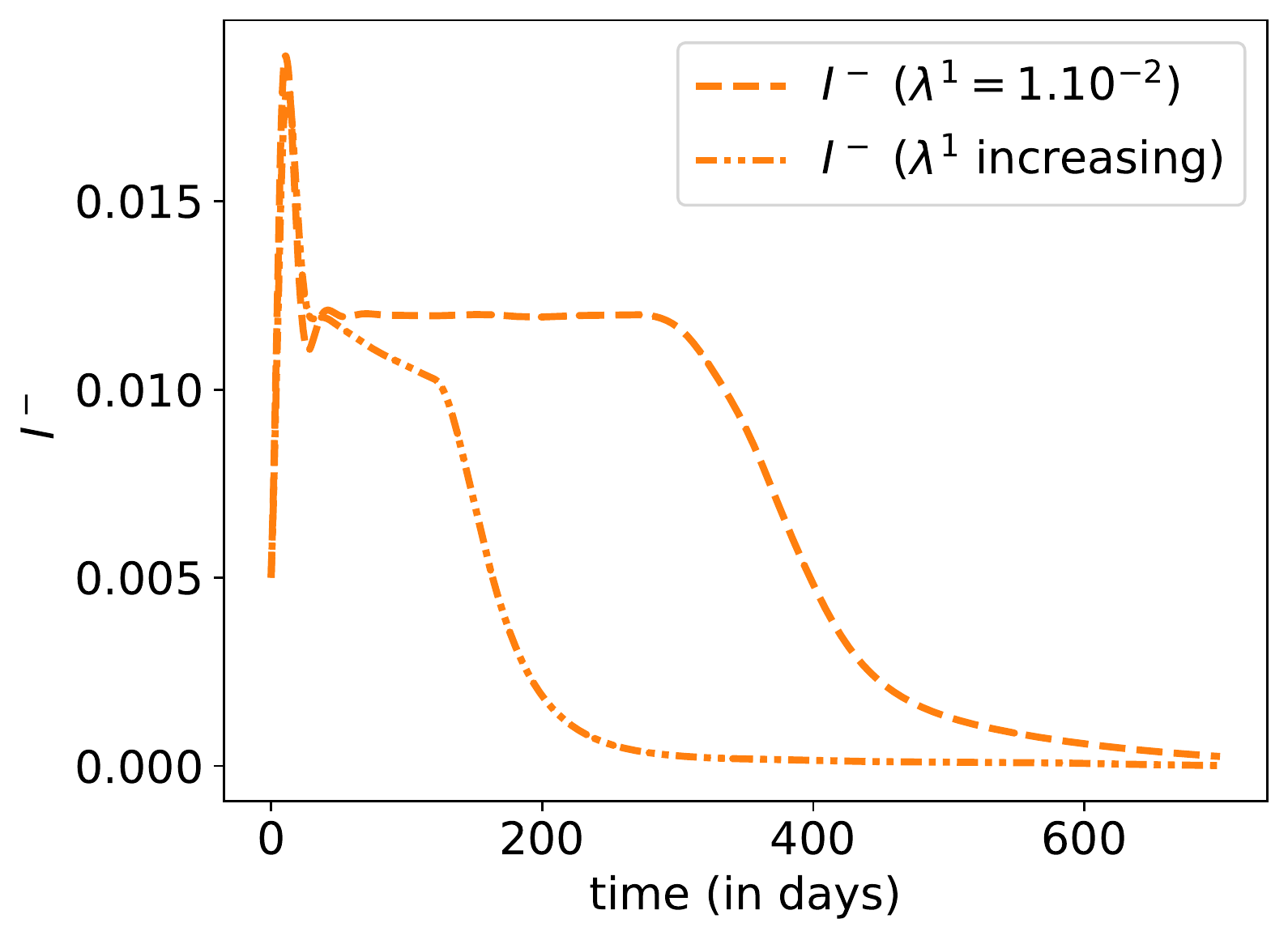}
	\end{subfigure}%
	\begin{subfigure}{.33\columnwidth}
		\centering 
		\includegraphics[width=\columnwidth]{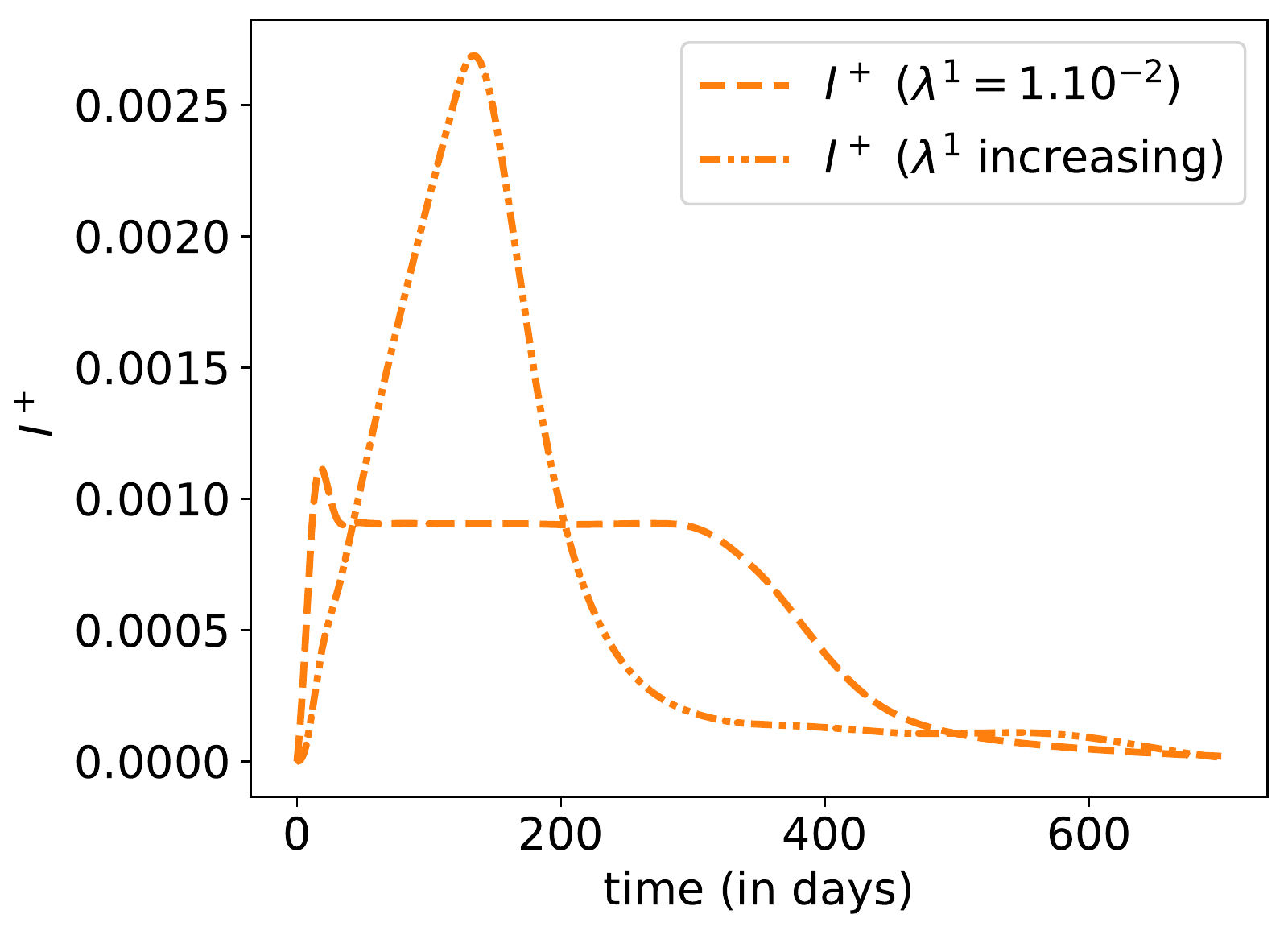}
	\end{subfigure}
	\begin{subfigure}{.33\columnwidth}
		\centering 
		\includegraphics[width=\columnwidth]{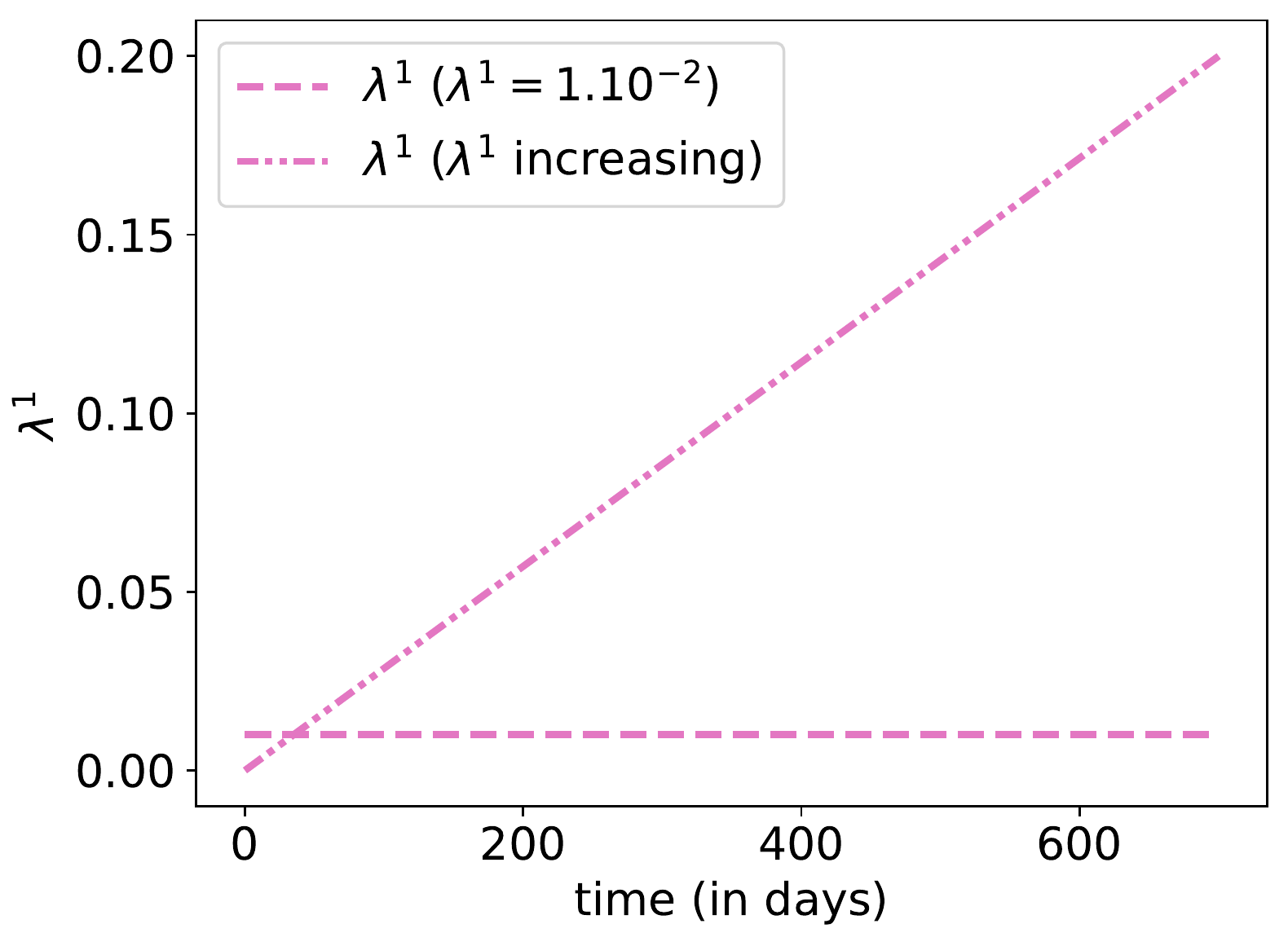}
	\end{subfigure}
	\caption{Evolution of states with optimal control $\delta$ for two different patterns for $\lambda^1$. The dotted line corresponds to the case $\lambda^1=1\%$ and the mixed line is the increasing scenarios for $\lambda^1_t$ (from 0\% and 20\% over two years).} 
 	\label{fig:Sensi_lambda1B5_linear}
\end{figure}

\newpage

\subsection{Optimizing over effort in virologic detection without lockdown intervention} \label{sec_opti_delta_lambda1_b6}

{
Here, we compare the situation without intervention at all and the situation with intervention through $\lambda^1$. The optimal control problem for which we compute an approximate solution numerically is:
\begin{eqnarray}
 \inf_{\lambda^1\in\tilde{\mathcal{A}}} \big\lbrace \tilde J_T(0,\lambda^1,0)\big\rbrace\;,
\end{eqnarray}
with $\tilde{\mathcal{A}}$ the set of measurable functions from $[0,T]$ to $[0,1]$.
}

\begin{figure}[h]
	\begin{subfigure}{.33\columnwidth}
		\centering
		\includegraphics[width=\columnwidth]{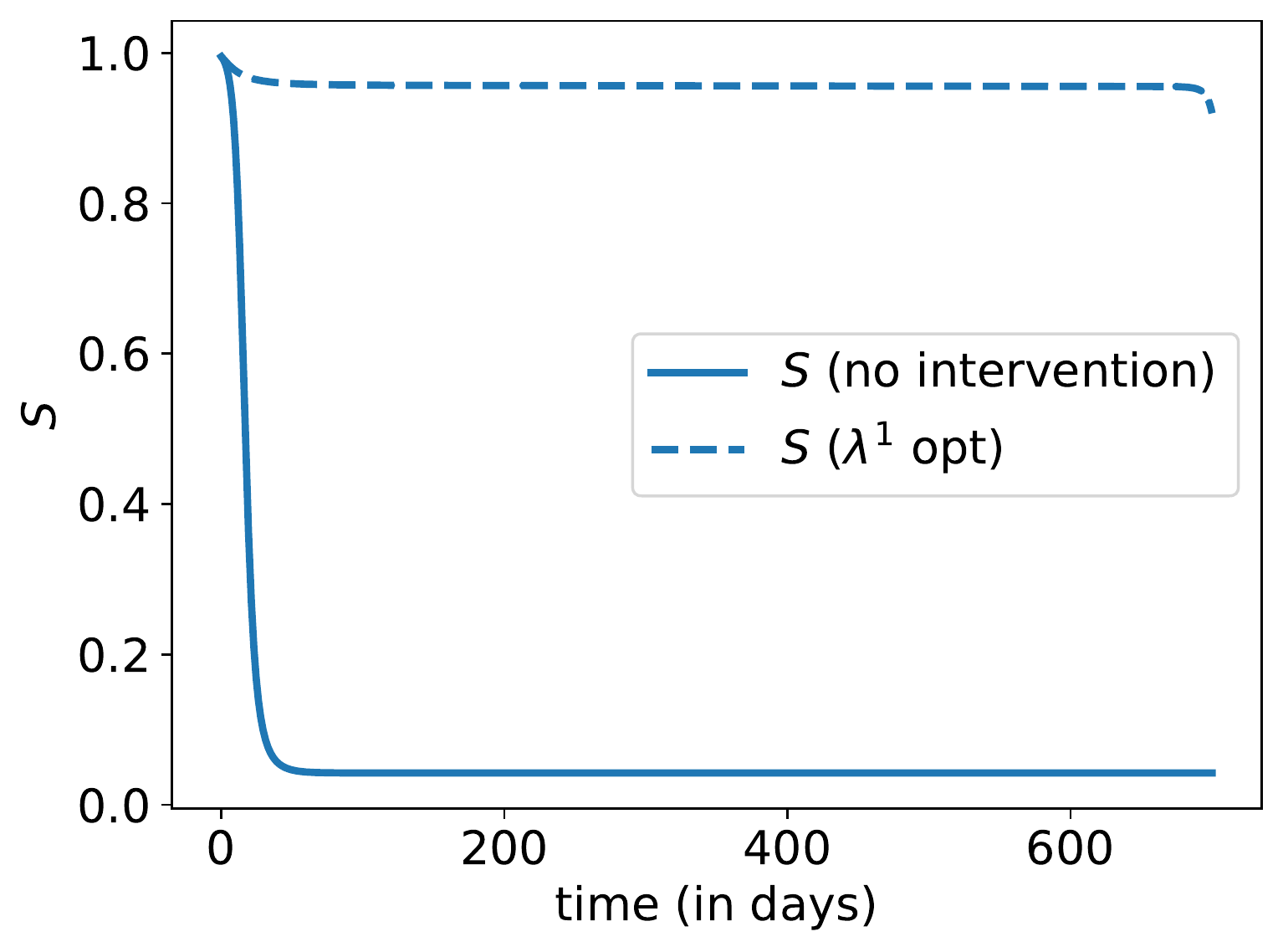}
	\end{subfigure}%
	\begin{subfigure}{.33\columnwidth}
		\centering 
		\includegraphics[width=\columnwidth]{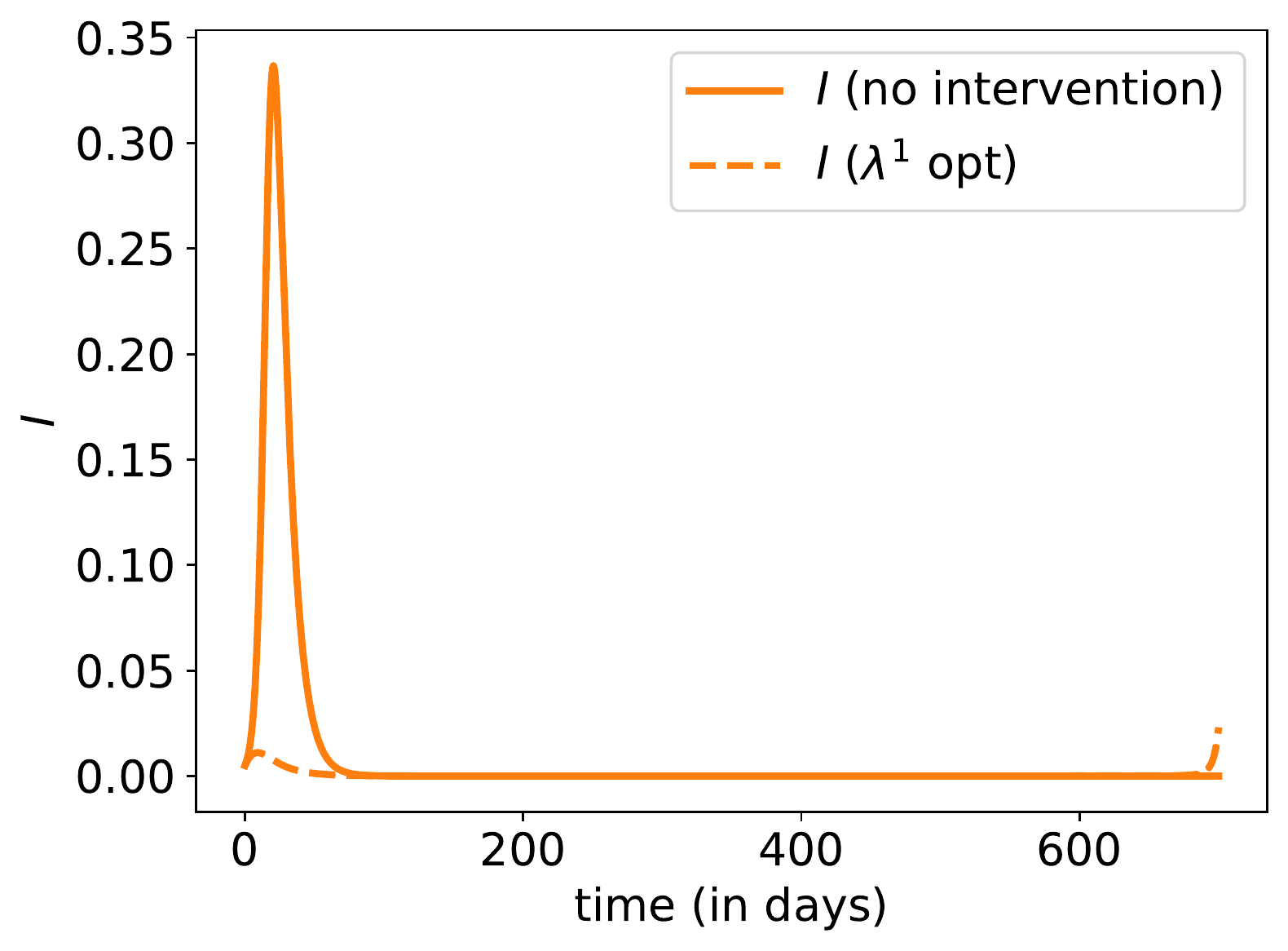}
	\end{subfigure}
	\begin{subfigure}{.33\columnwidth}
		\centering 
		\includegraphics[width=\columnwidth]{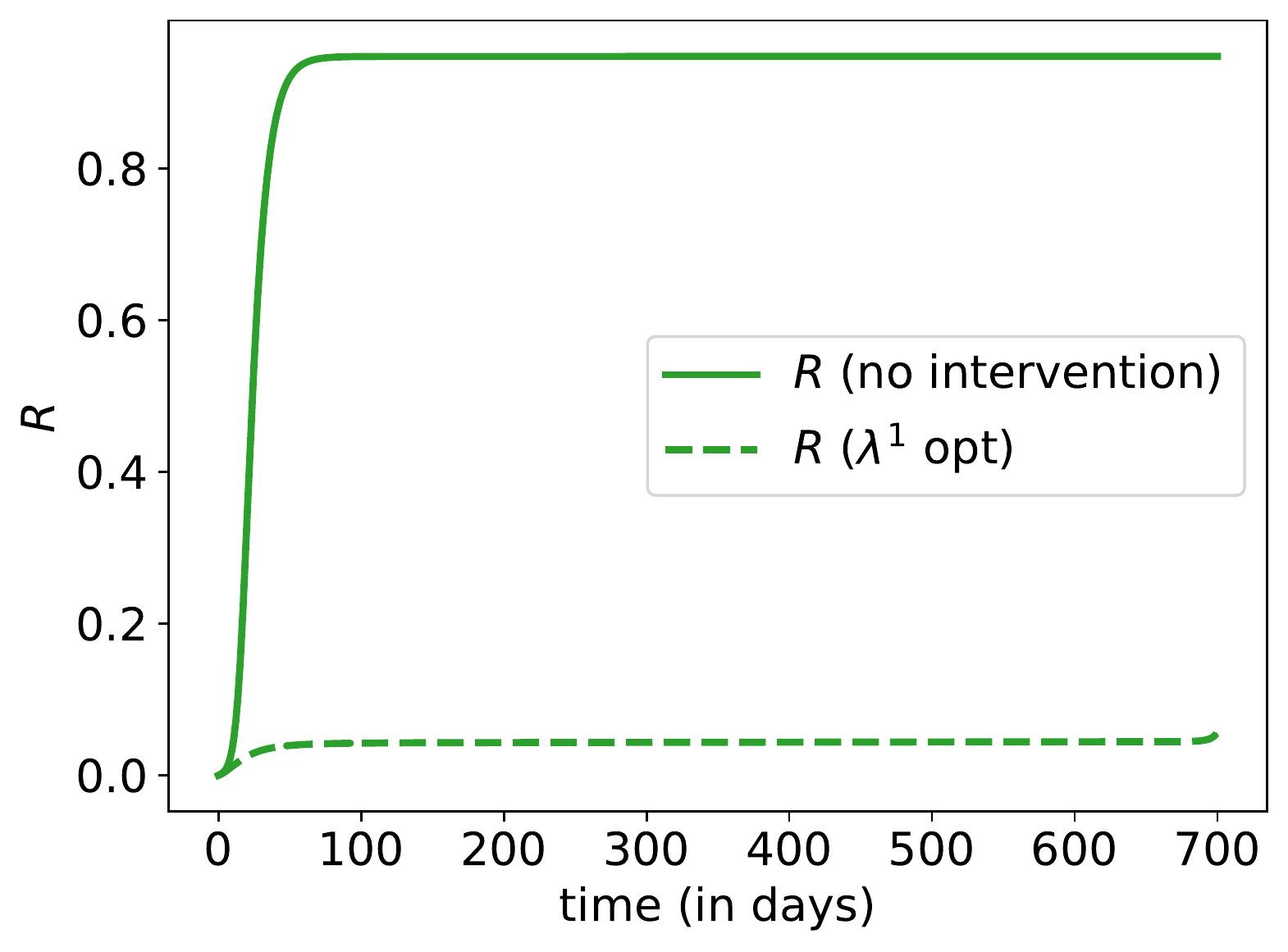}
	\end{subfigure}
	
	\begin{subfigure}{.33\columnwidth}
		\centering
		\includegraphics[width=\columnwidth]{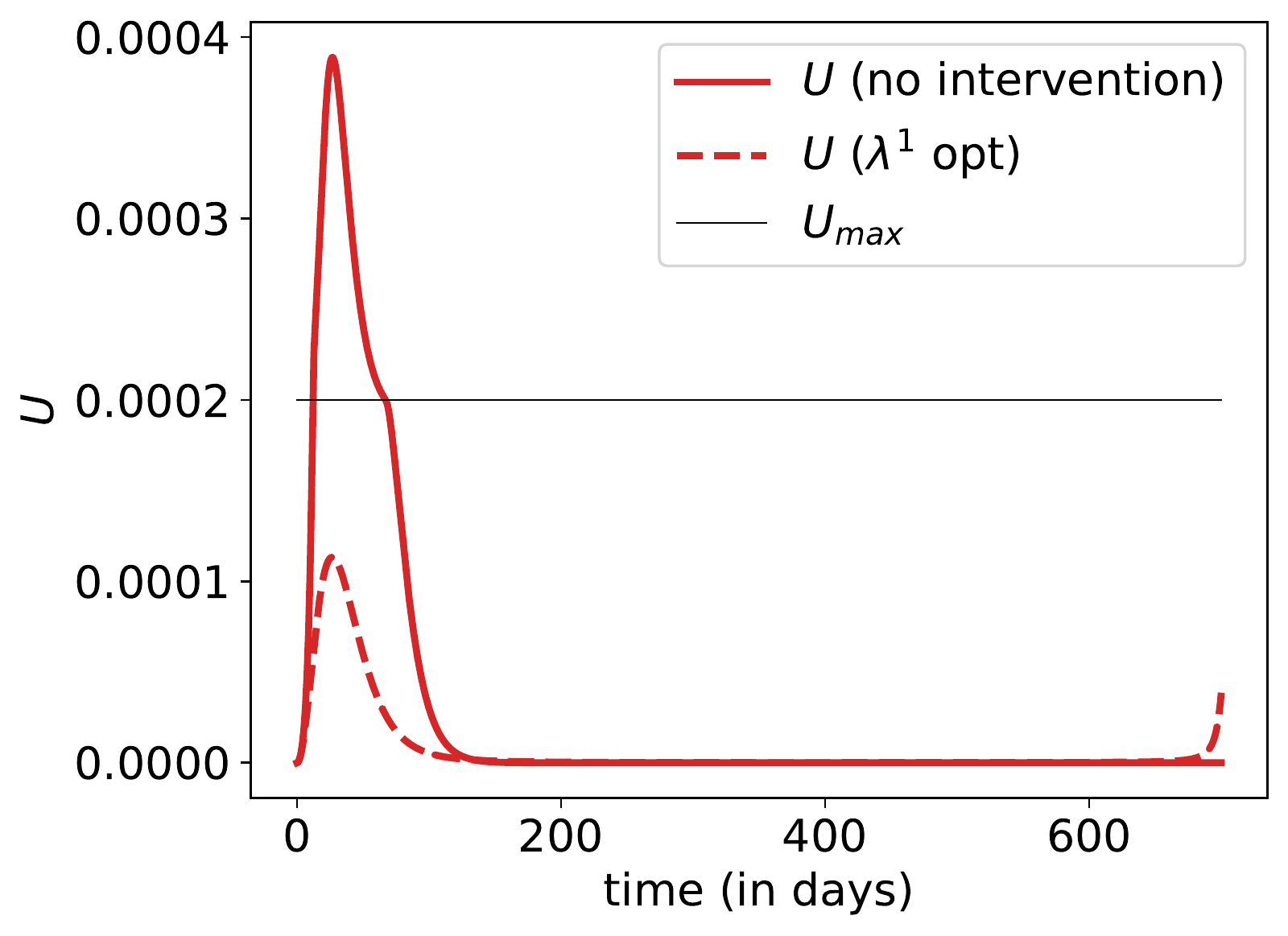}
	\end{subfigure}%
	\begin{subfigure}{.33\columnwidth}
		\centering 
		\includegraphics[width=\columnwidth]{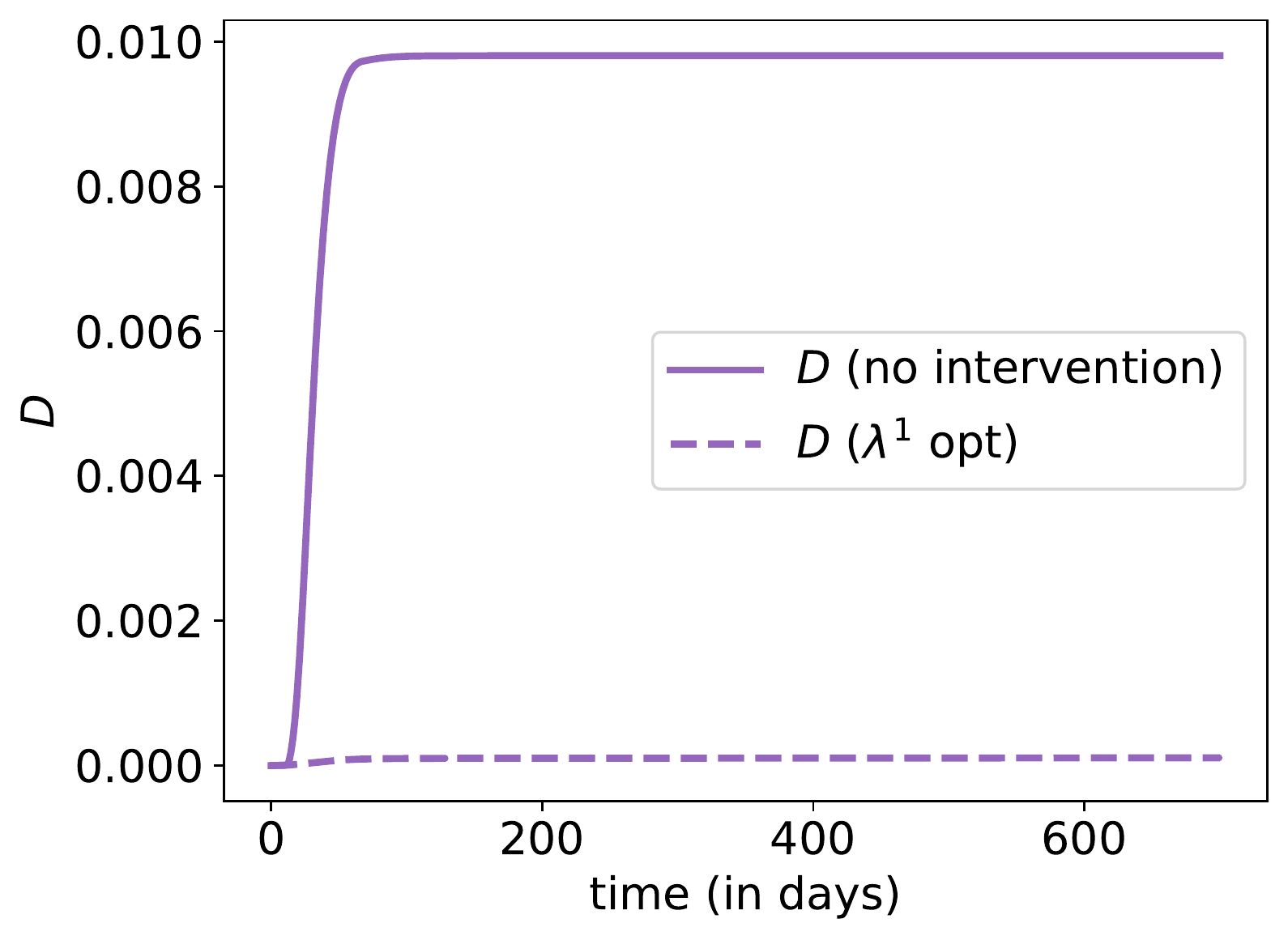}
	\end{subfigure}
	\begin{subfigure}{.33\columnwidth}
		\centering 
		\includegraphics[width=\columnwidth]{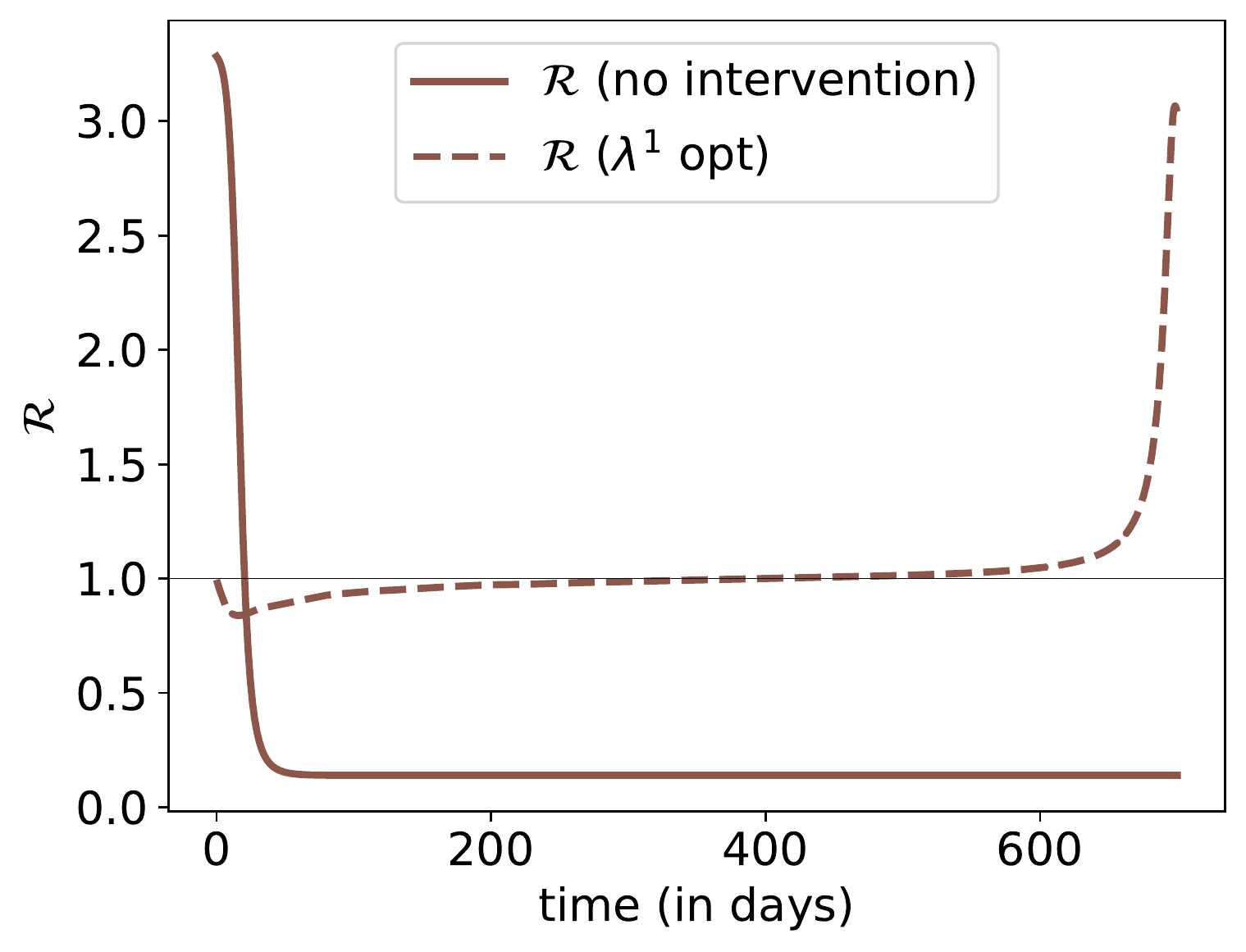}
	\end{subfigure}
	
	\begin{subfigure}{.33\columnwidth}
		\centering
		\includegraphics[width=\columnwidth]{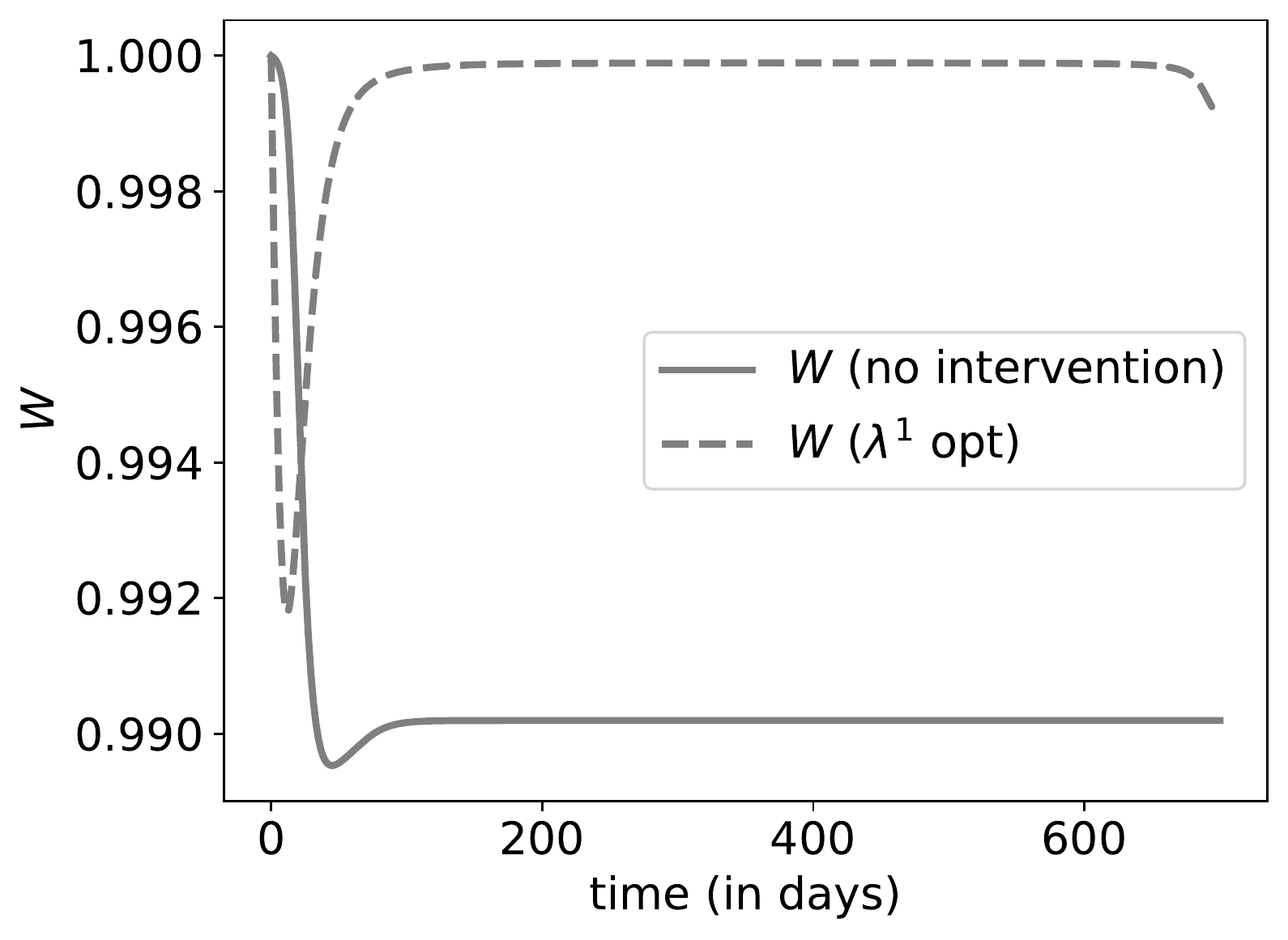}
	\end{subfigure}%
	\begin{subfigure}{.33\columnwidth}
		\centering 
		\includegraphics[width=\columnwidth]{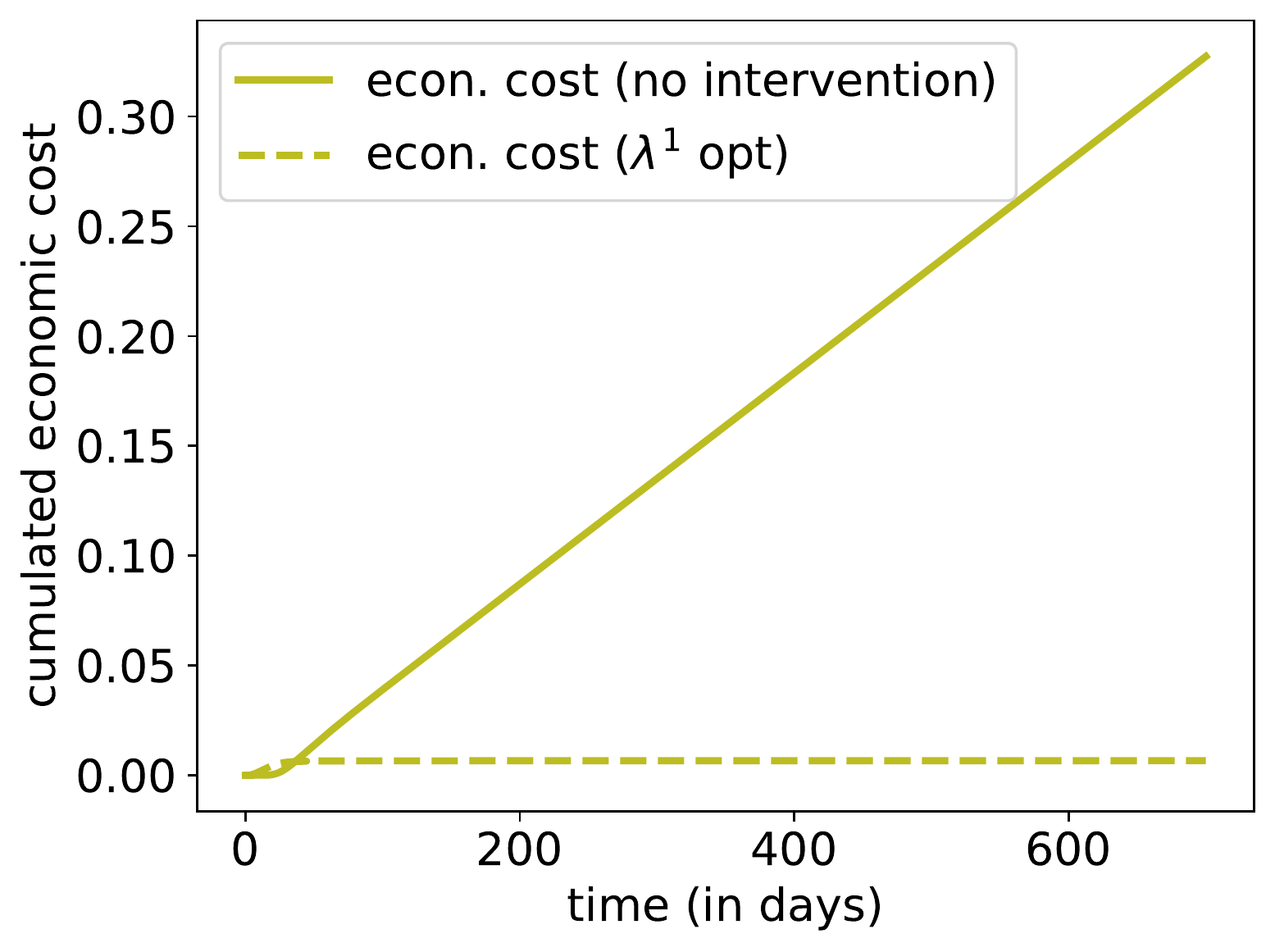}
	\end{subfigure}
	\begin{subfigure}{.33\columnwidth}
		\centering 
		\includegraphics[width=\columnwidth]{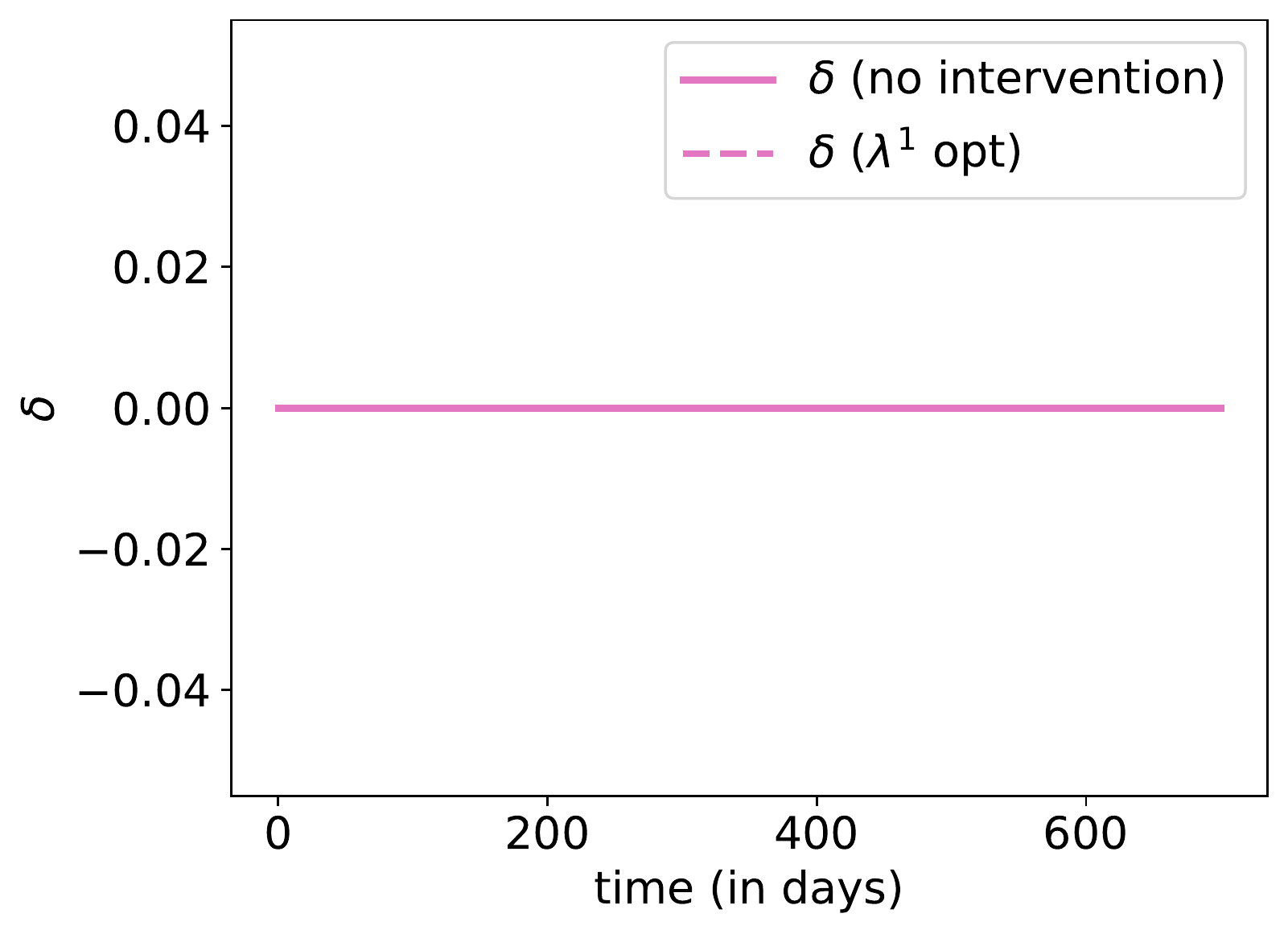}
	\end{subfigure}
	
	\begin{subfigure}{.33\columnwidth}
		\centering
		\includegraphics[width=\columnwidth]{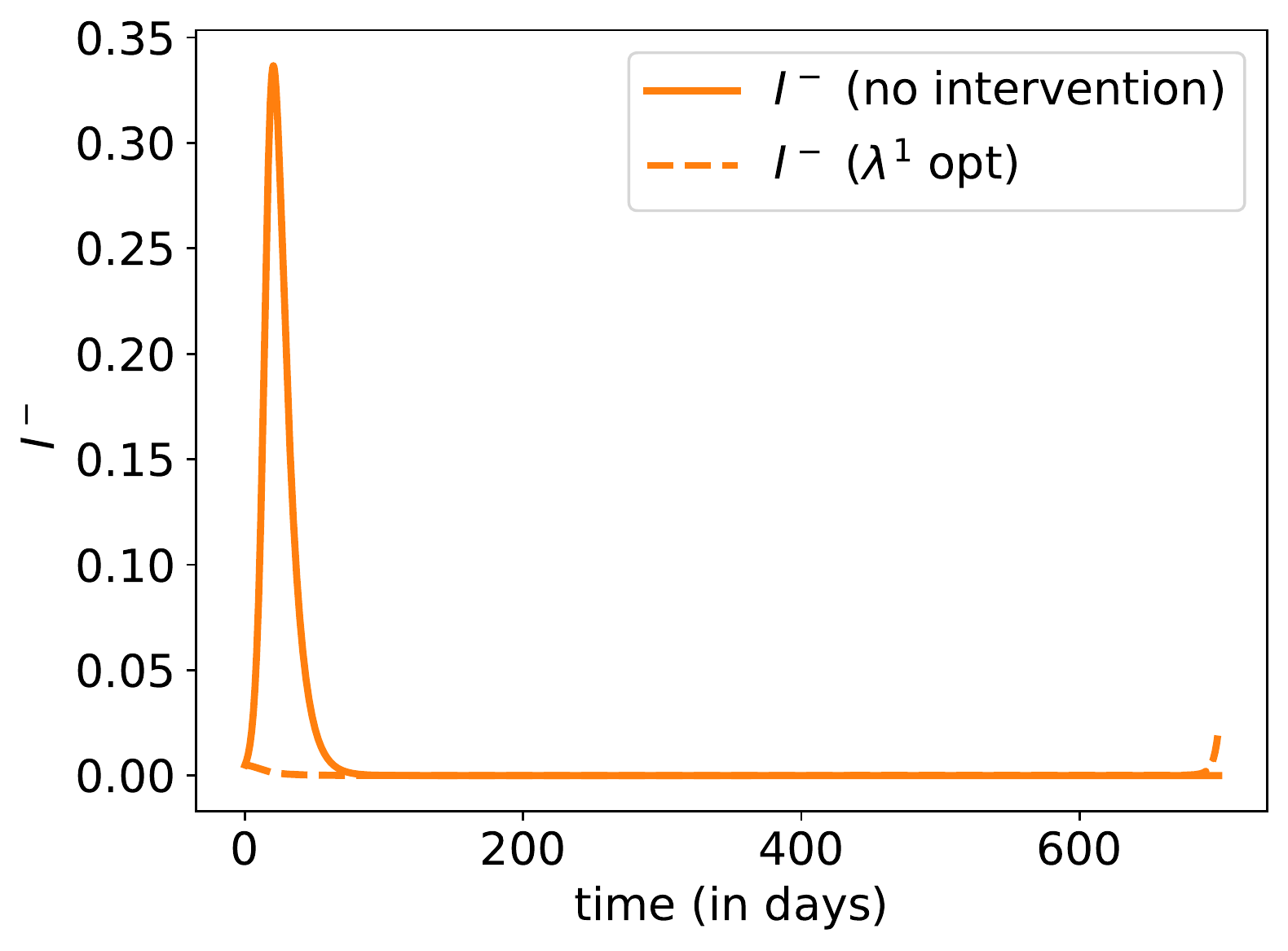}
	\end{subfigure}%
	\begin{subfigure}{.33\columnwidth}
		\centering 
		\includegraphics[width=\columnwidth]{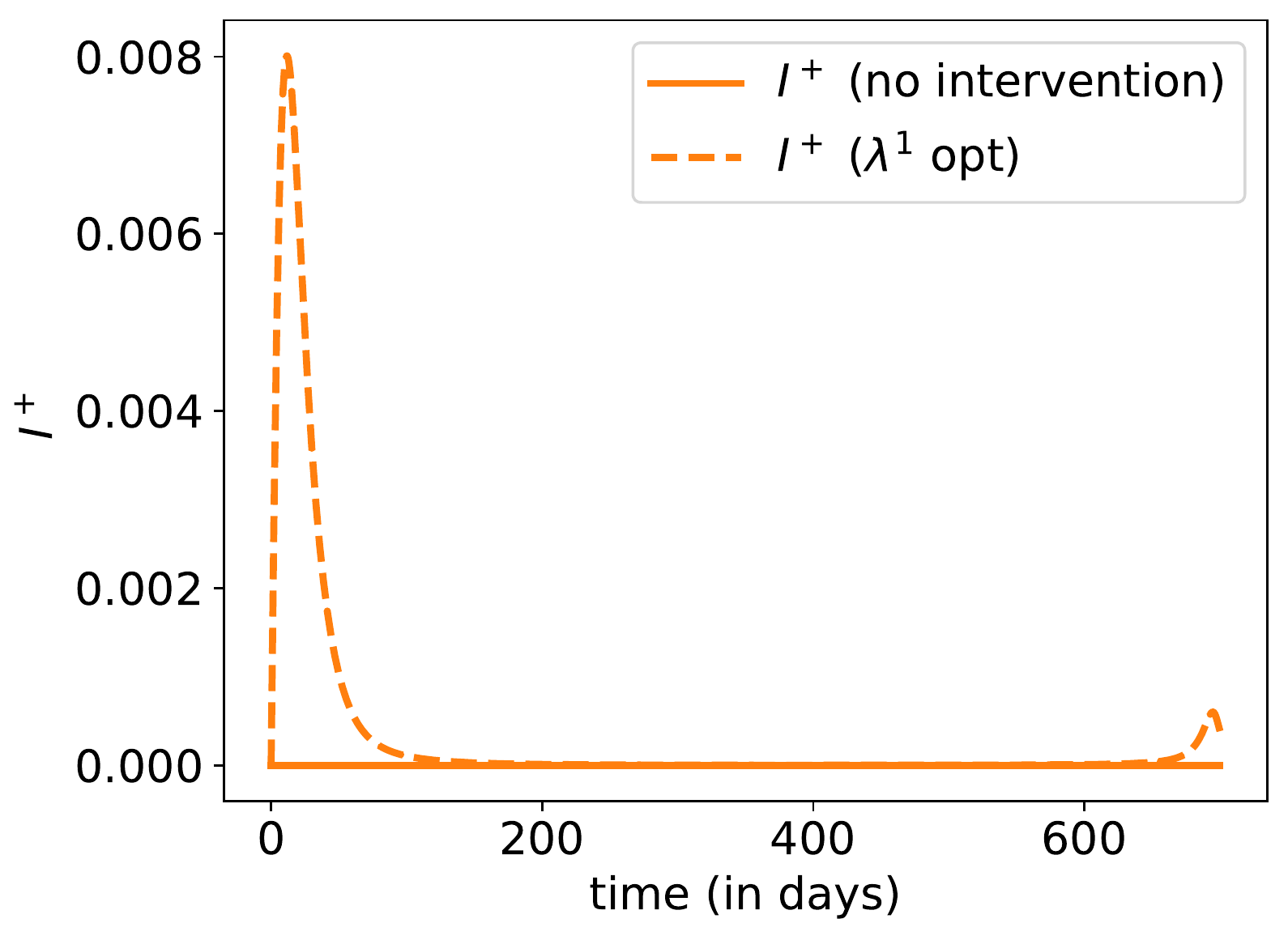}
	\end{subfigure}
	\begin{subfigure}{.33\columnwidth}
		\centering 
		\includegraphics[width=\columnwidth]{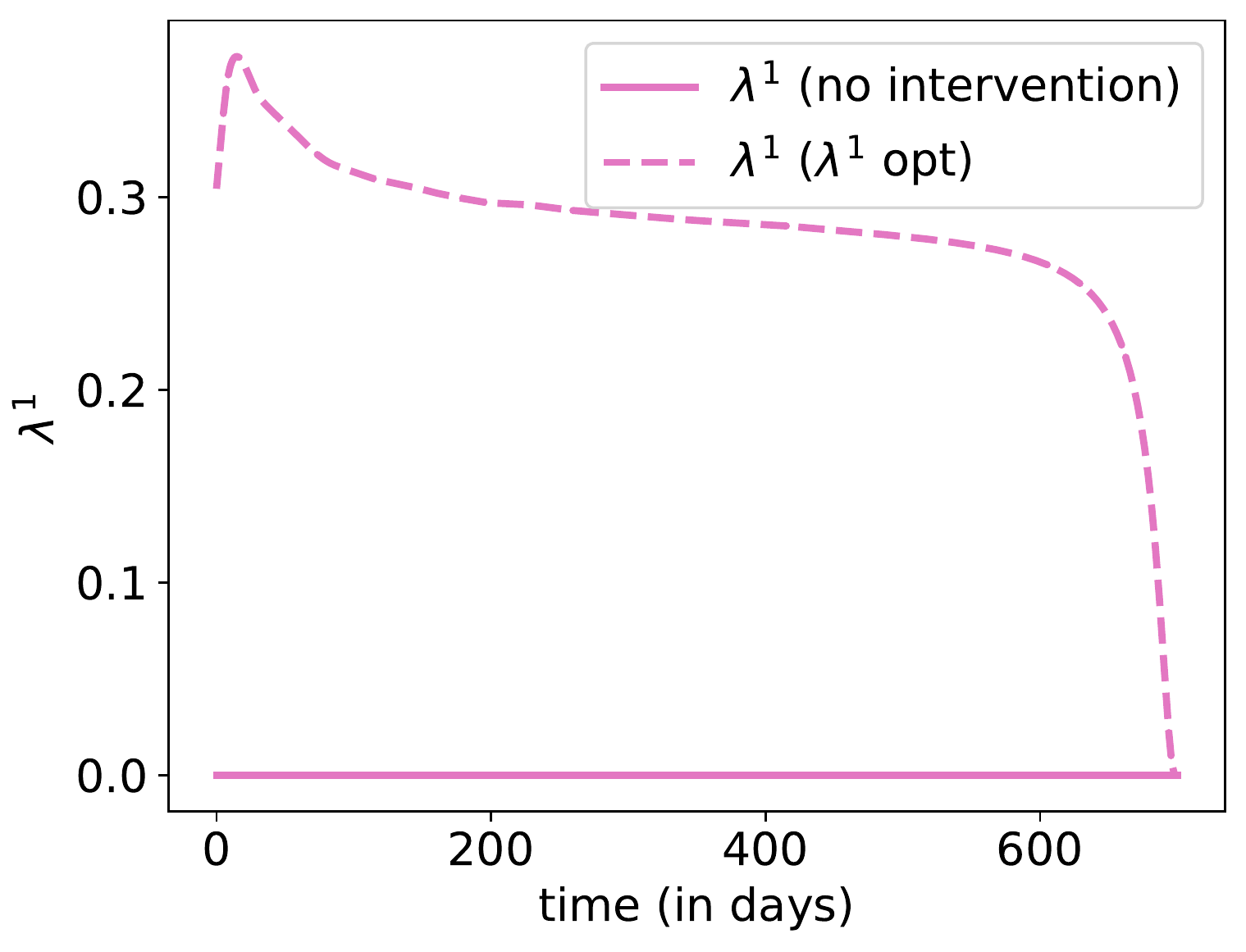}
	\end{subfigure}
	\caption{Evolution of states with optimal control $(0,\lambda^1,0)$, where the plain line corresponds to the benchmark scenario discussed in Section \ref{sec:3:1} (with no intervention $\delta=\lambda^1=\lambda^2=0$); and evolution with optimal control $\lambda^1$ (when $\delta=\lambda^2=0$).} 
 	\label{fig:Sensi_lambda1B6}
\end{figure}

\newpage 

\subsection{Optimizing over both lockdown intervention $\delta$ and effort in virologic detection $\lambda^1$} \label{sec_opti_delta_lambda1_b7}

{
Here, we compare the situation with optimal lockdown policy versus the situation in which we optimize over both lockdown and virologic effort. The latter corresponds to the following optimal control problem, for which we compute an approximate solution numerically:
\begin{eqnarray}
 \inf_{(\delta,\lambda^1)\in\tilde{\mathcal{A}} \times \tilde{\mathcal{A}}} \big\lbrace \tilde J_T(\delta,\lambda^1,0)\big\rbrace\;,
\end{eqnarray}
with $\tilde{\mathcal{A}}$ the set of measurable functions from $[0,T]$ to $[0,1]$.
}

\begin{figure}[h]
	\begin{subfigure}{.33\columnwidth}
		\centering
		\includegraphics[width=\columnwidth]{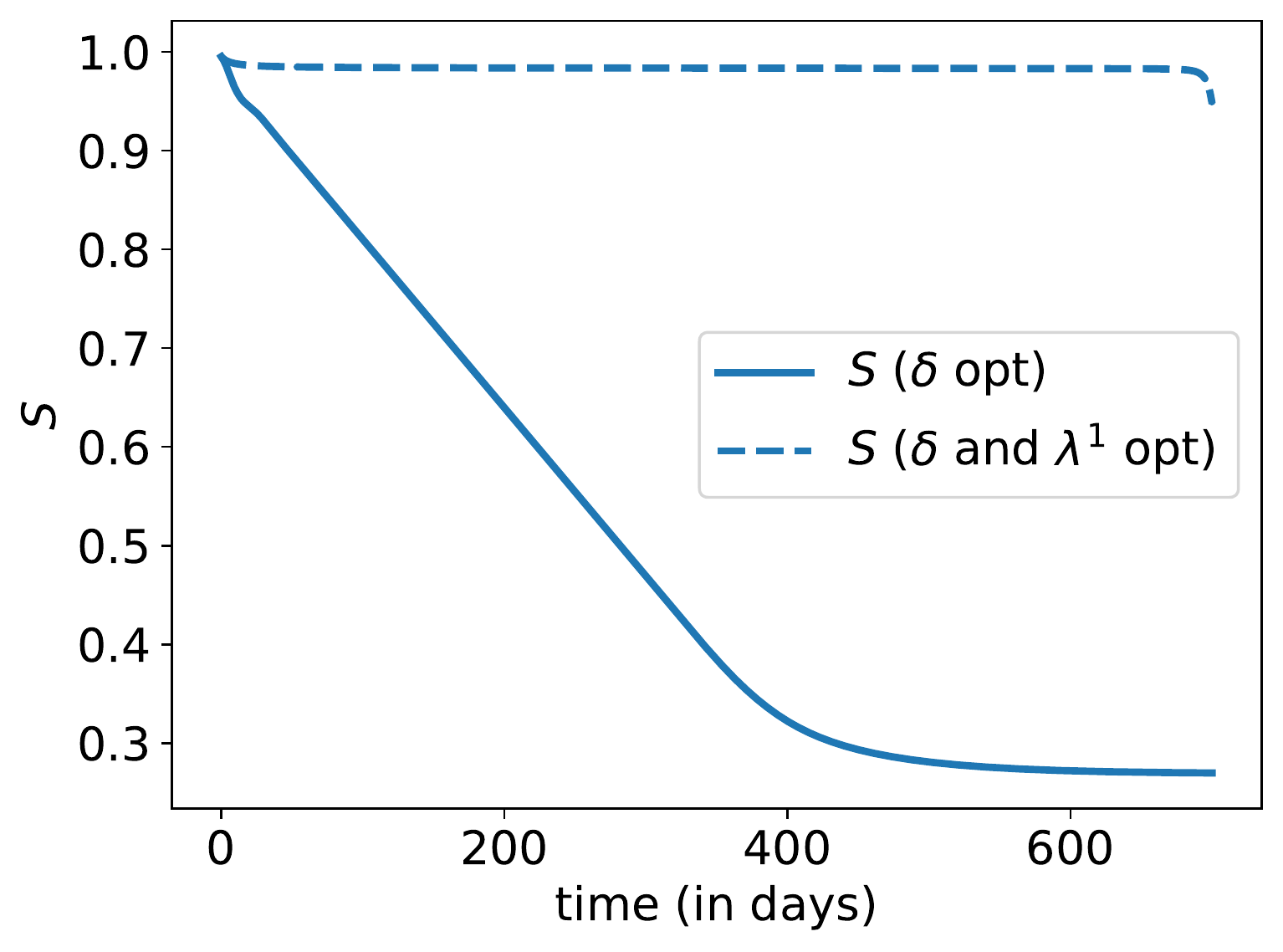}
	\end{subfigure}%
	\begin{subfigure}{.33\columnwidth}
		\centering 
		\includegraphics[width=\columnwidth]{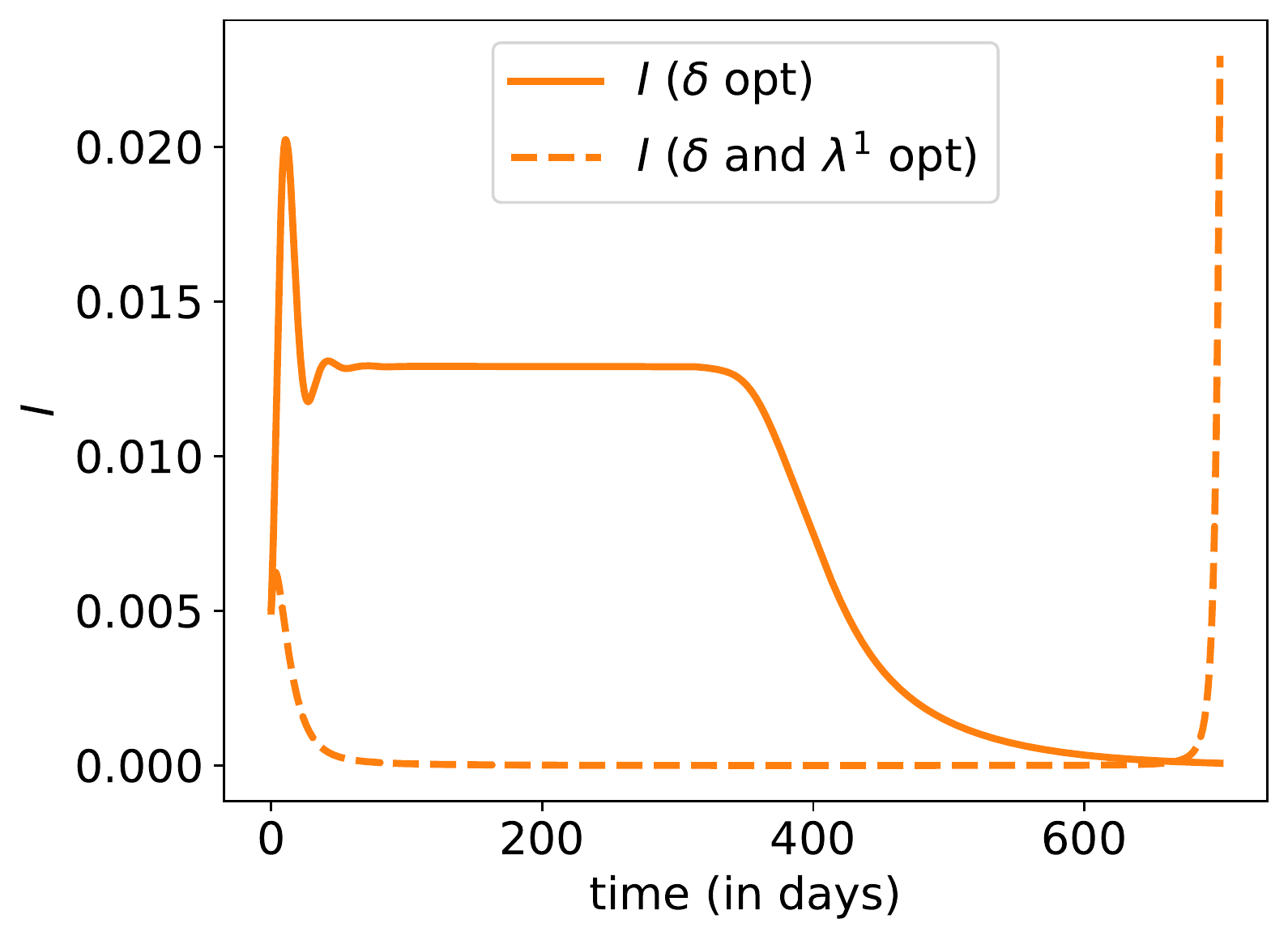}
	\end{subfigure}
	\begin{subfigure}{.33\columnwidth}
		\centering 
		\includegraphics[width=\columnwidth]{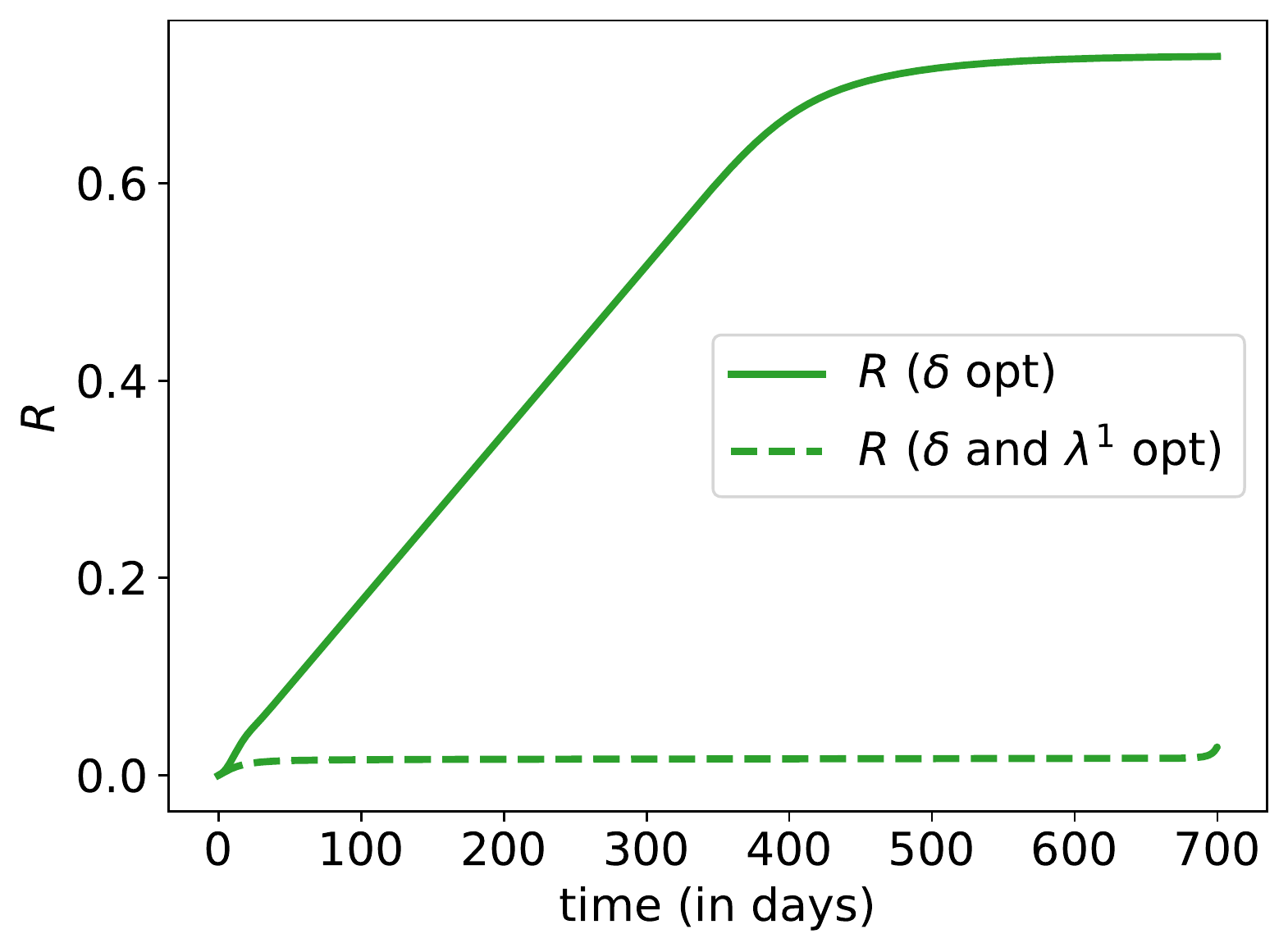}
	\end{subfigure}
	
	\begin{subfigure}{.33\columnwidth}
		\centering
		\includegraphics[width=\columnwidth]{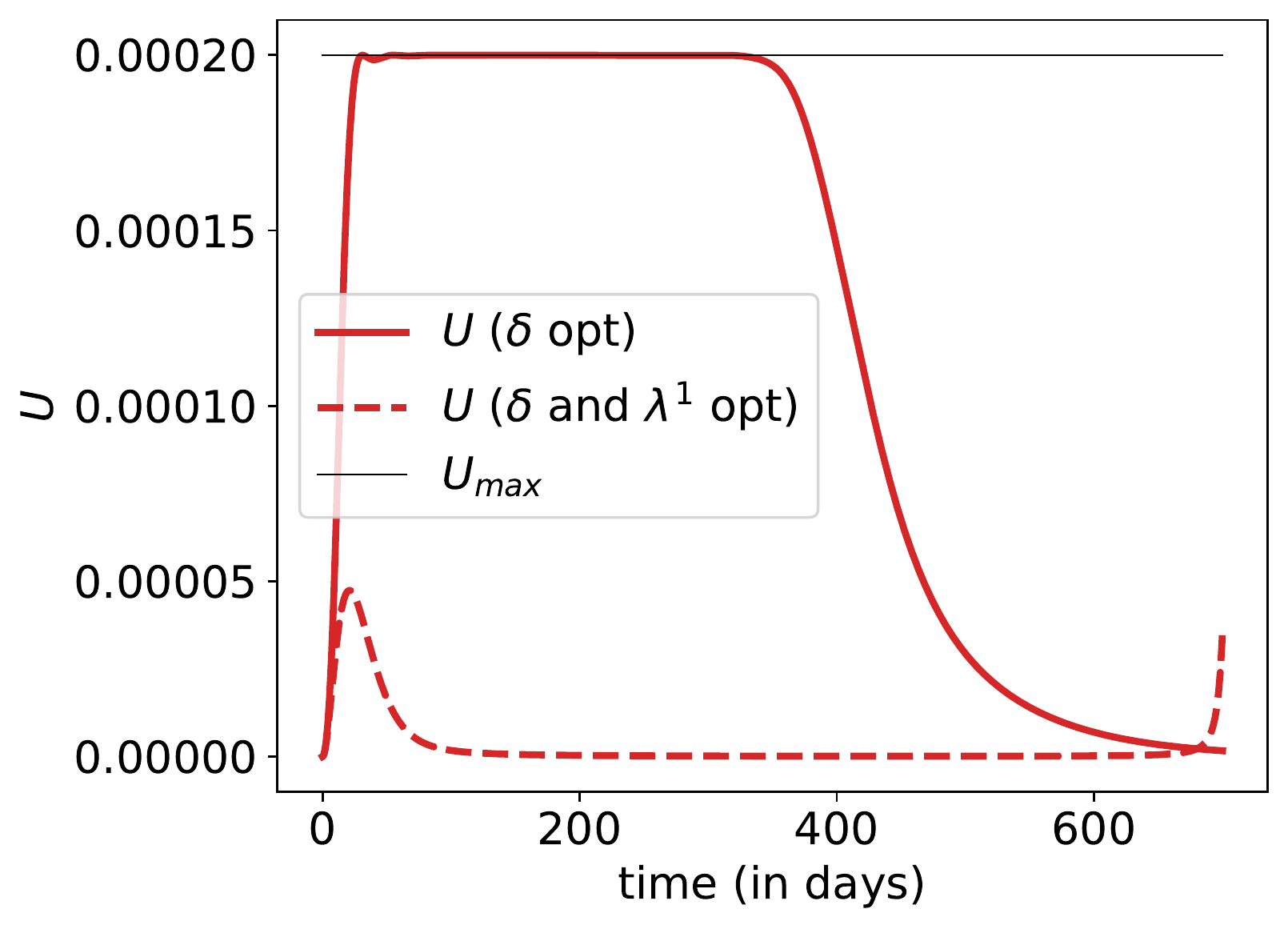}
	\end{subfigure}%
	\begin{subfigure}{.33\columnwidth}
		\centering 
		\includegraphics[width=\columnwidth]{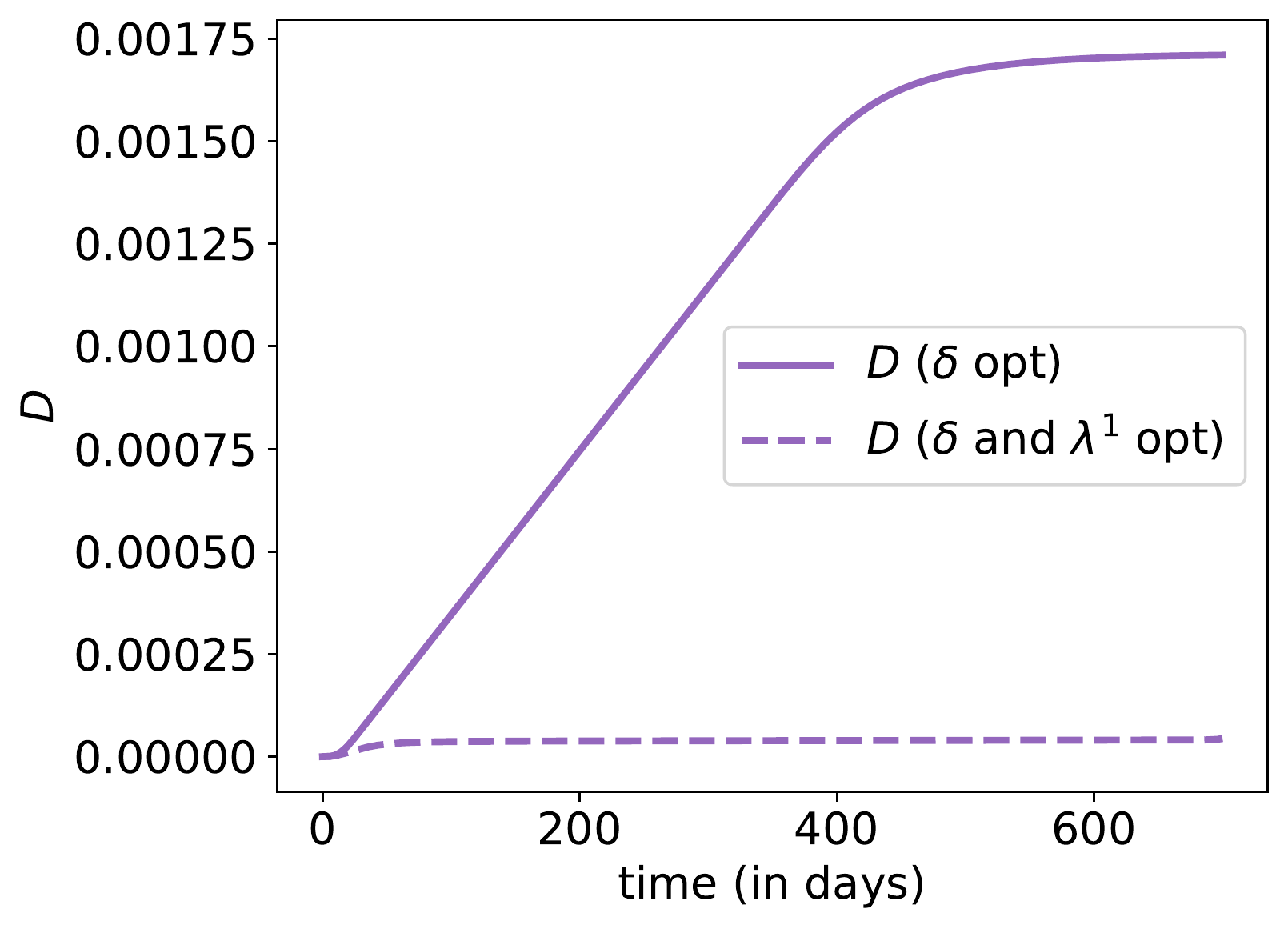}
	\end{subfigure}
	\begin{subfigure}{.33\columnwidth}
		\centering 
		\includegraphics[width=\columnwidth]{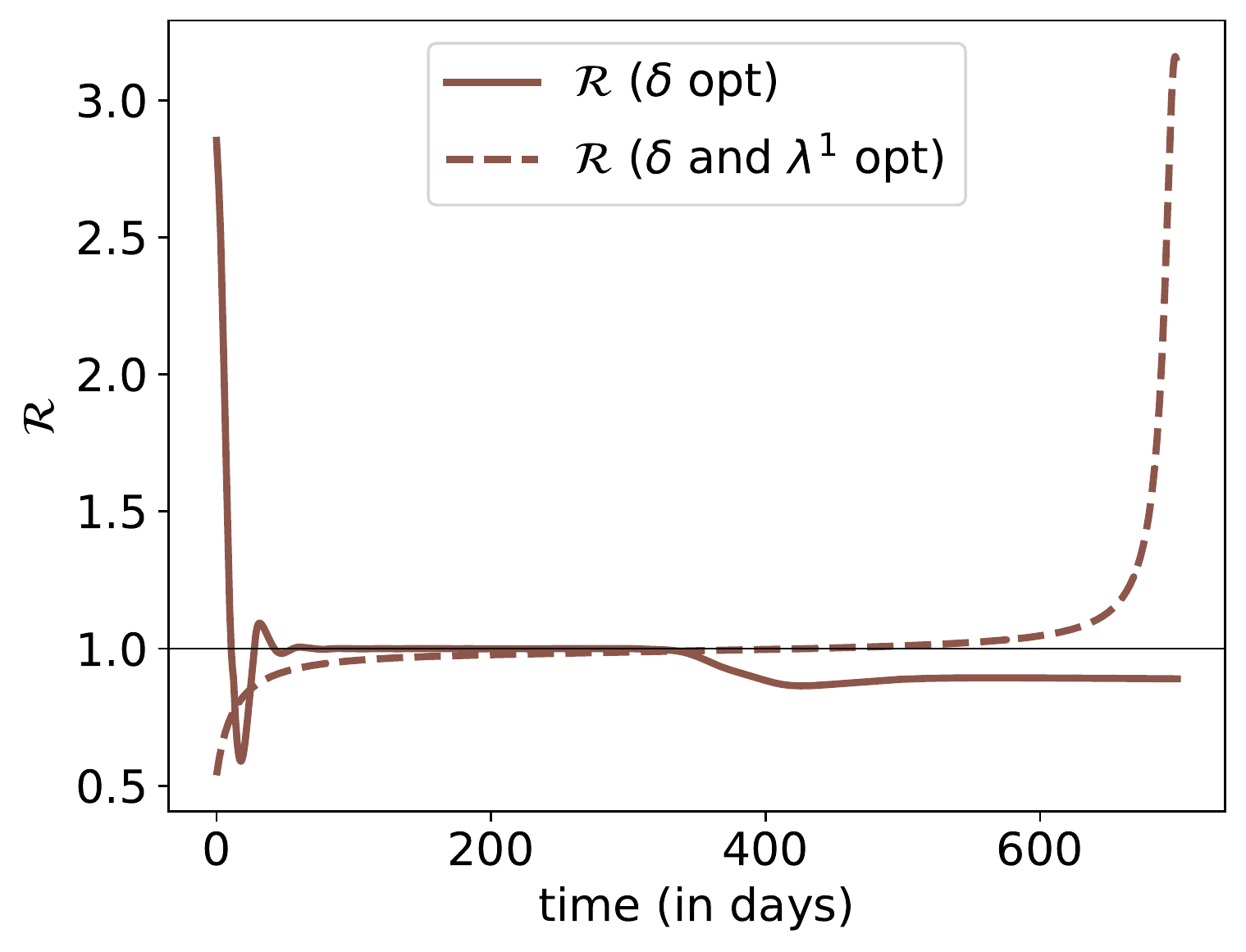}
	\end{subfigure}
	
	\begin{subfigure}{.33\columnwidth}
		\centering
		\includegraphics[width=\columnwidth]{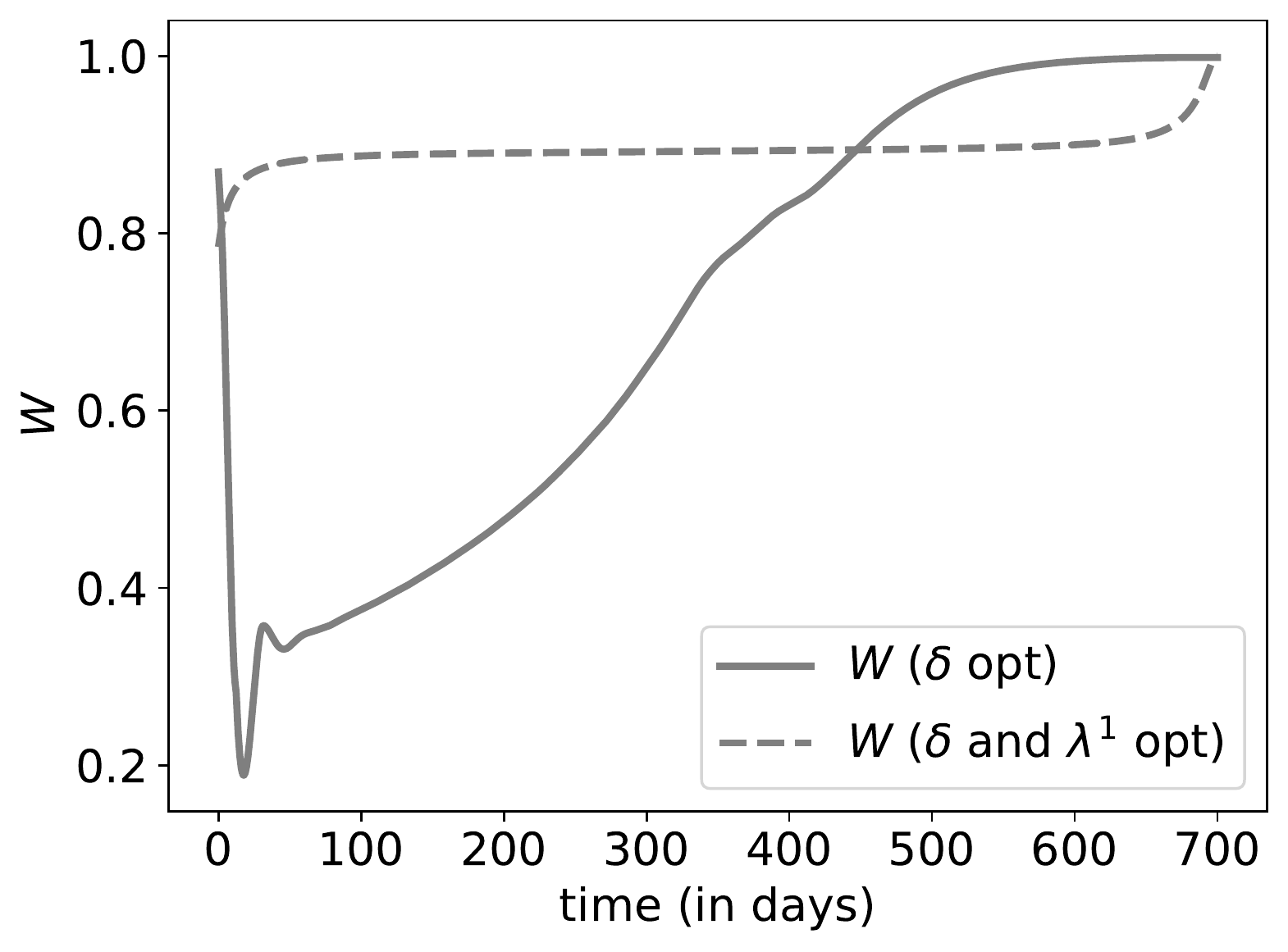}
	\end{subfigure}%
	\begin{subfigure}{.33\columnwidth}
		\centering 
		\includegraphics[width=\columnwidth]{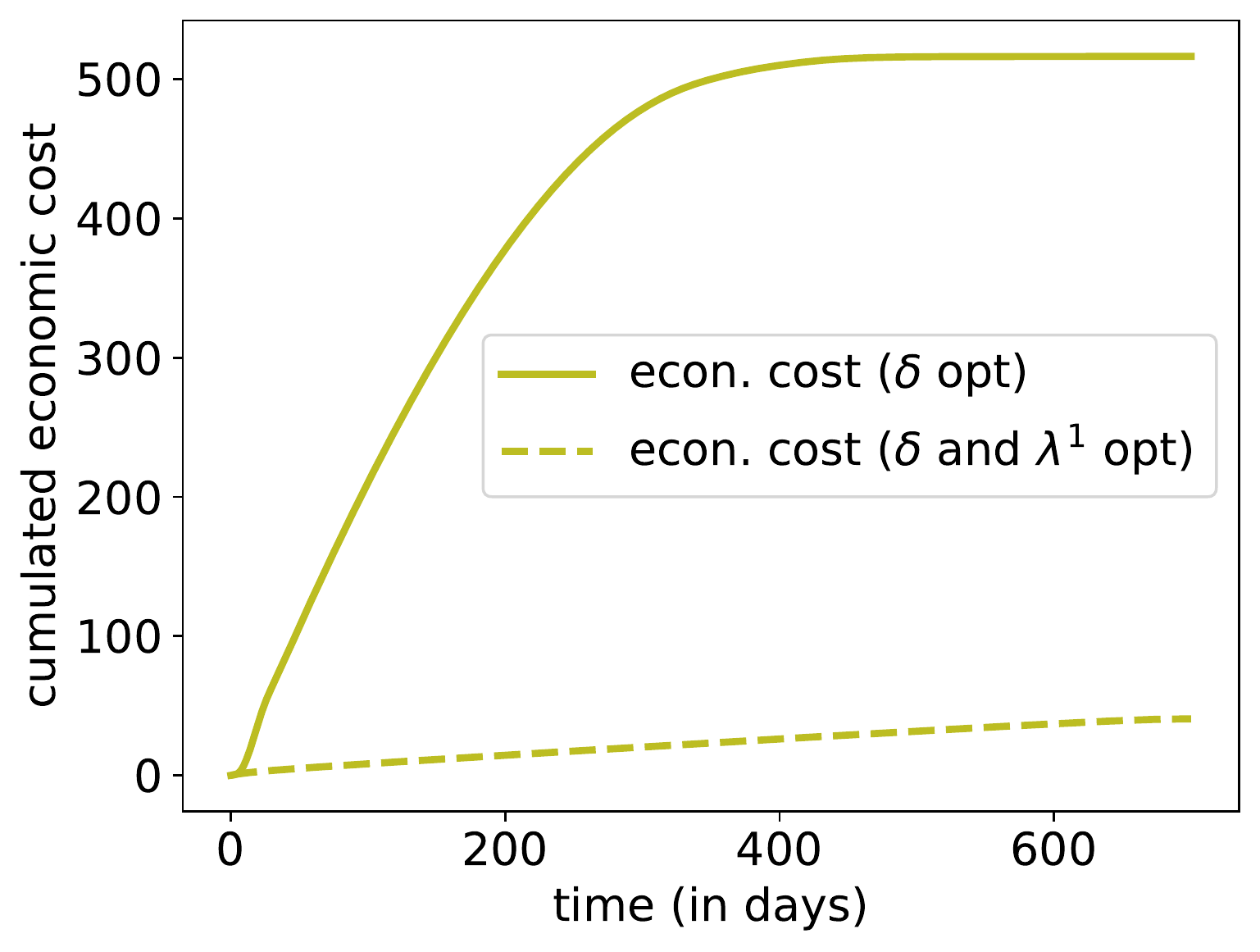}
	\end{subfigure}
	\begin{subfigure}{.33\columnwidth}
		\centering 
		\includegraphics[width=\columnwidth]{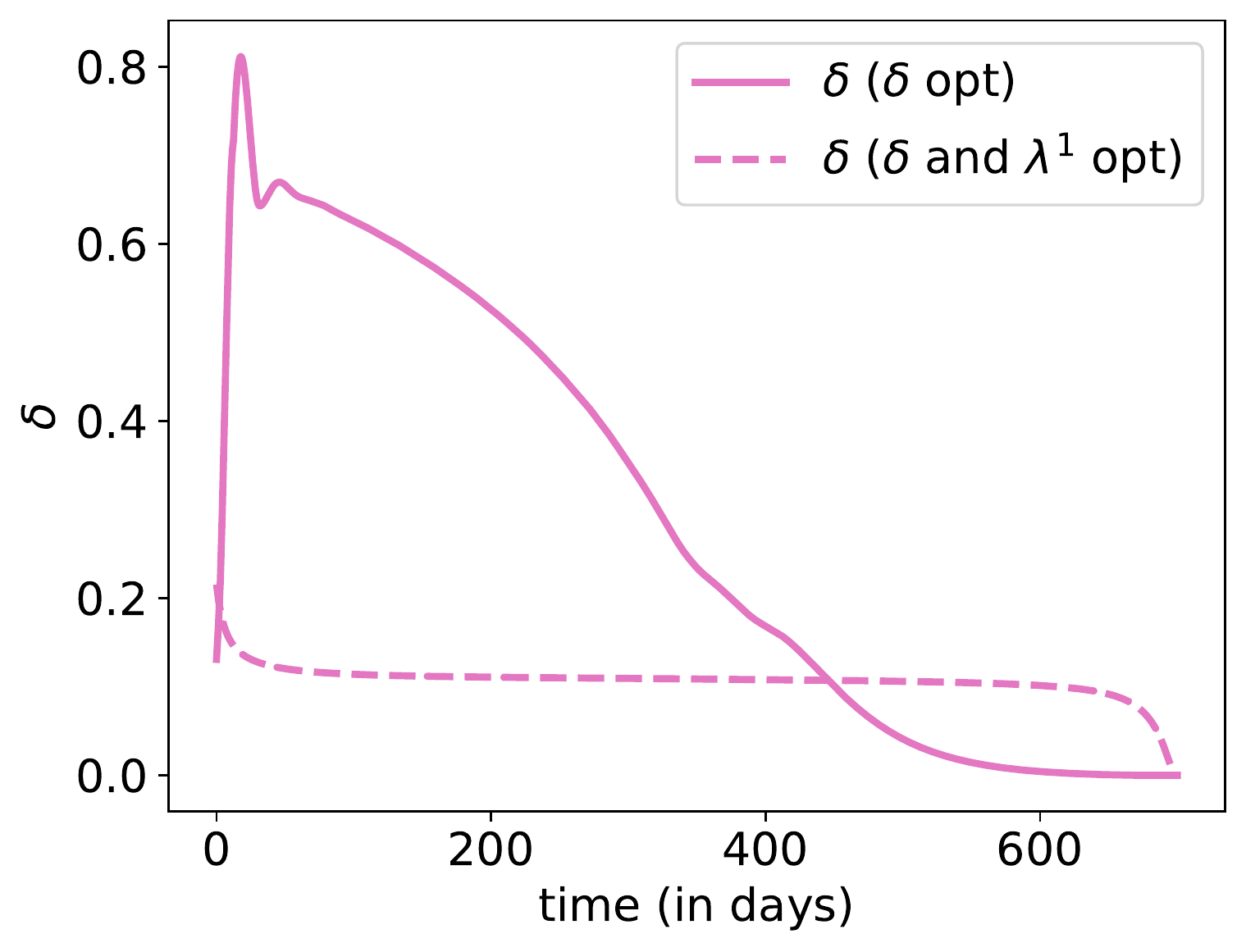}
	\end{subfigure}
	
	\begin{subfigure}{.33\columnwidth}
		\centering
		\includegraphics[width=\columnwidth]{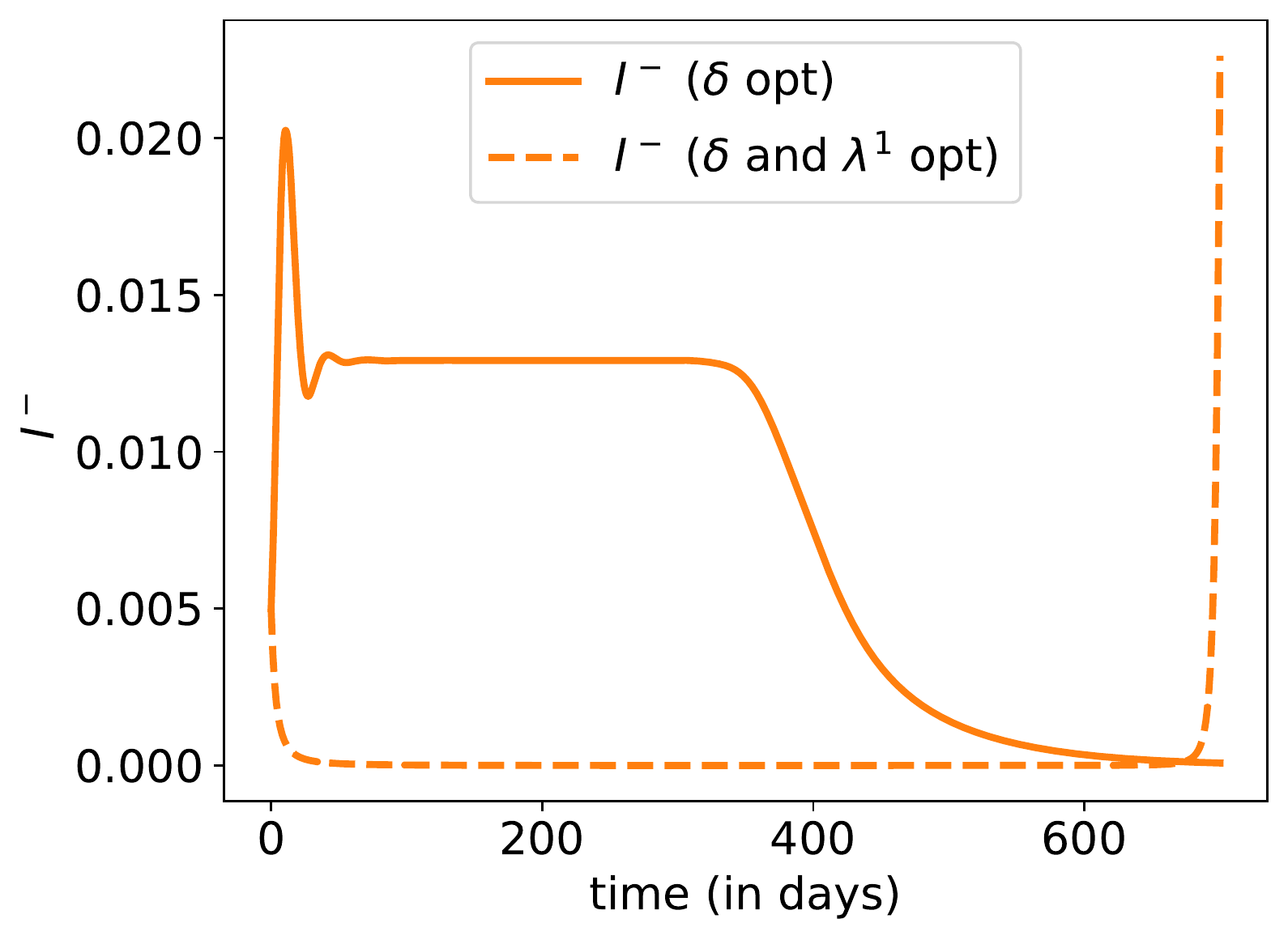}
	\end{subfigure}%
	\begin{subfigure}{.33\columnwidth}
		\centering 
		\includegraphics[width=\columnwidth]{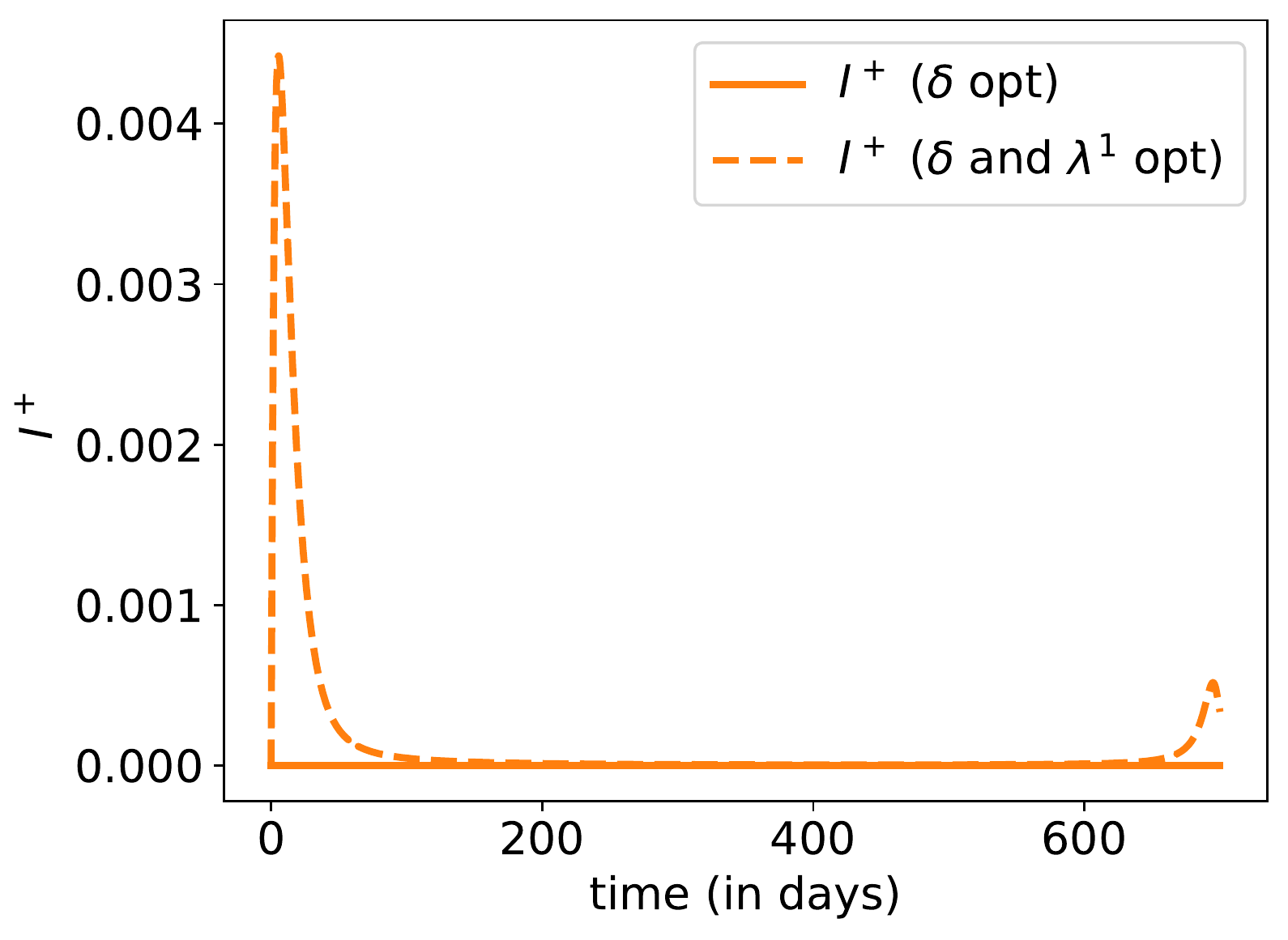}
	\end{subfigure}
	\begin{subfigure}{.33\columnwidth}
		\centering 
		\includegraphics[width=\columnwidth]{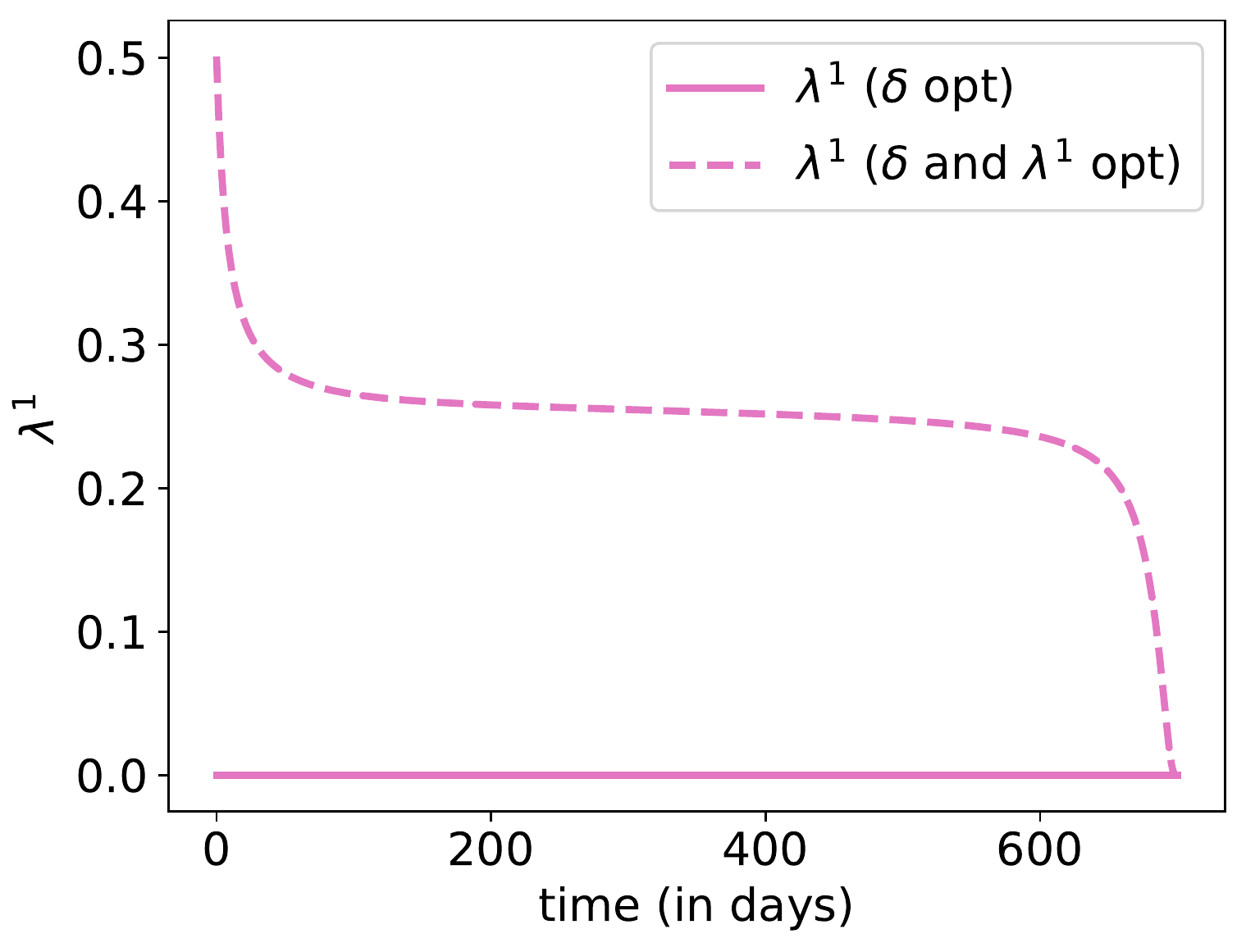}
	\end{subfigure}
	\caption{Evolution of states with optimal controls $(\delta,\lambda^1,0)$, where the plain line corresponds to the benchmark scenario $(\delta,0,0)$ discussed in Section \ref{subsec:optimal:policy} (plain line); and evolution with joint optimal controls $(\delta,\lambda^1,0)$ (dashed line).} 
 	\label{fig:Sensi_lambda1B7}
\end{figure}

\newpage

\subsection{Impact of additional effort in immunity detection $\lambda^2$} \label{sec_sensi_lambda2}

{
Here, we add virologic effort for a constant value of $\lambda^2$ and we compare the result of the optimization over $\delta$ for several values of $\lambda^2$. The optimal control problem for which we compute an approximate solution numerically is:
\begin{eqnarray}
 \inf_{\delta\in\tilde{\mathcal{A}}} \big\lbrace \tilde J_T(\delta,0,\lambda^2)\big\rbrace\;,
\end{eqnarray}
with $\tilde{\mathcal{A}}$ the set of measurable functions from $[0,T]$ to $[0,1]$.
}

\begin{figure}[h]
	\begin{subfigure}{.33\columnwidth}
		\centering
		\includegraphics[width=\columnwidth]{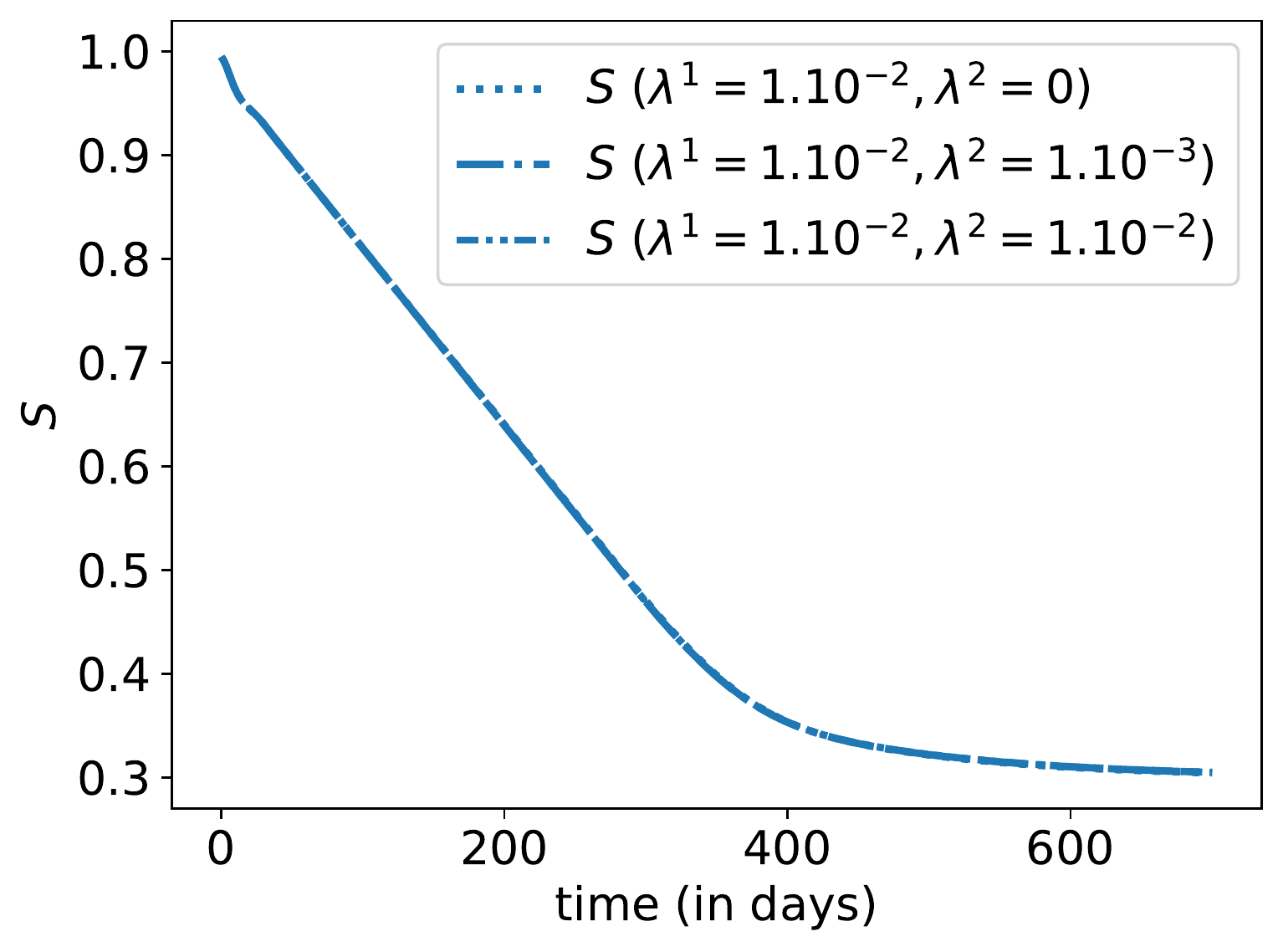}
	\end{subfigure}%
	\begin{subfigure}{.33\columnwidth}
		\centering 
		\includegraphics[width=\columnwidth]{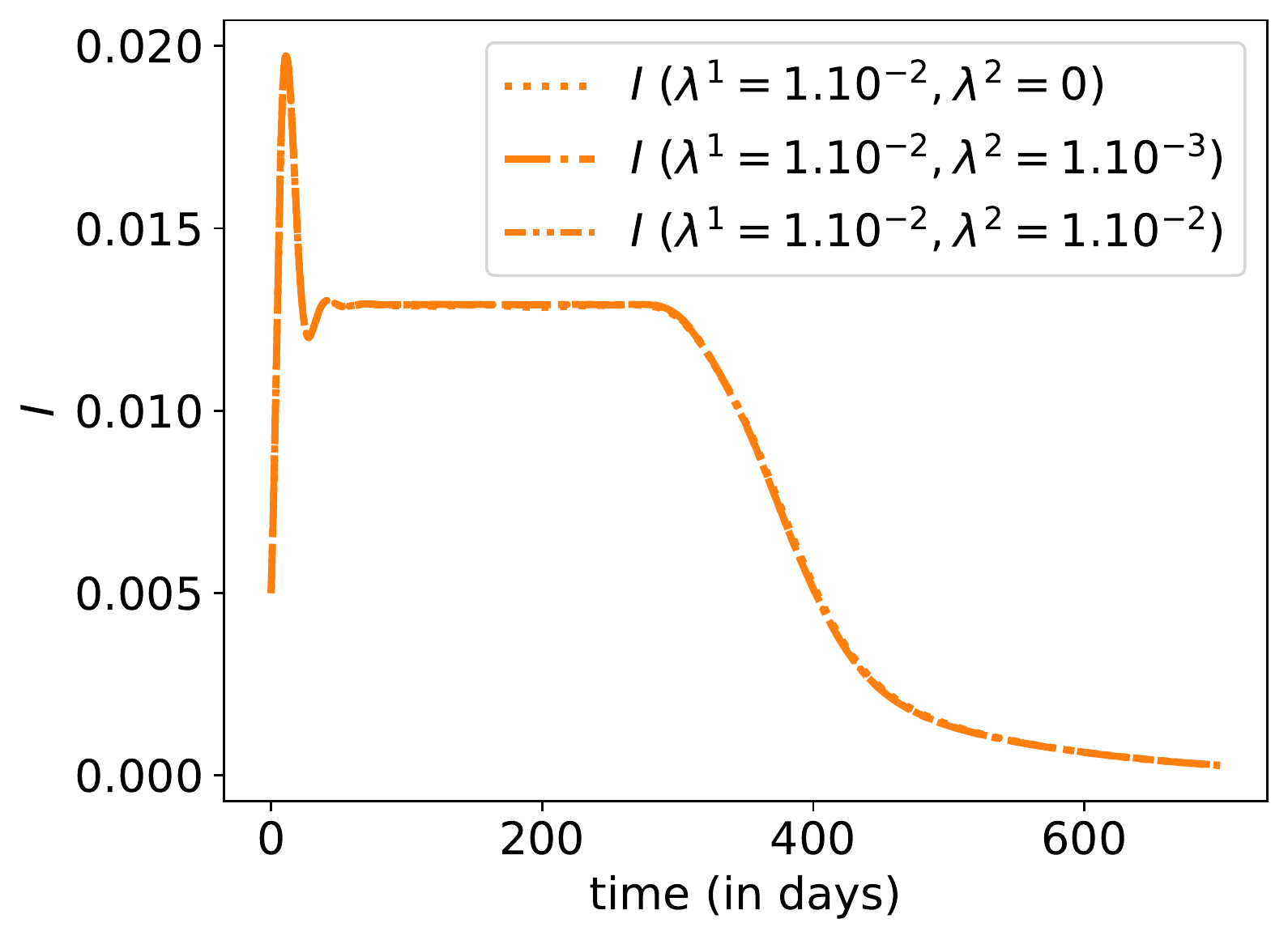}
	\end{subfigure}
	\begin{subfigure}{.33\columnwidth}
		\centering 
		\includegraphics[width=\columnwidth]{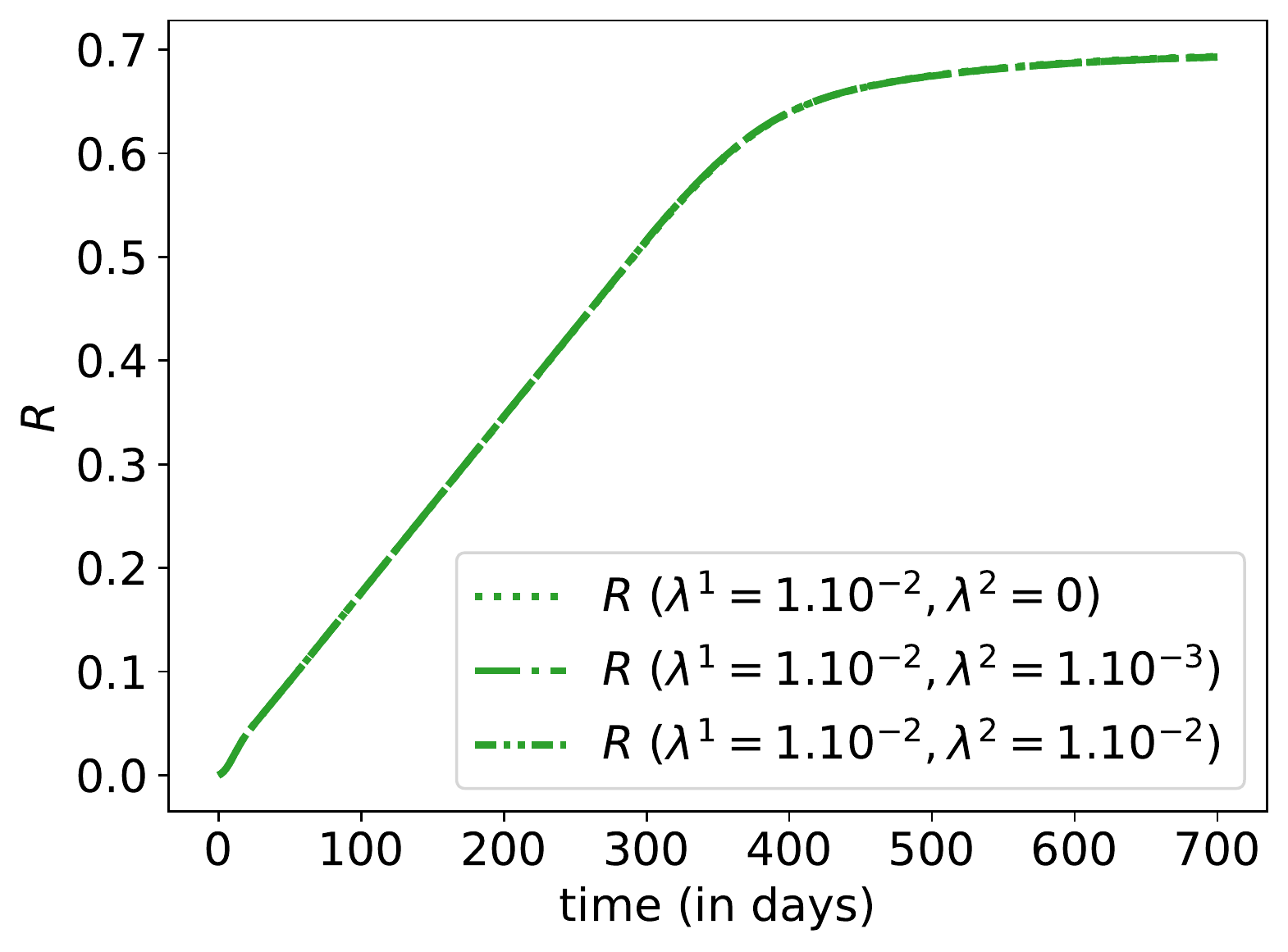}
	\end{subfigure}
	
	\begin{subfigure}{.33\columnwidth}
		\centering
		\includegraphics[width=\columnwidth]{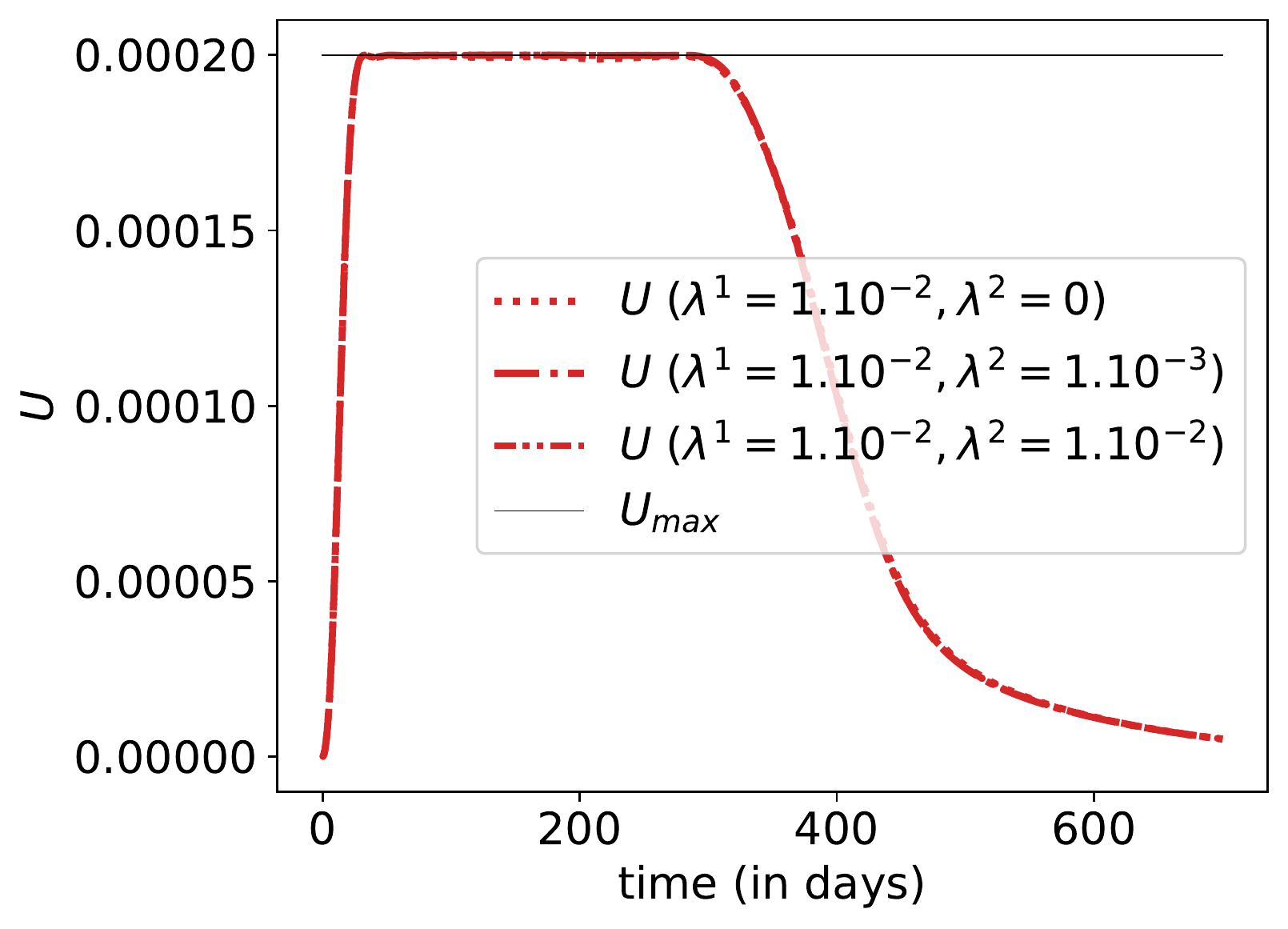}
	\end{subfigure}%
	\begin{subfigure}{.33\columnwidth}
		\centering 
		\includegraphics[width=\columnwidth]{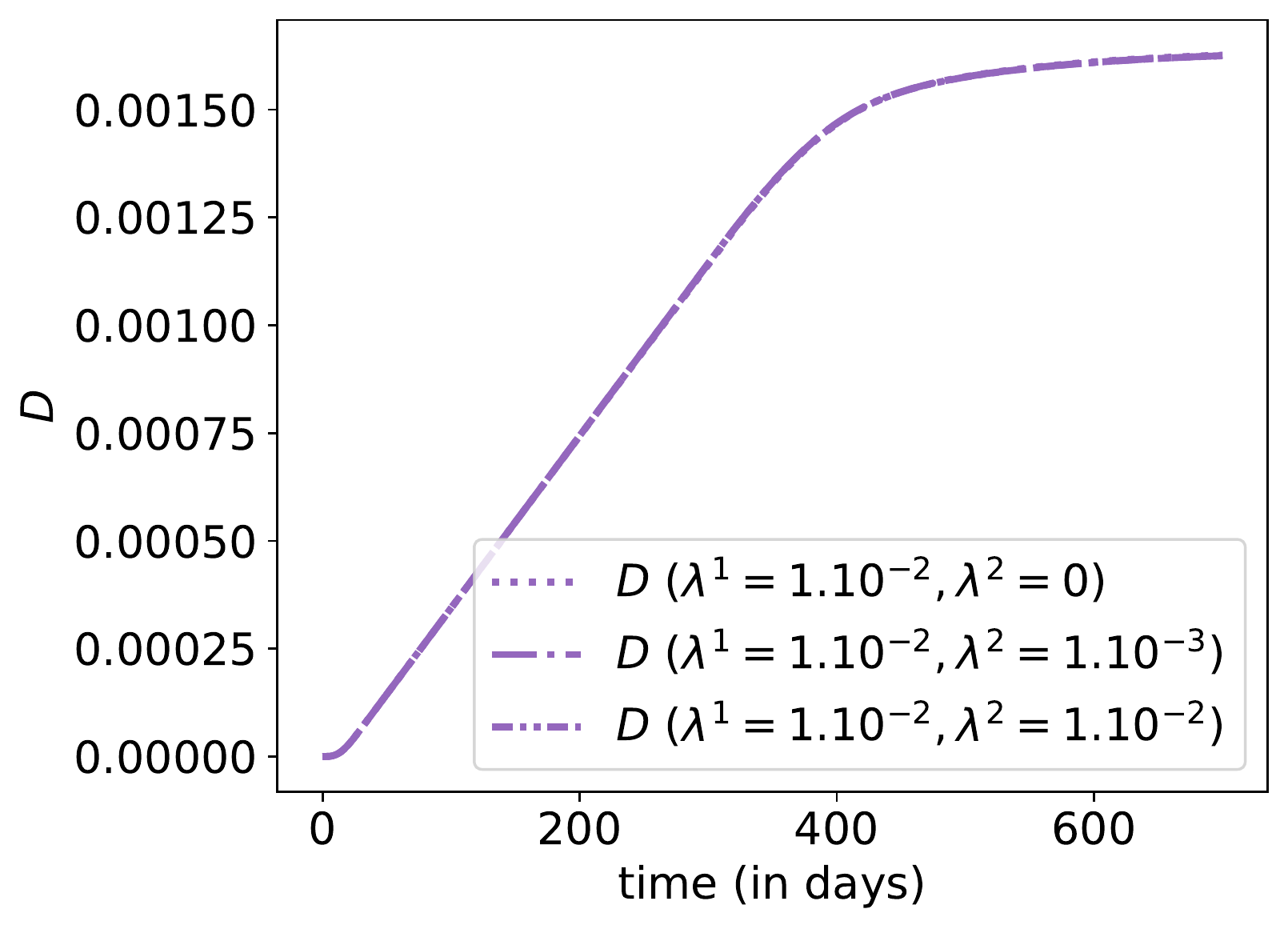}
	\end{subfigure}
	\begin{subfigure}{.33\columnwidth}
		\centering 
		\includegraphics[width=\columnwidth]{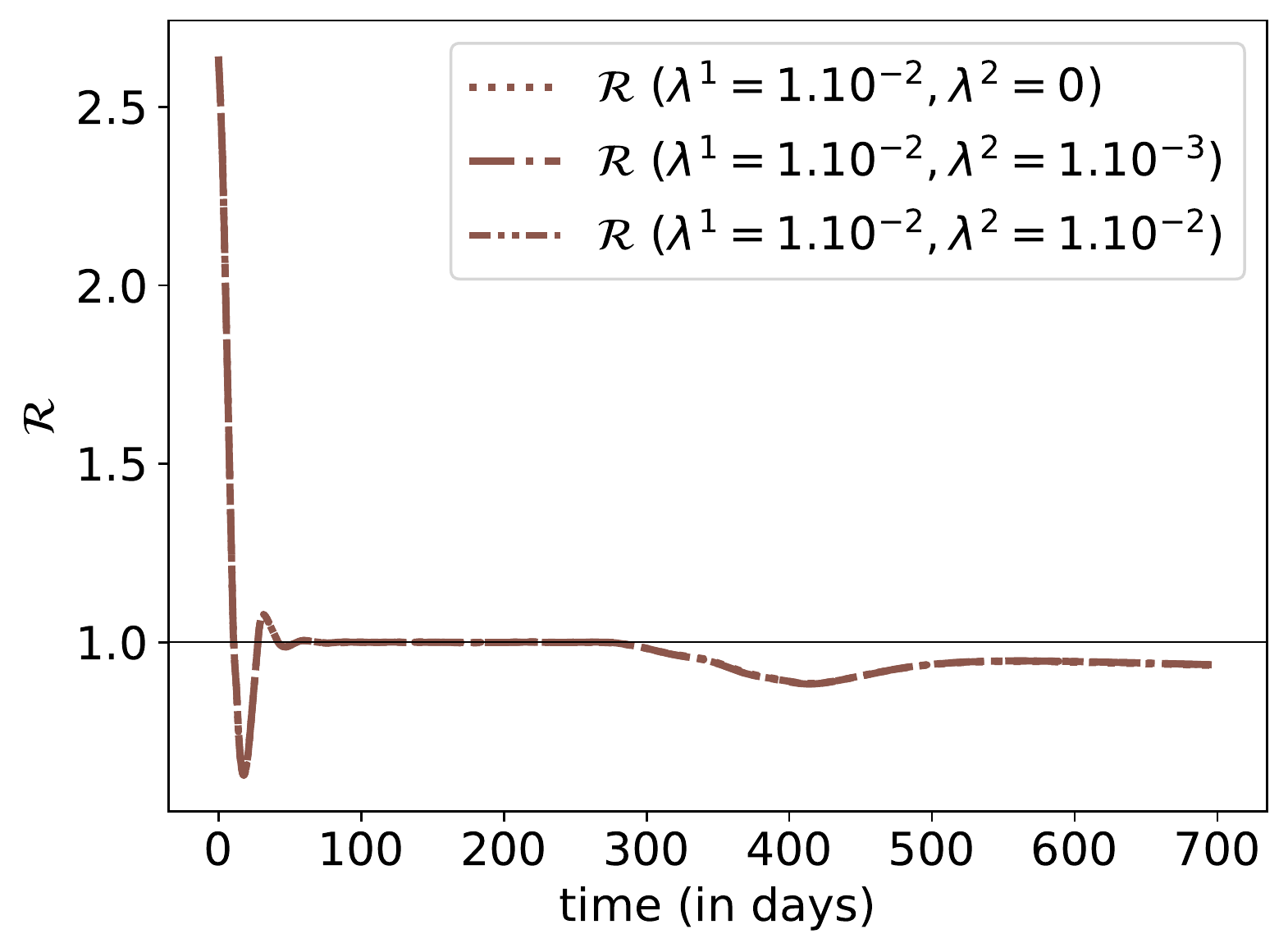}
	\end{subfigure}
	
	\begin{subfigure}{.33\columnwidth}
		\centering
		\includegraphics[width=\columnwidth]{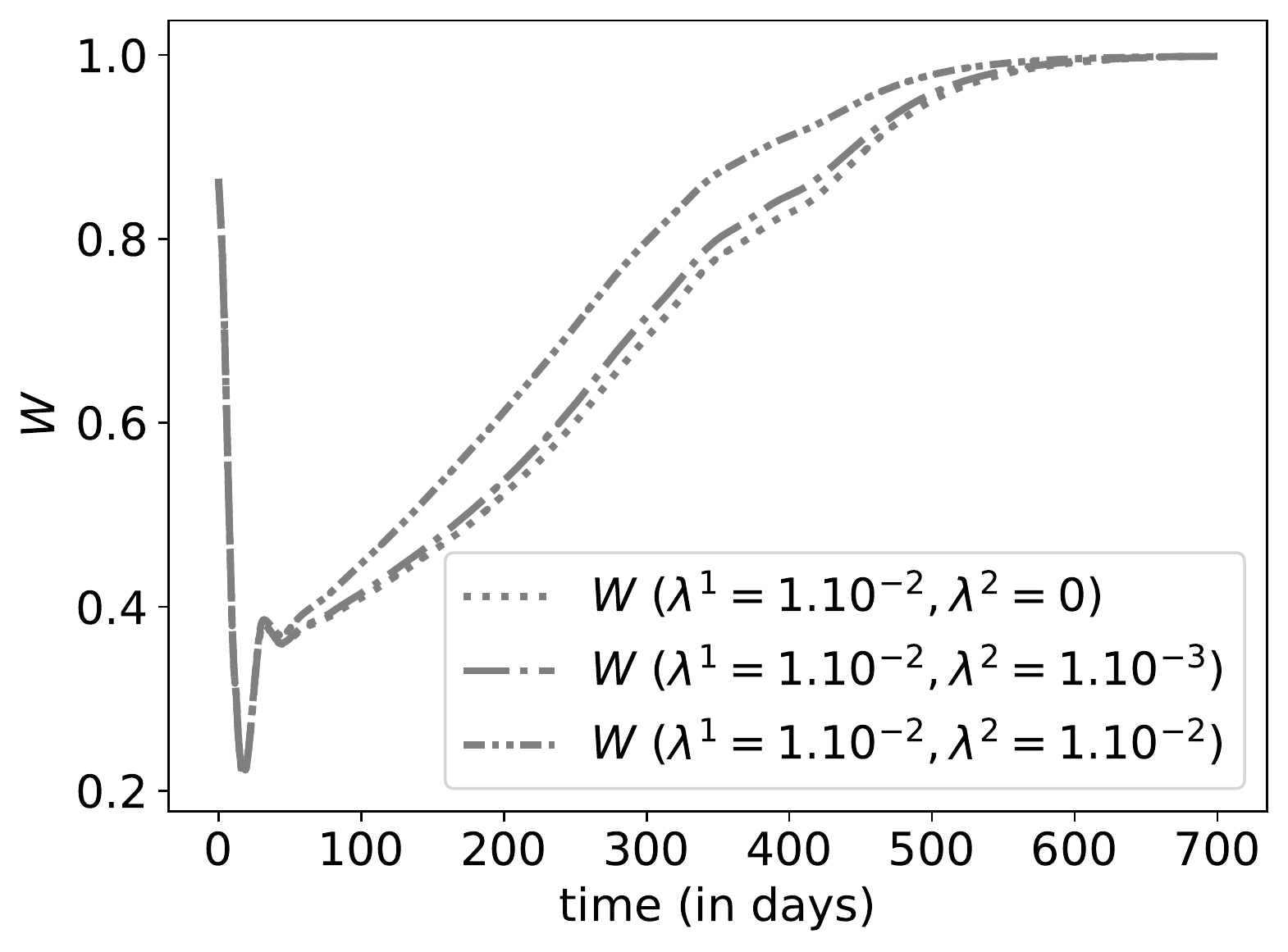}
	\end{subfigure}%
	\begin{subfigure}{.33\columnwidth}
		\centering 
		\includegraphics[width=\columnwidth]{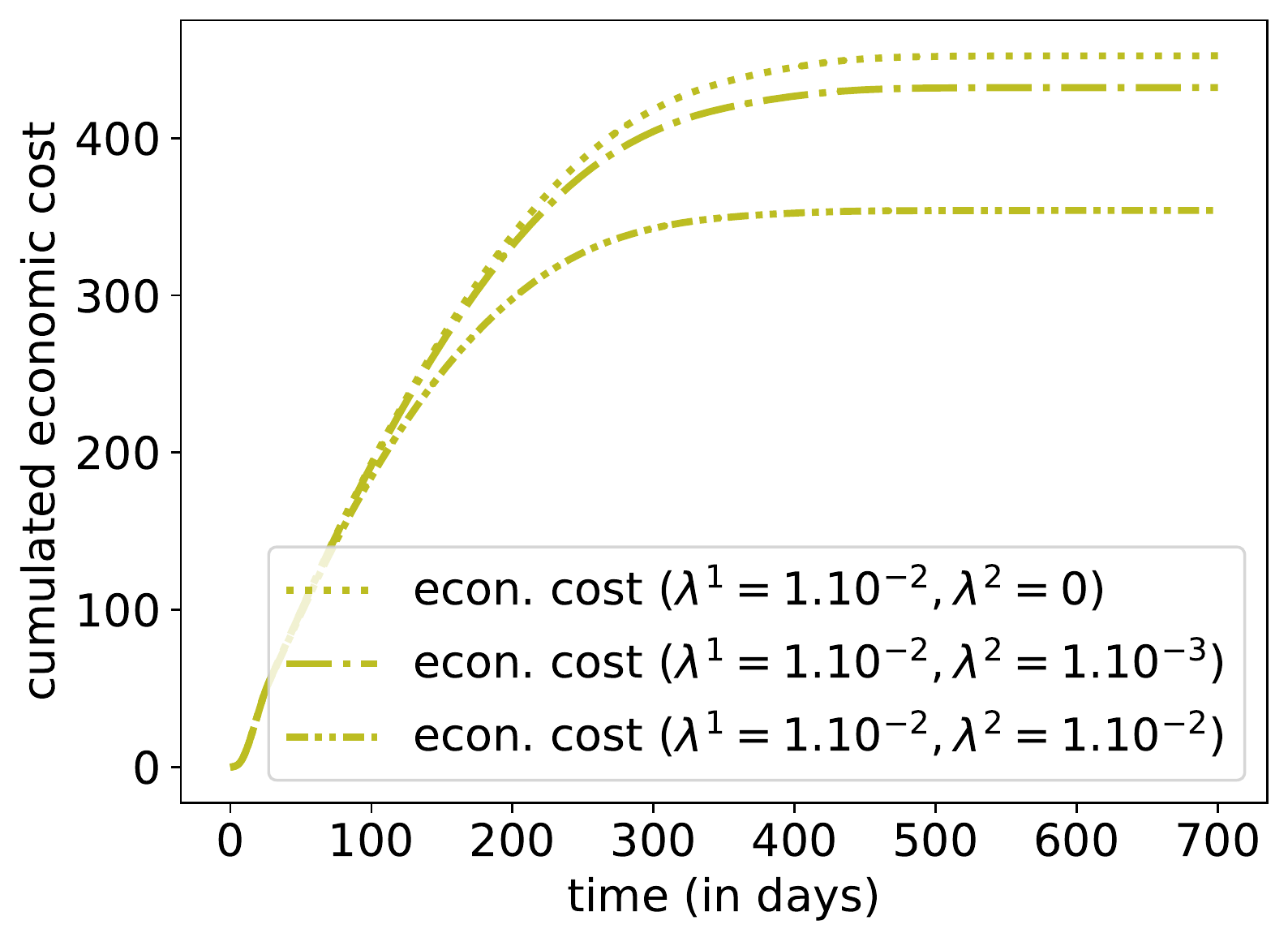}
	\end{subfigure}
	\begin{subfigure}{.33\columnwidth}
		\centering 
		\includegraphics[width=\columnwidth]{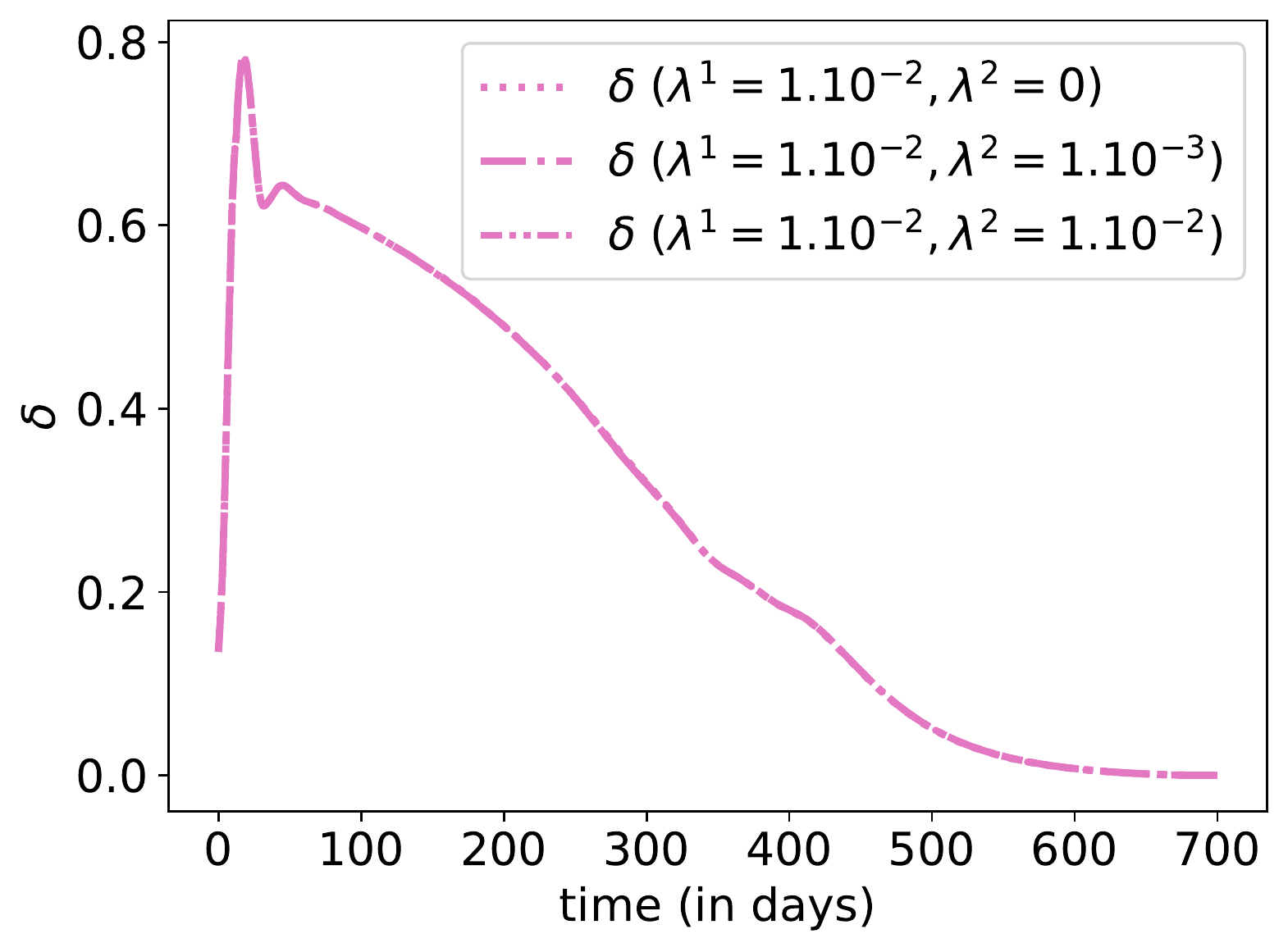}
	\end{subfigure}
	
	\begin{subfigure}{.33\columnwidth}
		\centering
		\includegraphics[width=\columnwidth]{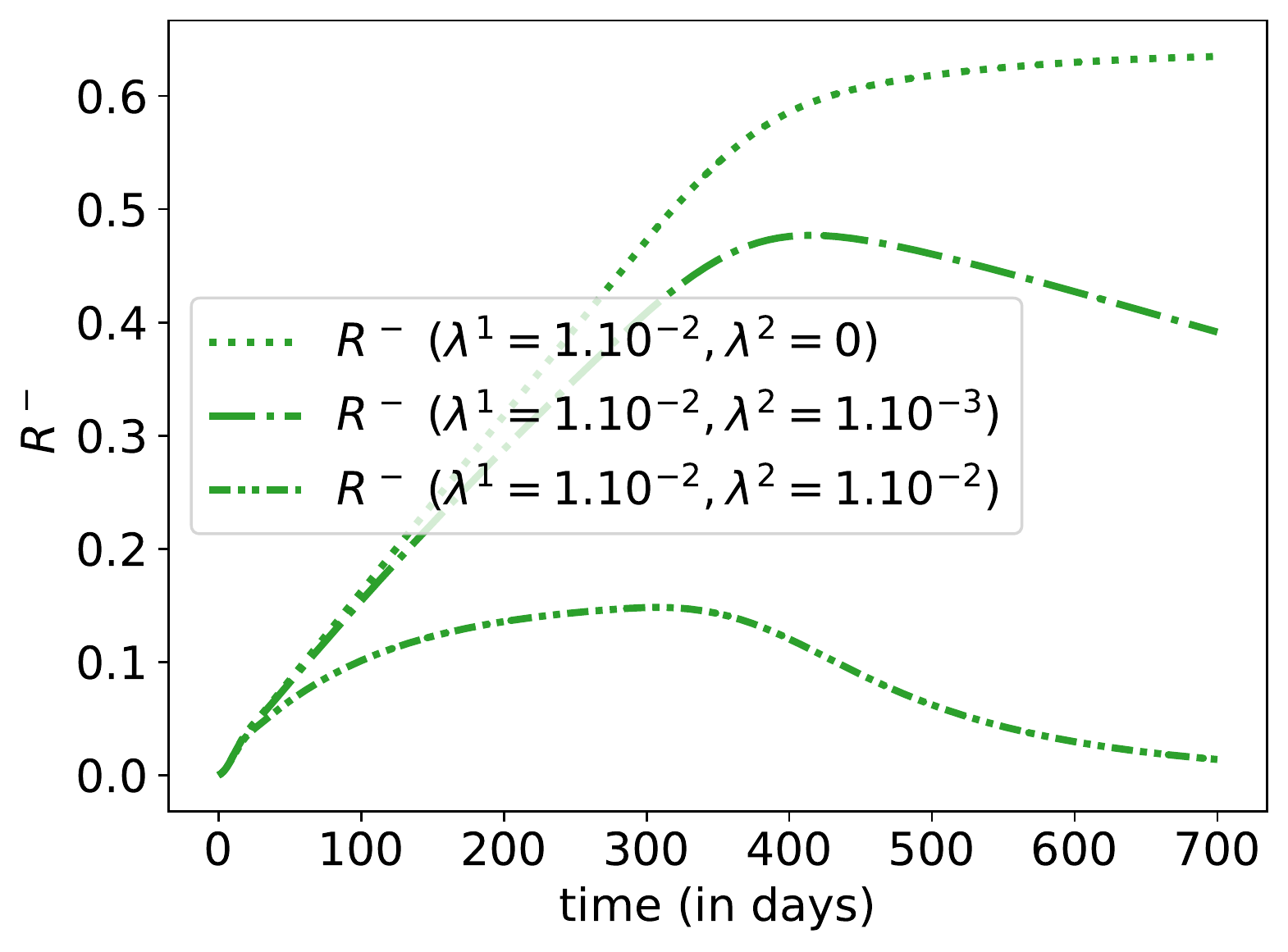}
	\end{subfigure}%
	\begin{subfigure}{.33\columnwidth}
		\centering 
		\includegraphics[width=\columnwidth]{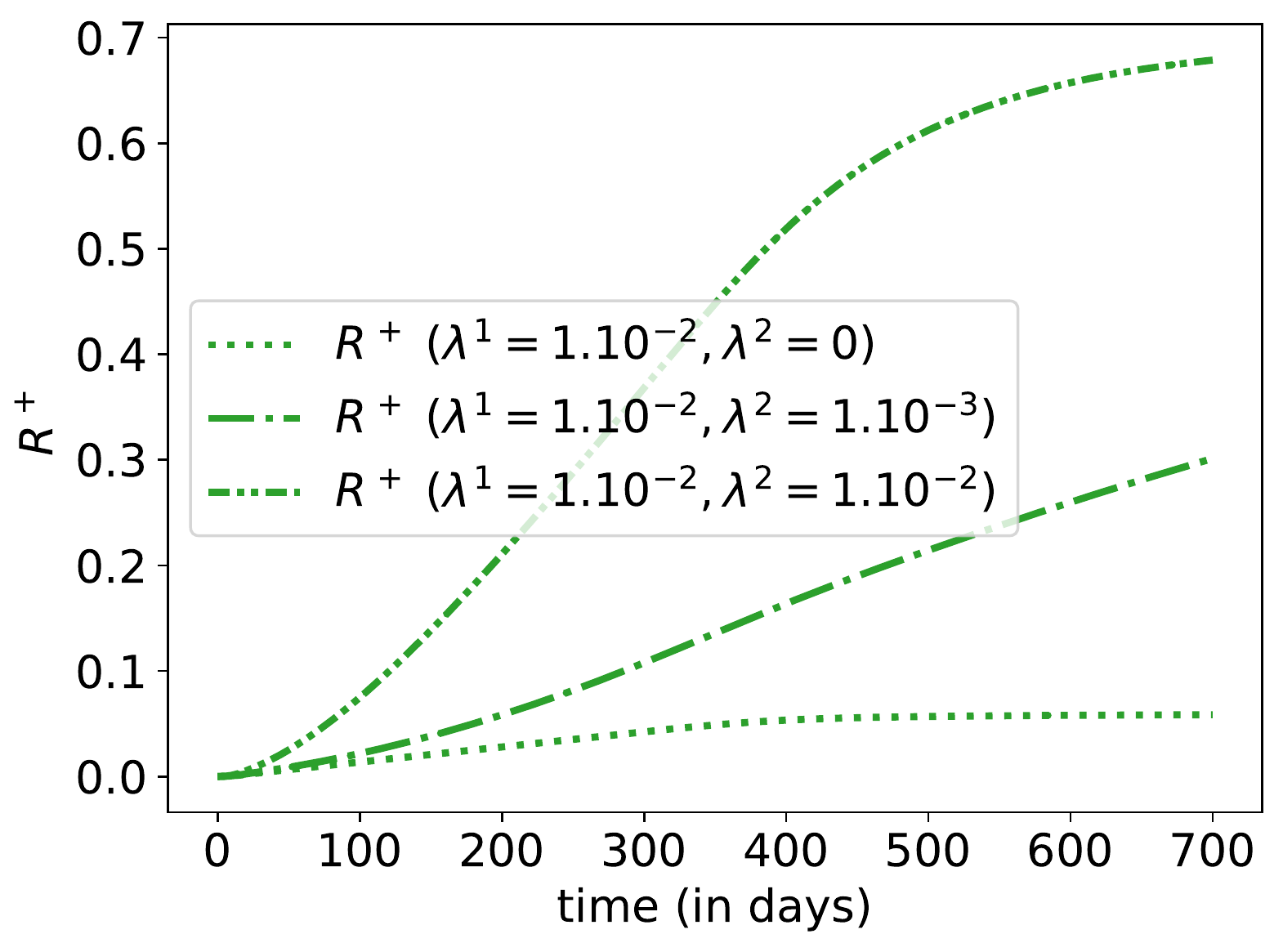}
	\end{subfigure}
	\begin{subfigure}{.33\columnwidth}
		\centering 
		\includegraphics[width=\columnwidth]{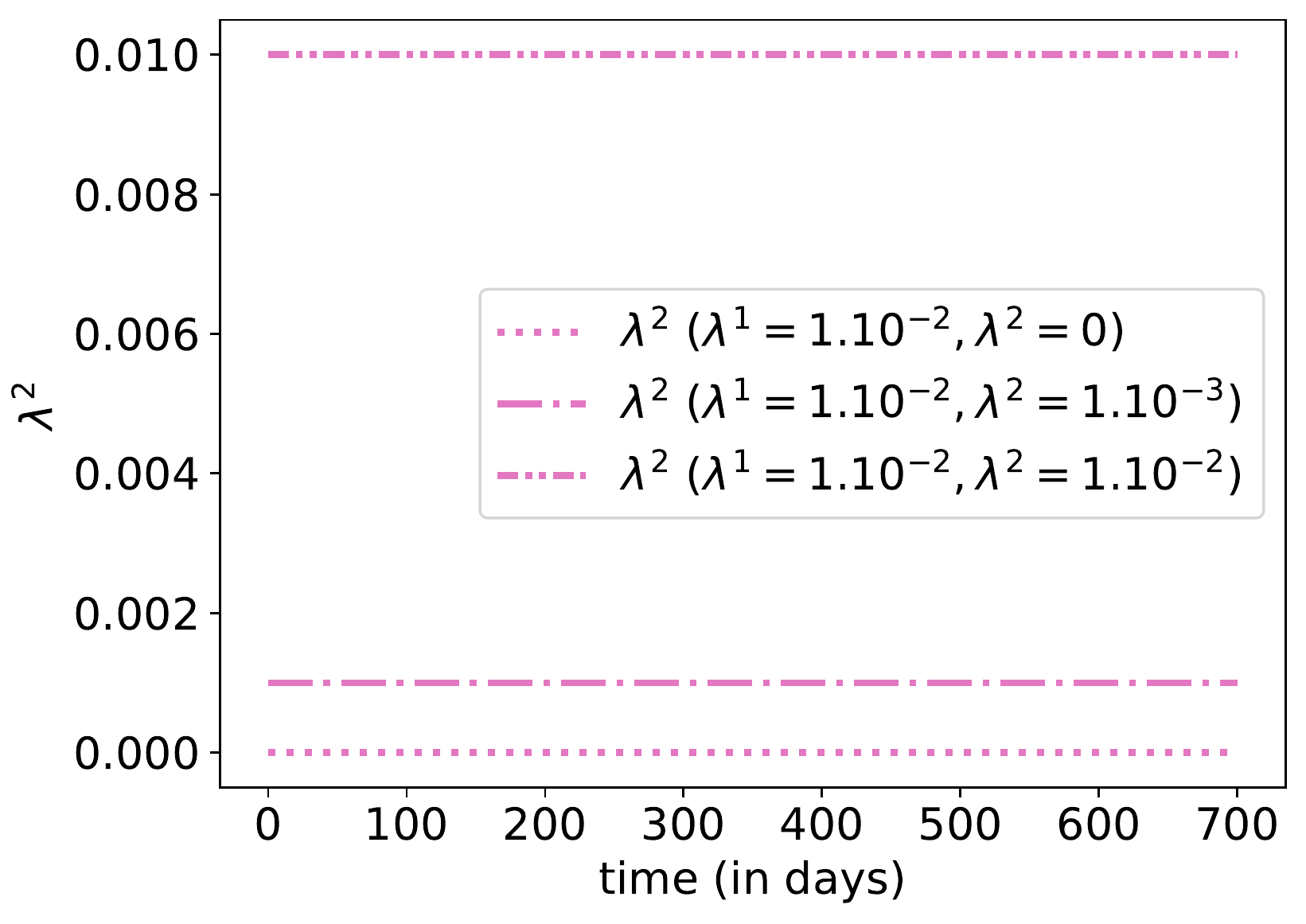}
	\end{subfigure}
	\caption{Evolution of states with optimal control $\delta$ for three different values of $\lambda^2$ given a fixed level of $\lambda^1$ (here $1\%$), with  $\lambda^2=0$ (dotted line), $\lambda^2=1$\textperthousand~(dashed line) and $\lambda^2=1\%$ (mixed line), i.e. $(\delta^*,\lambda_1,\lambda_2)$.
	}
 	\label{fig:Sensi_lambda2}
\end{figure}

\newpage

\subsection{Optimizing over effort in immunity detection $\lambda^2$ without lockdown intervention} \label{sec_opti_lambda2}

{
Here, we compare the situation without intervention at all and the situation with intervention through $\lambda^2$. The optimal control problem for which we compute an approximate solution numerically is:
\begin{eqnarray}
 \inf_{\lambda^2\in\tilde{\mathcal{A}}} \big\lbrace \tilde J_T(0,0,\lambda^2)\big\rbrace\;,
\end{eqnarray}
with $\tilde{\mathcal{A}}$ the set of measurable functions from $[0,T]$ to $[0,1]$.
}

\begin{figure}[h]
	\begin{subfigure}{.33\columnwidth}
		\centering
		\includegraphics[width=\columnwidth]{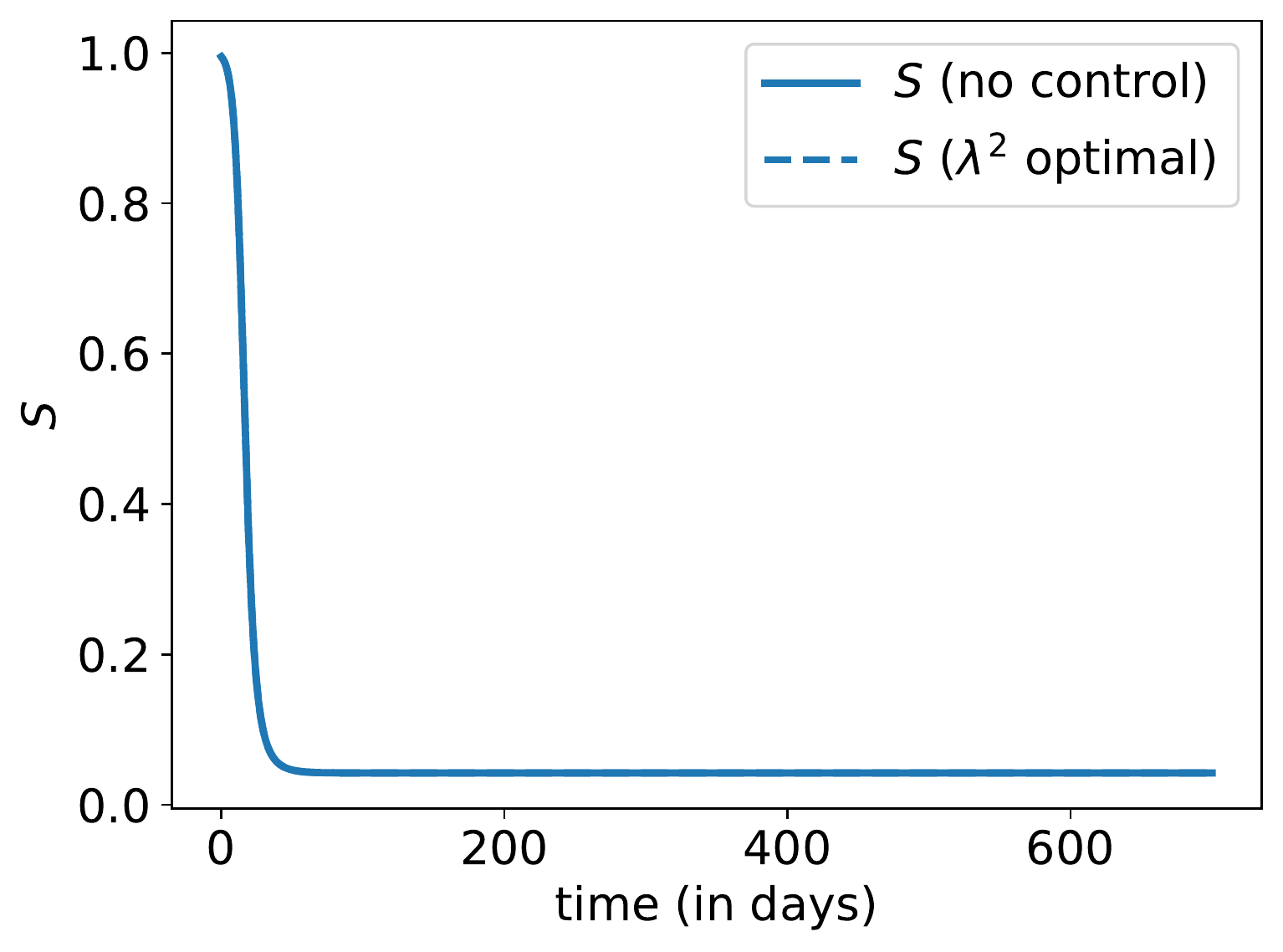}
	\end{subfigure}%
	\begin{subfigure}{.33\columnwidth}
		\centering 
		\includegraphics[width=\columnwidth]{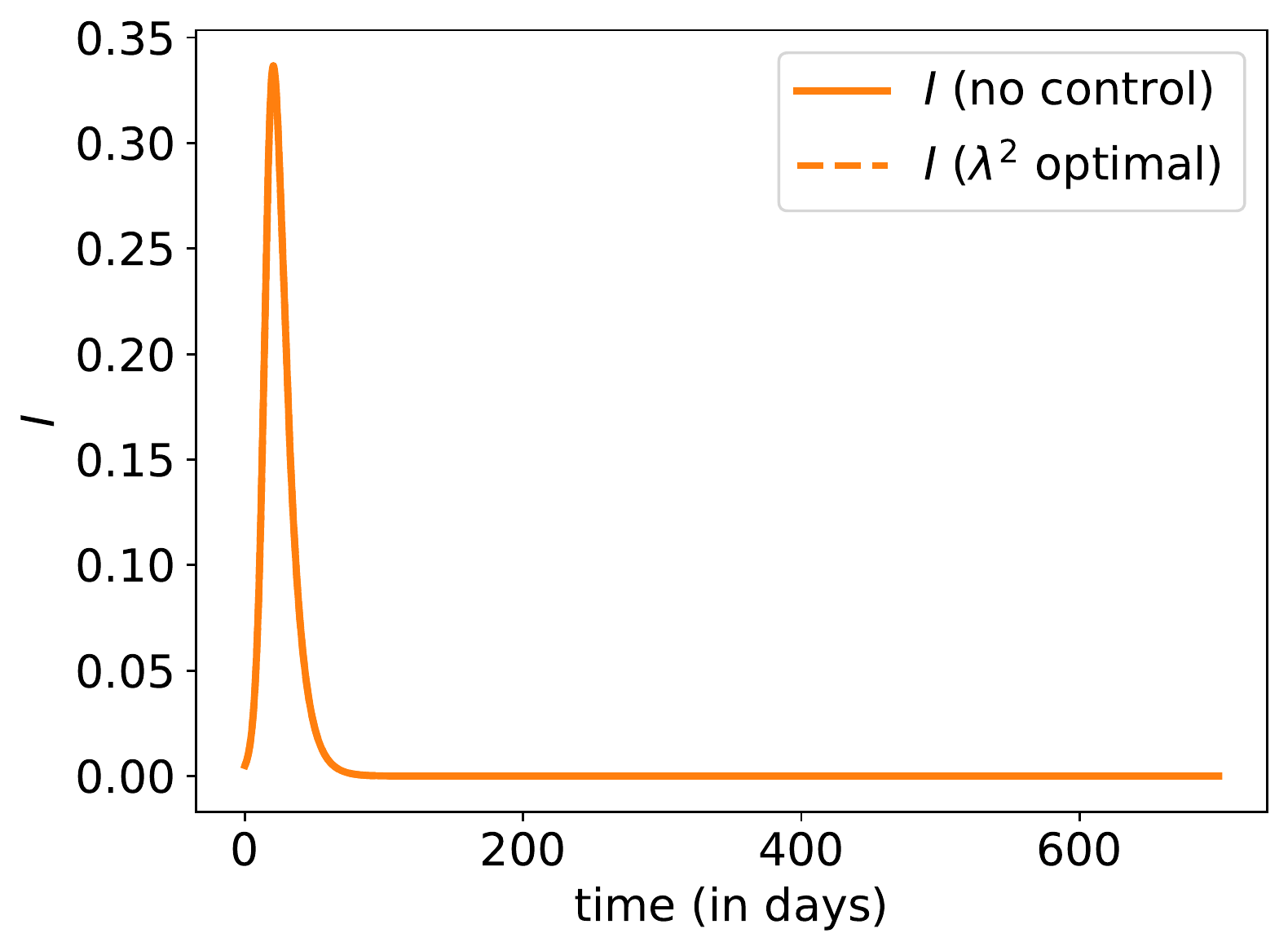}
	\end{subfigure}
	\begin{subfigure}{.33\columnwidth}
		\centering 
		\includegraphics[width=\columnwidth]{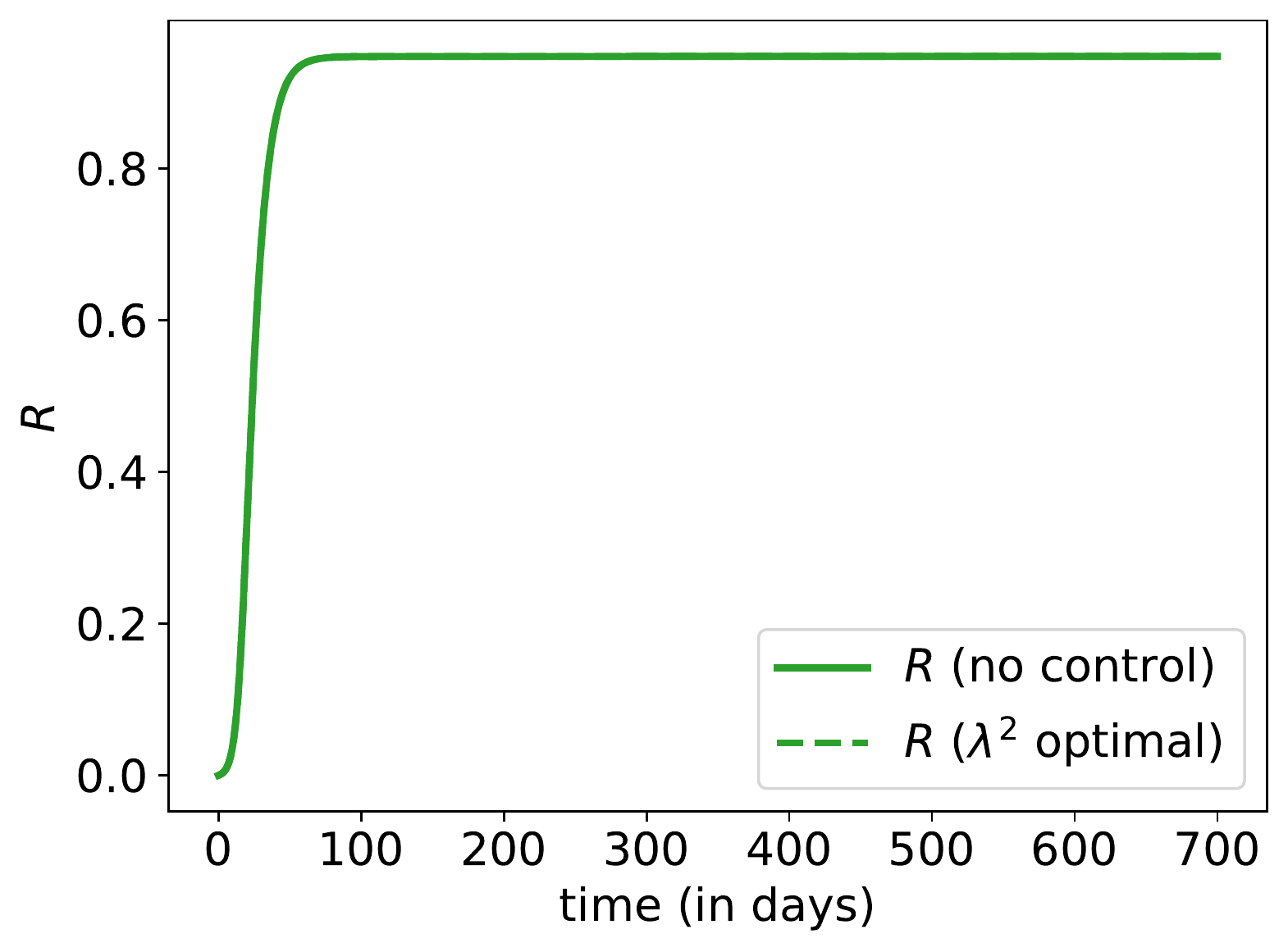}
	\end{subfigure}
	
	\begin{subfigure}{.33\columnwidth}
		\centering
		\includegraphics[width=\columnwidth]{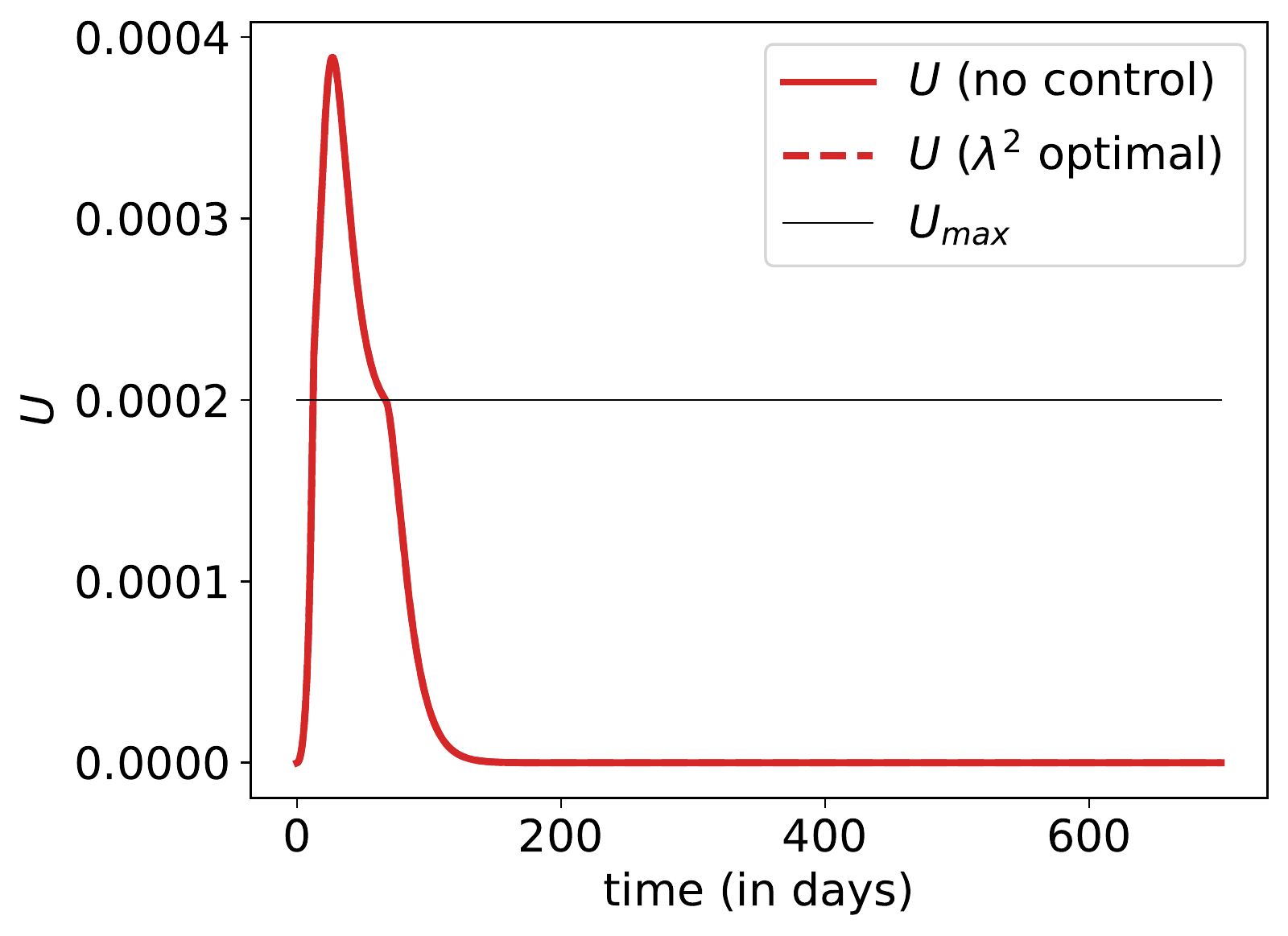}
	\end{subfigure}%
	\begin{subfigure}{.33\columnwidth}
		\centering 
		\includegraphics[width=\columnwidth]{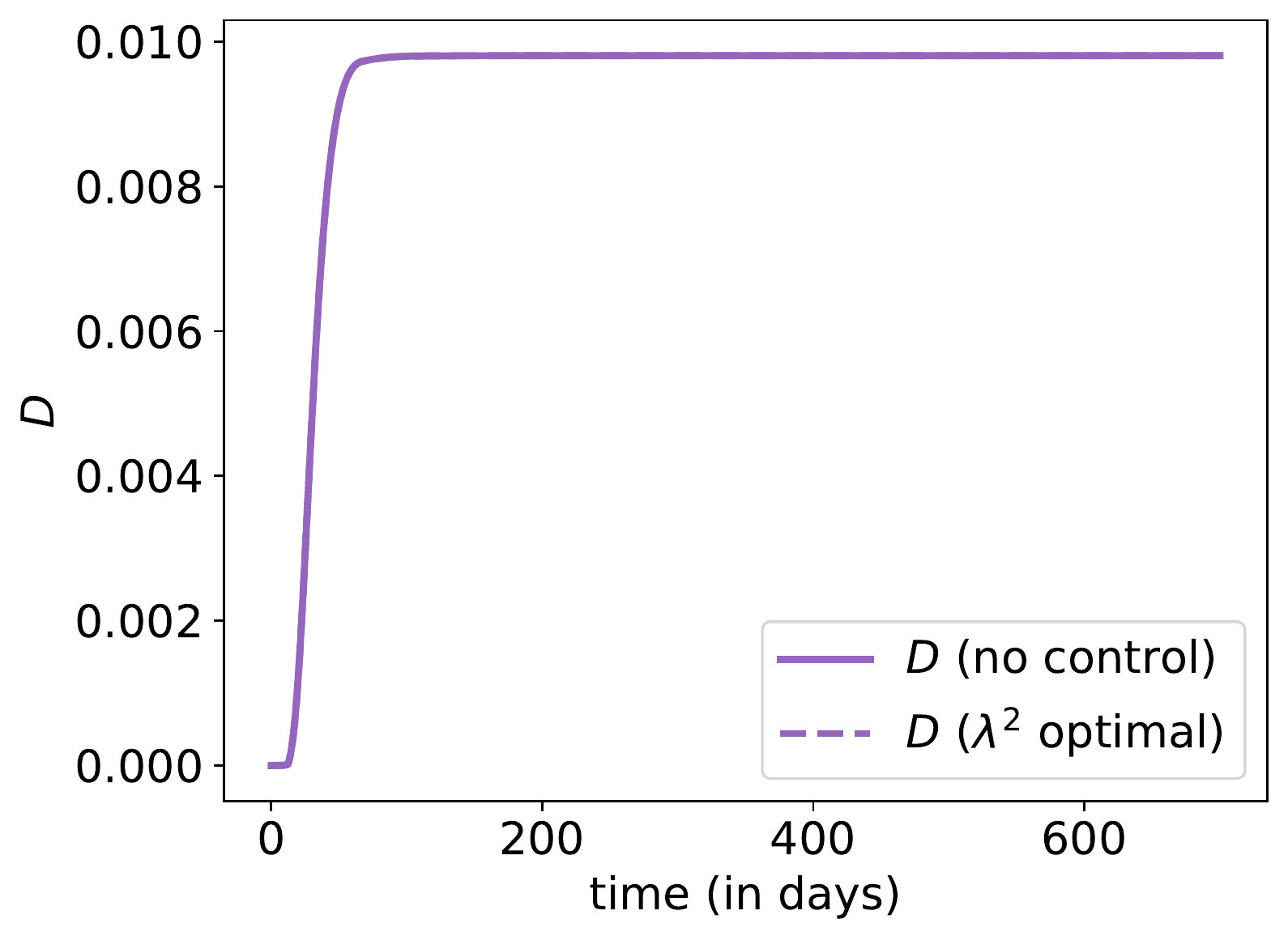}
	\end{subfigure}
	\begin{subfigure}{.33\columnwidth}
		\centering 
		\includegraphics[width=\columnwidth]{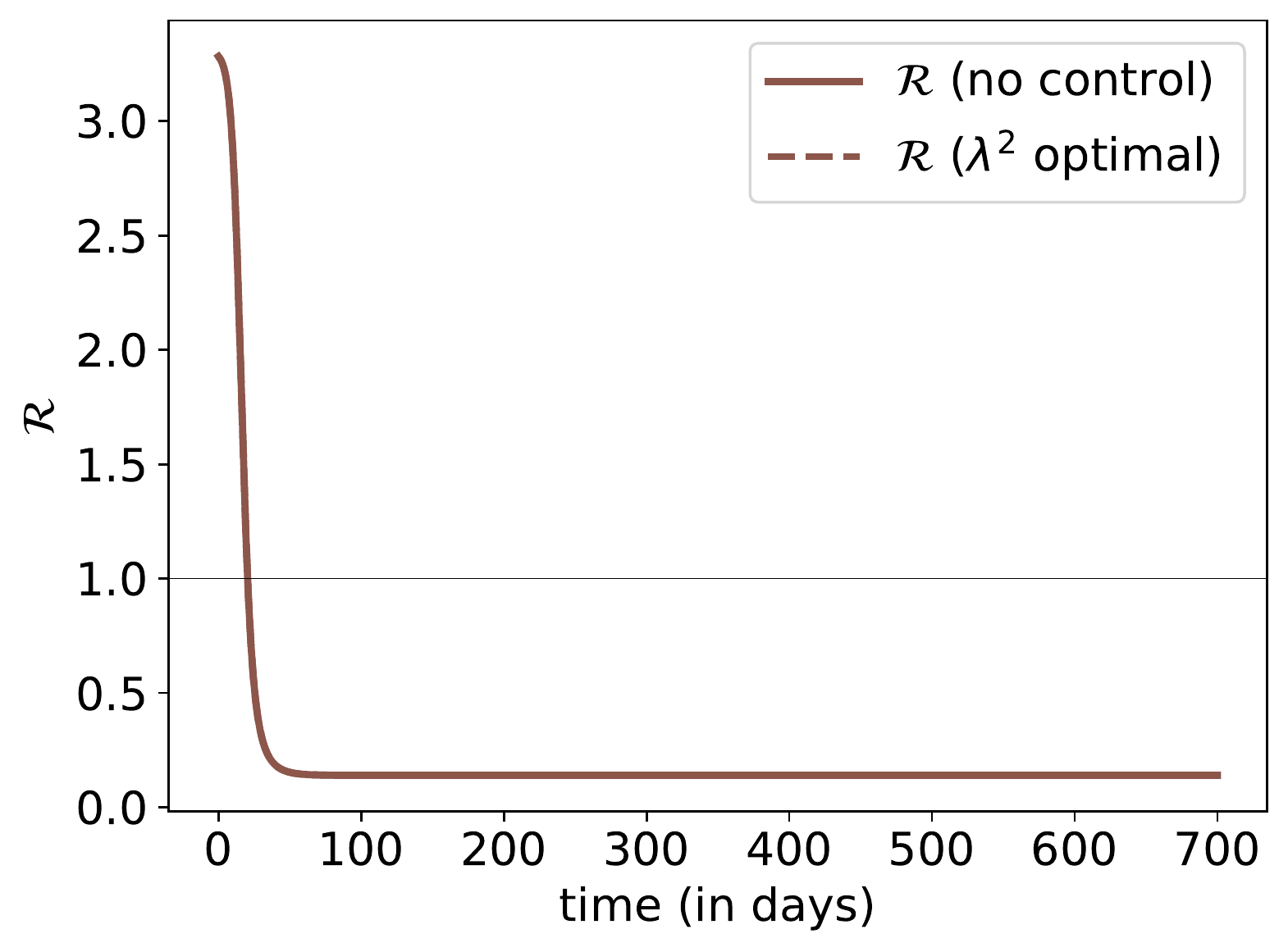}
	\end{subfigure}
	
	\begin{subfigure}{.33\columnwidth}
		\centering
		\includegraphics[width=\columnwidth]{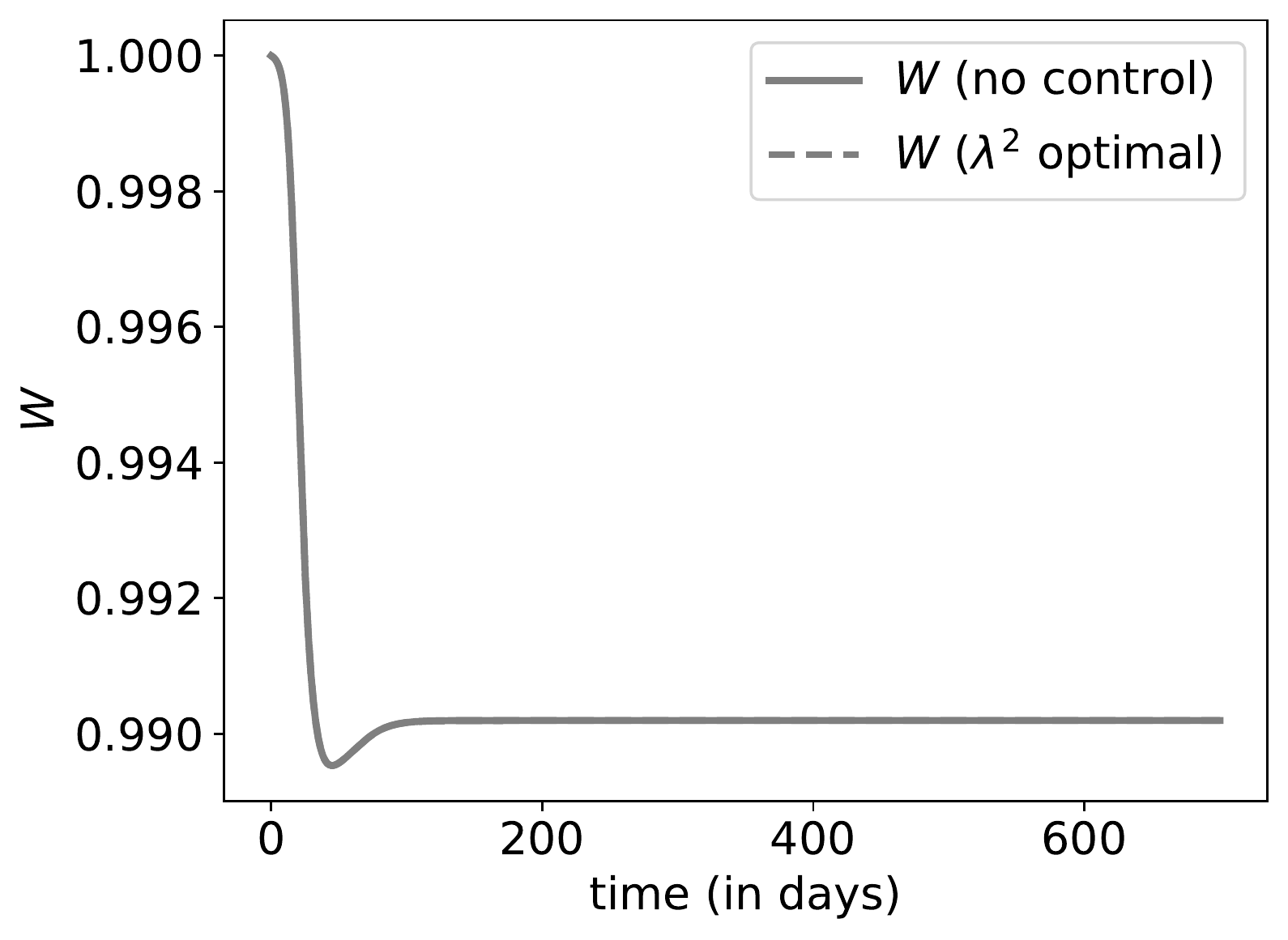}
	\end{subfigure}%
	\begin{subfigure}{.33\columnwidth}
		\centering 
		\includegraphics[width=\columnwidth]{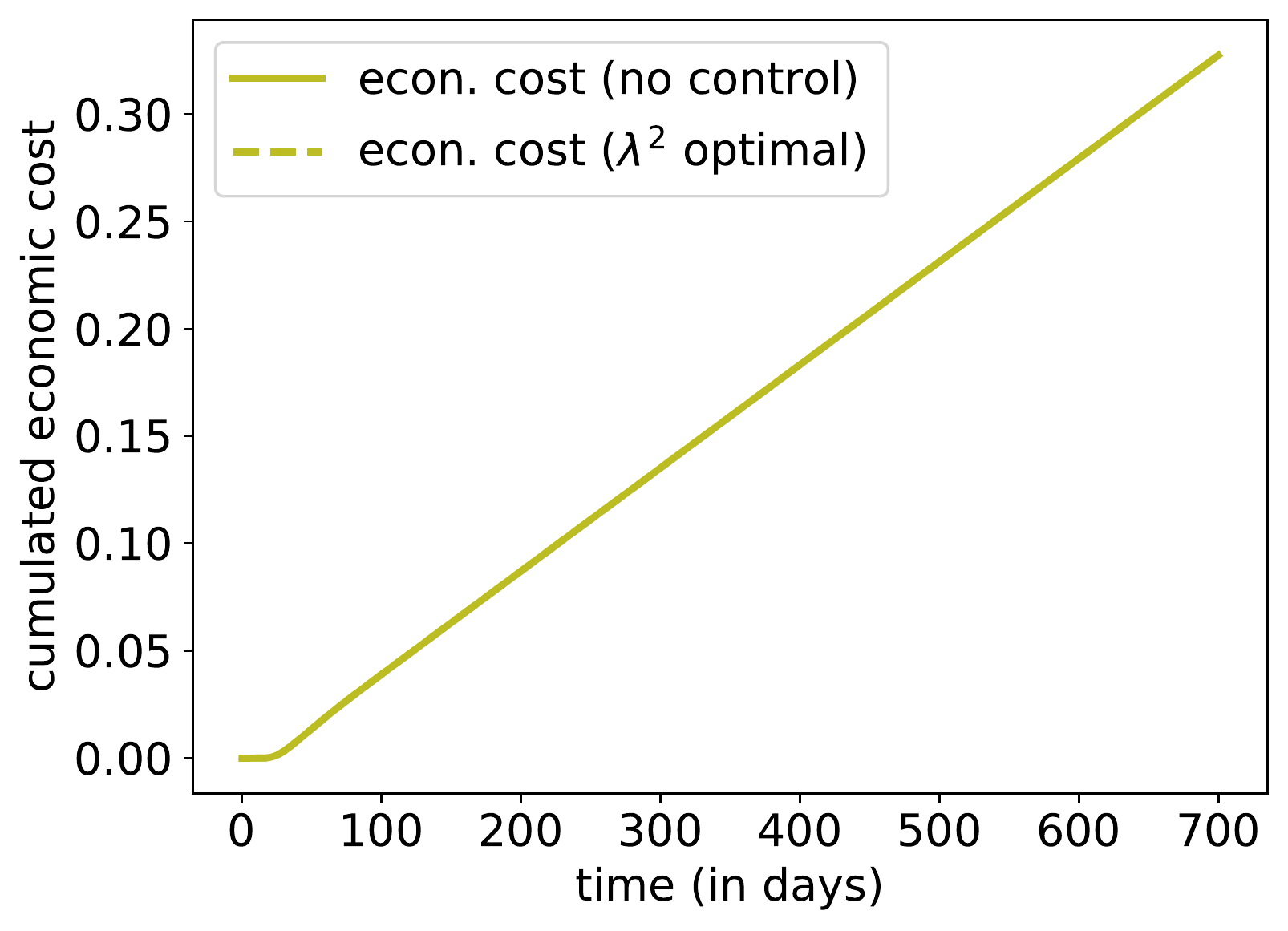}
	\end{subfigure}
	\begin{subfigure}{.33\columnwidth}
		\centering 
		\includegraphics[width=\columnwidth]{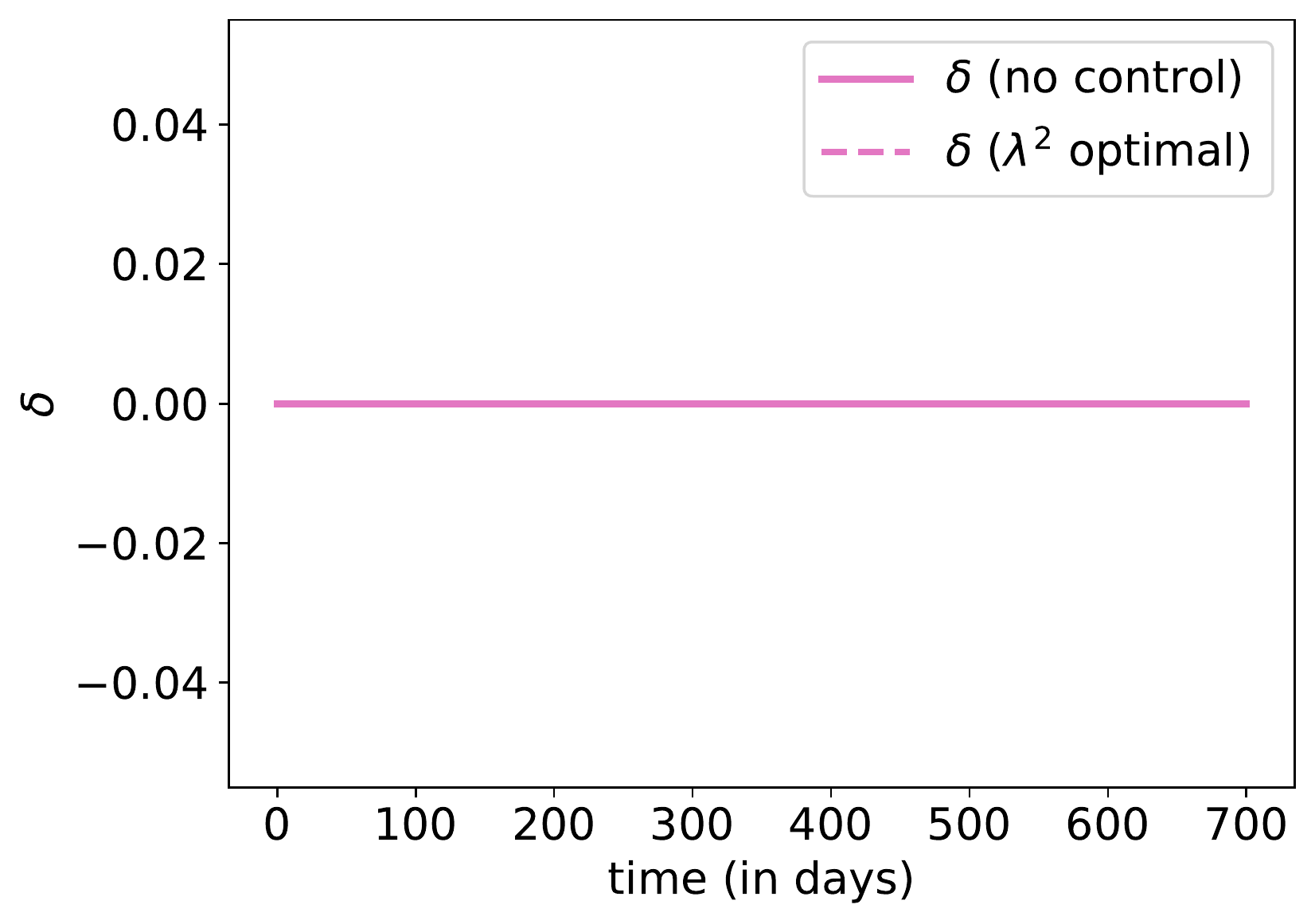}
	\end{subfigure}
	
	\begin{subfigure}{.33\columnwidth}
		\centering
		\includegraphics[width=\columnwidth]{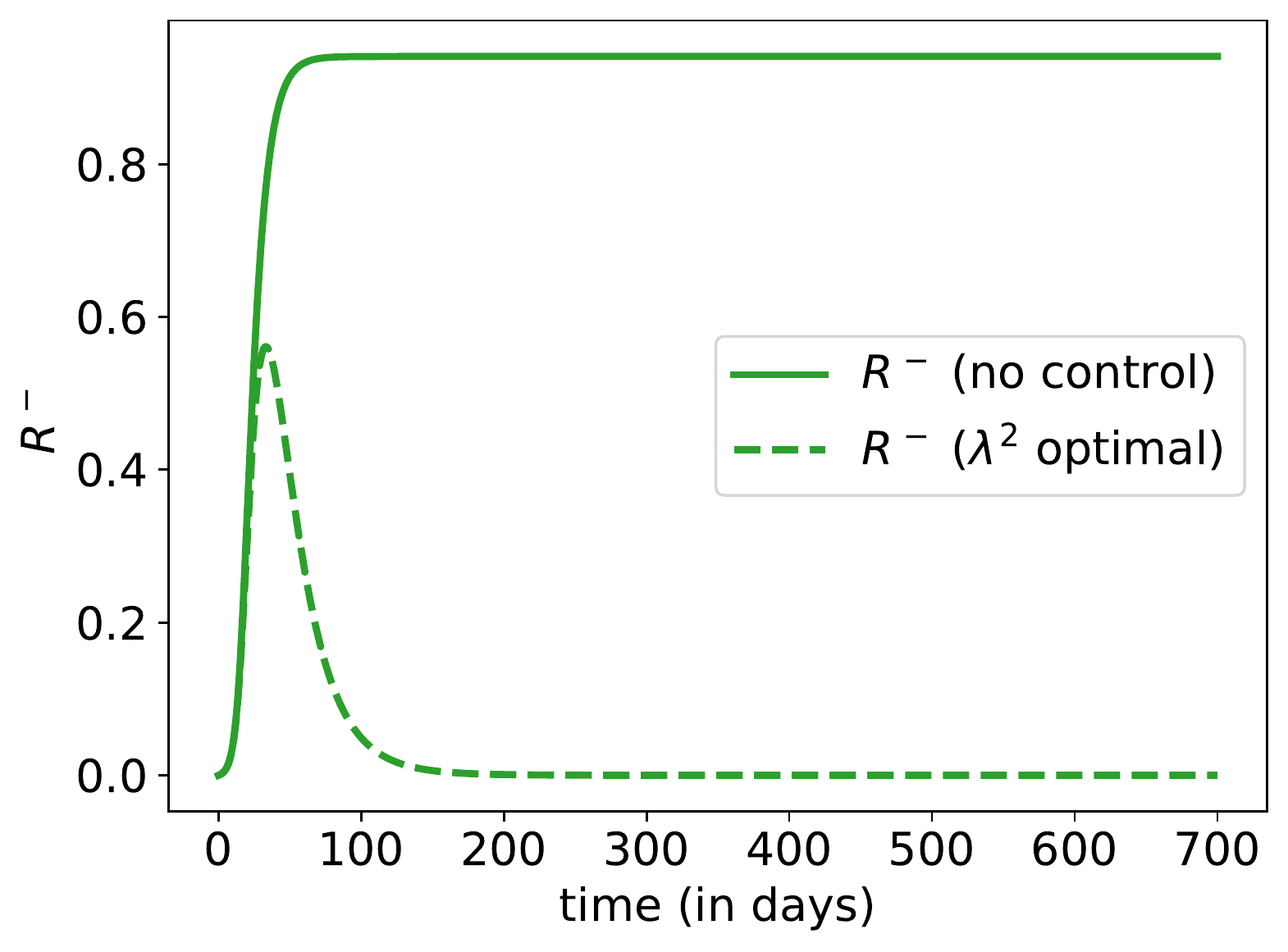}
	\end{subfigure}%
	\begin{subfigure}{.33\columnwidth}
		\centering 
		\includegraphics[width=\columnwidth]{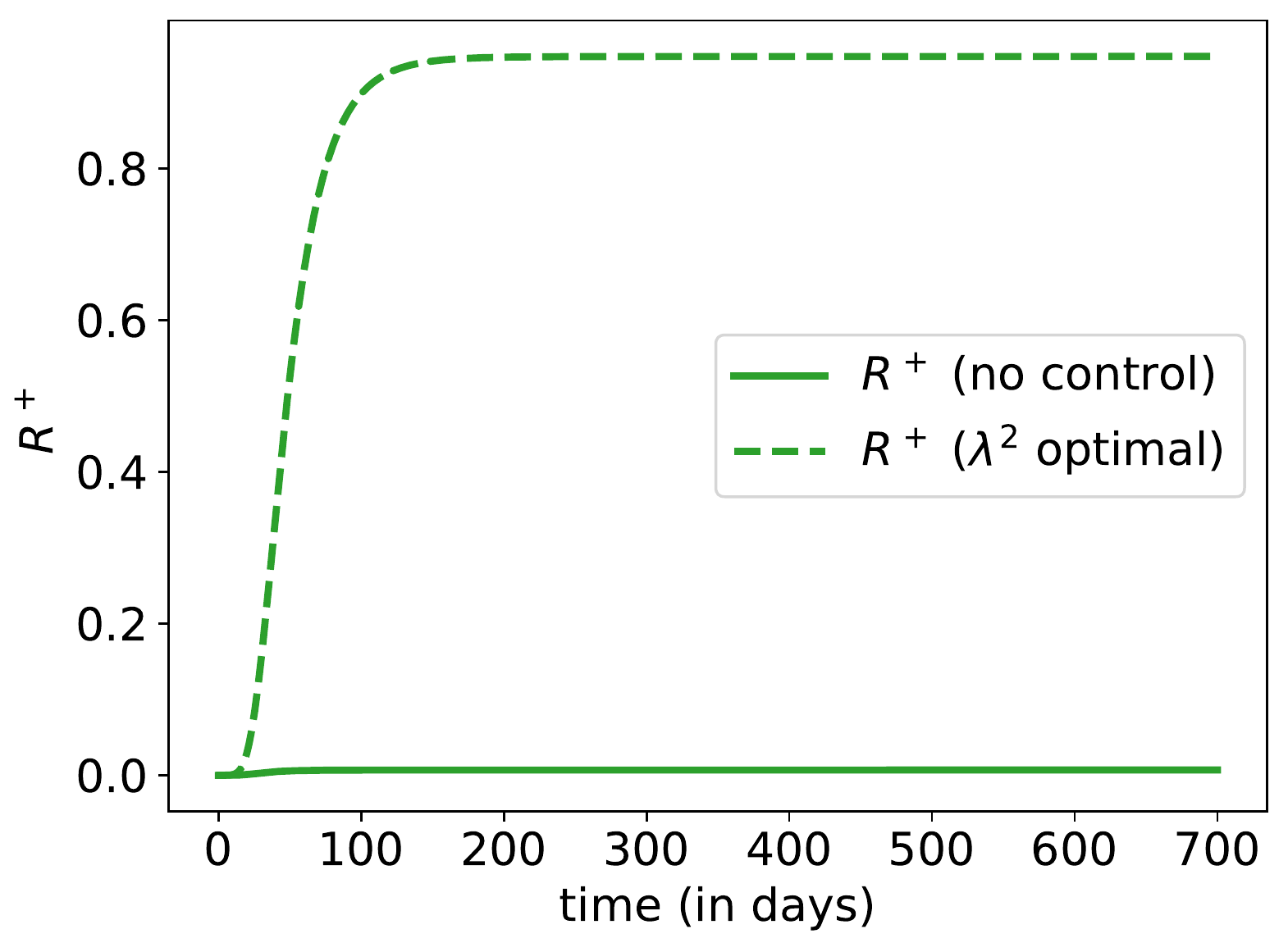}
	\end{subfigure}
	\begin{subfigure}{.33\columnwidth}
		\centering 
		\includegraphics[width=\columnwidth]{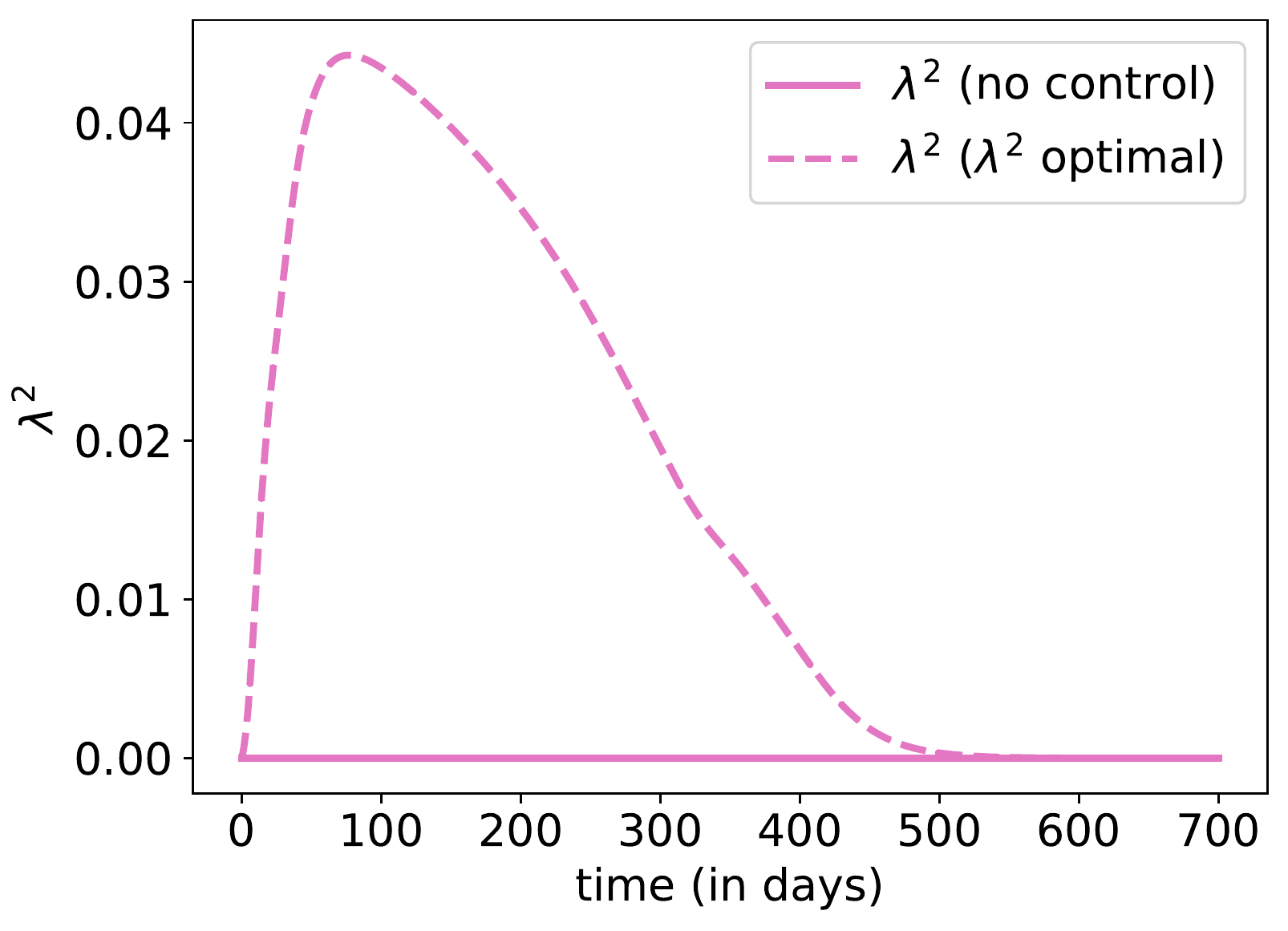}
	\end{subfigure}	\caption{Evolution of states with optimal controls $(0,0,\lambda^2)$, where the plain line corresponds to the benchmark scenario discussed in Section \ref{sec:3:1} (with no intervention $\delta=\lambda^1=\lambda^2=0$); and evolution with optimal control $\lambda^2$ (when $\delta=\lambda^1=0$). 
	}
 	\label{fig:Optim_delta_lambda2}
\end{figure}

\newpage 

\subsection{Optimizing over both lockdown intervention $\delta$ and effort in immunity detection $\lambda^2$} \label{sec_opti_delta_lambda2_b10}

{
Here, we compare the situation with optimal lockdown policy versus the situation in which we optimize over both lockdown and immunity detection. The latter corresponds to the following optimal control problem, for which we compute an approximate solution numerically:
\begin{eqnarray}
 \inf_{(\delta,\lambda^2)\in\tilde{\mathcal{A}} \times \tilde{\mathcal{A}}} \big\lbrace \tilde J_T(\delta,0,\lambda^2)\big\rbrace\;,
\end{eqnarray}
with $\tilde{\mathcal{A}}$ the set of measurable functions from $[0,T]$ to $[0,1]$.
}

\begin{figure}[h]
	\begin{subfigure}{.33\columnwidth}
		\centering
		\includegraphics[width=\columnwidth]{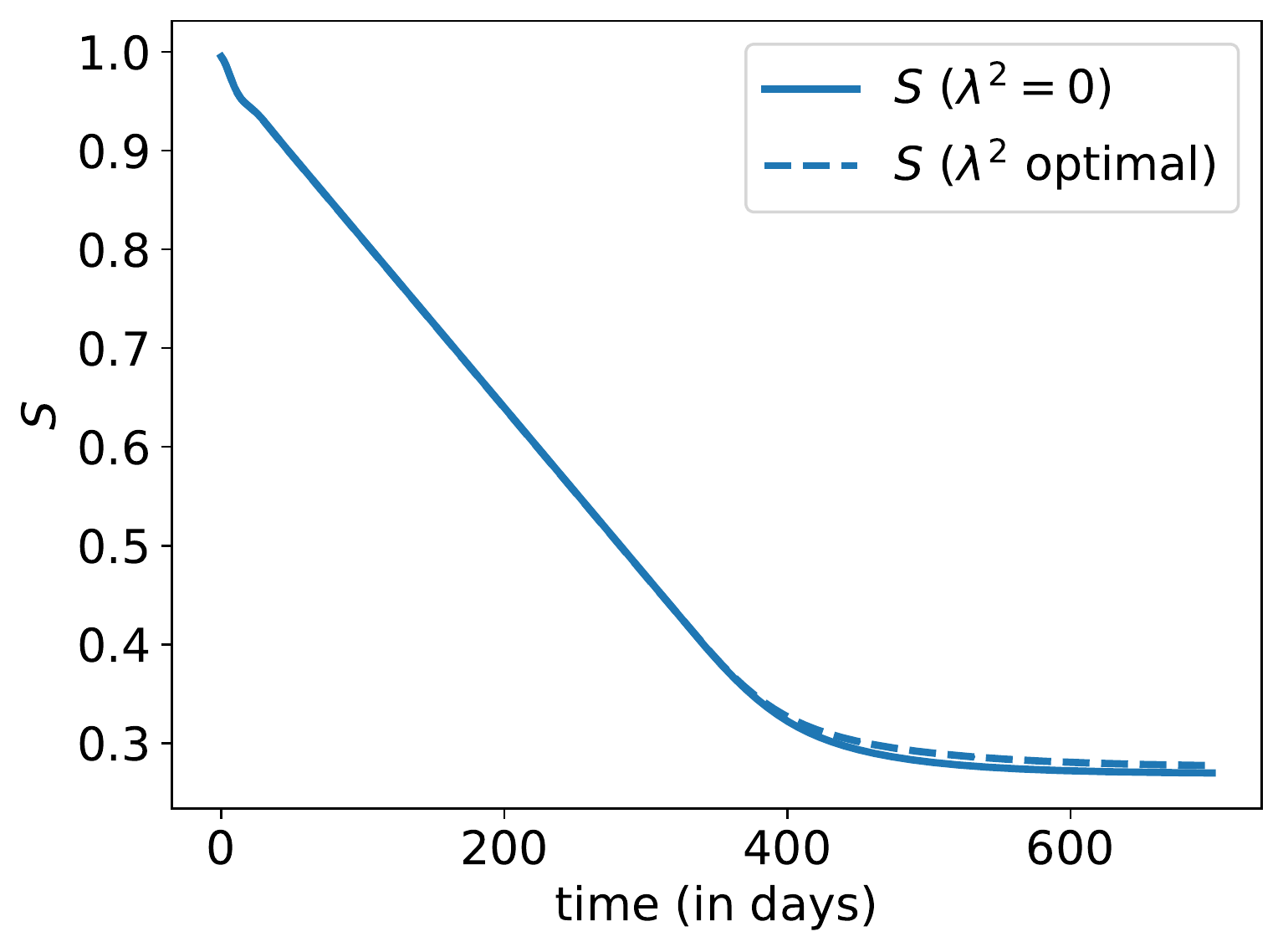}
	\end{subfigure}%
	\begin{subfigure}{.33\columnwidth}
		\centering 
		\includegraphics[width=\columnwidth]{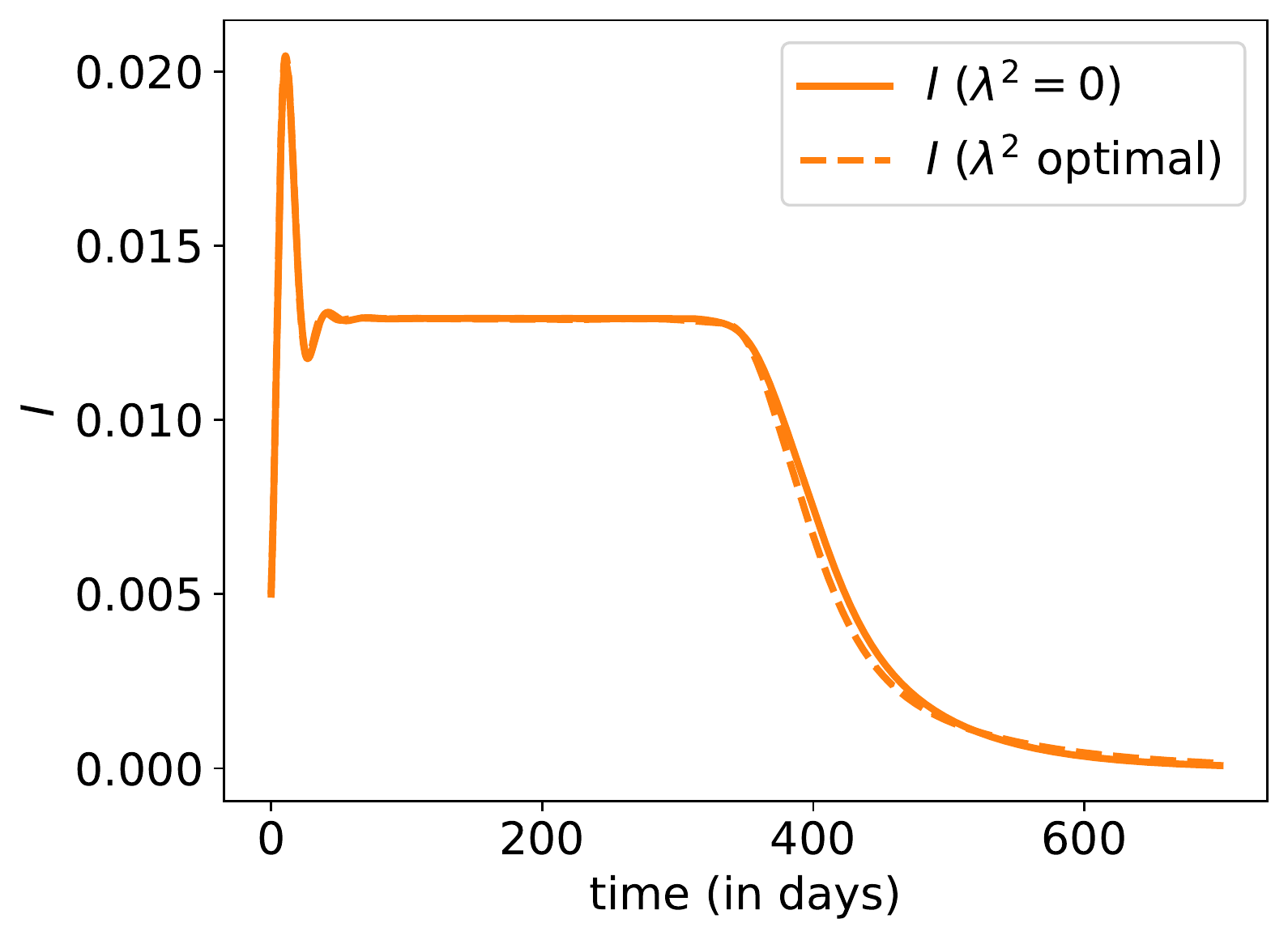}
	\end{subfigure}
	\begin{subfigure}{.33\columnwidth}
		\centering 
		\includegraphics[width=\columnwidth]{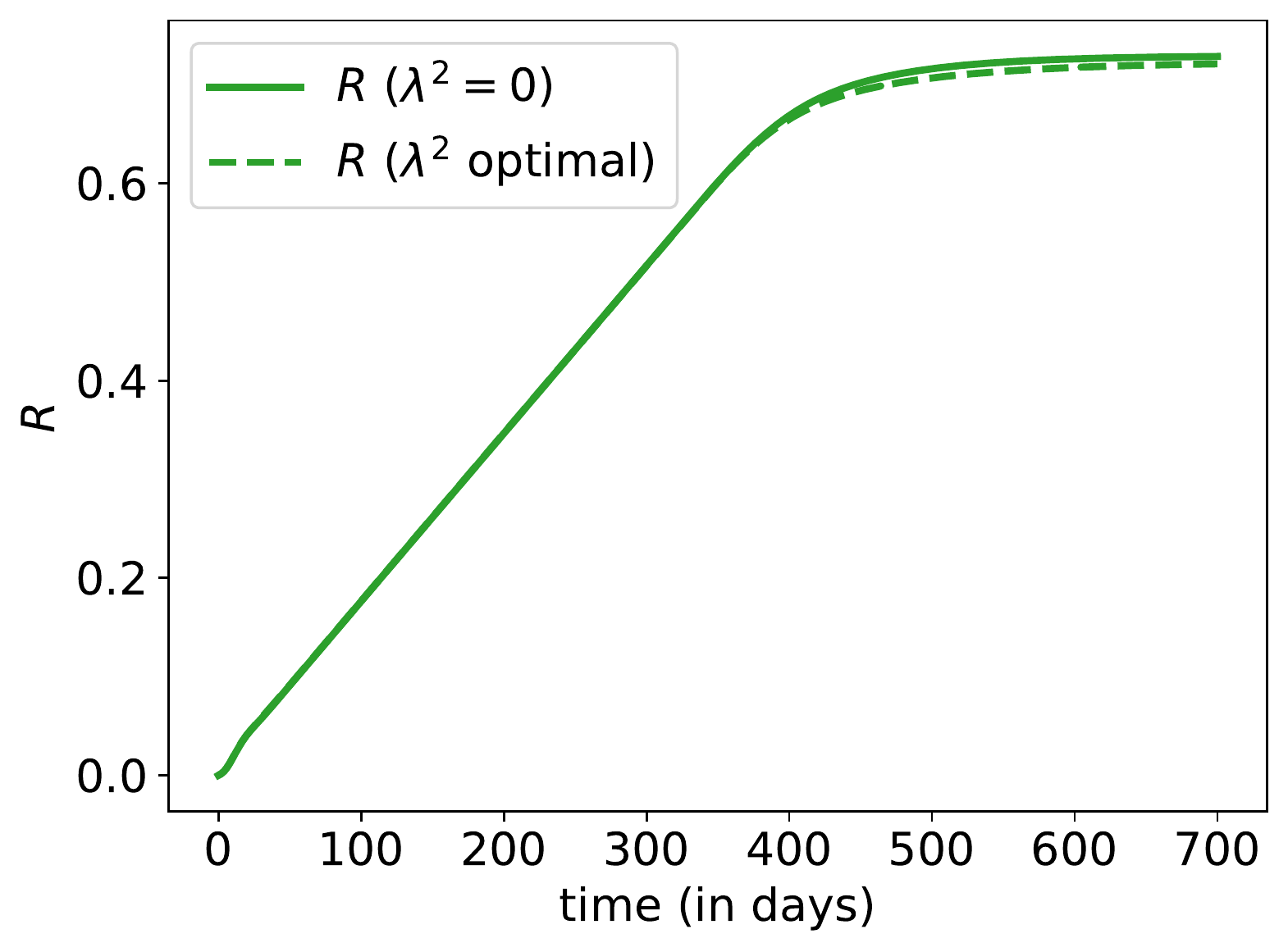}
	\end{subfigure}
	
	\begin{subfigure}{.33\columnwidth}
		\centering
		\includegraphics[width=\columnwidth]{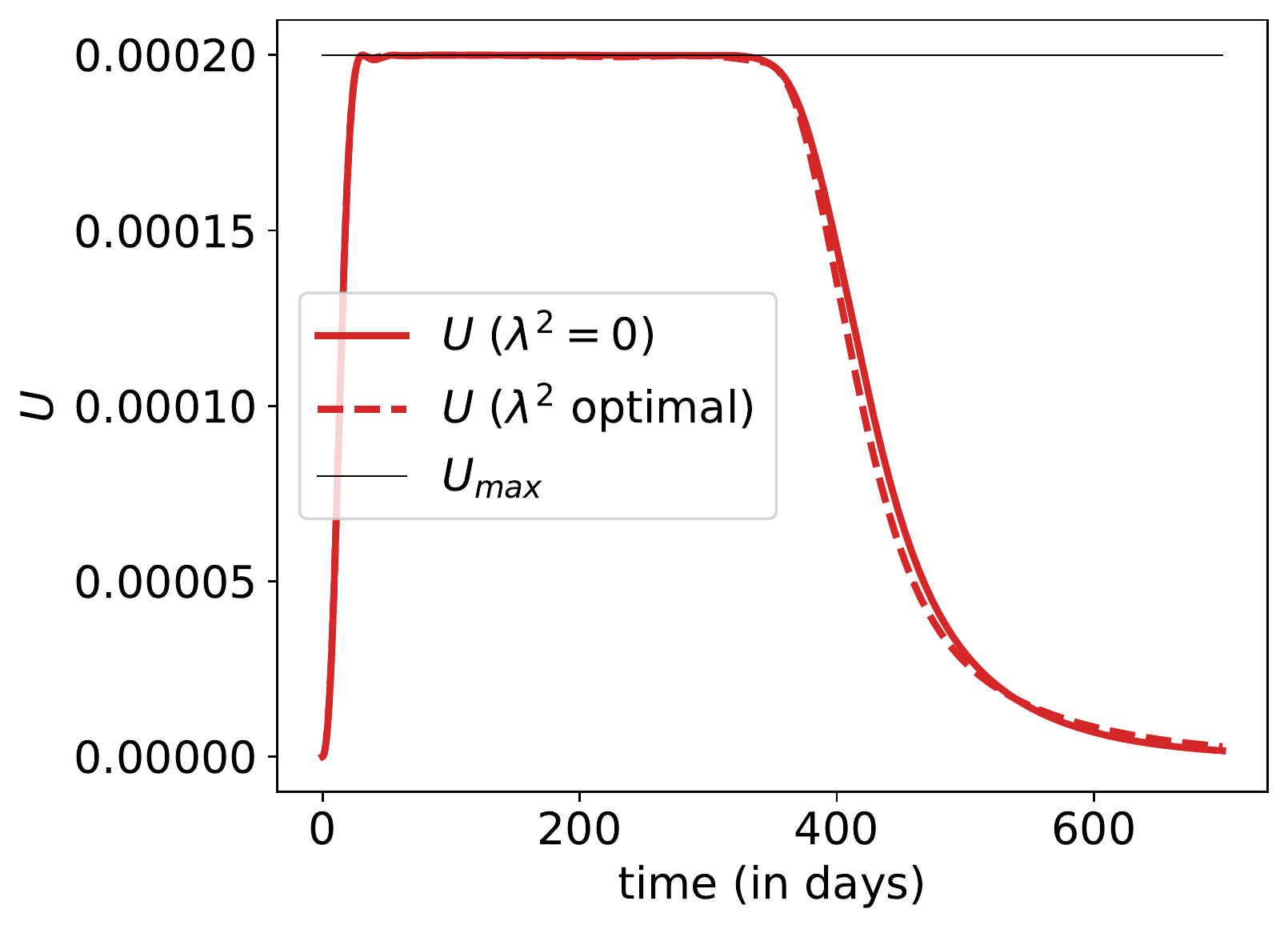}
	\end{subfigure}%
	\begin{subfigure}{.33\columnwidth}
		\centering 
		\includegraphics[width=\columnwidth]{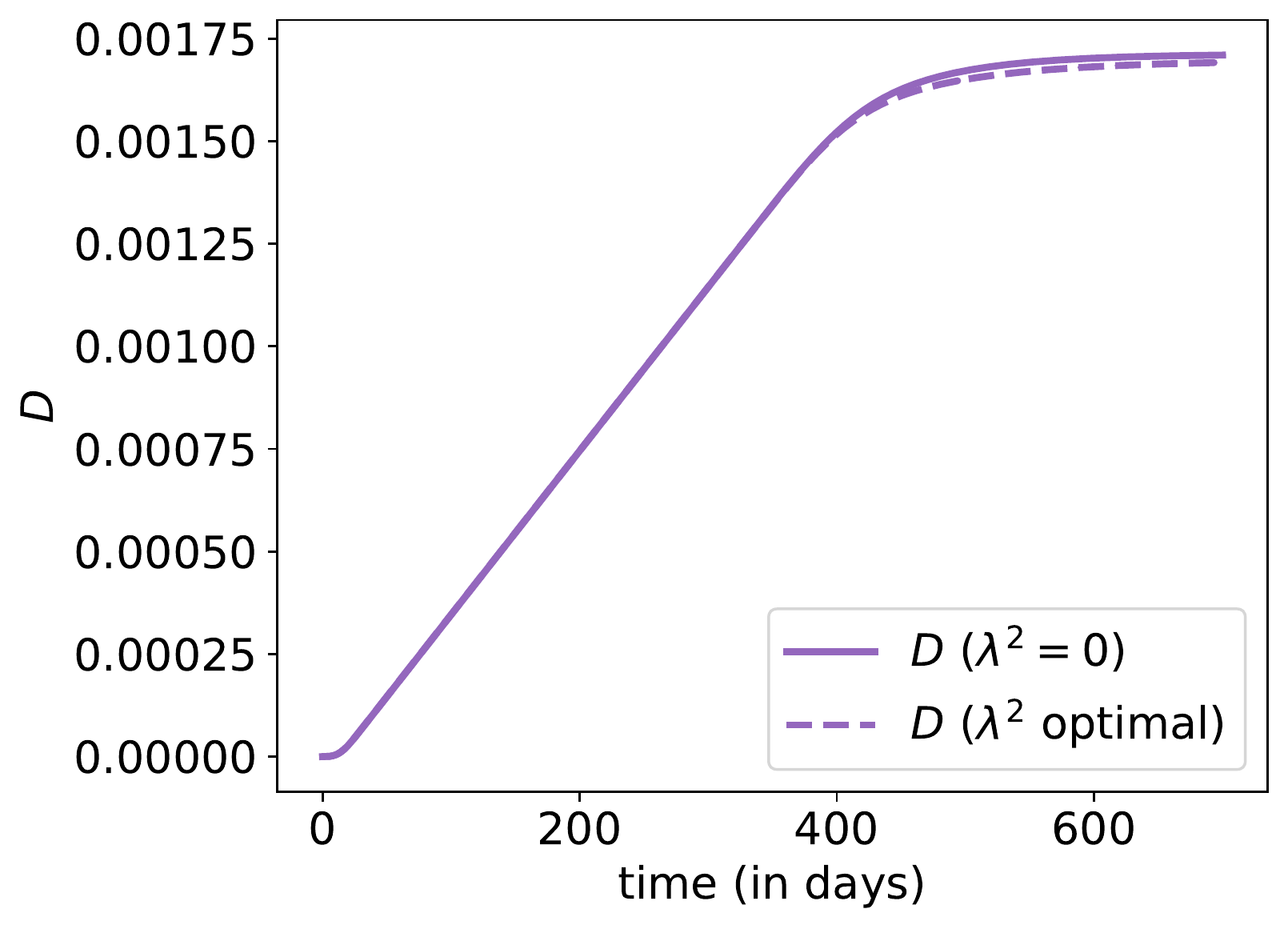}
	\end{subfigure}
	\begin{subfigure}{.33\columnwidth}
		\centering 
		\includegraphics[width=\columnwidth]{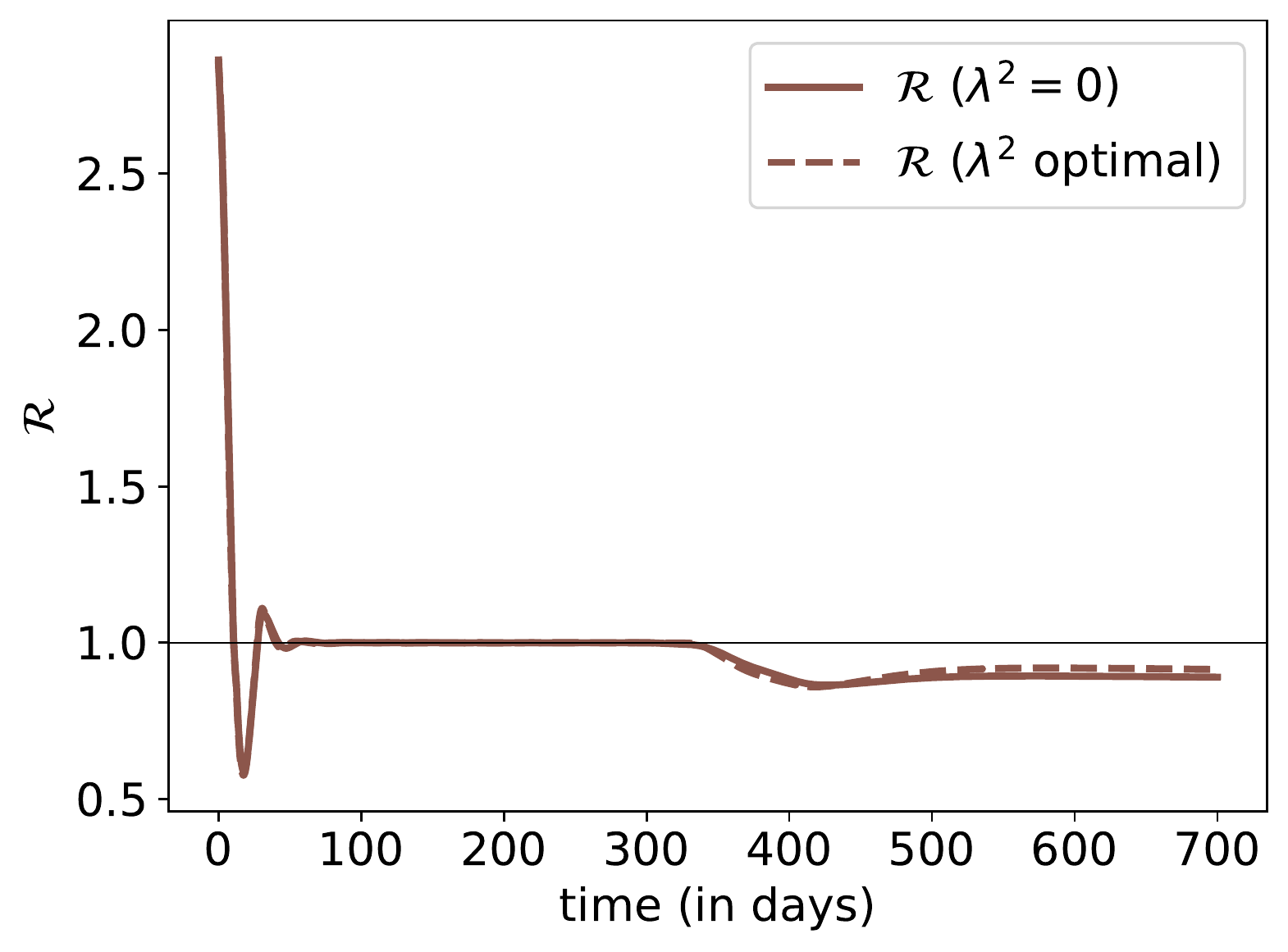}
	\end{subfigure}
	
	\begin{subfigure}{.33\columnwidth}
		\centering
		\includegraphics[width=\columnwidth]{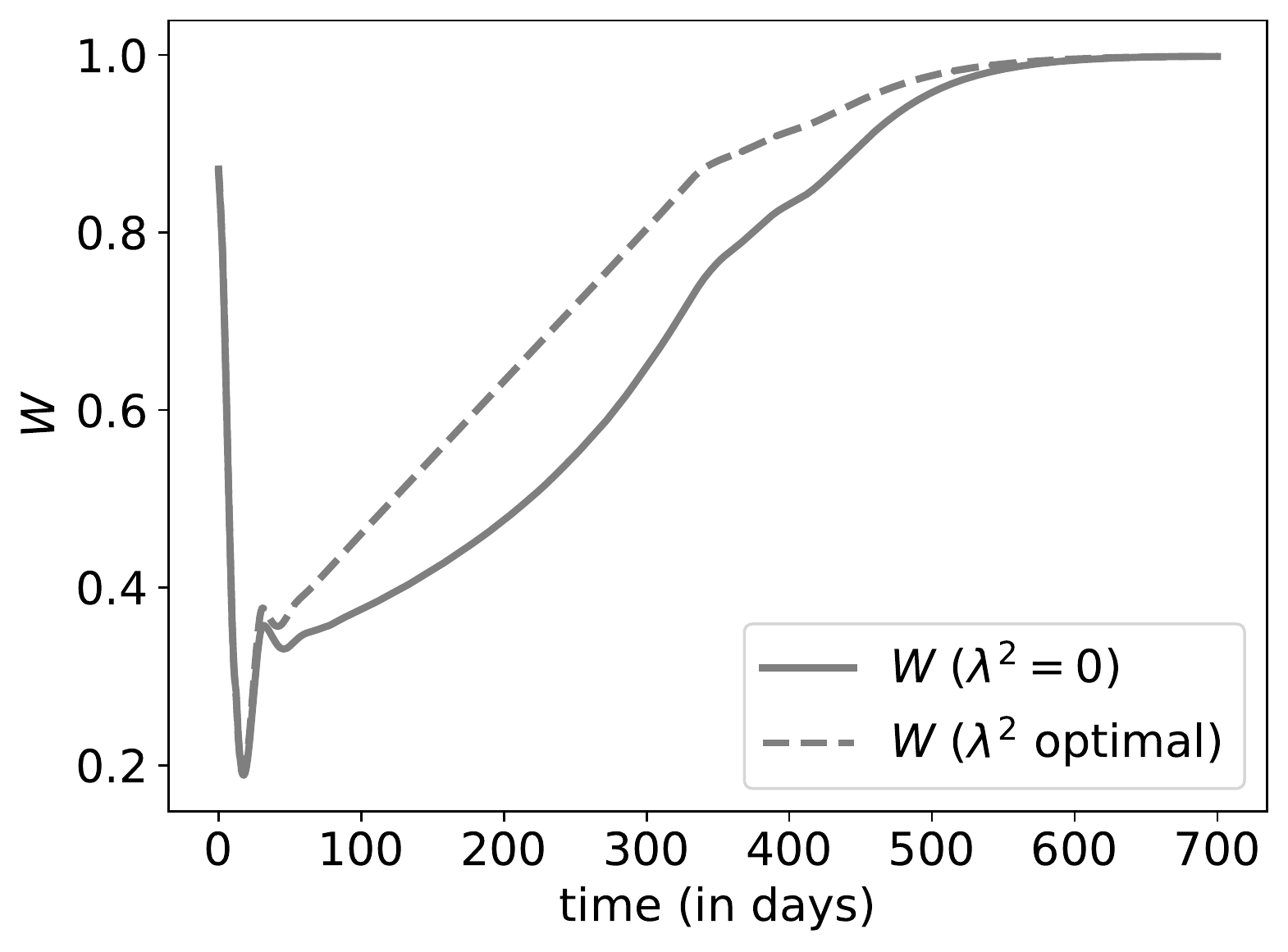}
	\end{subfigure}%
	\begin{subfigure}{.33\columnwidth}
		\centering 
		\includegraphics[width=\columnwidth]{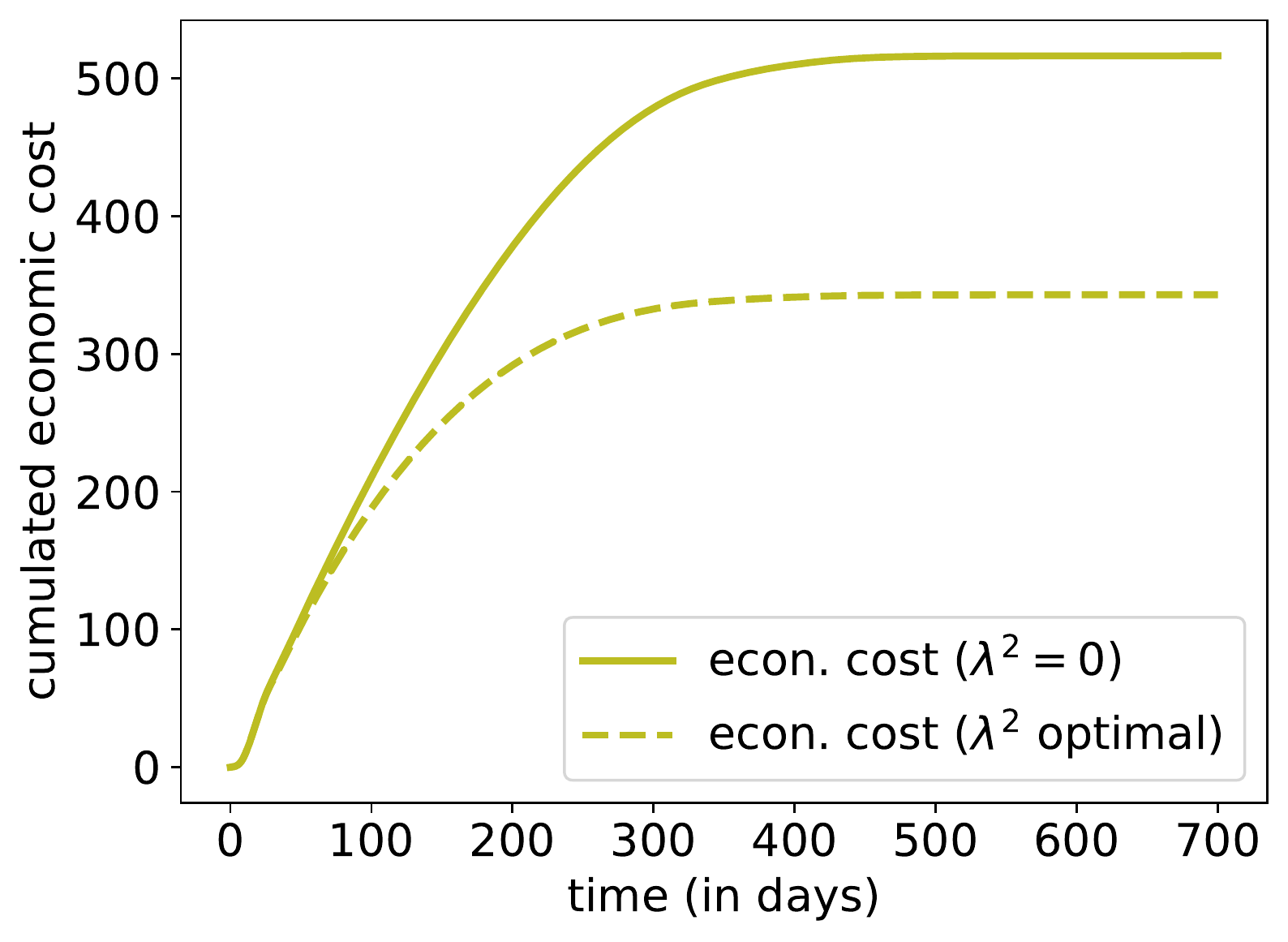}
	\end{subfigure}
	\begin{subfigure}{.33\columnwidth}
		\centering 
		\includegraphics[width=\columnwidth]{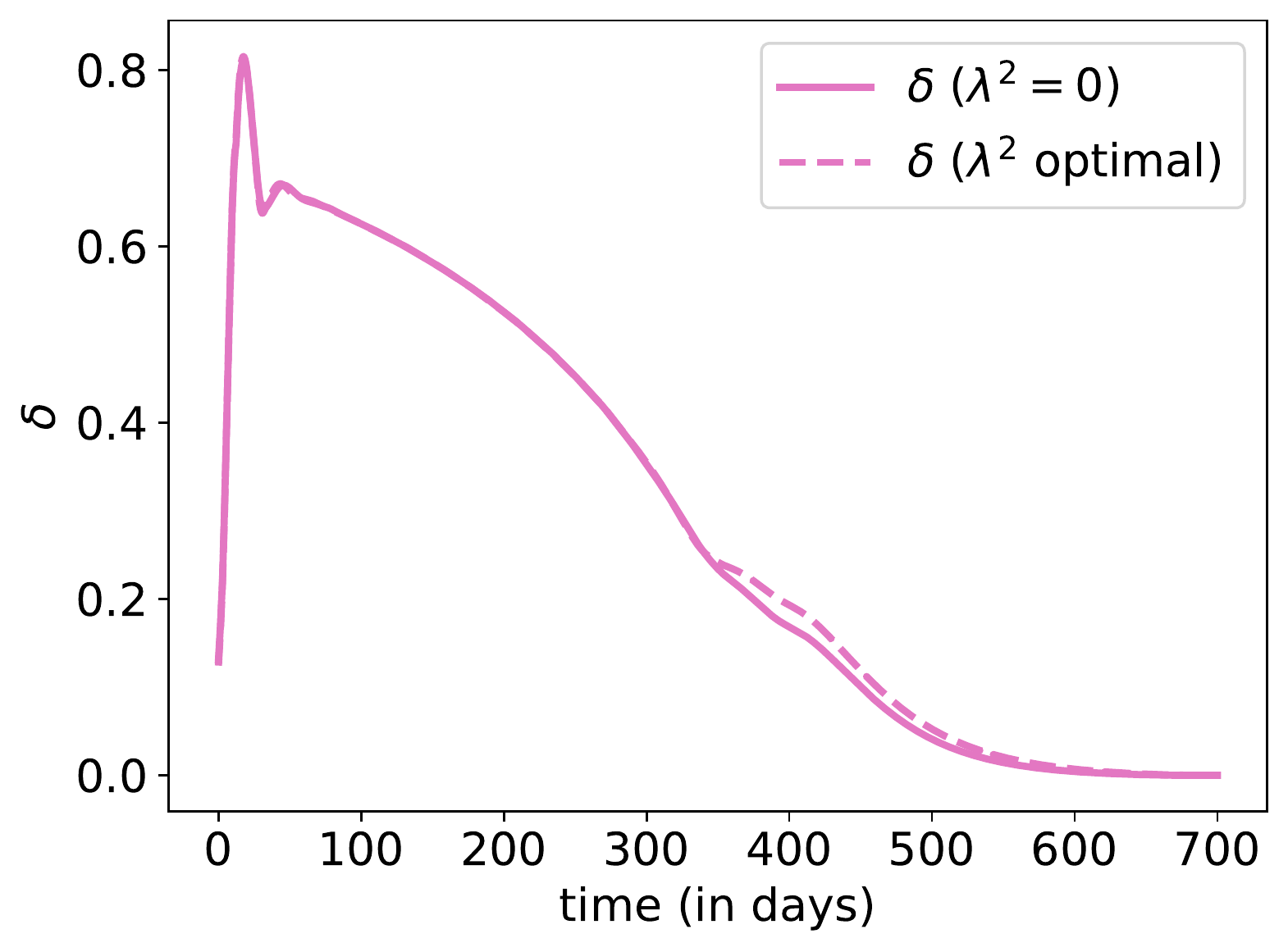}
	\end{subfigure}
	
	\begin{subfigure}{.33\columnwidth}
		\centering
		\includegraphics[width=\columnwidth]{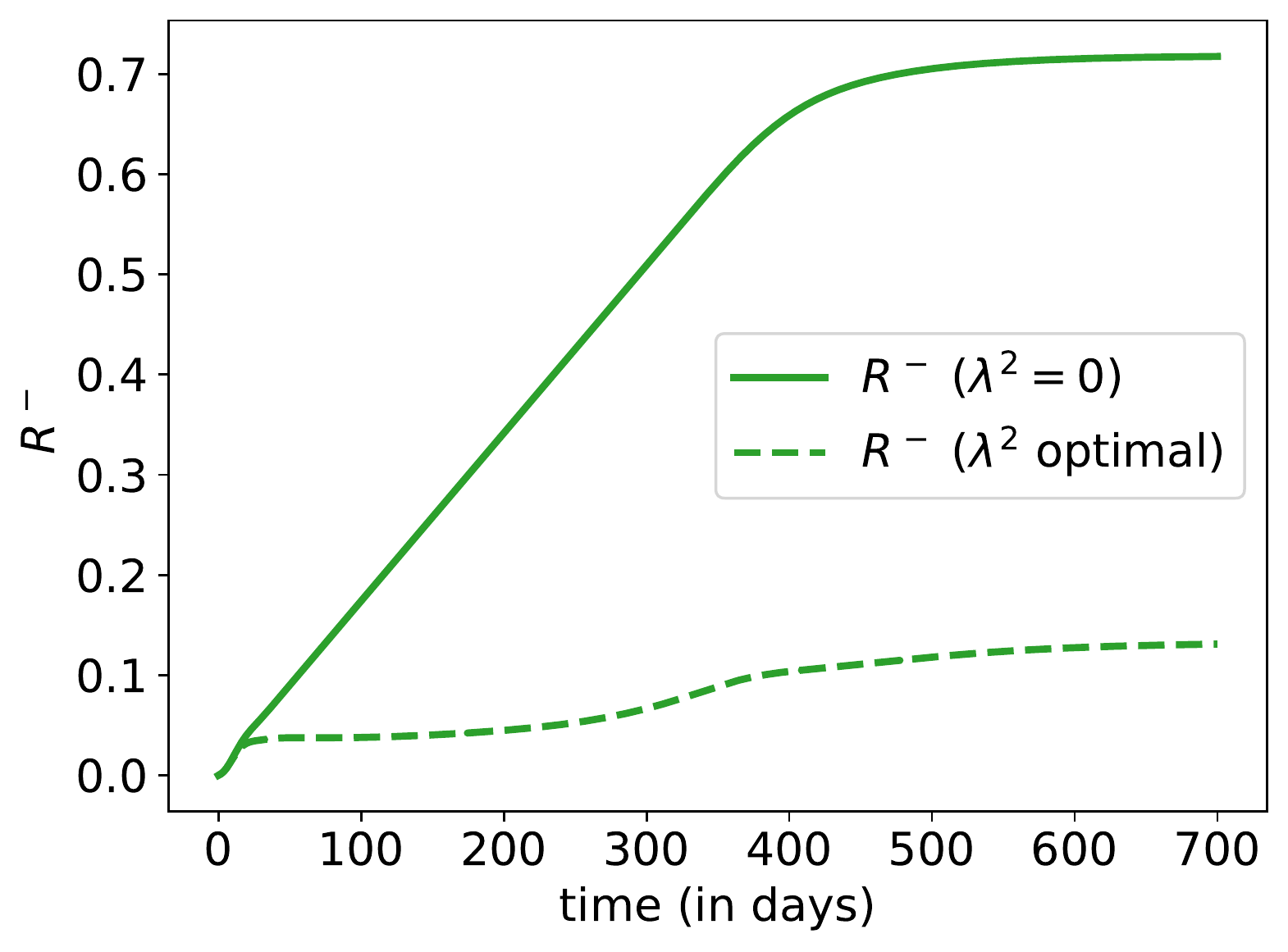}
	\end{subfigure}%
	\begin{subfigure}{.33\columnwidth}
		\centering 
		\includegraphics[width=\columnwidth]{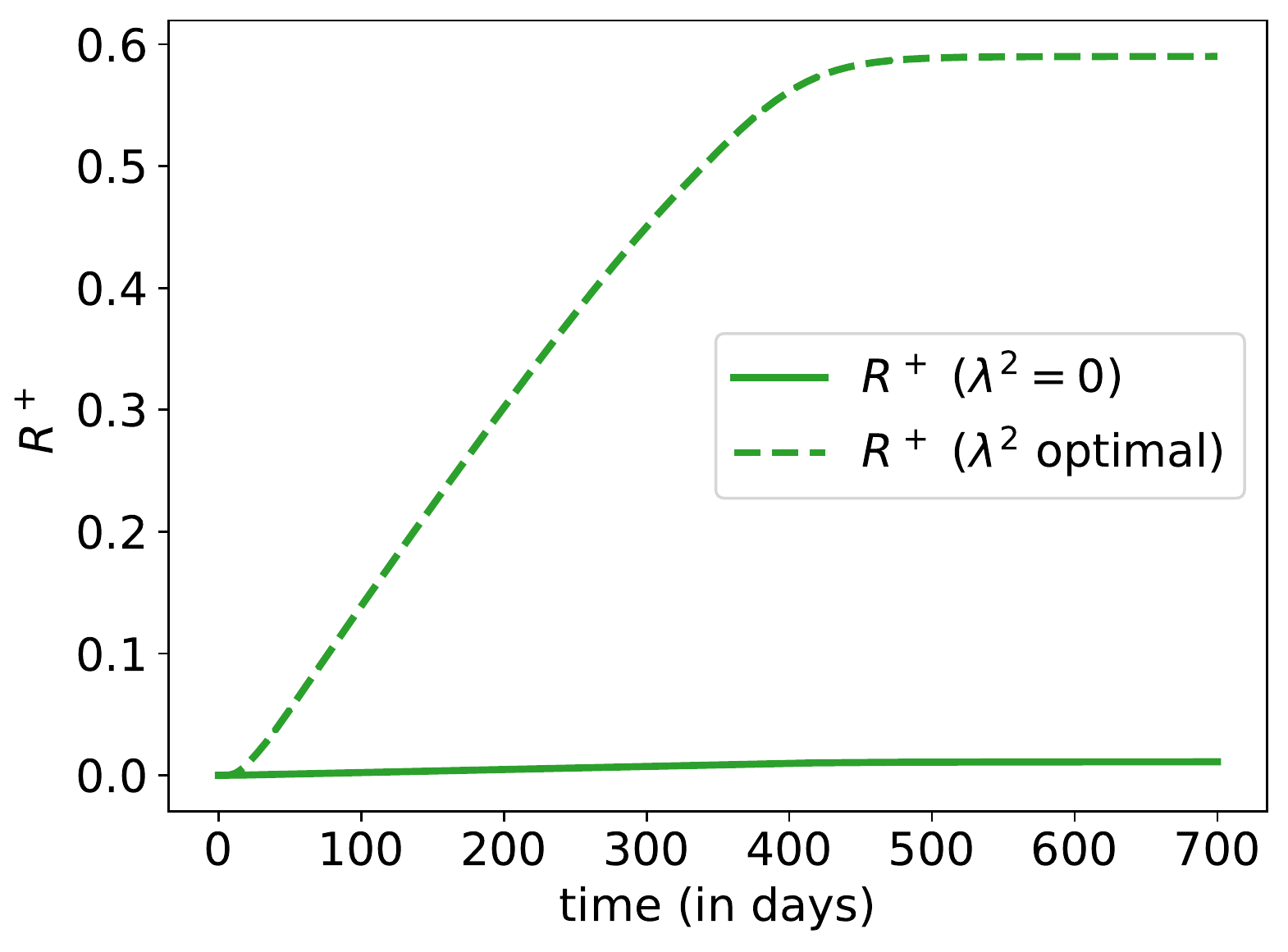}
	\end{subfigure}
	\begin{subfigure}{.33\columnwidth}
		\centering 
		\includegraphics[width=\columnwidth]{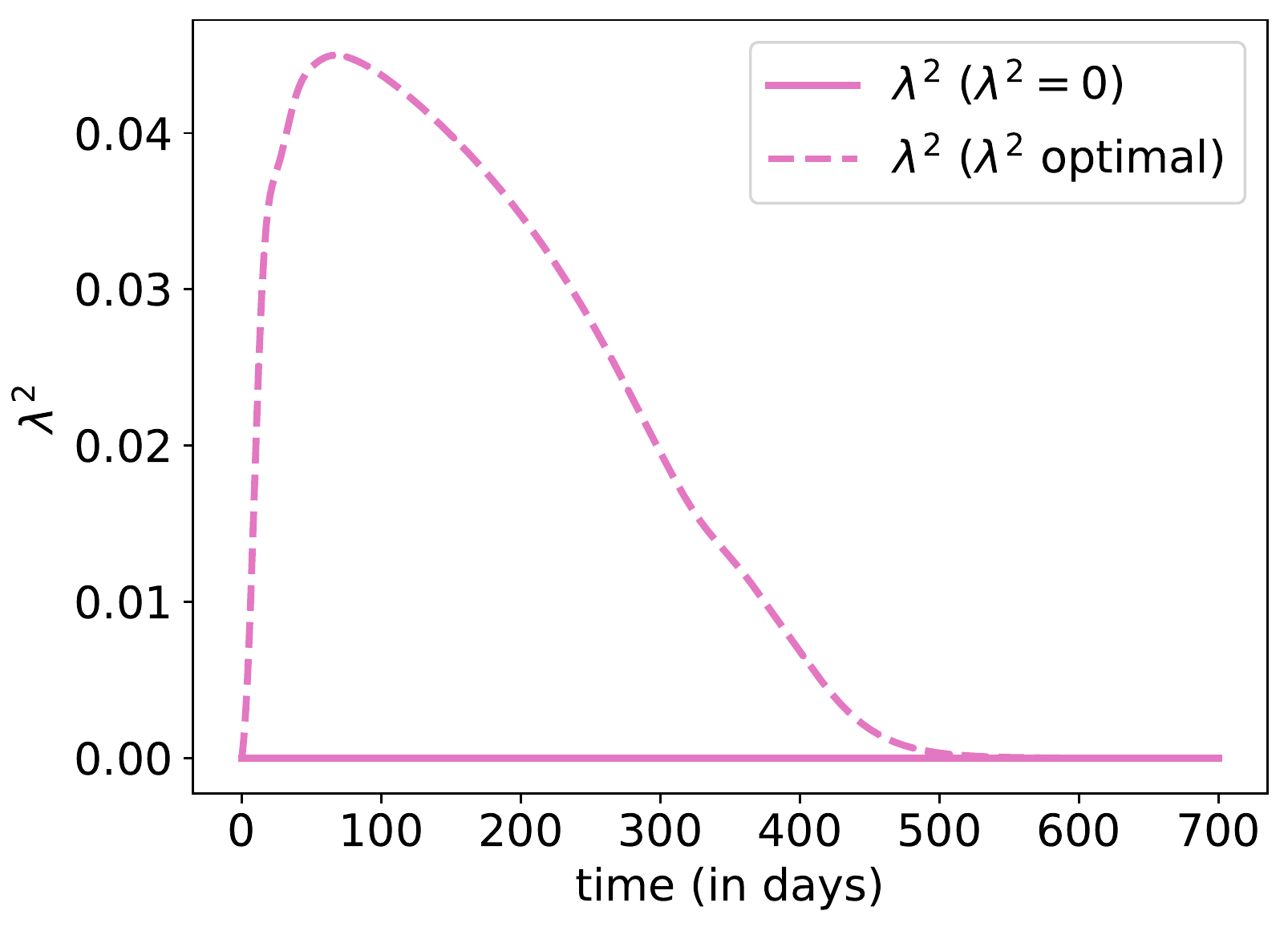}
	\end{subfigure}	\caption{Evolution of states with optimal controls $(\delta,0,\lambda^2)$, where the plain line corresponds to the benchmark scenario $(\delta,0,0)$ discussed in Section \ref{subsec:optimal:policy} (plain line); and evolution with joint optimal controls $(\delta,0,\lambda^2)$ (dashed line).
	}
 	\label{fig:Optim_delta_lambda2B10}
\end{figure}

\newpage

\subsection{Optimizing over lockdown intervention $\delta$,  and efforts in virologic detection $\lambda^1$ and immunity detection $\lambda^2$} \label{sec_opti_delta_lambda2_b11}

{Here, we compare the optimization on $\delta$ only versus the optimization~\eqref{Numeric_Control_Problem} over $(\delta,\lambda^1,\lambda^2) $. 
}

\begin{figure}[ht!]
	\begin{subfigure}{.33\columnwidth}
		\centering
		\includegraphics[width=\columnwidth]{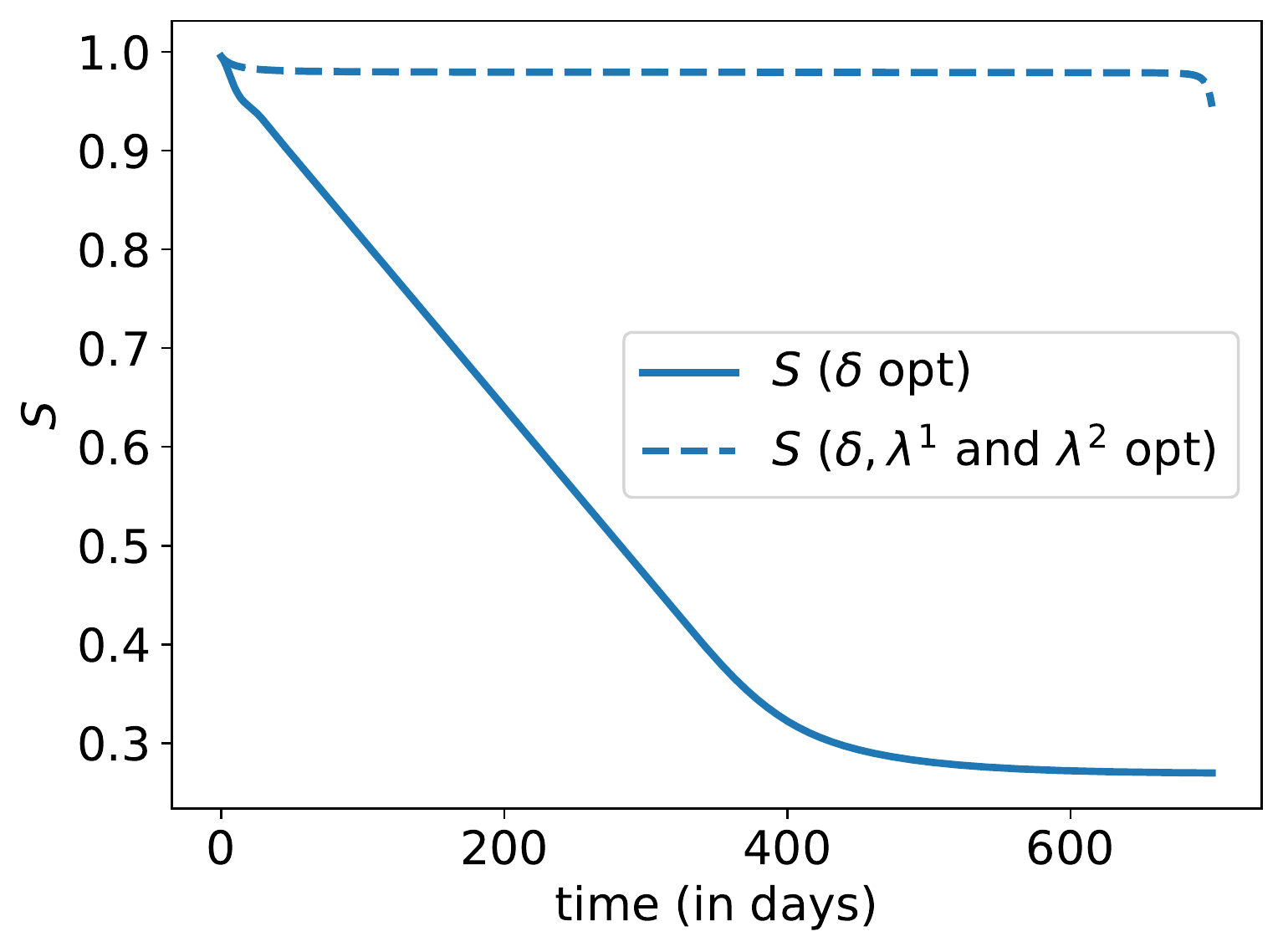}
	\end{subfigure}%
	\begin{subfigure}{.33\columnwidth}
		\centering 
		\includegraphics[width=\columnwidth]{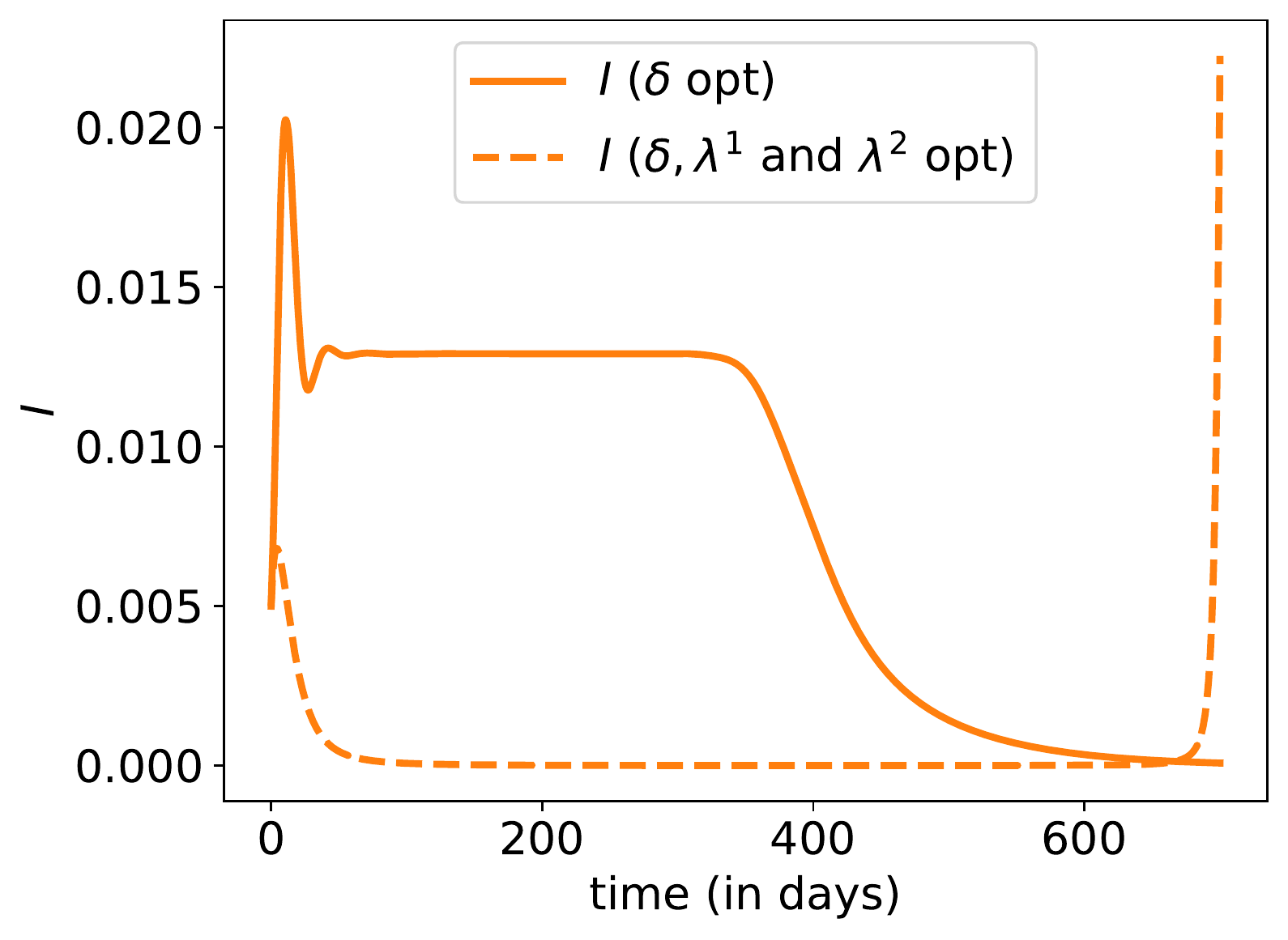}
	\end{subfigure}
	\begin{subfigure}{.33\columnwidth}
		\centering 
		\includegraphics[width=\columnwidth]{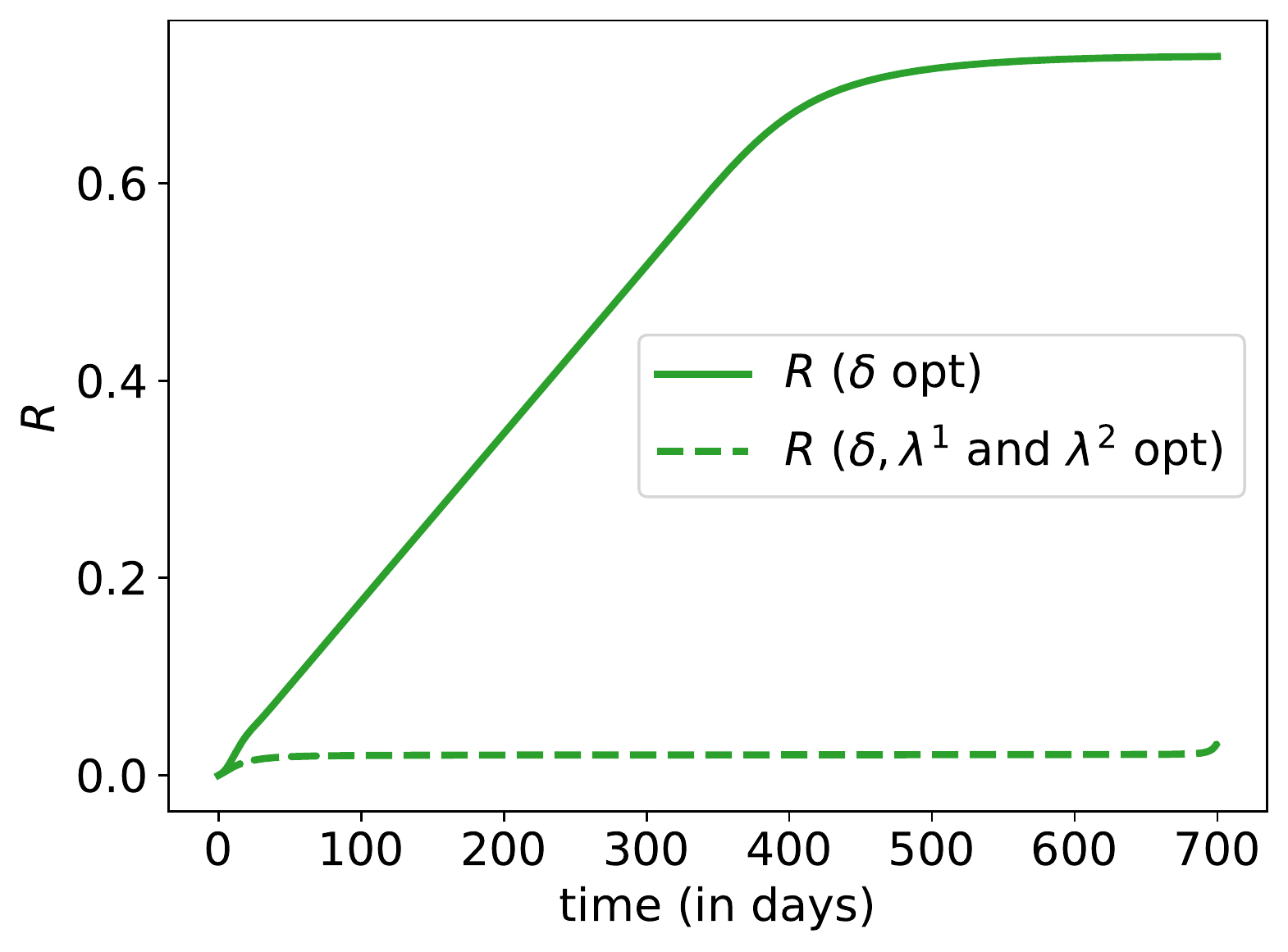}
	\end{subfigure}
	
	\begin{subfigure}{.33\columnwidth}
		\centering
		\includegraphics[width=\columnwidth]{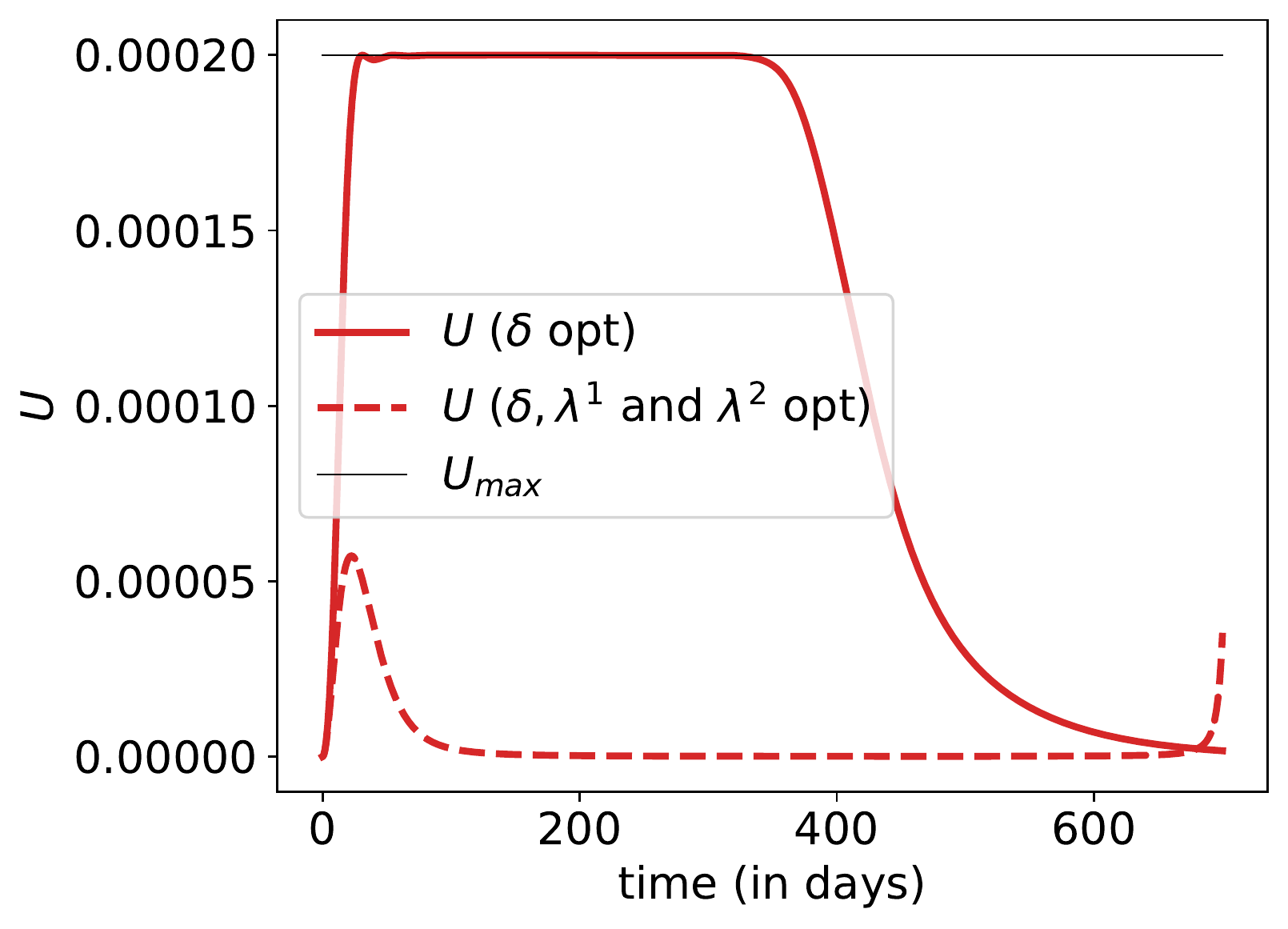}
	\end{subfigure}%
	\begin{subfigure}{.33\columnwidth}
		\centering 
		\includegraphics[width=\columnwidth]{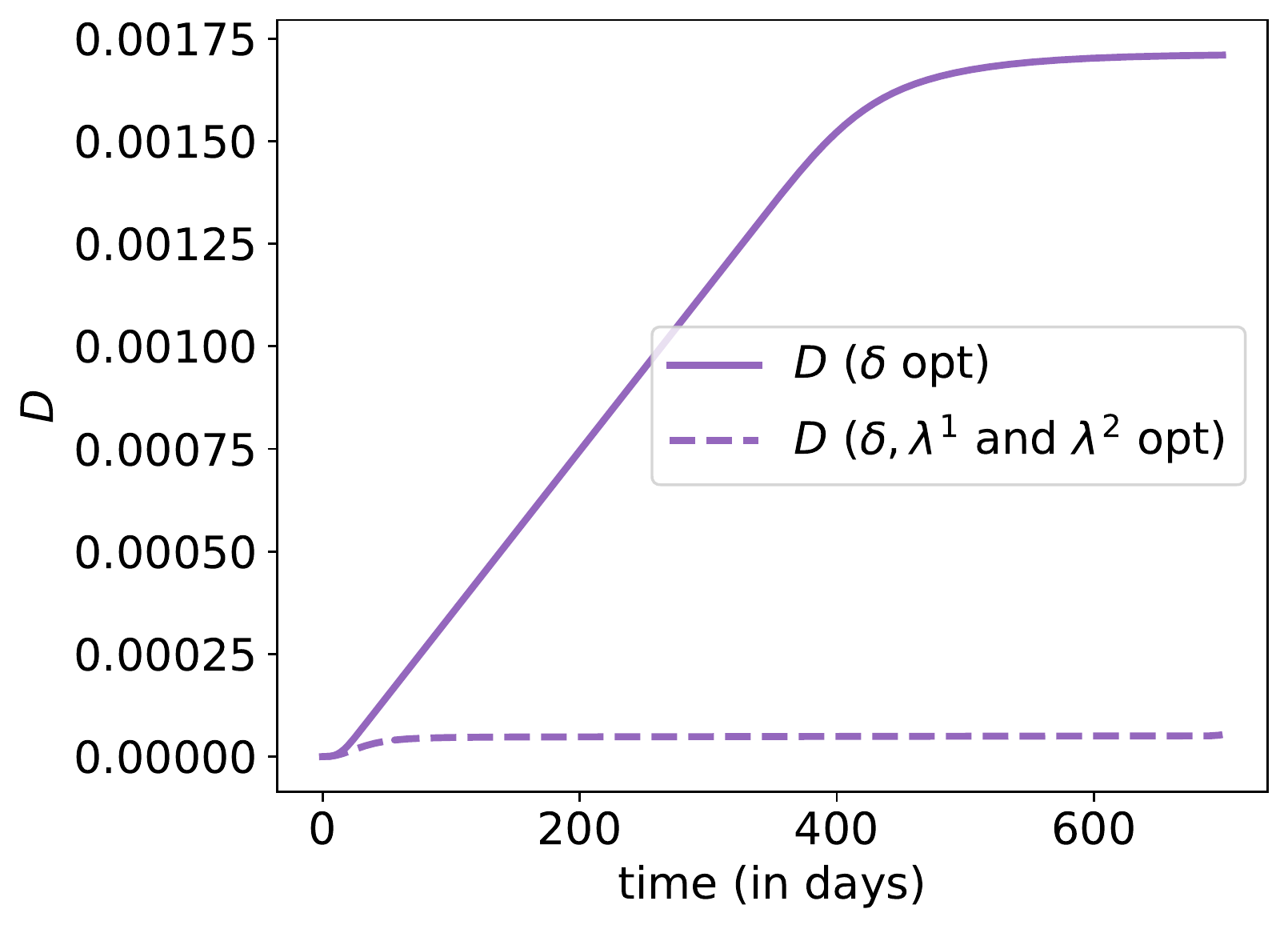}
	\end{subfigure}
	\begin{subfigure}{.33\columnwidth}
		\centering 
		\includegraphics[width=\columnwidth]{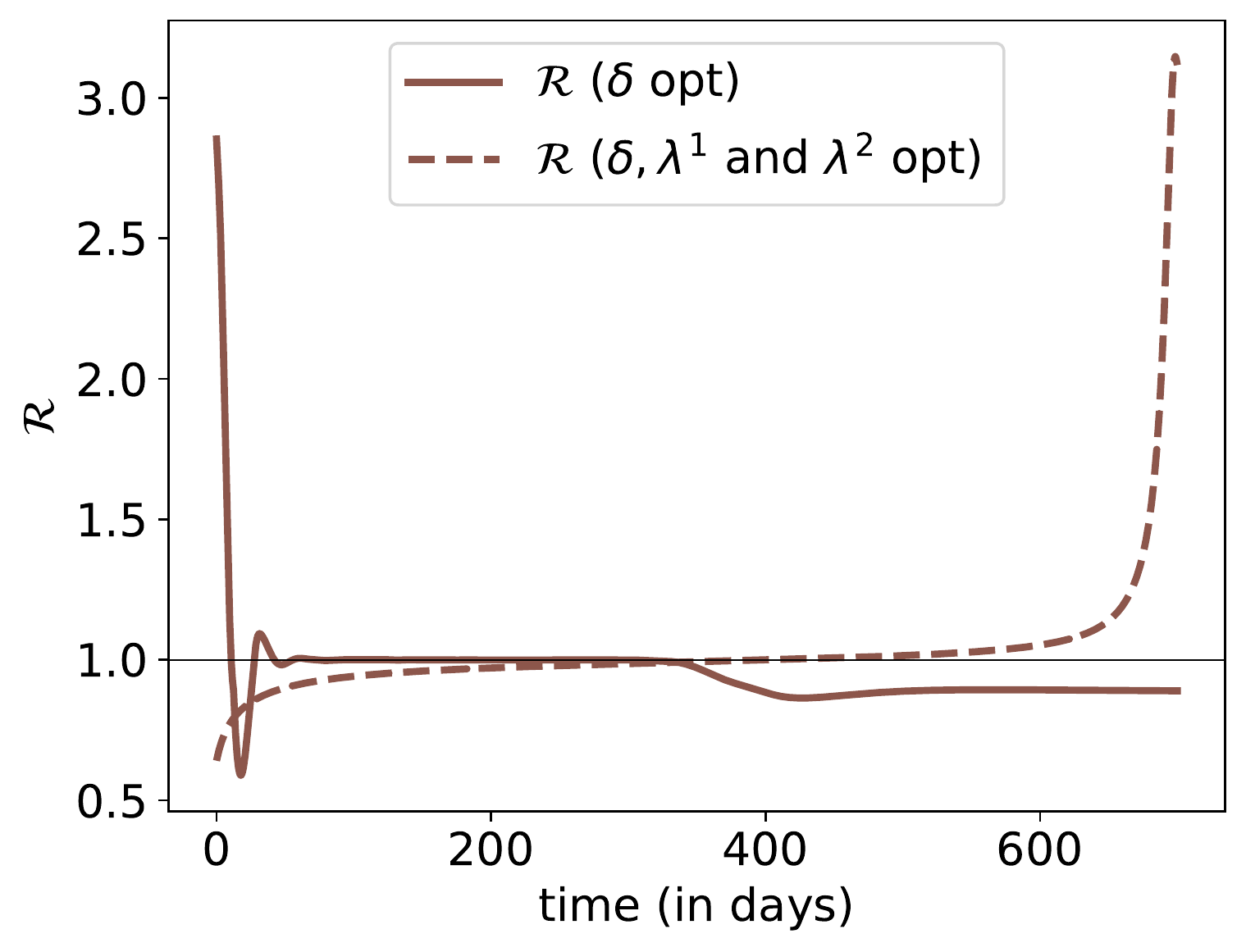}
	\end{subfigure}
	
	\begin{subfigure}{.33\columnwidth}
		\centering
		\includegraphics[width=\columnwidth]{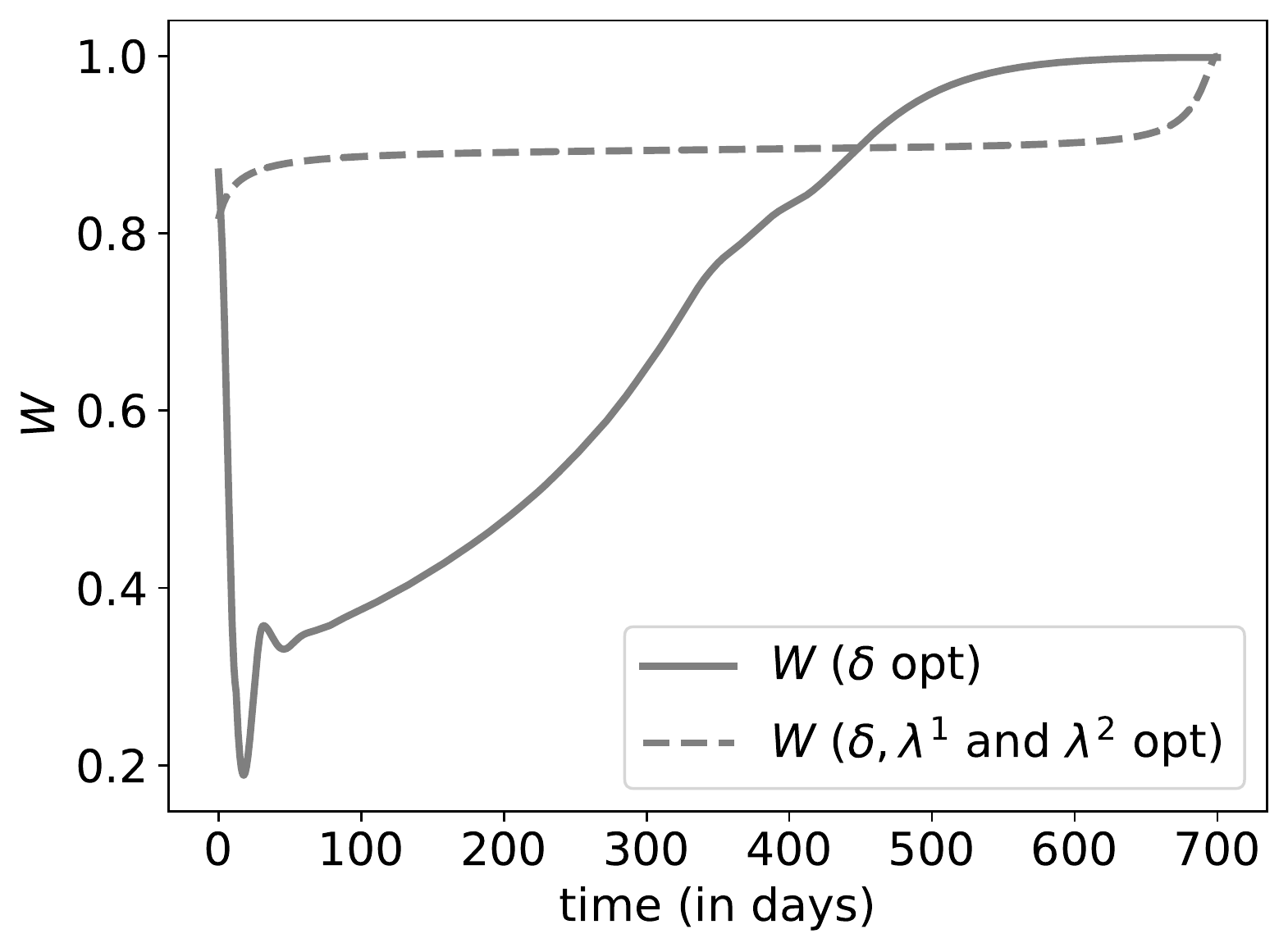}
	\end{subfigure}%
	\begin{subfigure}{.33\columnwidth}
		\centering 
		\includegraphics[width=\columnwidth]{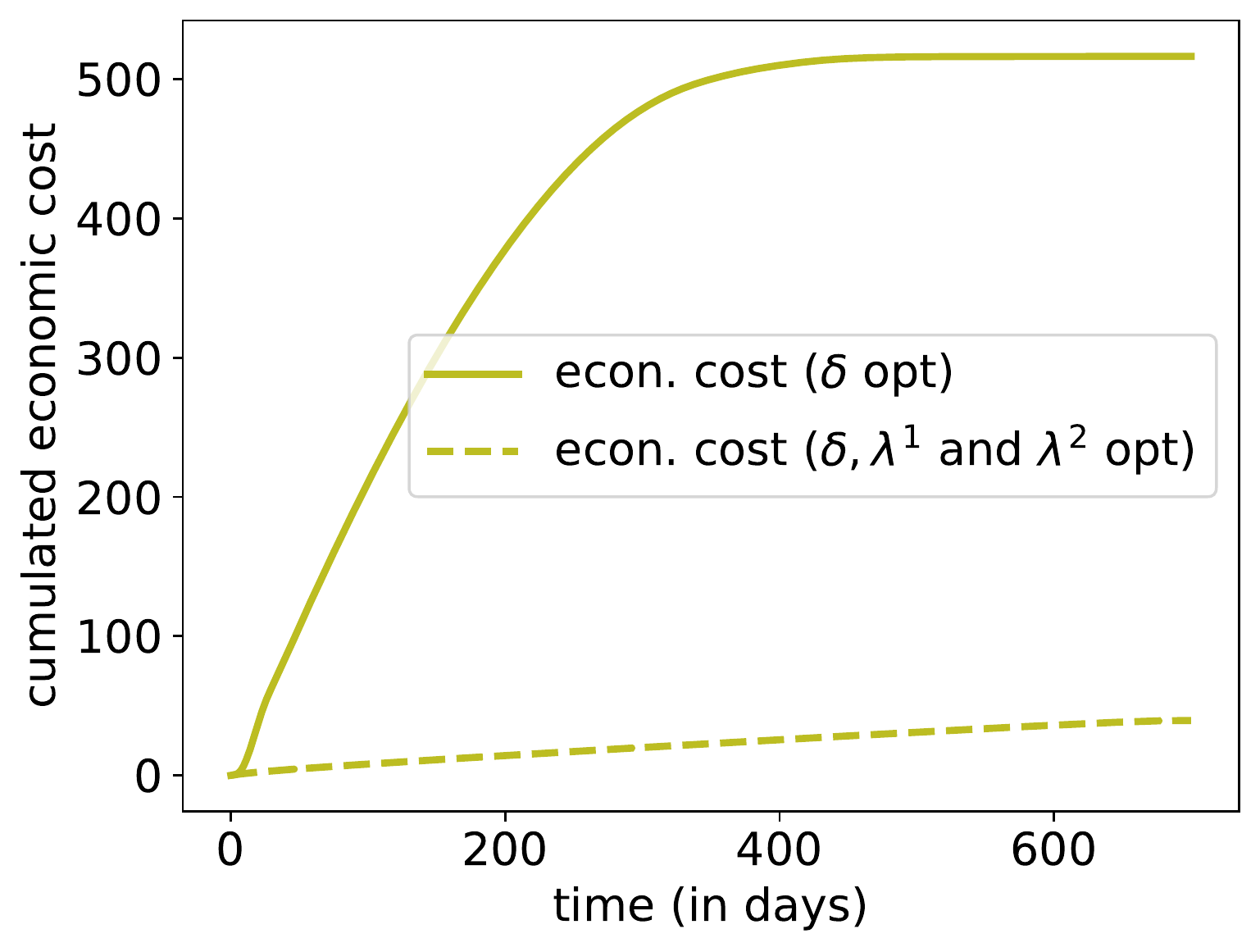}
	\end{subfigure}
	\begin{subfigure}{.33\columnwidth}
		\centering 
		\includegraphics[width=\columnwidth]{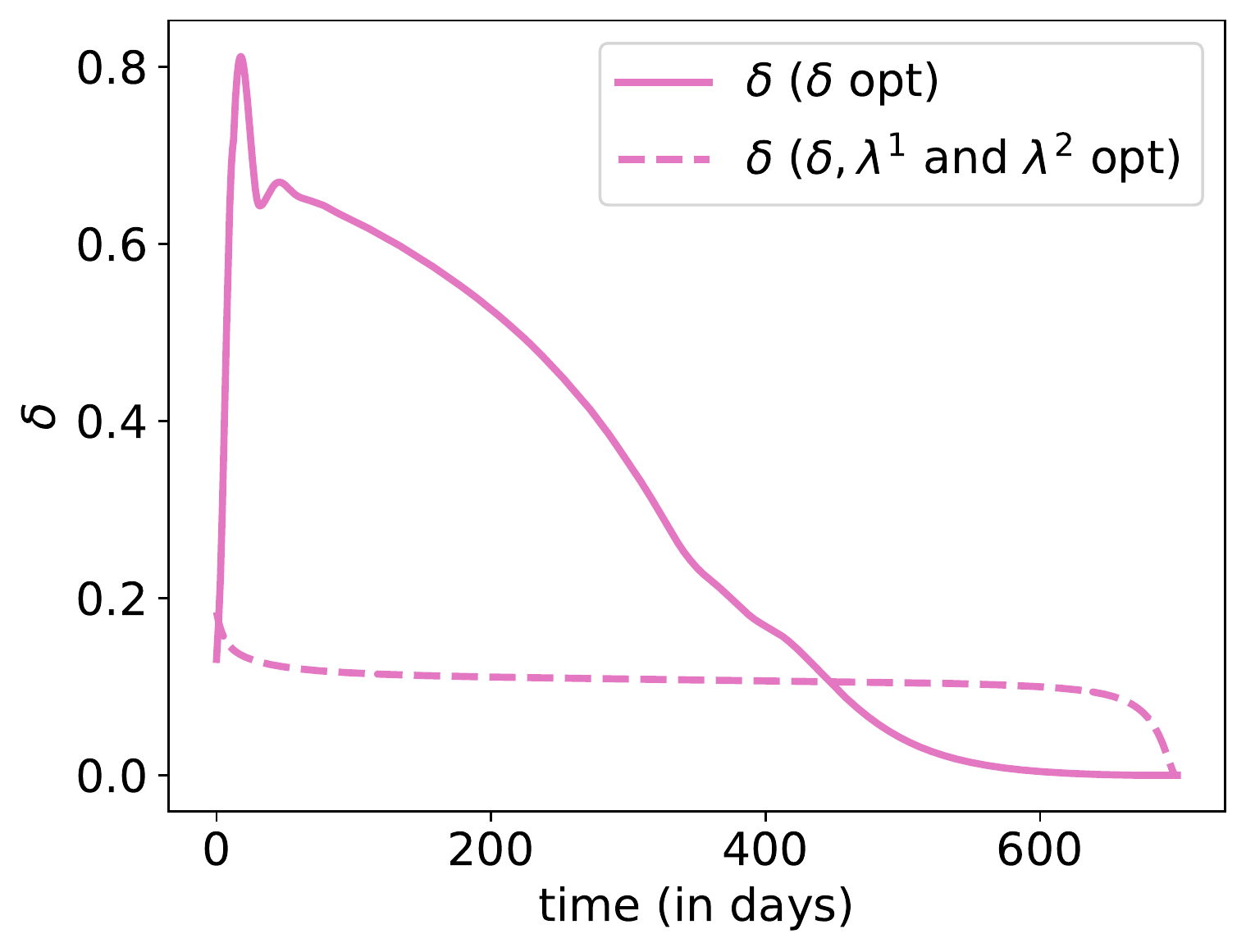}
	\end{subfigure}
	
	\begin{subfigure}{.33\columnwidth}
		\centering
		\includegraphics[width=\columnwidth]{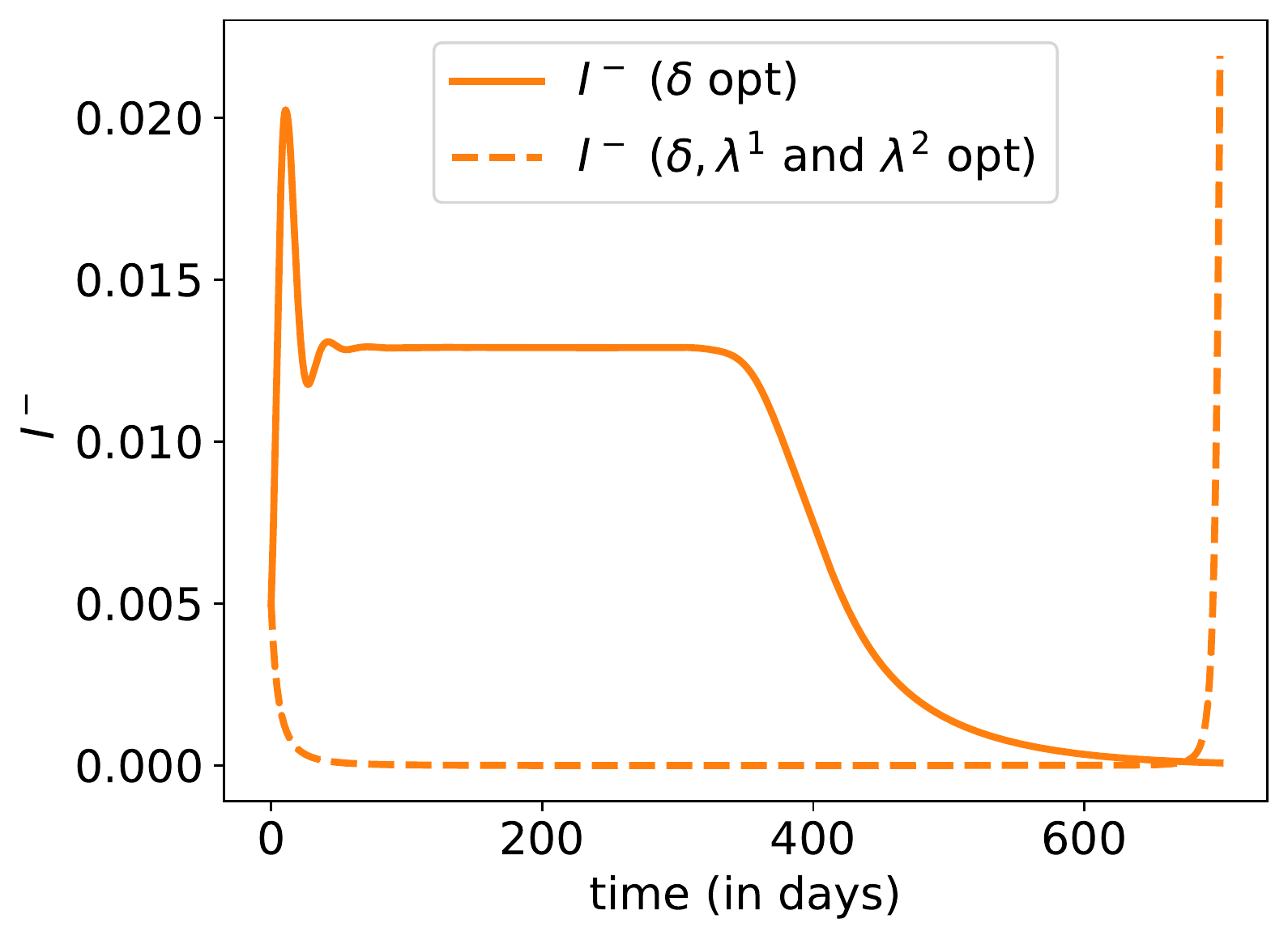}
	\end{subfigure}%
	\begin{subfigure}{.33\columnwidth}
		\centering 
		\includegraphics[width=\columnwidth]{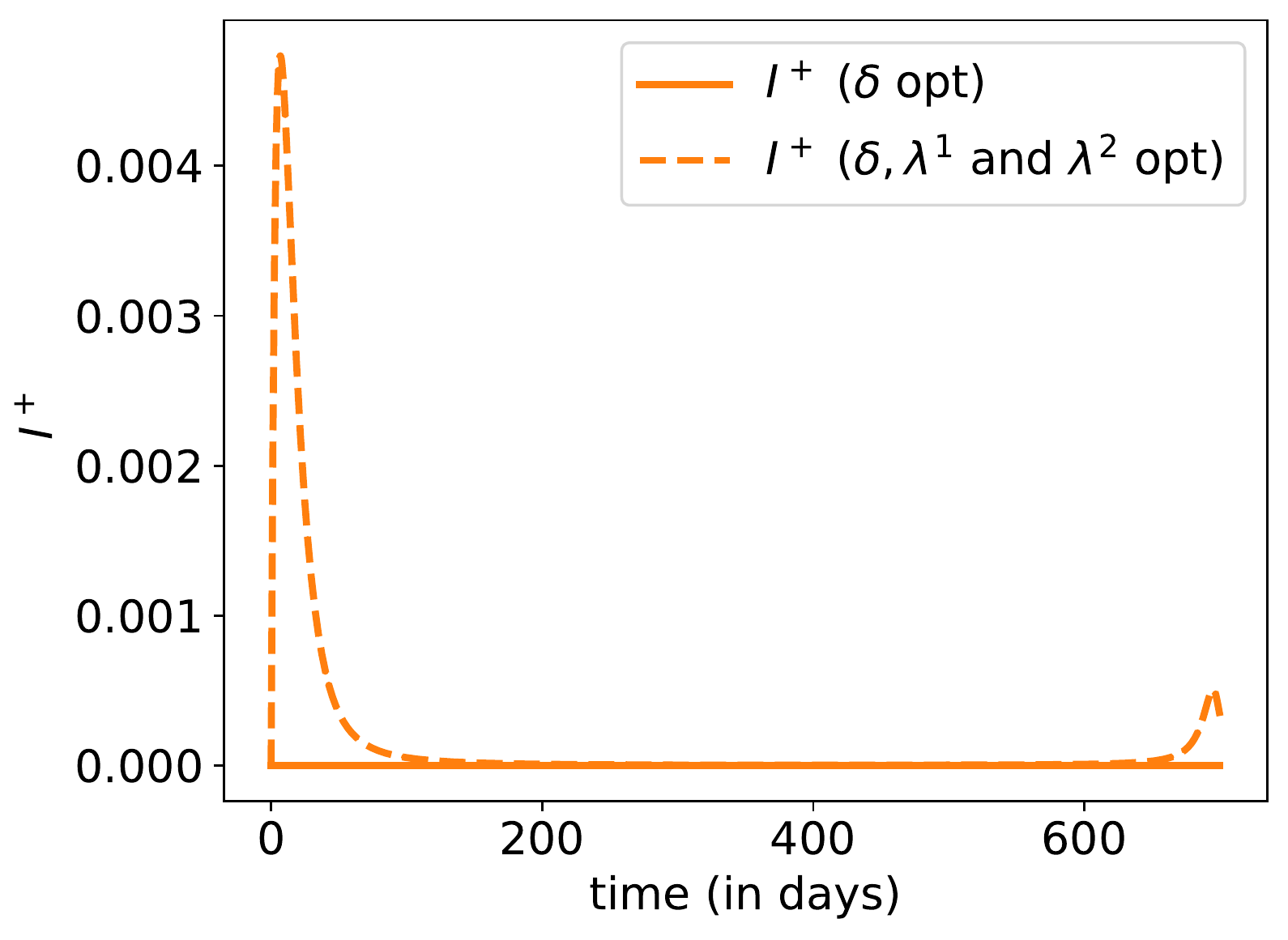}
	\end{subfigure}
	\begin{subfigure}{.33\columnwidth}
		\centering 
		\includegraphics[width=\columnwidth]{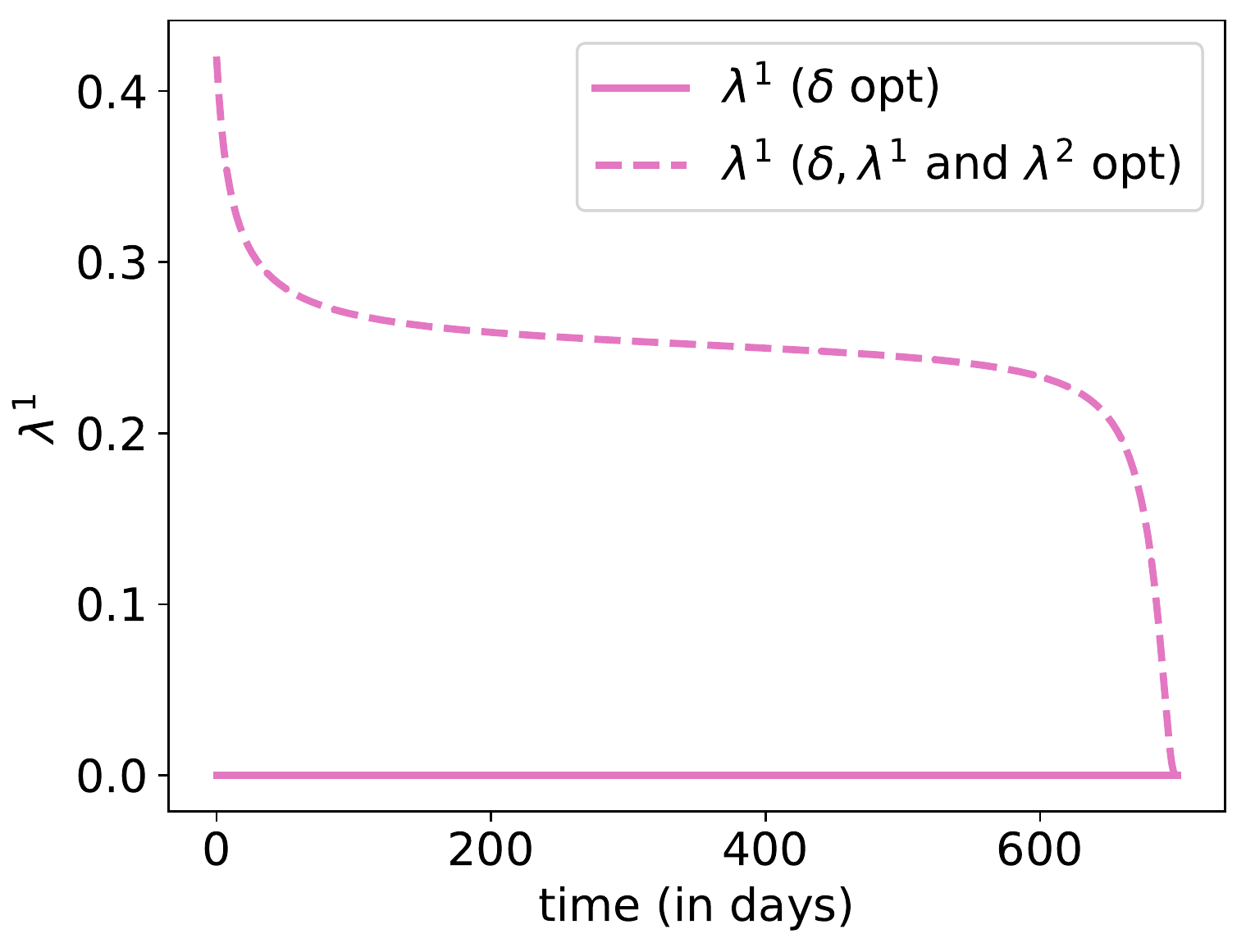}
	\end{subfigure}

	\begin{subfigure}{.33\columnwidth}
		\centering
		\includegraphics[width=\columnwidth]{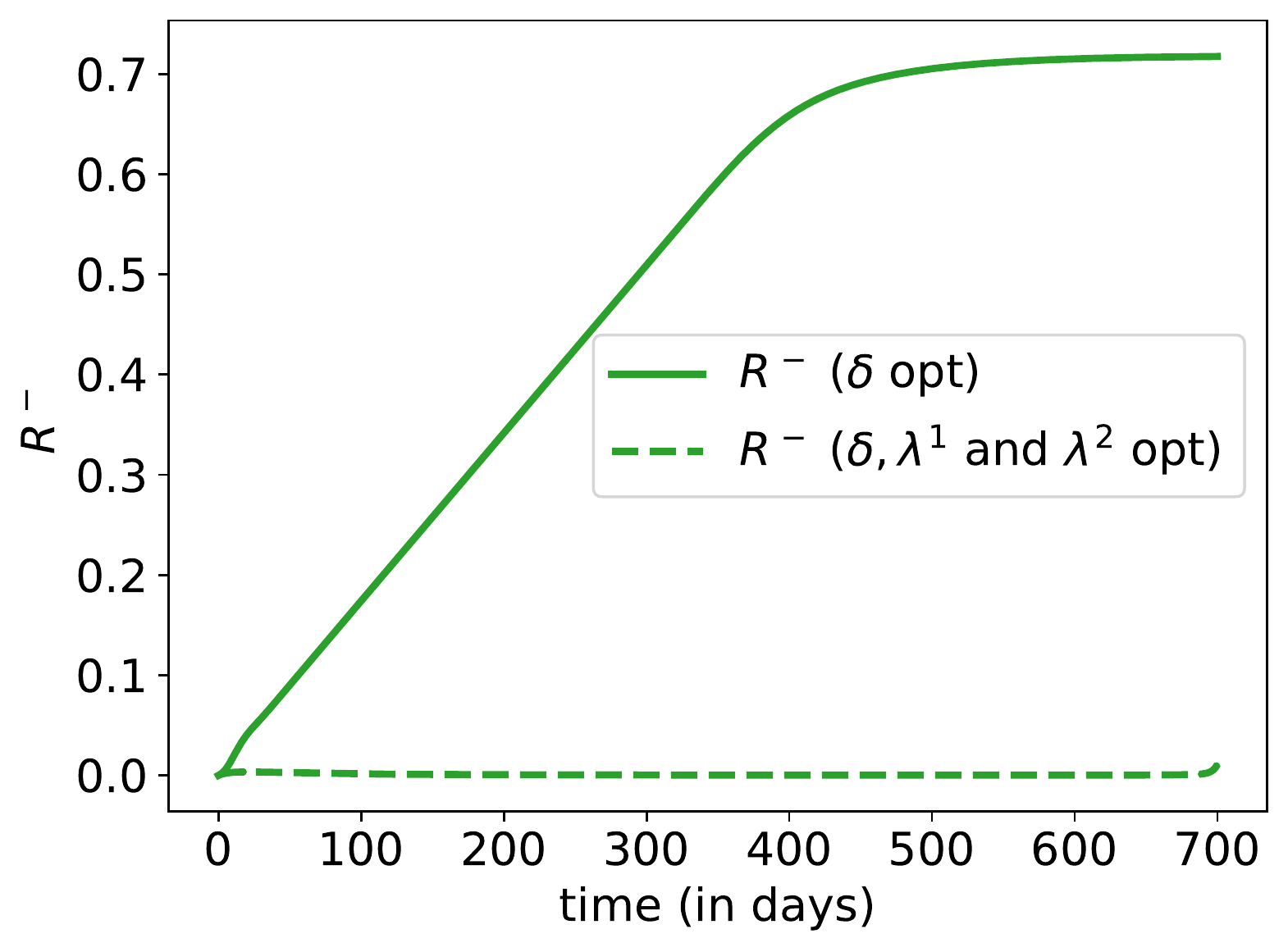}
	\end{subfigure}%
	\begin{subfigure}{.33\columnwidth}
		\centering 
		\includegraphics[width=\columnwidth]{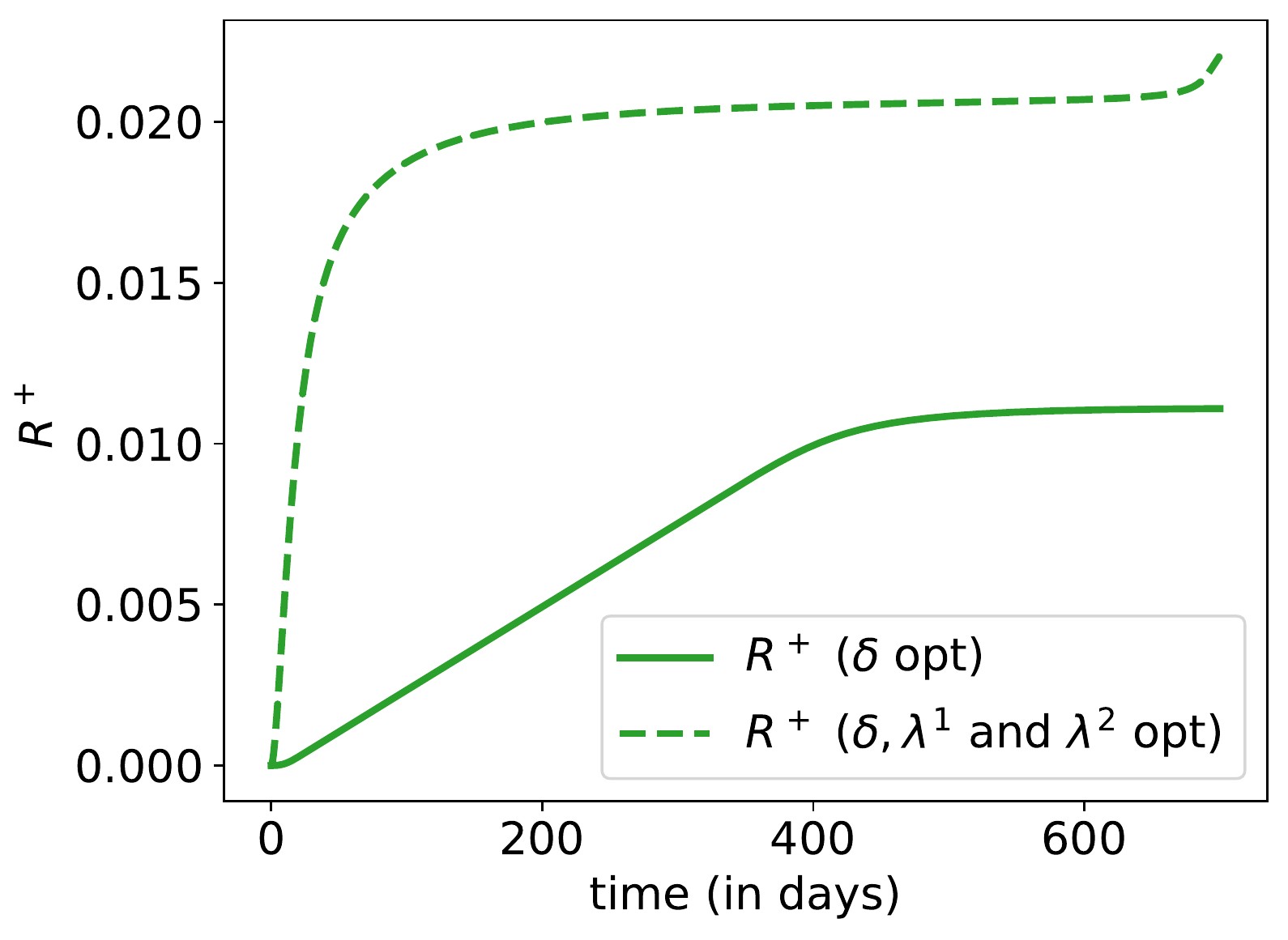}
	\end{subfigure}
	\begin{subfigure}{.33\columnwidth}
		\centering 
		\includegraphics[width=\columnwidth]{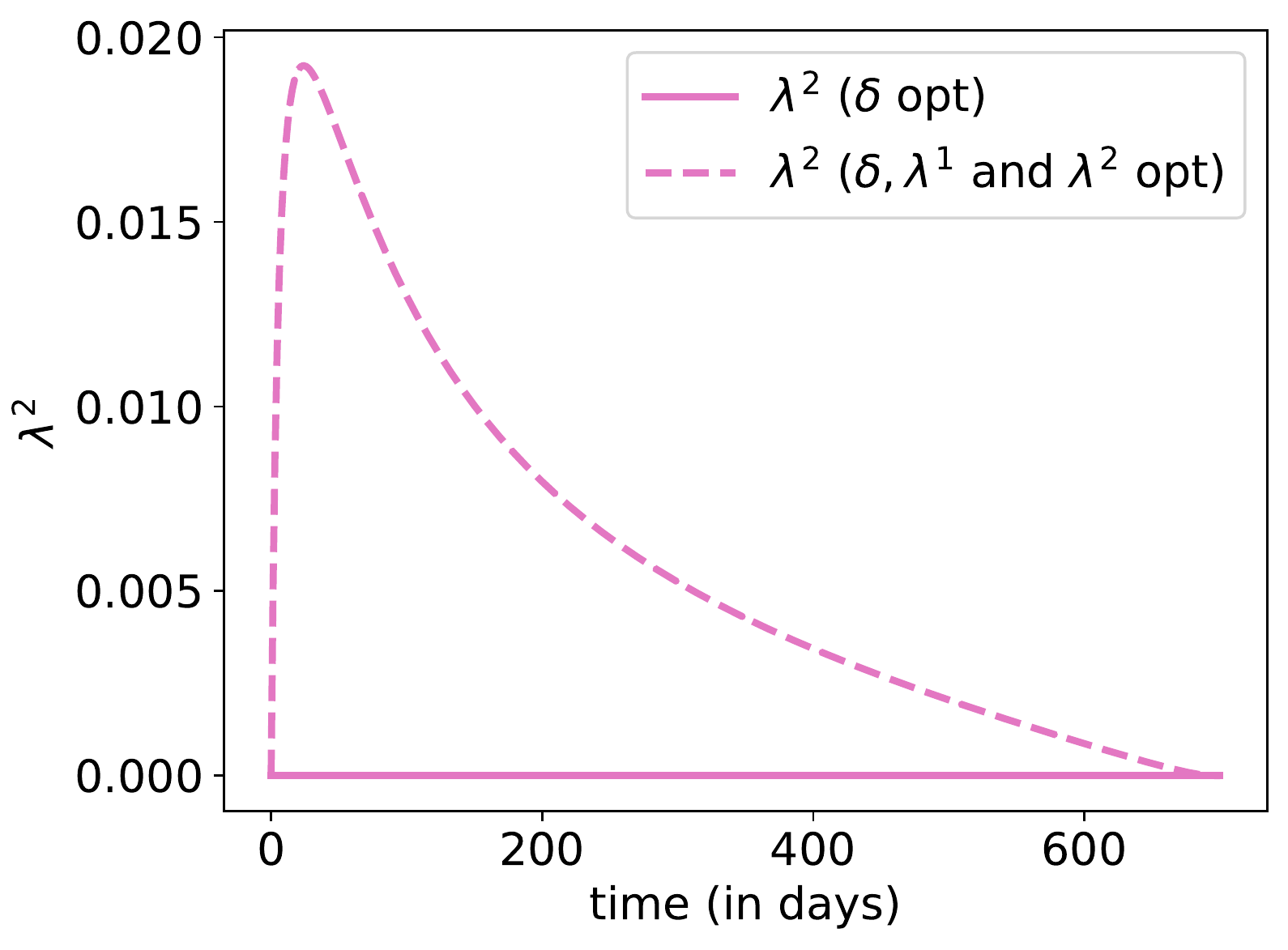}
	\end{subfigure}
	\caption{Evolution of states with optimal controls $(\delta,\lambda^1,\lambda^2)$, where the plain line corresponds to the benchmark scenario $(\delta,0,0)$ discussed in Section \ref{subsec:optimal:policy} (plain line); and evolution with joint optimal controls $(\delta,\lambda^1,\lambda^2)$ (dashed line).
	}
 	\label{fig:Optim_delta_lambda2B11}
\end{figure}

\newpage


\section{Numerical resolution of the optimal control problem}\label{sec_appendix_math}

We discuss in this section the numerical resolution of the optimal control problem  \eqref{Control_Problem} of interest. 

The original problem has infinite horizon, but for numerical purpose, we focus on the equivalent version over a sufficiently large time horizon $T$. The proximity between both optimal control problems is well established in the literature, see e.g. Section 4.1 in \cite{elie2008finite}, while we observe empirically very few impact by the choice of maturity $T$, as soon as $T$ is chosen large enough in order for the epidemic phase to be over (i.e. such that the proportion of infected people is small enough at time $T$). With respect to the random time of vaccination $\tau$, such approximation corresponds to optimizing over the time interval $[0,\tau\wedge T]$ instead of $[0,\tau]$.

Similarly, the problem \eqref{Control_Problem} is an optimal control problem with sate constraint on the ICU capacity, and we choose to replace for ease of numerical tractability the state constraint by a penalization cost with a sufficiently large weight $w_{\text{ICU}}>0$. Hence, we focus numerically on an approximate version of the original optimal control, which is given by:
\begin{eqnarray}\label{Numeric_Control_Problem}
 \inf_{(\delta,\lambda^1,\lambda^2)\in\mathcal{A}} ~  \big\lbrace\tilde J_T (\delta,\lambda^1,\lambda^2)\big\rbrace\;,
\end{eqnarray}
where the approximate objective function $\tilde J_T$ is given by 
 \begin{eqnarray}\label{eq_Global_cost_penalizationICU}
     \tilde J_T(\delta, \lambda^1,\lambda^2) &:=& w_{\text{sanitary}} \int_0^T e^{-\alpha t}  \drm D_t + w_{\text{econ}} \int_0^T e^{-\alpha t} (1-W_t)^2  \drm t\\ &+& w_{\text{prevalence}} \int_0^T e^{-\alpha t} |N^1_t|^2 \drm t + w_{\text{immunity}} \int_0^T e^{-\alpha t} |N^2_t|^2 \drm t,\nonumber\\
    &+& w_{\text{ICU}} \int_0^T e^{-\alpha t}  [U_t-U_{\max}]^+ \drm t\nonumber.
 \end{eqnarray}

\subsection{Optimality condition by Pontryagin's maximum principle}

To alleviate the notation and use more standrad notations from the litterature, the problem can be recast in the following form, where $a$ denotes the control and $X$ the state: Minimize
$$
    J(a) = \int_0^T f(t, X_t, a_t) \drm t + g(X_T)
$$
subject to
$$
    \dot X_t = b(X_t, a_t), \qquad X_0 = x_0 \hbox{ given}.
$$
We interpret $X_t = (S_t, I^-_t, I^+_t, R^-_t, R^+_t, H_t, U_t, D_t) \in [0,1]^8 \subset \R^8$, $a_t = (\delta, \lambda^1_t, \lambda^2_t) \in [0,1]^3$ and we take
\begin{align*}
    f(t, X, a) 
    &=  
    w_{\text{sanitary}}  e^{-\alpha t}  \gamma_{UD}(U) U
    + w_{\text{econ}} e^{-\alpha t} (1-W)^2 
    \\ 
    &\qquad + w_{\text{prevalence}}  e^{-\alpha t} |N^1|^2 
    + w_{\text{immunity}}  e^{-\alpha t} |N^2|^2 
    + w_{\text{ICU}} e^{-\alpha t} [U-U_{\max}]^+ \,,
\end{align*}
where 
$$
    Q := R^- + I^- + S, 
    \quad
    W := (1- \delta) Q + R^+,
    \quad
    N^1  := \lambda^1 Q + \gamma_{IH} I^-,
    \quad 
    N^2_t  := \lambda^2 Q.
$$

Moreover, we take
$$
    b(X, a)
    = 
    \begin{pmatrix}
    - \fgreen{(1-\delta)}\beta I^{-} S
    \\
     \fgreen{(1-\delta)}\beta I^{-}  S - \red{\lambda^1} I^{-} - (\gamma_{IR}+\gamma_{IH}) I^{-}
     \\
     \red{\lambda^1} I^{-}  - (\gamma_{IR}+\gamma_{IH}) I^{+}
    \\ 
    \gamma_{IR} I^{-}
                        - \blue{\lambda^2_{t}} R^{-}
     \\
     \gamma_{IR} I^{+}   + \blue{\lambda^2} R^{-} + \purple{\gamma_{HR}} H   +  \purple{\gamma_{UR}(U)} U
     \\
     \gamma_{IH} \big(I^{-}+I^{+}\big) - (\purple{\gamma_{HR} + \gamma_{HU})} H
     \\
     \purple{\gamma_{HU}} H   - (\purple{\gamma_{UR}(U)} + \purple{\gamma_{UD}(U)}) U 
     \\
     \purple{\gamma_{UD}(U)} U 
    \end{pmatrix}
    =
    \begin{pmatrix}
    - \fgreen{(1-\delta)}\beta I^{-} S
    \\
     \fgreen{(1-\delta)}\beta I^{-}  S - \red{\lambda^1} I^{-} - (\gamma_{IR}+\gamma_{IH}) I^{-}
     \\
     \red{\lambda^1} I^{-}  - (\gamma_{IR}+\gamma_{IH}) I^{+}
    \\ 
    \gamma_{IR} I^{-}
                        - \blue{\lambda^2_{t}} R^{-}
     \\
     \gamma_{IR} I^{+}   + \blue{\lambda^2} R^{-} + \purple{\gamma_{HR}} H   +  \purple{\rho_{UR}(U)}
     \\
     \gamma_{IH} \big(I^{-}+I^{+}\big) - (\purple{\gamma_{HR} + \gamma_{HU})} H
     \\
     \purple{\gamma_{HU}} H   - (\purple{\rho_{UR}(U)} + \purple{\rho_{UD}(U)}) 
     \\
     \purple{\rho_{UD}(U)} 
    \end{pmatrix} \, ,
$$
where we denoted $\rho_{UR}(u) = \gamma_{UR}(u)u$ and $\rho_{UD}(u) = \gamma_{UD}(u)u$ in order to alleviate the notation.

Let $H$ be the Hamiltonian defined by
$$
    H(t, x,y,a) = b(x,a) \cdot y + f(t, x,a).
$$

Pontryagin's maximum principle leads, at least informally, to the following optimality condition: if $\hat a=(\hat a_t)_t$ is optimal, then for every $t \in [0,T]$, 
\begin{equation*}
\left\{
\begin{split}
    0
    &= \partial_{\delta} H(\hat X_t, \hat Y_t, \hat a_t)
    \\
    &=
    2 w_{\text{econ}} e^{-\alpha t} |\hat Q_t|^2 \hat \delta_t
    + 2 w_{\text{econ}} e^{-\alpha t} \hat Q_t (1- \hat Q_t-\hat R^+_t)
    + \beta \hat I^{-}_t \hat S_t(\hat Y^S - \hat Y^{I^{-}}) \, ,
    \\
    0
    &= \partial_{\lambda^1} H(\hat X_t, \hat Y_t, \hat a_t)
    \\
    &=
    2 w_{\text{prevalence}} e^{-\alpha t} \hat Q_t(\hat Q_t \hat \lambda^1_t + \gamma_{IH} \hat I^-_t)  
    - \hat I^-_t(\hat Y^{I^-}_t - \hat Y^{I^+}_t) \, , 
    \\
    0
    &= \partial_{\lambda^2} H(\hat X_t, \hat Y_t, \hat a_t)
    \\
    &=
    2 w_{\text{immunity}} e^{-\alpha t} |\hat Q_t|^2 \hat \lambda^2_t 
    - \hat I^-_t(\hat Y^{I^-}_t - \hat Y^{I^+}_t) \, ,
\end{split}
\right.
\end{equation*}
where $(\hat X,\hat Y) = (S, I^-, I^+, R^-, R^+, H, U, D, Y^{S}, Y^{I^-}, Y^{I^+}, Y^{R^-}, Y^{R^+}, Y^{H}, Y^{U}, Y^{D})$ solve the forward-backward ODE system:
\begin{equation}
\label{FBODE-XY}
\begin{cases}
    \dot {\hat{X}}_t = b(\hat{X}_t, \hat a_t), 
    & \hat{X}_0 = x_0
    \\
    \dot {\hat{Y}}_t = - \partial_x H(t, \hat{X}_t, \hat{Y}_t, \hat a_t),
    & \hat{Y}_T = 0.
\end{cases}
\end{equation}

The backward equation can be written as follows (dropping the hats to alleviate the notation):
\begin{equation*}
\begin{cases}
    \dot {\hat{Y}}^{S}_t = - \Big[(1-\delta) \beta I^- (Y^{S}_t - Y^{I^-}_t) 
    \\
    \qquad\qquad - 2 w_{\text{econ}} e^{-\alpha t} (1-\delta) \big(1-W_t\big) + 2 w_{\text{prevalence}} e^{-\alpha t} \lambda^1 N^1_t + 2 w_{\text{immunity}} e^{-\alpha t} \lambda^2 N^2_t \Big]
    \\
    \dot {\hat{Y}}^{I^-}_t = -\Big[ (1-\delta) \beta S (Y^{S}_t - Y^{I^-}_t) + \lambda^1(Y^{I^+}_t - Y^{I^-}_t) + \gamma_{IR}(Y^{R^-}_t- Y^{I^-}_t) + \gamma_{IH}(Y^{H}_t - Y^{I^-}_t)
    \\
    \qquad\qquad - 2 w_{\text{econ}} e^{-\alpha t} (1-\delta) \big(1-W_t\big) + 2 w_{\text{prevalence}} e^{-\alpha t} (\lambda^1+\gamma_{IH}) N^1_t + 2 w_{\text{immunity}} e^{-\alpha t} \lambda^2 N^2_t \Big]
    \\
    \dot {\hat{Y}}^{I^+}_t = -\Big[ \gamma_{IR}(Y^{R^+}_t - Y^{I^+}_t) + \gamma_{IH}(Y^{H}_t - Y^{I^+}_t) \Big]
    \\
    \dot {\hat{Y}}^{R^-}_t = -\Big[ \lambda^2(Y^{R^+}_t - Y^{R^-}_t)
    \\
    \qquad\qquad - 2 w_{\text{econ}} e^{-\alpha t} (1-\delta) \big(1-W_t\big) + 2 w_{\text{prevalence}} e^{-\alpha t} \lambda^1 N^1_t + 2 w_{\text{immunity}} e^{-\alpha t} \lambda^2 N^2_t \Big]
    \\
    \dot {\hat{Y}}^{R^+}_t = -\Big[  - 2 w_{\text{econ}} e^{-\alpha t}  \big(1-W_t\big) \Big]
    \\
    \dot {\hat{Y}}^{H}_t = -\Big[ \gamma_{HR}(Y^{R^+}_t - Y^{H}_t) + \gamma_{HU}(Y^{U}_t - Y^{H}_t)\Big]
    \\
    \dot {\hat{Y}}^{U}_t = -\Big[ \rho_{UR}'(U)(Y^{R^+}_t - Y^{U}_t) + \rho_{UD}'(U)(Y^{D}_t - Y^{U}_t)
    \\
    \qquad\qquad + w_{\text{sanitary}} e^{-\alpha t} \rho_{UD}'(U_t) + w_{\text{ICU}} e^{-\alpha t} \mathbb{1}_{U_t > U_{\text{max}}} \Big]
    \\
    \dot {\hat{Y}}^{D}_t = 0 \, .
\end{cases}
\end{equation*}

\subsection{Numerical algorithm}\label{Sec_algo}

The numerical method is an iterative procedure to compute the optimal control $\hat a = (\hat \delta, \hat \lambda^1, \hat \lambda^2) = (\hat \delta_t, \hat \lambda^1_t, \hat \lambda^2_t)_{t \in [0T]}$. Starting from an initial guess $a^{(0)}$, at iteration $k \ge 0$, in order to compute $a^{(k+1)}$, we first compute the solution $(X^{(k)},Y^{(k)})$ to the forward-backward ODE system corresponding to~\eqref{FBODE-XY} but with control $a^{(k)}$ instead of $\hat a$, i.e.,
\begin{equation*}
\begin{cases}
    \dot X^{(k)}_t = b(X^{(k)}_t, a^{(k)}_t), 
    & X^{(k)}_0 = x_0
    \\
    \dot Y^{(k)}_t = - \partial_x H(X^{(k)}_t, Y^{(k)}_t, a^{(k)}_t),
    & Y^{(k)}_T = 0.
\end{cases}
\end{equation*}
We then set
\begin{equation*}
    \tilde a^{(k+1)}_t = a^{(k)}_t - \theta^{(k)} \partial_a H(X_t, Y_t, a^{(k)}_t) \, ,
\end{equation*}
where $\theta^{(k)}>0$ is a step size (which may depend on the iteration in order to adjust the convergence rate). 

Finally, we define, for each component $i$ of the control,
$$
    a^{i, (k+1)}_t = \pi_{[0,1]} (\tilde a^{i, (k+1)}_t)
$$
where $\pi_{[0,1]}$ denotes the projection on the interval $[0,1]$.

If the optimization is to be performed over only one component of the control (e.g. over $\delta$ only), then the updates are done only for this component instead of the whole vector of controls.

\newpage

\end{document}